\documentclass{article}
\usepackage{amssymb}

\usepackage{graphicx}
\usepackage{amsmath}

\input{tcilatex}

\begin{document}

\bigskip \rule[1in]{0in}{0.17in}

\begin{center}
{\Huge THE GENERAL STRUCTURE }

{\Huge \bigskip OF MATTER}
\end{center}

{\huge \vspace*{4in}}

\begin{center}
{\Huge M\'{a}rio Everaldo de Souza}
\end{center}

\pagebreak

\rule[1.5in]{0in}{0.17in}

{\LARGE THE GENERAL STRUCTURE }

\medskip \bigskip

{\LARGE \ OF MATTER}

{\LARGE \bigskip \rule[1in]{0in}{0.17in}}

{\Large M\'{a}rio Everaldo de Souza}

{\Large \medskip (e-mail mdesouza@ufs.br)}

{\Large Departamento de F\'{i}sica}

{\Large Universidade Federal de Sergipe}

\bigskip

\bigskip

\bigskip

\bigskip

\bigskip

\bigskip

\bigskip

\bigskip

\bigskip

\bigskip

\bigskip

\bigskip

\bigskip

\bigskip

\bigskip

\bigskip

\bigskip

\bigskip

\bigskip

\bigskip

\bigskip

\bigskip

\bigskip

\bigskip

\bigskip

\bigskip

\bigskip

\bigskip

\bigskip

\bigskip

\bigskip

\bigskip

\bigskip

\bigskip

\bigskip

\bigskip

\bigskip

\bigskip

\bigskip

\bigskip

\bigskip

\bigskip

\bigskip

\bigskip

\bigskip

\bigskip

\bigskip

\bigskip

\bigskip

\bigskip

\bigskip

\bigskip

\bigskip

\bigskip

\bigskip

\bigskip \rule[1.5in]{0in}{0.17in}

{\small Reproduction of this book, limited to a small number}

{\small of copies, is allowed for scientific purposes and forbidden }

{\small for other purposes. Further information and requests}

{\small should be addressed to the author at}

{\small Departamento de F\'{i}sica}

{\small Universidade Federal de Sergipe }

{\small Campus Universit\'{a}rio}

{\small 49100-000 S\~{a}o Crist\'{o}v\~{a}o, Sergipe}

{\small Brasil}

{\small e-mail mdesouza@sergipe.ufs.br}

\bigskip

\bigskip

\bigskip

{\small Cataloging by Main Library of Universidade Federal de Sergipe}

{\small \bigskip }

{\small S719s\qquad Souza, M\'{a}rio Everaldo de, 1954 - }

{\small \qquad \qquad \qquad \qquad The General Structure of Matter}

{\small \qquad \qquad\ \ \ - S\~{a}o Crist\'{o}v\~{a}o: Departamento de
F\'{i}sica, }

{\small \qquad \qquad\ \ \ Universidade Federal de Sergipe, 2001.}

{\small \bigskip }

{\small \qquad \qquad\ \ \ 1. General Principles of Physics. 2. Fundamental}

{\small \qquad \qquad\ \ \ \ Interactions. 3. Hadrons, Quarks and Prequarks. 
}

{\small \qquad \qquad\ \ \ 4. Quark Masses. 5. Masses of Hadrons. }

{\small \qquad \qquad\ \ \ 6. Structure of the Nucleon. 7. The Nucleon Sea. }

{\small \qquad \qquad\ \ \ 8. Nuclear Forces. 9. Galactic Evolution. }

{\small \qquad \qquad\ \ \ 10. Rotation of Galaxies. 11. Universal Expansion.%
}

{\small \qquad \qquad\ \ \ 12. Structure Formation. 13. The Fifth Force.}

{\small \bigskip }

\bigskip

\ \ \ \ \ \ \ \ \ \ \ \ \ 

\bigskip

\bigskip

\bigskip

\bigskip

\bigskip

\bigskip

\bigskip

\bigskip

\bigskip

\bigskip

\bigskip

\bigskip

\bigskip

\bigskip

\ \rule[1.5in]{0in}{0.17in}\ \ \ \ \ \ \ \ \ \ 

{\large Dedicated to the memory of Sir Isaac Newton}

{\large \ \ \ \ \ \ \ \ \ for his genius and for his humility.}

{\Large \bigskip }

\bigskip \bigskip

\bigskip

{\Large ``I do not know what I may appear to the world; }

{\Large but to myself \ I seem to have been only like a boy}

{\Large playing on the seashore, and diverting myself in }

{\Large now and then finding a smoother pebble or prettier}

{\Large shell than ordinary, while the great ocean of truth}

{\Large lay all undiscovered before me ''.}

{\Large \ \ \ \ \ \ \ \ \ \ \ \ \ \ \ \ \ \ \ \ \ \ \ \ \ \ \ \ \ \ \ \ \ \
\ \ \ \ \ \ \ \ \ \ \ \ Sir Isaac Newton}

{\Large \bigskip }

\bigskip

\bigskip \pagebreak

\bigskip

\vspace*{1in}

\ \ \ \ \ \ \ \ \ \ \ \ \ \ \ \ \ \ \ \ \ \ \ \ \ \ \ \ \ \ \ \ \ \ \ \ \ \ 
{\LARGE \ \ Preface}

\bigskip

This work intends to be just the tip of the iceberg that is concealed
beneath the present theoretical status quo. I do apologize beforehand if any
due reference was omitted. If it was, it was not done by mischievousness,
but by ignorance. Last year (Science, April 2000) and this year (Science,
January 2001) the Boo-merang experiment reported that the Universe began
with an initial mass. It is important to have in mind that in 1995, in the
work The Six Fundamental Forces of Nature, on top of p. 22 , I said that \ 
\textbf{``Each cycle of the Universe begins(t=0) with a certain volume of
neutrons, protons and electrons at a temperature of about 1 MeV, which is
necessary for the primordial formation of the light elements''.} Also, in
the work The Superstrong, Strong and Superweak Interactions, of April 2000,
on p. 47, 2nd paragraph I said \textbf{``It is quite remarkable the
similarity between a supernova explosion and the Big Bang. In supernova
debris we find sheets and filaments of gas, and underdense and overdense
regions. We find the same in the large scale structure of the Universe:
sheets, filaments and voids. There are other similarities. In supernova
debris we find shells of gas expanding at speeds in the range }$%
(10^{3}-10^{4})km/s$\textbf{. There are also shells in the Universe. As di
Nella and Paturel\ show ``The distribution of galaxies up to a distance of
200Mpc (650 million light-years) is flat and shows a strucutre like a shell
roughly centered on the Local Supercluster (Virgo Cluster). This result
clearly confirms the existence of the hypergalactic large scale structure
noted in 1988. This is presently the largest strucuture ever seen''. This is
so because both explosions, either in supernovae or in the Universe, are
caused by the same force: THE SUPERSTRONG FORCE''.}

The fundamental idea of this work is that Nature has six fundamental forces.
This implies that a quark is a composite structure formed by primons. It is
shown in this work that primons form quarks, supergluons, gluons and Higgs
bosons. They are thus related to the deepest essence of matter. With them we
can explain the sizes of baryons, the sizes of quarks and the nucleon sea.
And we can understand the origin of the harmonic effective potentials used
in the calculation of hadron spectra. And finally the proposal makes the
bridge from quarks to nuclear physics in a rather consistent way.

In this work almost all energy levels of baryons are calculated in an easy
way taking into account angular momentum, and the bound states of heavy
mesons are explained. Other considerations on mesons are included.



Some important results on formation and evolution of galaxies are presented
and their connection with the superstrong interaction is stressed. Some
cosmic and planetary evidences are shown on the existence of such
interaction. It is shown that this interaction avoids the formation of black
holes and plays a role in the formation of planets and stars.

It is clearly shown that nature has a generalized structured state which is
manifested in four different ways, and is characterized by a fermion charge
carrier (not interaction carrier which is a boson) and is always formed by
two different fundamental forces. From this we arrive at very important
results concerning the quantization of gravity.

In 1999 and 2000 several works were published on quintessence which is
actually a disguise for another field. Since 1991 I have proposed that
Fischbach's fifth force acts among galaxies and together with gravity makes
the galactic medium\textit{.} The present book reinforces this proposal and
shows the connection between neutrinos and the superweak interaction.

I would like to pay homage to David Schramm because I began to construct
Tables 1.1 and 1.2 (which gave origin to the rest of this work) just after a
talk that he gave at the University of Illinois at Chicago in 1989. In the
talk he showed the work of de Lapparent et al. of 1986 and the fresh work,
at that moment, of Broadhurst et al., both on the distribution of galaxies.
He stressed then that both works were very disturbing. After the talk I
suggested that there could be another force acting between galaxies and he
replied in a humble way: ``It is possible. It is a possible line of
thought.'' Unfortunately, this open-mindedness is rare in physics. Thus,
this work is also dedicated to his memory.

I emphasize once more that this work is just the tip of the iceberg and
therefore it has some qualitative results simply because we still have to
develop a theory on the new interactions.

It is quite odd that some critics do not accept the possibility of Nature
having six forces but accept, instead, extra dimensions, dark matter and its
array of strange particles, unification at any cost, inflation, neutrino
oscillations, and so on.

Aracaju, July 2001.\bigskip

\bigskip

\ \ \ \ \ \ \ \ \ \ \qquad \qquad \qquad \qquad \qquad M\'{a}rio Everaldo de
Souza \ \ 

\bigskip \qquad \qquad \qquad \qquad \qquad \qquad\ \ \ e-mail
mdesouza@ufs.br

\bigskip

\bigskip

\bigskip

\bigskip

\bigskip

\bigskip

\bigskip

\bigskip

\bigskip

\bigskip

\bigskip

\bigskip

\bigskip \pagebreak

\bigskip \bigskip \rule[1.8in]{0in}{0.17in}

\bigskip

\bigskip {\Huge Contents}

\bigskip \bigskip

{\huge 1}{\large \ \ }{\Large General Properties of Matter}

1.1 \ General Classification of Matter \ \ \ \ \ \ \ \ \ \ \ \ \ \ \ \ \ \ \
\ \ \ \ \ \ \ \ \ \ \ \ \ \ \ \ \ \ \ \ \ \ \ \ \ \ \ \ \ \ \ \ \ \ 10

1.2 \ Size and Number \ \ \ \ \ \ \ \ \ \ \ \ \ \ \ \ \ \ \ \ \ \ \ \ \ \ \
\ \ \ \ \ \ \ \ \ \ \ \ \ \ \ \ \ \ \ \ \ \ \ \ \ \ \ \ \ \ \ \ \ \ \ \ \ \
\ \ \ \ \ \ \ \ \ 13

\bigskip

{\huge 2}{\large \ \ }{\Large Prequarks, Quarks and Nucleons}

2.1 \ Prequarks and the Number of Quarks \ \ \ \ \ \ \ \ \ \ \ \ \ \ \ \ \ \
\ \ \ \ \ \ \ \ \ \ \ \ \ \ \ \ \ \ \ \ \ \ \ \ \ 15

2.2 \ The Structure of Nucleons, the Sizes of Quarks {\large u} and {\large d%
},

\ \ \ \ \ \ \ \ and the Stability of \ the Proton \ \ \ \ \ \ \ \ \ \ \ \ \
\ \ \ \ \ \ \ \ \ \ \ \ \ \ \ \ \ \ \ \ \ \ \ \ \ \ \ \ \ \ \ \ \ \ \ \ \ \
\ \ \ 18

2.3 \ The Sizes of Quarks, Primon Mass, and Generation

\ \ \ \ \ \ \ \ of Quark Mass (by Higgs Bosons) \ \ \ \ \ \ \ \ \ \ \ \ \ \
\ \ \ \ \ \ \ \ \ \ \ \ \ \ \ \ \ \ \ \ \ \ \ \ \ \ \ \ \ \ \ \ \ \ 20

2.4 \ The Seas and Sizes of Nucleons \ \ \ \ \ \ \ \ \ \ \ \ \ \ \ \ \ \ \ \
\ \ \ \ \ \ \ \ \ \ \ \ \ \ \ \ \ \ \ \ \ \ \ \ \ \ \ \ \ \ \ \ 27

2.4.1 The Proton Sea Content \ \ \ \ \ \ \ \ \ \ \ \ \ \ \ \ \ \ \ \ \ \ \ \
\ \ \ \ \ \ \ \ \ \ \ \ \ \ \ \ \ \ \ \ \ \ \ \ \ \ \ \ \ \ \ \ \ \ \ \ \ \
27

2.4.2 The Neutron Sea Content \ \ \ \ \ \ \ \ \ \ \ \ \ \ \ \ \ \ \ \ \ \ \
\ \ \ \ \ \ \ \ \ \ \ \ \ \ \ \ \ \ \ \ \ \ \ \ \ \ \ \ \ \ \ \ \ \ \ \ \ 31

2.4.3 The Contribution of Both Seas to the Structure Functions \ \ \ \ \ \ \
\ \ \ \ \ \ 31

\bigskip

{\huge 3}{\large \ \ }{\Large The Strong, Superstrong and Higgs Bosons,}

{\Large \ \ \ \ Gluons and Interactions \ Between Primons}

3.1 \ The Higgs Bosons \ \ \ \ \ \ \ \ \ \ \ \ \ \ \ \ \ \ \ \ \ \ \ \ \ \ \
\ \ \ \ \ \ \ \ \ \ \ \ \ \ \ \ \ \ \ \ \ \ \ \ \ \ \ \ \ \ \ \ \ \ \ \ \ \
\ \ \ \ \ \ \ \ \ \ 35

3.2 \ The Bosons of the Strong and Superstrong Interactions \ \ \ \ \ \ \ \
\ \ \ \ \ \ \ \ \ \ 36

3.3 \ The Nature of Gluons \ \ \ \ \ \ \ \ \ \ \ \ \ \ \ \ \ \ \ \ \ \ \ \ \
\ \ \ \ \ \ \ \ \ \ \ \ \ \ \ \ \ \ \ \ \ \ \ \ \ \ \ \ \ \ \ \ \ \ \ \ \ \
\ \ \ \ \ 38

3.4 \ The Interaction Matrix Between Primons of Different \ \ \ \ \ \ \ \ \
\ \ \ \ \ \ \ \ \ \ \ \ \ 

\ \ \ \ \ \ \ \ Quarks with Different Supercolors \ \ \ \ \ \ \ \ \ \ \ \ \
\ \ \ \ \ \ \ \ \ \ \ \ \ \ \ \ \ \ \ \ \ \ \ \ \ \ \ \ \ \ \ \ \ \ \ \ 40

3.5 \ The Lagrangian of Quantum Superchromodynamics \ \ \ \ \ \ \ \ \ \ \ \
\ \ \ \ \ \ \ \ \ \ \ \ 42

3.6 \ Primons and Weak Interactions \ \ \ \ \ \ \ \ \ \ \ \ \ \ \ \ \ \ \ \
\ \ \ \ \ \ \ \ \ \ \ \ \ \ \ \ \ \ \ \ \ \ \ \ \ \ \ \ \ \ \ \ \ \ 42

\bigskip \pagebreak

\rule[1.0in]{0in}{0.17in}{\huge 4}{\large \ \ }{\Large Some Topics of QCD}

4.1 \ The Potential of a Quark Pair and the Usual QCD Potential \ \ \ \ \ \
\ \ \ \ \ 44

4.2 \ The Confining Term of the Usual QCD Potential \ \ \ \ \ \ \ \ \ \ \ \
\ \ \ \ \ \ \ \ \ \ \ \ \ \ \ \ 46

4.3 \ Asymptotic Freedom \ \ \ \ \ \ \ \ \ \ \ \ \ \ \ \ \ \ \ \ \ \ \ \ \ \
\ \ \ \ \ \ \ \ \ \ \ \ \ \ \ \ \ \ \ \ \ \ \ \ \ \ \ \ \ \ \ \ \ \ \ \ \ \
\ \ \ \ \ 47

\bigskip

{\huge 5}{\large \ \ }{\Large The Energies and Sizes of Hadrons }

5.1 \ The Energies of Baryons \ \ \ \ \ \ \ \ \ \ \ \ \ \ \ \ \ \ \ \ \ \ \
\ \ \ \ \ \ \ \ \ \ \ \ \ \ \ \ \ \ \ \ \ \ \ \ \ \ \ \ \ \ \ \ \ \ \ \ \ \
\ \ 51

5.1.1 In Cartesian Coordinates \ \ \ \ \ \ \ \ \ \ \ \ \ \ \ \ \ \ \ \ \ \ \
\ \ \ \ \ \ \ \ \ \ \ \ \ \ \ \ \ \ \ \ \ \ \ \ \ \ \ \ \ \ \ \ \ \ \ \ \ 52

5.1.2 In Polar Coordinates \ \ \ \ \ \ \ \ \ \ \ \ \ \ \ \ \ \ \ \ \ \ \ \ \
\ \ \ \ \ \ \ \ \ \ \ \ \ \ \ \ \ \ \ \ \ \ \ \ \ \ \ \ \ \ \ \ \ \ \ \ \ \
\ \ \ 61

5.1.3 \ Relation Between Energy and Angular Momentum \ \ \ \ \ \ \ \ \ \ \ \
\ \ \ \ \ \ \ \ \ \ 68

5.1.4 \ The Sizes of Baryons \ \ \ \ \ \ \ \ \ \ \ \ \ \ \ \ \ \ \ \ \ \ \ \
\ \ \ \ \ \ \ \ \ \ \ \ \ \ \ \ \ \ \ \ \ \ \ \ \ \ \ \ \ \ \ \ \ \ \ \ \ \
\ \ \ 68

5.1.5 \ Spin-Orbit Interaction \ \ \ \ \ \ \ \ \ \ \ \ \ \ \ \ \ \ \ \ \ \ \
\ \ \ \ \ \ \ \ \ \ \ \ \ \ \ \ \ \ \ \ \ \ \ \ \ \ \ \ \ \ \ \ \ \ \ \ \ \
\ \ \ 71

5.2 \ Generalization of the Gell-Mann-Okubo Mass Formula \ \ \ \ \ \ \ \ \ \
\ \ \ \ \ \ \ \ 72

5.3 \ The Bound States of Mesons \ \ \ \ \ \ \ \ \ \ \ \ \ \ \ \ \ \ \ \ \ \
\ \ \ \ \ \ \ \ \ \ \ \ \ \ \ \ \ \ \ \ \ \ \ \ \ \ \ \ \ \ \ \ \ \ 76

5.3.1 \ Heavy Mesons \ \ \ \ \ \ \ \ \ \ \ \ \ \ \ \ \ \ \ \ \ \ \ \ \ \ \ \
\ \ \ \ \ \ \ \ \ \ \ \ \ \ \ \ \ \ \ \ \ \ \ \ \ \ \ \ \ \ \ \ \ \ \ \ \ \
\ \ \ \ \ \ \ \ 77

5.3.2 \ Light Mesons \ \ \ \ \ \ \ \ \ \ \ \ \ \ \ \ \ \ \ \ \ \ \ \ \ \ \ \
\ \ \ \ \ \ \ \ \ \ \ \ \ \ \ \ \ \ \ \ \ \ \ \ \ \ \ \ \ \ \ \ \ \ \ \ \ \
\ \ \ \ \ \ \ \ \ \ \ 80

5.3.3 \ The Sizes of Mesons \ \ \ \ \ \ \ \ \ \ \ \ \ \ \ \ \ \ \ \ \ \ \ \
\ \ \ \ \ \ \ \ \ \ \ \ \ \ \ \ \ \ \ \ \ \ \ \ \ \ \ \ \ \ \ \ \ \ \ \ \ \
\ \ \ 84

\bigskip

{\huge 6}{\large \ \ }{\Large The Superstrong Force and the Universe}

6.1 \ The Supernovae Evidence for the Superstrong Interaction \ \ \ \ \ \ \
\ \ \ \ \ \ \ 87

6.2 \ The Formation and Evolution of Galaxies \ \ \ \ \ \ \ \ \ \ \ \ \ \ \
\ \ \ \ \ \ \ \ \ \ \ \ \ \ \ \ \ \ \ \ \ \ \ \ 89

6.2.1 \ The Formation of Galaxies and Primordial Stars \ \ \ \ \ \ \ \ \ \ \
\ \ \ \ \ \ \ \ \ \ \ \ \ \ \ 89

6.2.2 \ The Evolution of Galaxies \ \ \ \ \ \ \ \ \ \ \ \ \ \ \ \ \ \ \ \ \
\ \ \ \ \ \ \ \ \ \ \ \ \ \ \ \ \ \ \ \ \ \ \ \ \ \ \ \ \ \ \ \ \ \ \ \ \ \
\ 91

6.3 \ The Formation of\ Structure: Voids, Sheets and Clusters \ \ \ \ \ \ \
\ \ \ \ \ \ \ \ \ \ \ 96

6.4 \ No need of \ (for) Dark Matter \ \ \ \ \ \ \ \ \ \ \ \ \ \ \ \ \ \ \ \
\ \ \ \ \ \ \ \ \ \ \ \ \ \ \ \ \ \ \ \ \ \ \ \ \ \ \ \ \ \ \ \ \ \ \ \ 96

6.5 \ The Expansion of the Universe and the Background \ \ \ \ \ \ 

\ \ \ \ \ \ \ Radiation \ \ \ \ \ \ \ \ \ \ \ \ \ \ \ \ \ \ \ \ \ \ \ \ \ \
\ \ \ \ \ \ \ \ \ \ \ \ \ \ \ \ \ \ \ \ \ \ \ \ \ \ \ \ \ \ \ \ \ \ \ \ \ \
\ \ \ \ \ \ \ \ \ \ \ \ \ \ \ \ \ \ \ \ \ \ \ 97

6.6 \ The Planetary Evidence for the Superstrong Interaction \ \ \ \ \ \ \ \
\ \ \ \ \ \ \ \ \ \ \ 98

6.7 \ The Rotation of Spiral Galaxies \ \ \ \ \ \ \ \ \ \ \ \ \ \ \ \ \ \ \
\ \ \ \ \ \ \ \ \ \ \ \ \ \ \ \ \ \ \ \ \ \ \ \ \ \ \ \ \ \ \ \ \ \ \ \ 98

6.8 \ Black Holes Do Not Exist \ \ \ \ \ \ \ \ \ \ \ \ \ \ \ \ \ \ \ \ \ \ \
\ \ \ \ \ \ \ \ \ \ \ \ \ \ \ \ \ \ \ \ \ \ \ \ \ \ \ \ \ \ \ \ \ \ \ \ \ \
\ 101

\bigskip \pagebreak

{\huge 7} \ {\Large Associated Fermions and the Hidden Realm}

{\Large \ \ \ \ \ of Gravity}

7.1 \ Associated Fermions and the Dual Role of Neutrinos \ \ \ \ \ \ \ \ \ \
\ \ \ \ \ \ \ \ \ \ \ \ 104

7.2 \ The Hidden Realm of Gravity \ \ \ \ \ \ \ \ \ \ \ \ \ \ \ \ \ \ \ \ \
\ \ \ \ \ \ \ \ \ \ \ \ \ \ \ \ \ \ \ \ \ \ \ \ \ \ \ \ \ \ \ \ \ \ \ \ 108

\bigskip

{\huge 8 }{\Large Properties of the Galactic Structured State}

8.1 \ The Superweak Force and its Connection to Neutrinos \ \ \ \ \ \ \ \ \
\ \ \ \ \ \ \ \ \ 112

8.2 \ The Galactic Medium is a Strange \ `Metal of\ Clusters \ \ 

\ \ \ \ \ \ \ of Galaxies and Neutrinos' \ \ \ \ \ \ \ \ \ \ \ \ \ \ \ \ \ \
\ \ \ \ \ \ \ \ \ \ \ \ \ \ \ \ \ \ \ \ \ \ \ \ \ \ \ \ \ \ \ \ \ \ \ \ \ \
\ \ \ \ \ 115

8.3 \ Properties of the Neutrino Gas of the Universe \ \ \ \ \ \ \ \ \ \ \ \
\ \ \ \ \ \ \ \ \ \ \ \ \ \ \ \ \ \ \ 116

8.4 \ Neutrino Levels in a Weak Periodic Potential \ \ \ \ \ \ \ \ \ \ \ \ \
\ \ \ \ \ \ \ \ \ \ \ \ \ \ \ \ \ \ \ \ 118

8.4.1 \ Neutrino Energy Bands in One Dimension \ \ \ \ \ \ \ \ \ \ \ \ \ \ \
\ \ \ \ \ \ \ \ \ \ \ \ \ \ \ \ \ \ \ \ 121

\bigskip

{\huge 9 }{\Large Another Solution to the Solar Neutrino Problem}

9.1 The Solar Neutrino Problem and its Current Solution \ \ \ \ \ \ \ \ \ \
\ \ \ \ \ \ \ \ \ \ \ \ 125

9.2 Another Solution to the Solar Neutrino Problem \ \ \ \ \ \ \ \ \ \ \ \ \
\ \ \ \ \ \ \ \ \ \ \ \ \ \ \ \ 126

\bigskip

{\huge 10}{\Large \ Some Topics in Nuclear Physics}

10.1 \ The Nuclear Potential and the Stability of the Deuteron,

\ \ \ \ \ \ \ \ \ Triton and Alpha Particle \ \ \ \ \ \ \ \ \ \ \ \ \ \ \ \
\ \ \ \ \ \ \ \ \ \ \ \ \ \ \ \ \ \ \ \ \ \ \ \ \ \ \ \ \ \ \ \ \ \ \ \ \ \
\ \ \ \ \ \ 130

10.2 \ The Absence of Nuclides with A=5

\ \ \ \ \ \ \ \ \ \ and the Instability of Be$^{8}$ \ \ \ \ \ \ \ \ \ \ \ \
\ \ \ \ \ \ \ \ \ \ \ \ \ \ \ \ \ \ \ \ \ \ \ \ \ \ \ \ \ \ \ \ \ \ \ \ \ \
\ \ \ \ \ \ \ \ \ \ \ 134

\bigskip

{\huge Appendix: }{\Large \ Brief Vita} \ \ \ \ \ \ \ \ \ \ \ \ \ \ \ \ \ \
\ \ \ \ \ \ \ \ \ \ \ \ \ \ \ \ \ \ \ \ \ \ \ \ \ \ \ \ \ \ \ \ \ \ \ \ \ \
135

\bigskip\ \ \ \ \ \ \ \bigskip \bigskip

\bigskip \bigskip \pagebreak

\bigskip\ \ \ \ \ \ \ \ \ \ \ \ \ \ \ \ \ \ \ \ \ \ \ \ \ \ \ \ \ \ 

\bigskip \vspace*{0.5in}

\section{{\protect\LARGE General Properties of Matter\ }{\protect\Huge \ \ \
\ \ \ \ \ \ \ }}

\bigskip \bigskip \rule[1in]{0in}{0.17in}

\subsection{\protect\bigskip {\protect\Large General Classification of Matter%
}}

Science has utilized specific empirical classifications of matter which have
revealed hidden laws and symmetries. Two of the most known classifications
are the Periodic Table of the Elements and Gell-Mann's classification of
particles(which paved the way towards the quark model). Let us go on the
footsteps of Mendeleev and let us attempt to achieve a general
classification of matter, including all kinds of matter formed along the
universal expansion, and by doing so we may find the links between the
elementary particles and the large bodies of the universe.

It is well known that the different kinds of matter appeared at different
epochs of the universal expansion and that they are imprints of the
different sizes of the universe along the expansion. Taking a closer look at
the different kinds of matter we may classify them as belonging to two
distinct general states. One state is characterized by a single unit with
angular momentum. The angular momentum may either be the intrinsic angular
momentum, spin, or the orbital angular momentum. The other state is
characterized by some degree of correlation among the interacting particles
and may be called the structured state. The angular momentum may(or may not)
be present in this state. The fundamental units of matter make the
structured states, that is, they are the building blocks of everything, 
\textit{stepwise}. In what follows we will not talk about the weak force
since it does not form any stable matter and is rather related to
instability in matter. Along the universal expansion nature made different
building blocks and different media to fill space. The weak force did not
form any building block and is out of our discussion. As is well known this
force is special in many other ways. For example, it violates parity and has
no ``effective potential''(or static potential) as the other interactions
do. Besides, the weak force is known to be left-handed, that is, particles
experience this force only when their spin direction is anti-aligned with
their momentum. Right-handed particles appear to experience no weak
interaction, although, if they have electric charge, they may still interact
electromagnetically. Later on we will include the weak force into the
discussion. Each structured state is mainly formed by two types of
fundamental forces. Due to the interactions among the units one expects
other kinds of forces in the structured state. In this fashion we can form a
chain from the quarks to the galactic superstructures and extrapolate at the
two ends towards the constituents of quarks and towards the whole Universe.

The units of matter are the nucleons, the atom, the galaxies, etc. The `et
cetera' will become clearer later on in this work. In the structured state
one finds the quarks, the nuclei, the gasses, liquids and solids, and the
galactic liquid. Let us, for example, examine the sequence
nucleon-nucleus-atom. As is well known a nucleon is made out of quarks and
held together by means of the strong force. The atom is made out of the
nucleus and the electron(we will talk about the electron later), and is held
together by means of the electromagnetic force. The nucleus, which is in the
middle of the sequence, is held together by the strong force(attraction
among nucleons) and by the electromagnetic force(repulsion among protons).
In other words, we may say that the nucleus is the result of a compromise
between these two forces. Let us, now, turn to the sequence
atom-(gas,liquid,solid)-galaxy. The gasses, liquids and solids are also
formed by two forces, namely, the electromagnetic and the gravitational
forces. Because the gravitational force is $10^{39}$ weaker than the
electromagnetic force the polarization in gasses, liquids and solids is
achieved by the sole action of the electromagnetic force because it has two
signs. But it is well known that large masses of gasses, liquids and solids
are unstable configurations of matter in the absence of gravity. Therefore,
they are formed by the electromagnetic and gravitational forces. Large
amounts of nucleons(and electrons) at some time in the history of the
universe gave origin to galaxies which are the biggest individual units of
creation. We arrive again at a single fundamental force that holds a galaxy
together, which is the gravitational force. There is always the same
pattern: one goes from one fundamental force which holds a single
unit(nucleon, atom, galaxy) together to two fundamental forces which coexist
in a medium. The interactions in the medium form a new unit in which the
action of another fundamental force appears. We are not talking any more
about the previous unit which exists inside the new unit(such as the
nucleons in the nucleus of an atom).

Actually, we can also argue that according to Noether's theorem$^{1}$ there
should exist a force connected to baryon number conservation. This is the
basis for the proposal on the fifth force. And all experiments do show that
the proton does not decay.

By placing all kinds of matter together in a table in the order of the
universal expansion we can construct the two tables below, one for the
states and units of matter and another one for the fundamental forces.

In order to make an atom we need the electron besides the nucleus.
Therefore, just the clumping of nucleons is not enough in this case. Let us
just borrow the electron for now. Therefore, it looks like that the electron
belongs to a separate class and is an elementary particle. That is, the
electron itself is not one of the units. Therefore, we can complete Table 1
with the prequark and with the Universe. The above considerations may be
summarized by the following: \textit{the different kinds of building blocks
of the Universe(at different times of the expansion) are intimately related
to the idea of filling space.} That is, depending on its size, the Universe
is filled with different units.

Following the same reasoning we can say that there should exist a force,
other than the strong force, acting between any two prequarks. We call it
superstrong force. Also, for the `galactic liquid' there must be another
fundamental force at play. Because it must be much weaker than the
gravitational force(otherwise, it would already have been found on Earth) we
expect it to be a very weak force. Let us call it the superweak force.

Actually, in a nucleus, there is also the action of the superstrong force
for very small distances between the nucleons.

Summing up all fundamental forces we arrive at \textit{six forces for
nature: the superstrong, the strong, the electromagnetic, the gravitational,
the superweak and the weak forces}. We will see later on that these two
tables are very important and reveal a very important role of the stable
fermions which has not been taken care of until now.


\vspace*{0.5in}

\begin{center}
\begin{tabular}{ccccccc}
\hline\hline
&  &  &  &  &  &  \\ 
& ? &  & quark &  & nucleon &  \\ 
&  &  &  &  &  &  \\ 
& nucleon &  & nucleus &  & atom &  \\ 
&  &  &  &  &  &  \\ 
& atom &  & gas &  & galaxy &  \\ 
&  &  & liquid &  &  &  \\ 
&  &  & solid &  &  &  \\ 
&  &  &  &  &  &  \\ 
& galaxy &  & galactic medium &  & ? &  \\ 
&  &  &  &  &  &  \\ \hline\hline
&  &  &  &  &  & 
\end{tabular}
\end{center}

\vskip .1in

\begin{center}
\parbox{4in}
{Table 1.1. The table is arranged in such a way to show the links between the 
structured states and the units of creation. The interrogation marks above 
imply the existence of prequarks and of the Universe itself as units of
creation}
\end{center}

\vspace*{0.5in}

\begin{center}
\begin{tabular}{ccc}
\hline\hline
&  &  \\ 
? & ? &  \\ 
& strong force & strong force \\ 
&  &  \\ 
strong force & strong force &  \\ 
& electromagnetic force & electromagnetic force \\ 
&  &  \\ 
electromagnetic force & electromagnetic force &  \\ 
& gravitational force & gravitational force \\ 
&  &  \\ 
gravitational force & gravitational force & ? \\ 
& ? &  \\ 
&  &  \\ \hline\hline
&  & 
\end{tabular}
\end{center}

\vskip .1in

\begin{center}
\parbox{4.5in}
{Table 1.2. Three of the fundamental forces of nature. Each force 
appears twice and is linked to another force by means of a structured 
state. The interrogation marks suggest the existence of two other
fundamental forces. Compare with Table 1.1.}
\end{center}

\vspace*{0.5in}

\subsection{\noindent {\protect\Large Size and Number}}

We know that the sizes of things are related to relations between the
constituent forces. For instance, the size of a mountain is related to the
relation between the gravitational and electromagnetic forces because the
matter at the base of the mountain can not be smashed by the weight of the
matter above the base. Applying the same logic to all forces we have the
following. Since primons are very light fermions (baryons) they should have
the size of an electron around $10^{-17}$ m. The 4 primons combine and form
the 6 quarks. Thus, the strong and superstrong interactions should have
comparable strengths. And that is so, for as we will see the superstrong
force strength is about 10 times the strong force strength. We can also see
this in the formation of baryons. With the 6 quarks we form about 50
baryons. Going to the next pair of forces we notice that the strong force is
about 100 stronger than the electromagnetic force. This is in line with the
numbers of nucleons and nuclei, for when we combine the nucleons we form
about $10^{3}$ nuclei (or atoms). Following the same reasoning we find that
with the electromagnetic and gravitational forces we can form an enormous
quantity of things because their relative strengths is about $10^{37}$.
Also, with the gravitational and superweak forces we make all kinds of
structures. This means that the superweak force is much weaker than the
gravitational force.

The essence of all this is the following: When the two forces have
comparable strengths they compete and limit the number of things done with
them. Conversely, when one is much stronger than the other one the number of
things is enormous due to the weak competition between them.

The sizes of things are also related to the strengths between the two forces
of each pair. For example, when we put together a small number of atoms they
can have arbitrary shapes because the gravitational force is much weaker
than the electromagnetic force. When the number of atoms increases the
macroscopic body becomes more and more spherical due to the increasing
influence of the gravitational force. Let us, then, take a look at the sizes
of the units in Table 1.2. They are \emph{primon, nucleon, atom, galaxy,
Universe}. Their sizes in meters are about $<10^{-17}$, $10^{-15}$, $%
10^{-10} $, $10^{21}$, $>10^{26}$, respectively. Hence, we notice once more
that the superweak force should be much weaker than the gravitational force,
and that the upper limit for the strength of the superstrong force with
respect to the strong force is about 100. From the precision of the
equivalence principle we have that the superweak force has to be about $%
10^{11}$ weaker than the gravitational force. As we will see in chapter 9
the cross section of the superweak interaction is extremely small. \bigskip

\noindent {\Large References}

\bigskip

\noindent 1) \ E. Noether, in \textit{Nach. Ges. Wiss. G\"{o}ttingen,} 171,
1918.

\bigskip

\bigskip \pagebreak

\bigskip

\bigskip \vspace*{0.5in}

\section{\protect\LARGE Prequarks, Quarks and \protect\nolinebreak \protect%
\nolinebreak Nucleons}

{\LARGE \bigskip \rule[1.5in]{0in}{0.17in}}

\subsection{\protect\Large Prequarks and the Number of Quarks}

{\normalsize \textbf{\ }}

As we saw in chapter 1 it has been proposed by De Souza$%
^{(1,2,3,4,5,6,7,8,9,10,11,12)}$ that Nature has six fundamental forces. One
of the new forces, called superstrong force, acts between any two quarks and
between quarks constituents. Actually, quark composition is an old idea,
although it has been proposed on different grounds$^{(13,14,15,16)}$. A
major distinction is that in this work leptons are supposed to be elementary
particles. This is actually consistent with the smallness of the electron
mass which is already too small for a particle with a very small radius$%
^{(17)}$.

In order to distinguish the model proposed in this work from other models of
the literature we will name these prequarks with a different name. We may
call them primons, a word derived from the Latin word primus which means
first.

Let us develop some preliminary ideas which will help us towards the
understanding of the superstrong interaction. Since a baryon is composed of
three quarks it is reasonable to consider that a quark is composed of two
primons. The new interaction between them exists by means of the exchange of
new bosons.

In order to reproduce the spectrum of 6 quarks and their colors we need 4
primons in 3 supercolor states. Each color is formed by the two supercolors
of two different primons that form a particular quark. Therefore, the
symmetry group associated with the supercolor filed is SU(2). As to the
charge, one has charge (+5/6)e and any other one has charge (-1/6)e. And
what about spin? How can we have prequarks with spins equal to 1/2 and also
have quarks with spin equal to 1/2? There are two solutions to this
question. One is to consider that at the prequark level Planck's constant is
redefined as ${\hbar }/2$. I adopted this solution in a previous version of
this work. It leads to some problems. One of them is that in the end we will
have to deal with anyons. But anyons violate P and T while the strong
interaction does not. I believe that the other solution is more plausibe
although it depends on a postulate which may be expressed in the following
way:\textbf{\ Primons are fermions with spins equal to 1/2 but each spin (z
component) makes an angle of }$\pi /3$\textbf{\ with the direction of the
hadron spin (z component), so that the total spin of the quark is 1/2}. This
means that the system of primons in a baryon is a very cooperative system in
the sense that primons interact in such a way as to maintain their spins (z
component) making either $\pi /3$ or $2\pi /3$ with the baryon's spin
direction. The total angle between the two spins of the two primons of a
quark should always be $2\pi /3$. Of course, we are saying that primons are
special fermions because their spins can not be aligned by a magnetic field
due to their mutual interaction that couples each pair making a rigid angle
between their spins, and only the total z component of each pair is aligned
by the field. This means that primon spin behaves as a normal vector. Thus,
with respect to spin such a system is highly ordered. This spin picture
sheds some light onto the proton spin puzzle and is in line with the work of
Srivastava and Widom on the spin of the proton$^{18}.$ As we will see
shortly in the next sections this is the only spin arrangement that
maintains primons as fermions and allows the exchange of scalar and
vectorial bosons between primons.

Let us choose the +Z direction as the direction of the proton's spin. Each
primon spin contributes with $(1/2)\cos (\pi /3)=(1/4)$ along the Z
direction ( see Fig 2.1). Thus, each quark has a spin equal to $1/2$. We
also see that the spins of the two primons in a quark can rotate freely
around the Z axis but they have to rotate at the same time so that the two
components in the XY plane cancel out. Therefore, it is possible to have the
exchange of scalar and vector bosons between primons of different quarks.

This is quite in line with the known properties of the nuclear potential
which may be described with terms due to the exchange of pions as well as
the exchange of vector mesons such as $\rho $ and $\omega $. Let us consider
that the superstrong field is mediated by the exchange of vector bosons.

As we see above a baryon is a very complex system. That is why the above
considerations are quite qualitative and incomplete and deserve further
investigation.

Taking into account the above considerations on spin and charge we have the
following tables for primons(Tables 2.1, 2.2, 2.3). According to Table 2.2
the maximum number of quarks is six. There should exist similar tables for
the corresponding antiparticles.

As we will see many different bosons may mediate the strong and superstrong
interactions among primons. And there are colored and colorless mesons.

\vspace*{0.3in}

\begin{center}
\begin{tabular}{||ccc||ccc|ccc|ccc||}
\hline\hline
&  &  &  &  &  &  &  &  &  &  &  \\ 
&  &  &  & $\alpha $ &  &  & $\beta $ &  &  & $\gamma $ &  \\ 
&  &  &  &  &  &  &  &  &  &  &  \\ \hline\hline
&  &  &  &  &  &  &  &  &  &  &  \\ 
& $\alpha $ &  &  &  &  &  & blue &  &  & green &  \\ 
&  &  &  &  &  &  &  &  &  &  &  \\ \hline
&  &  &  &  &  &  &  &  &  &  &  \\ 
& $\beta $ &  &  & blue &  &  &  &  &  & red &  \\ 
&  &  &  &  &  &  &  &  &  &  &  \\ \hline
&  &  &  &  &  &  &  &  &  &  &  \\ 
& $\gamma $ &  &  & green &  &  & red &  &  &  &  \\ 
&  &  &  &  &  &  &  &  &  &  &  \\ \hline\hline
\end{tabular}
\end{center}

\vskip .1in

\begin{center}
\parbox{3in}
{Table 2.1. Generation of colors from supercolors}
\end{center}

\bigskip

\begin{center}
\begin{tabular}{||ccc||ccc|ccc||}
\hline\hline
&  &  &  &  &  &  &  &  \\ 
& superflavor &  &  & charge &  &  & spin &  \\ 
&  &  &  &  &  &  &  &  \\ \hline\hline
&  &  &  &  &  &  &  &  \\ 
& $p_{1} $ &  &  & + $\frac{5}{6} $ &  &  & $\frac{1}{2} $ &  \\ 
&  &  &  &  &  &  &  &  \\ \hline
&  &  &  &  &  &  &  &  \\ 
& $p_{2} $ &  &  & - $\frac{1}{6} $ &  &  & $\frac{1}{2} $ &  \\ 
&  &  &  &  &  &  &  &  \\ \hline
&  &  &  &  &  &  &  &  \\ 
& $p_{3} $ &  &  & - $\frac{1}{6} $ &  &  & $\frac{1}{2} $ &  \\ 
&  &  &  &  &  &  &  &  \\ \hline
&  &  &  &  &  &  &  &  \\ 
& $p_{4} $ &  &  & - $\frac{1}{6} $ &  &  & $\frac{1}{2} $ &  \\ 
&  &  &  &  &  &  &  &  \\ \hline\hline
\end{tabular}
\end{center}

\vskip .1in

\begin{center}
\parbox{3in}
{Table 2.2. Electric charges and spins of primons}
\end{center}

\bigskip \pagebreak

\bigskip \vspace*{0.5in}

\bigskip

\begin{center}
\begin{tabular}{||ccc||ccc|ccc|ccc|ccc||}
\hline\hline
&  &  &  &  &  &  &  &  &  &  &  &  &  &  \\ 
&  &  &  & $p_{1}$ &  &  & $p_{2}$ &  &  & $p_{3}$ &  &  & $p_{4}$ &  \\ 
&  &  &  &  &  &  &  &  &  &  &  &  &  &  \\ \hline\hline
&  &  &  &  &  &  &  &  &  &  &  &  &  &  \\ 
& $p_{1}$ &  &  &  &  &  & u &  &  & c &  &  & t &  \\ 
&  &  &  &  &  &  &  &  &  &  &  &  &  &  \\ \hline
&  &  &  &  &  &  &  &  &  &  &  &  &  &  \\ 
& $p_{2}$ &  &  & u &  &  &  &  &  & d &  &  & s &  \\ 
&  &  &  &  &  &  &  &  &  &  &  &  &  &  \\ \hline
&  &  &  &  &  &  &  &  &  &  &  &  &  &  \\ 
& $p_{3}$ &  &  & c &  &  & d &  &  &  &  &  & b &  \\ 
&  &  &  &  &  &  &  &  &  &  &  &  &  &  \\ \hline
&  &  &  &  &  &  &  &  &  &  &  &  &  &  \\ 
& $p_{4}$ &  &  & t &  &  & s &  &  & b &  &  &  &  \\ 
&  &  &  &  &  &  &  &  &  &  &  &  &  &  \\ \hline\hline
\end{tabular}
\end{center}

\vskip .2in

\begin{center}
\parbox{3in}
{Table 2.3. Composition of quark flavors}
\end{center}

\bigskip

\bigskip

\subsection{\protect\Large The Structure of Nucleons, the Sizes of Quarks u
and d, and the Stability of the Proton }

Deep inelastic electron scattering$^{(19,20)}$ has shown that the
distributions of electric charge in the nucleons are represented by the two
graphs below(Figs. 2.2a and 2.2b). These distributions have inspired the
pion cloud model of the nucleon which has been quite sucessful at explaining
many of its properties.

Analyzing these two figures one easily sees that shells of electric charge
exist in both nucleons. The proton has two shells with mean radii of about
0.2fm and 0.7fm and the neutron has three shells with radii of about 0.15fm,
0.65fm, and 1.8fm. Let us disregard the outermost shell of the neutron.
Therefore, each nucleon has two shells of primons located at about 0.16fm
and 0.67fm from the center. We can only explain the existence of these
shells if we admit that quarks are composite and formed of prequarks. The
two shells are, then, prequark shells, showing that a quark is composed of
two prequarks. Considering what was presented above primons with the same
supercolors tend to stay away from each other and primons with different
supercolors attract each other. Therefore, primons are arranged inside the
proton as is shown in Fig. 2.3. The charge of each one of the two
shells(inner and outer shells) is +1/2. In terms of primon shells we can say
that the proton has the configuration 
\begin{equation*}
(p_{1}^{\alpha }p_{2}^{\beta }p_{3}^{\gamma })^{1}(p_{2}^{\beta
}p_{1}^{\gamma }p_{2}^{\alpha })^{2}.
\end{equation*}
\noindent The superscripts 1 and 2 mean the inner and outer shells,
respectively. Let us dispose the primons of the inner shell clockwise. A
primon of one shell with the closest primon of the other shell forms a
quark. In each shell there is a plane of primons. The two planes are linked
by the three strong bonds, that is, by the three quarks. A primon of a quark
with a primon of another quark forms a weak bond when they are different and
have different supercolors. The three quarks of the inner shell of the
proton, for example, are linked by weak bonds. Due to the exchange of gluons
the colors change, and therefore the weak bonds change all the time, but the
lowest potential energy of the inner shell should happen when it has three
different supercolors since equal colors repel each other. Thus, all
possible configurations of the proton are:

$(p_{1}^{\alpha }p_{2}^{\beta }p_{3}^{\gamma })^{1}(p_{2}^{\beta
}p_{1}^{\gamma }p_{2}^{\alpha })^{2}$; $(p_{1}^{\alpha }p_{2}^{\beta
}p_{3}^{\gamma })^{1}(p_{2}^{\gamma }p_{1}^{\alpha }p_{2}^{\beta })^{2}$; $%
(p_{1}^{\alpha }p_{2}^{\gamma }p_{3}^{\beta })^{1}(p_{2}^{\beta
}p_{1}^{\alpha }p_{2}^{\gamma })^{2}$;

$(p_{1}^{\alpha }p_{2}^{\gamma }p_{3}^{\beta })^{1}(p_{2}^{\gamma
}p_{1}^{\beta }p_{2}^{\alpha })^{2}$; $(p_{1}^{\beta }p_{2}^{\alpha
}p_{3}^{\gamma })^{1}(p_{2}^{\gamma }p_{1}^{\beta }p_{2}^{\alpha })^{2}$; $%
(p_{1}^{\beta }p_{2}^{\alpha }p_{3}^{\gamma })^{1}(p_{2}^{\alpha
}p_{1}^{\gamma }p_{2}^{\beta })^{2}$;

$(p_{1}^{\beta }p_{2}^{\gamma }p_{3}^{\alpha })^{1}(p_{2}^{\alpha
}p_{1}^{\beta }p_{2}^{\gamma })^{2}$; $(p_{1}^{\beta }p_{2}^{\gamma
}p_{3}^{\alpha })^{1}(p_{2}^{\gamma }p_{1}^{\alpha }p_{2}^{\beta })^{2}$; $%
(p_{1}^{\gamma }p_{2}^{\alpha }p_{3}^{\beta })^{1}(p_{2}^{\beta
}p_{1}^{\gamma }p_{2}^{\alpha })^{2}$;

$(p_{1}^{\gamma }p_{2}^{\alpha }p_{3}^{\beta })^{1}(p_{2}^{\alpha
}p_{1}^{\beta }p_{2}^{\gamma })^{2}$; $(p_{1}^{\gamma }p_{2}^{\beta
}p_{3}^{\alpha })^{1}(p_{2}^{\alpha }p_{1}^{\gamma }p_{2}^{\beta })^{2}$ and 
$(p_{1}^{\gamma }p_{2}^{\beta }p_{3}^{\alpha })^{1}(p_{2}^{\beta
}p_{1}^{\alpha }p_{2}^{\gamma })^{2}$.

Since the $u$ quark does not decay $p_{1}$ and $p_{2}$ have to be stable and
since $d$ decays $p_{3}$ has to decay as $p_{3}\rightarrow p_{1}e^{-}\bar{%
\nu _{e}}$. Why then does not the proton decay since it contains a $d$
quark? The outer shell of the proton contains the primons $p_{1}$ and $p_{2}$
which are stable. Since the proton does not decay the inner shell which is
composed of the primons \ $p_{1}$, \ $p_{2}$ and \ $p_{3}$ has to be very
stable. This means that these three primons are in a deep potential well.
The neutron, on the other hand, has a $p_{3}$ primon in the outer shell, and
therefore, may decay. That is, if the $p_{3}$ of the outer shell is in a
well it must be so shallow that this primon may not be bound(that is, there
is no bound state). A quark is not, then, a pointlike particle as a lepton
is. It is an extended object. The $u$ and $d$ have average sizes of about
0.5fm and their sizes may be as large as 1fm.

Following the same reasoning the configuration of primons in the neutron
should be as shown in Fig. 2.4. The charge of the inner shell is +1/2 and
the charge of the outer shell is -1/2. As we saw above the inner shell of
the neutron should be equal to that of the proton. The configuration of the
neutron is $(p_{1}p_{2}p_{3})^{1}(p_{2}p_{3}p_{2})^{2}$ which differs from
the proton's in the outer shell.

As the neutron decays via the weak interaction into $n\rightarrow p^{+}e^{-}%
\bar{\nu _{e}}$ the primon $p_{3}$ should decay accordingly as $%
p_{3}\rightarrow p_{1}e^{-}\bar{\nu _{e}}$. In the case of the neutron the $%
p_{3}$ which decays is that of the outer shell. The primon $p_{4}$ should
also be unstable against weak decay.

Figures 2.3 and 2.4 are planar displays of three-dimensional spatial
configurations. In this way we reconcile the pion cloud vision of the
nucleon with the quark model. We easily see that a bare nucleon is a nucleon
without its outer shell since it is this shell that makes the difference
between nucleons.

We may idenfify primons as partons$^{11}$ which are supposed to be pointlike
and with spin equal to 1/2. But primons are supposed to be almost massless.
Therefore, they do not carry much momenta. Only their pairwise combinations,
that is, quarks, carry momenta. That is why it is very difficult to see
primons. At very high $Q^{2}$ they are not seen simply because they are very
light.

Therefore, it looks like that nature has been fooling us since a long time
ago, at least since the sixties: \textbf{The pointlike particles that we
have observed in the nucleon are not quarks, they are prequarks.}

A very important quantity that corroborates the arrangement of primons in
the nucleons is the value of the electric dipole moment(EDM) in each
nucleon. The values of the EDM for the proton and the neutron are$^{(21)}$ $%
d=(-4\pm 6)\times 10^{-23}$ecm and $d<1.1\times 10^{-25}$ecm, respectively.
According to the above picture we expect that the neutron EDM should be
smaller than the proton's because the outer layer of the neutron is $%
(p_{2}p_{3}p_{2})$ while the proton's is $(p_{2}p_{1}p_{2})$. Therefore,
since the primons $p_{2}$ and $p_{3}$ have the same charge(-1/6), while the
primons $p_{1}$ and $p_{2}$ have quite different charges(+5/6 and -1/6), the
outer layer of the neutron should be more spherical than the proton's. And
since the inner layer is the same for both nucleons, the neutron EDM should
be smaller than the proton's.

This picture also means that each quark(i.e., each pair of primons) does not
rotate much about its center of mass.

\vspace*{0.5in}

\subsection{\protect\Large The Sizes of Quarks, Primon Mass, and \newline
Generation of Quark Mass}

Having in mind what was developed above it is reasonable, therefore, to
consider that the two primons that form a quark(inside a baryon) are bound
by the combination of the strong and superstrong interactions. Therefore,
they should generate an effective potential well. Each well has to have just
one bound state which is the mass of each quark. Hopefully in the near
future we may have more information on such potential.

The heavier a quark is the deeper should be the well generated by the two
primons. Also, the well should be narrower because the heavier a quark is
the more it must be bound. For a given quantum number, $n$, the energy of a
well increases as it narrows. The potential of the top quark is extremely
deep since it is much more massive than the other quarks are. We are able,
then, to understand the decays of quarks. The lowest level is, of course,
the ground state of the $u$ quark. The ground state of the $d$ quark is
slightly above that of the $u$ quark, and the ground state of the $s$ quark
is above the ground state of the $d$ quark. This should also happen for the
other heavier quarks. Therefore, we expect that the potentials of all quarks
should be as shown in Fig.2.5 (the well of the $d$ quark is not shown).
These potentials and bound states are in line with the observed decay chain $%
b\rightarrow c\rightarrow s\rightarrow u$ and with the decays $d\rightarrow
ue\bar{\nu}_{e}$, $b\rightarrow s\gamma $. What about the masses of primons?
Since quarks $u$ and $d$ have about the same mass of 0.3GeV, we expect that $%
p_{1}$, $p_{2}$ and $p_{3}$ have the same mass. But, since the combination $%
p_{1}p_{3}$ generates the $c$ quark which is much heavier(about 1.5GeV) than 
$u$, we can infer that the different masses of quarks come from the strong
and superstrong interactions. Thus, we may suspect that all primons have the
same mass which is a sort of primitive, inherent mass, which may be of the
same kind of the mass that leptons have.

Let us now see what is behind quark masses. Several researchers have tried
to relate their masses to something more fundamental. In order to do this
let us approximate each well of Fig. 2.5 by an infinite potential well, that
is, the mass of each quark(in units of energy), $E_{q}$, should be given by 
\begin{equation}
E_{q}=\frac{{\hbar }^{2}{\pi }^{2}}{8ma^{2}}
\end{equation}
\noindent where $m$ is the mass of a primon, and $a$ is the average size of
each quark. Since each mass corresponds to a single level in each well, and
considering that primons have approximately the same mass, we obtain that
each quark mass should be related to the average distance between each pair
of primons, that is, to the width of each well. Therefore, we should have
the approximate relations: 
\begin{equation}
0.3=\frac{C}{R_{u}^{2}};\;0.5=\frac{C}{R_{s}^{2}};\;1.5=\frac{C}{R_{c}^{2}}%
;\;5.0=\frac{C}{R_{b}^{2}};\;150=\frac{C}{R_{t}^{2}}.
\end{equation}

\bigskip

\noindent where $C$ is a constant and $R_{u}$, $R_{s}$,$R_{c}$,$R_{b}$ and $%
R_{t}$ are the widths of the wells. As we discussed in section 2, $%
R_{u}\approx 0.5$F. We may assume that $R_{u}=R_{d}$. We arrive at the very
important relations about quark sizes: 
\begin{equation}
R_{s}^{2}=\frac{3}{5}R_{u}^{2}=0.6R_{u}^{2}{\approx }0.15F^{2},
\end{equation}
\begin{equation}
R_{c}^{2}=\frac{5}{15}R_{s}^{2}=\frac{5}{15}\frac{3}{5}R_{u}^{2}=0.2R_{u}^{2}%
{\approx }0.05F^{2},
\end{equation}
\begin{equation}
R_{b}^{2}=\frac{15}{50}R_{c}^{2}=\frac{15}{50}\frac{5}{15}\frac{3}{5}%
R_{u}^{2}=0.06R_{u}^{2}{\approx }0.015F^{2},
\end{equation}
\begin{equation}
R_{t}^{2}=\frac{5}{150}R_{b}^{2}=\frac{50}{1500}\frac{15}{50}\frac{5}{15}%
\frac{3}{5}R_{u}^{2}{\approx }0.002R_{u}^{2}{\approx }0.0005F^{2}.
\end{equation}

It is quite interesting that there are some very fascinating relations. A
very important one is: 
\begin{equation}
\frac{m_{d}}{m_{u}}=\frac{R_{u}^{2}}{R_{d}^{2}}=1=3^{0}
\end{equation}
\begin{equation}
\frac{m_{c}}{m_{s}}=\frac{R_{s}^{2}}{R_{c}^{2}}=3=3^{1}
\end{equation}
\begin{equation}
\frac{m_{t}}{m_{b}}=\frac{R_{b}^{2}}{R_{t}^{2}}=30{\approx }3^{3},
\end{equation}
\noindent and, thus, there is a factor of 10 between the last two relations.
Other quite important relations are: 
\begin{equation}
\frac{m_{b}}{m_{s}}=\frac{R_{s}^{2}}{R_{b}^{2}}=10{\approx }3^{2}
\end{equation}
\begin{equation}
\frac{m_{t}}{m_{c}}=\frac{R_{c}^{2}}{R_{t}^{2}}=100{\approx }3^{4},
\end{equation}

\noindent which has the same factor of 10. Therefore, the heavier a quark is
the smaller it is. The approximation $100\approx 3^{4}$ is completely
justified because we approximated the finite well by an infinite well and
also made the top quark mass approximately equal to 150GeV.

The above results agree quite well with the work of Povh and H\"{u}fner$%
^{22} $ that have found $<r^{2}>_{u,d}=0.36$ $fm^{2}$ and $<r^{2}>_{s}=0.16$ 
$fm^{2}$ as effective radii of the constituent quarks. Also Povh$^{23}$
reports the following hadronic radii: $<r_{h}^{2}>=0.72$ $fm^{2}$, for
proton, $<r_{h}^{2}>=0.62$ $fm^{2}$, for $\Sigma ^{-}$, $<r_{h}^{2}>=0.54$ $%
fm^{2}$, for $\Xi ^{-}$, $<r_{h}^{2}>=0.43$ $fm^{2}$, for $\pi ^{-}$, and $%
<r_{h}^{2}>=0.37$ $fm^{2}$, for $K^{-}$. These radii clearly indicate that
the $s$ quark is smaller than the $u$ quark.

In order to have very light primons we can consider that every pair of
primons of a quark are bound by means of a very strong spring. Since every
potential well has just one level we have the mass of each quark equal to 
\begin{equation}
m_{q}c^{2}\approx \frac{{\hbar \omega }}{2}=\frac{{\hbar }}{2}\sqrt{\frac{k}{%
\mu _{p}}}
\end{equation}

\noindent in which $\mu _{p}$ is the reduced mass of the pair of primons and 
$k$ is the effective constant of the spring between them. It is worth
mentioning that a quite similar idea is used for explaining quark
confinement and based on it a term $Kr$ is introduced in the effective
potential. For the $u$ quark, for example, we have $m_{u}c^{2}\approx 0.3$%
GeV. On the other hand if we consider a harmonic potential we have 
\begin{equation}
m_{q}c^{2}\approx \frac{1}{2}k_{u}(R_{q})^{2}
\end{equation}
where $R_{q}$ is the size of the quark $q$. For $u$ we obtain $k_{u}\approx
10^{20}J/m^{2}\approx 2GeV/fm^{2}.$ Using this figure above we obtain $\mu
_{p}\approx 10^{-28}kg$ which is about the proton mass. Therefore, in order
to have light primons the effective well has to have a larger dependence
with the distance between the two primons. Considering that the potential is
symmetrical about the equilibrium position we may try to use the potential 
\begin{equation}
V(x)=\alpha _{u}x^{4}.
\end{equation}
The energy levels of the potential $V(x)=\alpha x^{\upsilon }$ are given by$%
^{24}$%
\begin{equation}
E_{n}=\left[ \sqrt{\frac{\pi }{2\mu }}\nu \hbar a^{1/\nu }\frac{\Gamma (%
\frac{3}{2}+\frac{1}{\nu })}{\Gamma (\frac{1}{\nu })}\right] ^{2\nu /(2+\nu
)}(n+\frac{1}{2})^{2\nu /(2+\nu )}.
\end{equation}
Thus, for $\nu =4$ and $n=0$ we have

\begin{equation}
E_{0}=\left[ \sqrt{\frac{\pi }{2\mu _{p}}}4\hbar a_{u}^{1/4}\frac{\Gamma (%
\frac{3}{2}+\frac{1}{4})}{\Gamma (\frac{1}{4})}\right] ^{4/3}(0+\frac{1}{2}%
)^{4/3}
\end{equation}
and then we obtain (making $E_{0}=m_{q}c^{2}$) 
\begin{equation}
\mu _{p}\sim 0.25\hbar ^{2}\sqrt{\frac{a_{u}}{(m_{q}c^{2})^{3}}}
\end{equation}
which can be extremely light depending on the value of $a_{u}.$ \ The above
figure should be taken with caution because it is a result of
nonrelativistic quantum mechanics but it does not change the fact that
primons may have a very small mass.

Thus, if primons interact via a very strong potential such as $V(x)=\alpha
_{u}x^{4}$ they can be extremely light fermions. We can then propose a more
general effective potential of the form $V(x)=\frac{1}{4}\alpha _{u}x^{4}-%
\frac{1}{2}k_{u}x^{2}$ where the last term is chosen negative. Generalizing
the coordinate $x$ we can consider the \ ``potential energy '' 
\begin{equation}
V_{0}(\phi )=\frac{1}{4}\lambda ^{2}\phi ^{4}-\frac{1}{2}\mu ^{2}\phi ^{2}
\end{equation}
where $\phi $ is a field related to the presence of the two primons (or of
other primons of the same baryon) and is the scalar interaction between
them, and $\mu $ and $\lambda $ are real constants. Hence we can propose the
Lagrangian 
\begin{equation}
\mathcal{L}_{0}\mathcal{=}\frac{1}{2}(\partial _{\nu }\phi )(\partial ^{\nu
}\phi )+\frac{1}{2}\mu ^{2}\phi ^{2}-\frac{1}{4}\lambda ^{2}\phi ^{4}
\end{equation}
between the two primons of a quark. Since a quark only exists by means of
the combination of the two primons we may consider that its initial mass is
very small. The above Lagrangian is symmetric in $\phi $ but let us recall
that primons can interact by other means, electromagnetically, for example.
Therefore, we can make the transformation $\phi \longrightarrow \phi +\eta
_{ev}$ where $\eta _{ev}$ is a deviation caused by the electromagnetic field
and vacuum. The new potential energy up to second power in $\eta _{ev}$ is 
\begin{equation}
V(\phi ,\eta _{ev})=V_{0}-\mu ^{2}\phi \eta _{ev}-\frac{1}{2}\mu ^{2}\eta
_{ev}^{2}+\lambda ^{2}\phi ^{3}\eta _{ev}+\frac{3}{2}\lambda ^{2}\phi
^{2}\eta _{ev}^{2}.
\end{equation}

\noindent \bigskip\ $V(\phi ,\eta _{ev})$ has a minimum at 
\begin{equation}
\eta _{ev}(\phi )=\frac{-\mu ^{2}\phi +\lambda ^{2}\phi ^{3}}{\mu
^{2}-3\lambda ^{2}\phi ^{2}}.
\end{equation}

\noindent Since $\eta _{ev}$ is small let us make $\mu ^{2}\phi -\lambda
^{2}\phi ^{3}=\delta $ (a small quantity). Then we can make $\phi \approx
\pm \frac{\mu }{\lambda }+\epsilon $ and obtain $\epsilon \approx -\frac{%
\delta }{2\mu ^{2}}$ and thus \ 
\begin{equation}
\phi \approx \pm \frac{\mu }{\lambda }-\frac{\delta }{2\mu ^{2}}
\end{equation}
and the symmetry has disappeared. But it is not spontaneously broken, it is
broken by the perturbation $\eta _{ev}$. Substituting the above value of $%
\phi $ into Eq 20 we have

\begin{equation}
U(\eta _{ev})=V(\phi ,\eta _{ev})-V_{0}\approx \mu ^{2}\eta _{ev}^{2}
\end{equation}
and the approximate Lagrangian is 
\begin{equation}
\mathcal{L=}\frac{1}{2}(\partial _{\nu }\eta _{ev})(\partial ^{\nu }\eta
_{ev})-\mu ^{2}\eta _{ev}^{2}
\end{equation}
which is a Klein-Gordon Lagrangian with mass 
\begin{equation}
m=\sqrt{2}\mu \hbar /c
\end{equation}
which may be an effective mass. This is in complete agreement with the ideas
above discussed on primons for as we saw the two primons of a quark have to
interact by means of a scalar field because the z components of their spins
are neither parallel nor antiparallel. Taking a look at Table 2.3 we observe
that we need three scalar bosons, $\eta _{ev}^{+}$, $\eta _{ev}^{-}$ and $%
\eta _{ev}^{0}$. The first and second \ particles are exchanged between the
primons of the quarks $p_{1}p_{2}(u)$, $p_{1}p_{3}(c)$, and $p_{1}p_{4}(t)$,
and the neutral boson is exchanged between the primons of the quarks $%
p_{2}p_{3}(d)$, $p_{2}p_{4}(s)$, and $p_{3}p_{4}(b)$. Therefore, three Higgs
bosons produce the masses of quarks. This is also in line with the recent
observations about the beginning of the Universe which have shown that the
Universe expanded from an initial mass$^{25}$.

It is quite interesting that we should have a triplet of scalar bosons. And
we notice immediately a very important trend: The charged bosons produce
masses larger\ than those produced by the neutral boson, considering the
quark generations

\begin{center}
\bigskip $\left( 
\begin{array}{c}
u \\ 
d
\end{array}
\right) ,\left( 
\begin{array}{c}
c \\ 
s
\end{array}
\right) ,\left( 
\begin{array}{c}
t \\ 
b
\end{array}
\right) .$
\end{center}

{\large Therefore, the origin of mass in quarks is also linked to the origin
of charge.}

\bigskip \pagebreak

\vspace*{0.5in}

This is summarized below in Table 2.4.

\bigskip

\begin{center}
\begin{tabular}{||ccc||ccc|ccc|ccc||}
\hline\hline
&  &  &  &  &  &  &  &  &  &  &  \\ 
& Quark &  &  & Mass(GeV) &  &  & Charge &  &  & Mass Generator &  \\ 
&  &  &  &  &  &  &  &  &  & (Higgs Bosons) &  \\ \hline\hline
&  &  &  &  &  &  &  &  &  &  &  \\ 
& $u(p_{1}p_{2})$ &  &  & $0.3$ &  &  & +$\frac{2}{3}$ &  &  & $\eta
_{ev}^{+}$, $\eta _{ev}^{-}$ &  \\ 
&  &  &  &  &  &  &  &  &  &  &  \\ \hline
&  &  &  &  &  &  &  &  &  &  &  \\ 
& $c(p_{1}p_{3})$ &  &  & $1.5$ &  &  & +$\frac{2}{3}$ &  &  & $\eta
_{ev}^{+}$, $\eta _{ev}^{-}$ &  \\ 
&  &  &  &  &  &  &  &  &  &  &  \\ \hline
&  &  &  &  &  &  &  &  &  &  &  \\ 
& $t(p_{1}p_{4})$ &  &  & $170$ &  &  & +$\frac{2}{3}$ &  &  & $\eta
_{ev}^{+}$, $\eta _{ev}^{-}$ &  \\ 
&  &  &  &  &  &  &  &  &  &  &  \\ \hline
&  &  &  &  &  &  &  &  &  &  &  \\ 
& $d(p_{2}p_{3})$ &  &  & $\gtrapprox 0.3$ &  &  & -$\frac{1}{3}$ &  &  & $%
\eta _{ev}^{0}$ &  \\ 
&  &  &  &  &  &  &  &  &  &  &  \\ \hline
&  &  &  &  &  &  &  &  &  &  &  \\ 
& $s(p_{2}p_{4})$ &  &  & $0.5$ &  &  & -$\frac{1}{3}$ &  &  & $\eta
_{ev}^{0}$ &  \\ 
&  &  &  &  &  &  &  &  &  &  &  \\ \hline
&  &  &  &  &  &  &  &  &  &  &  \\ 
& $b(p_{3}p_{4})$ &  &  & $4.5$ &  &  & -$\frac{1}{3}$ &  &  & $\eta
_{ev}^{0}$ &  \\ 
&  &  &  &  &  &  &  &  &  &  &  \\ \hline\hline
\end{tabular}

\vspace*{0.2in}

\parbox{4in}
{Table 2.4. The masses of quarks and their generators. As is known the mass of 
the $d$ quark is slightly larger than that of the $u$ quark. There is a clear
division between the three first quarks and the other three quarks. The quarks
generated by the charged bosons have larger masses and larger charges and 
those generated by $\eta_{ev}^{0}$ have smaller masses and smaller charges}.
\end{center}

\bigskip

\noindent Table 2.3 shows that quark masses come from the interaction term,
that is, we should have

\bigskip 
\begin{equation}
m_{12}=<p_{1}|\mu ^{2}\eta_{ev}^{2}|p_{2}>=m_{u}c^{2}=0.3GeV
\end{equation}

\begin{equation}
m_{13}=<p_{1}|\mu ^{2}\eta_{ev}^{2}|p_{3}>=m_{c}c^{2}=1.5GeV
\end{equation}

\begin{equation}
m_{14}=<p_{1}|\mu ^{2}\eta_{ev}^{2}|p_{4}>=m_{t}c^{2}=170GeV
\end{equation}

\begin{equation}
m_{23}=<p_{2}|\mu ^{2}\eta_{ev}^{2}|p_{3}>=m_{d}c^{2}=0.3GeV
\end{equation}

\begin{equation}
m_{24}=<p_{2}|\mu ^{2}\eta_{ev}^{2}|p_{4}>=m_{s}c^{2}=0.5GeV
\end{equation}

\begin{equation}
m_{34}=<p_{3}|\mu ^{2}\eta_{ev}^{2}|p_{4}>=m_{b}c^{2}=4.5GeV
\end{equation}

\noindent and the following mass matrix:

\bigskip

\begin{eqnarray}
M^{q} &=&\left( 
\begin{array}{cccc}
0 & m_{12} & m_{13} & m_{14} \\ 
m_{12} & 0 & m_{23} & m_{24} \\ 
m_{13} & m_{23} & 0 & m_{34} \\ 
m_{14} & m_{24} & m_{34} & 0
\end{array}
\right) =\left( 
\begin{array}{cccc}
0 & m_{u} & m_{c} & m_{t} \\ 
m_{u} & 0 & m_{d} & m_{s} \\ 
m_{c} & m_{d} & 0 & m_{b} \\ 
m_{t} & m_{s} & m_{b} & 0
\end{array}
\right) =  \notag \\
&=&\left( 
\begin{array}{cccc}
0 & 0.3 & 1.5 & 170 \\ 
0.3 & 0 & 0.3 & 0.5 \\ 
1.5 & 0.3 & 0 & 4.5 \\ 
170 & 0.5 & 4.5 & 0
\end{array}
\right) GeV
\end{eqnarray}

\vspace*{0.5in}

We cannot diagonalize this matrix because masses come from terms of
interaction between different primons, which are off-diagonal terms. This
means that the effective potential should contain odd powers of $r$ which
make the potential asymmetric. It also means that each mass is an
interaction mass and not an intrinsic mass as the mass of an electron. This
is quite in line with the idea of a ``bare mass'' and of a ``constituent
mass'' for quarks although their bare mass have actually to be much smaller.

As we will see in the next section the three bosons $\eta _{ev}^{+}$, $\eta
_{ev}^{-}$ and $\eta _{ev}^{0}$ are scalar colored combinations of $p%
\overline{p}$ pairs. For example, in the $u$ quark the primons $p_{1}$ and $%
p_{2}$ exchange the $\eta _{ev}^{-}=\overline{p_{1}p_{3}}p_{3}p_{2}+%
\overline{p_{1}p_{4}}p_{4}p_{2}$ and the $\eta _{ev}^{+}=\overline{p_{2}p_{3}%
}p_{3}p_{1}+\overline{p_{2}p_{4}}p_{4}p_{1}$ bosons. And in the same way in
the $c$ quark the primons $p_{1}$ and $p_{3}$ exchange the $\eta _{ev}^{-}=%
\overline{p_{1}p_{2}}p_{2}p_{3}+\overline{p_{1}p_{4}}p_{4}p_{2}$ and the $%
\eta _{ev}^{+}=\overline{p_{3}p_{2}}p_{2}p_{1}+\overline{p_{3}p_{4}}%
p_{4}p_{1}$ bosons. This means that $\overline{p_{1}p_{3}}p_{3}p_{2}+%
\overline{p_{1}p_{4}}p_{4}p_{2}$ and $\overline{p_{1}p_{2}}p_{2}p_{3}+%
\overline{p_{1}p_{4}}p_{4}p_{2}$ are expressed mathematically by the same
expression.

Some important questions continue lurking about. We may ask for example: Do
the masses of primons have an origin at all? It may be an electromagnetic
mass, such as the electron's. As we will see later on in section 7.2 gravity
is a very strange field and has a huge surprise for us that is revealed
thanks to Tables 1 and 2.

\subsection{\protect\bigskip {\protect\Large \noindent The Seas and Sizes of
Nucleons}}

\subsubsection{\protect\large The Proton Sea Content}

In order to find all the interactions let us consider three parts:
interactions in the outer shell, interactions in the inner shell and
interactions between the two shells, and let us first calculate the
transitions in Fig. 2.3. Of course, we are supposing that vacuum(in the
presence of matter) creates que $p\bar{p}$ pairs. In order to identify the
interactions let us \ take a look at all the possible vertices. Later on we
will identify the bosons that are involved in the interactions.

\noindent i) Interactions in the outer shell 
\begin{eqnarray*}
p_{1}^{\alpha } &+&\overline{p_{1}^{\alpha }p_{3}^{\gamma }}p_{3}^{\gamma
}p_{2}^{\beta }{\rightarrow }p_{2}^{\beta } \\
p_{1}^{\alpha } &+&\overline{p_{1}^{\alpha }p_{4}^{\gamma }}p_{4}^{\gamma
}p_{2}^{\beta }{\rightarrow }p_{2}^{\beta } \\
p_{2}^{\beta } &+&\overline{p_{2}^{\beta }p_{3}^{\gamma }}p_{3}^{\gamma
}p_{1}^{\alpha }{\rightarrow }p_{1}^{\alpha } \\
p_{2}^{\beta } &+&\overline{p_{2}^{\beta }p_{4}^{\gamma }}p_{4}^{\gamma
}p_{1}^{\alpha }{\rightarrow }p_{1}^{\alpha } \\
p_{1}^{\alpha } &+&\overline{p_{1}^{\alpha }p_{3}^{\beta }}p_{3}^{\beta
}p_{2}^{\gamma }{\rightarrow }p_{2}^{\gamma } \\
p_{1}^{\alpha } &+&\overline{p_{1}^{\alpha }p_{4}^{\beta }}p_{4}^{\beta
}p_{2}^{\gamma }{\rightarrow }p_{2}^{\gamma } \\
p_{2}^{\gamma } &+&\overline{p_{2}^{\gamma }p_{3}^{\beta }}p_{3}^{\beta
}p_{1}^{\alpha }{\rightarrow }p_{1}^{\alpha } \\
p_{2}^{\gamma } &+&\overline{p_{2}^{\gamma }p_{4}^{\beta }}p_{4}^{\beta
}p_{1}^{\alpha }{\rightarrow }p_{1}^{\alpha } \\
p_{2}^{\gamma } &+&\overline{p_{2}^{\gamma }p_{1}^{\alpha }}p_{1}^{\alpha
}p_{2}^{\beta }{\rightarrow }p_{2}^{\beta } \\
p_{2}^{\gamma } &+&\overline{p_{2}^{\gamma }p_{3}^{\alpha }}p_{3}^{\alpha
}p_{2}^{\beta }{\rightarrow }p_{2}^{\beta } \\
p_{2}^{\gamma } &+&\overline{p_{2}^{\gamma }p_{4}^{\alpha }}p_{4}^{\alpha
}p_{2}^{\beta }{\rightarrow }p_{2}^{\beta } \\
p_{2}^{\beta } &+&\overline{p_{2}^{\beta }p_{1}^{\alpha }}p_{1}^{\alpha
}p_{2}^{\gamma }{\rightarrow }p_{2}^{\gamma } \\
p_{2}^{\beta } &+&\overline{p_{2}^{\beta }p_{3}^{\alpha }}p_{3}^{\alpha
}p_{2}^{\gamma }{\rightarrow }p_{2}^{\gamma } \\
p_{2}^{\beta } &+&\overline{p_{2}^{\beta }p_{4}^{\alpha }}p_{4}^{\alpha
}p_{2}^{\gamma }{\rightarrow }p_{2}^{\gamma }
\end{eqnarray*}

\noindent ii) Interactions in the inner shell 
\begin{eqnarray*}
p_{1}^{\gamma } &+&\overline{p_{1}^{\gamma }p_{3}^{\alpha }}p_{3}^{\alpha
}p_{2}^{\beta }{\rightarrow }p_{2}^{\beta } \\
p_{1}^{\gamma } &+&\overline{p_{1}^{\gamma }p_{4}^{\alpha }}p_{4}^{\alpha
}p_{2}^{\beta }{\rightarrow }p_{2}^{\beta } \\
p_{2}^{\beta } &+&\overline{p_{2}^{\beta }p_{3}^{\alpha }}p_{3}^{\alpha
}p_{1}^{\gamma }{\rightarrow }p_{1}^{\gamma } \\
p_{2}^{\beta } &+&\overline{p_{2}^{\beta }p_{4}^{\alpha }}p_{4}^{\alpha
}p_{1}^{\gamma }{\rightarrow }p_{1}^{\gamma } \\
p_{1}^{\gamma } &+&\overline{p_{1}^{\gamma }p_{2}^{\beta }}p_{2}^{\beta
}p_{3}^{\alpha }{\rightarrow }p_{3}^{\alpha } \\
p_{1}^{\gamma } &+&\overline{p_{1}^{\gamma }p_{4}^{\beta }}p_{4}^{\beta
}p_{3}^{\alpha }{\rightarrow }p_{3}^{\alpha } \\
p_{3}^{\alpha } &+&\overline{p_{3}^{\alpha }p_{2}^{\beta }}p_{2}^{\beta
}p_{1}^{\gamma }{\rightarrow }p_{1}^{\gamma }
\end{eqnarray*}

\begin{eqnarray*}
p_{3}^{\alpha } &+&\overline{p_{3}^{\alpha }p_{4}^{\beta }}p_{4}^{\beta
}p_{1}^{\gamma }{\rightarrow }p_{1}^{\gamma } \\
p_{2}^{\beta } &+&\overline{p_{2}^{\beta }p_{1}^{\gamma }}p_{1}^{\gamma
}p_{3}^{\alpha }{\rightarrow }p_{3}^{\alpha } \\
p_{2}^{\beta } &+&\overline{p_{2}^{\beta }p_{4}^{\gamma }}p_{4}^{\gamma
}p_{3}^{\alpha }{\rightarrow }p_{3}^{\alpha } \\
p_{3}^{\alpha } &+&\overline{p_{3}^{\alpha }p_{1}^{\gamma }}p_{1}^{\gamma
}p_{2}^{\beta }{\rightarrow }p_{2}^{\beta } \\
p_{3}^{\alpha } &+&\overline{p_{3}^{\alpha }p_{4}^{\gamma }}p_{4}^{\gamma
}p_{2}^{\beta }{\rightarrow }p_{2}^{\beta }
\end{eqnarray*}

\bigskip

\noindent iii) Interactions between the two shells 
\begin{eqnarray*}
p_{1}^{\gamma } &+&\overline{p_{1}^{\gamma }p_{3}^{\alpha }}p_{3}^{\alpha
}p_{2}^{\beta }{\rightarrow }p_{2}^{\beta } \\
p_{1}^{\gamma } &+&\overline{p_{1}^{\gamma }p_{4}^{\alpha }}p_{4}^{\alpha
}p_{2}^{\beta }{\rightarrow }p_{2}^{\beta } \\
p_{2}^{\beta } &+&\overline{p_{2}^{\beta }p_{3}^{\alpha }}p_{3}^{\alpha
}p_{1}^{\gamma }{\rightarrow }p_{1}^{\gamma } \\
p_{2}^{\beta } &+&\overline{p_{2}^{\beta }p_{4}^{\alpha }}p_{4}^{\alpha
}p_{1}^{\gamma }{\rightarrow }p_{1}^{\gamma } \\
p_{1}^{\alpha } &+&\overline{p_{1}^{\alpha }p_{2}^{\beta }}p_{2}^{\beta
}p_{1}^{\gamma }{\rightarrow }p_{1}^{\gamma } \\
p_{1}^{\alpha } &+&\overline{p_{1}^{\alpha }p_{3}^{\beta }}p_{3}^{\beta
}p_{1}^{\gamma }{\rightarrow }p_{1}^{\gamma } \\
p_{1}^{\alpha } &+&\overline{p_{1}^{\alpha }p_{4}^{\beta }}p_{4}^{\beta
}p_{1}^{\gamma }{\rightarrow }p_{1}^{\gamma } \\
p_{1}^{\gamma } &+&\overline{p_{1}^{\gamma }p_{2}^{\beta }}p_{2}^{\beta
}p_{1}^{\alpha }{\rightarrow }p_{1}^{\alpha } \\
p_{1}^{\gamma } &+&\overline{p_{1}^{\gamma }p_{3}^{\beta }}p_{3}^{\beta
}p_{1}^{\alpha }{\rightarrow }p_{1}^{\alpha } \\
p_{1}^{\gamma } &+&\overline{p_{1}^{\gamma }p_{4}^{\beta }}p_{4}^{\beta
}p_{1}^{\alpha }{\rightarrow }p_{1}^{\alpha } \\
p_{2}^{\gamma } &+&\overline{p_{2}^{\gamma }p_{1}^{\alpha }}p_{1}^{\alpha
}p_{2}^{\beta }{\rightarrow }p_{2}^{\beta } \\
p_{2}^{\gamma } &+&\overline{p_{2}^{\gamma }p_{3}^{\alpha }}p_{3}^{\alpha
}p_{2}^{\beta }{\rightarrow }p_{2}^{\beta } \\
p_{2}^{\gamma } &+&\overline{p_{2}^{\gamma }p_{4}^{\alpha }}p_{4}^{\alpha
}p_{2}^{\beta }{\rightarrow }p_{2}^{\beta } \\
p_{2}^{\beta } &+&\overline{p_{2}^{\beta }p_{1}^{\alpha }}p_{1}^{\alpha
}p_{2}^{\gamma }{\rightarrow }p_{2}^{\gamma } \\
p_{2}^{\beta } &+&\overline{p_{2}^{\beta }p_{3}^{\alpha }}p_{3}^{\alpha
}p_{2}^{\gamma }{\rightarrow }p_{2}^{\gamma } \\
p_{2}^{\beta } &+&\overline{p_{2}^{\beta }p_{4}^{\alpha }}p_{4}^{\alpha
}p_{2}^{\gamma }{\rightarrow }p_{2}^{\gamma } \\
p_{3}^{\alpha } &+&\overline{p_{3}^{\alpha }p_{1}^{\gamma }}p_{1}^{\gamma
}p_{2}^{\beta }{\rightarrow }p_{2}^{\beta } \\
p_{3}^{\alpha } &+&\overline{p_{3}^{\alpha }p_{4}^{\gamma }}p_{4}^{\gamma
}p_{2}^{\beta }{\rightarrow }p_{2}^{\beta } \\
p_{2}^{\beta } &+&\overline{p_{2}^{\beta }p_{1}^{\gamma }}p_{1}^{\gamma
}p_{3}^{\alpha }{\rightarrow }p_{3}^{\alpha } \\
p_{2}^{\beta } &+&\overline{p_{2}^{\beta }p_{4}^{\gamma }}p_{4}^{\gamma
}p_{3}^{\alpha }{\rightarrow }p_{3}^{\alpha } \\
p_{1}^{\alpha } &+&\overline{p_{1}^{\alpha }p_{2}^{\beta }}p_{2}^{\beta
}p_{3}^{\alpha }{\rightarrow }p_{3}^{\alpha } \\
p_{1}^{\alpha } &+&\overline{p_{1}^{\alpha }p_{2}^{\gamma }}p_{2}^{\gamma
}p_{3}^{\alpha }{\rightarrow }p_{3}^{\alpha }
\end{eqnarray*}

\pagebreak

\begin{eqnarray*}
p_{1}^{\alpha } &+&\overline{p_{1}^{\alpha }p_{4}^{\beta }}p_{4}^{\beta
}p_{3}^{\alpha }{\rightarrow }p_{3}^{\alpha } \\
p_{1}^{\alpha } &+&\overline{p_{1}^{\alpha }p_{4}^{\gamma }}p_{4}^{\gamma
}p_{3}^{\alpha }{\rightarrow }p_{3}^{\alpha } \\
p_{3}^{\alpha } &+&\overline{p_{3}^{\alpha }p_{2}^{\beta }}p_{2}^{\beta
}p_{1}^{\alpha }{\rightarrow }p_{1}^{\alpha } \\
p_{3}^{\alpha } &+&\overline{p_{3}^{\alpha }p_{2}^{\gamma }}p_{2}^{\gamma
}p_{1}^{\alpha }{\rightarrow }p_{1}^{\alpha } \\
p_{3}^{\alpha } &+&\overline{p_{3}^{\alpha }p_{4}^{\beta }}p_{4}^{\beta
}p_{1}^{\alpha }{\rightarrow }p_{1}^{\alpha } \\
p_{3}^{\alpha } &+&\overline{p_{3}^{\alpha }p_{4}^{\gamma }}p_{4}^{\gamma
}p_{1}^{\alpha }{\rightarrow }p_{1}^{\alpha } \\
p_{1}^{\gamma } &+&\overline{p_{1}^{\gamma }p_{3}^{\alpha }}p_{3}^{\alpha
}p_{2}^{\gamma }{\rightarrow }p_{2}^{\gamma } \\
p_{1}^{\gamma } &+&\overline{p_{1}^{\gamma }p_{3}^{\beta }}p_{3}^{\beta
}p_{2}^{\gamma }{\rightarrow }p_{2}^{\gamma } \\
p_{1}^{\gamma } &+&\overline{p_{1}^{\gamma }p_{4}^{\alpha }}p_{4}^{\alpha
}p_{2}^{\gamma }{\rightarrow }p_{2}^{\gamma } \\
p_{1}^{\gamma } &+&\overline{p_{1}^{\gamma }p_{4}^{\beta }}p_{4}^{\beta
}p_{2}^{\gamma }{\rightarrow }p_{2}^{\gamma } \\
p_{2}^{\gamma } &+&\overline{p_{2}^{\gamma }p_{3}^{\alpha }}p_{3}^{\alpha
}p_{1}^{\gamma }{\rightarrow }p_{1}^{\gamma } \\
p_{2}^{\gamma } &+&\overline{p_{2}^{\gamma }p_{3}^{\beta }}p_{3}^{\beta
}p_{1}^{\gamma }{\rightarrow }p_{1}^{\gamma } \\
p_{2}^{\gamma } &+&\overline{p_{2}^{\gamma }p_{4}^{\alpha }}p_{4}^{\alpha
}p_{1}^{\gamma }{\rightarrow }p_{1}^{\gamma } \\
p_{2}^{\gamma } &+&\overline{p_{2}^{\gamma }p_{4}^{\beta }}p_{4}^{\beta
}p_{1}^{\gamma }{\rightarrow }p_{1}^{\gamma } \\
p_{2}^{\beta } &+&\overline{p_{2}^{\beta }p_{1}^{\alpha }}p_{1}^{\alpha
}p_{2}^{\beta }{\rightarrow }p_{2}^{\beta } \\
p_{2}^{\beta } &+&\overline{p_{2}^{\beta }p_{1}^{\gamma }}p_{1}^{\gamma
}p_{2}^{\beta }{\rightarrow }p_{2}^{\beta } \\
p_{2}^{\beta } &+&\overline{p_{2}^{\beta }p_{3}^{\alpha }}p_{3}^{\alpha
}p_{2}^{\beta }{\rightarrow }p_{2}^{\beta } \\
p_{2}^{\beta } &+&\overline{p_{2}^{\beta }p_{3}^{\gamma }}p_{3}^{\gamma
}p_{2}^{\beta }{\rightarrow }p_{2}^{\beta } \\
p_{2}^{\beta } &+&\overline{p_{2}^{\beta }p_{4}^{\alpha }}p_{4}^{\alpha
}p_{2}^{\beta }{\rightarrow }p_{2}^{\beta } \\
p_{2}^{\beta } &+&\overline{p_{2}^{\beta }p_{4}^{\gamma }}p_{4}^{\gamma
}p_{2}^{\beta }{\rightarrow }p_{2}^{\beta }.
\end{eqnarray*}

\bigskip

\noindent The two primons of a quark do not interact by means of vector
bosons because their spins (the z components) make an angle (see Fig. 2.1).
They can only exchange scalar bosons which are the bosons that produce their
masses. And since quarks are colored these bosons are also colored. This
does not contradict QCD because QCD gluons are exchanged between quarks. As
has been shown experimentally by PETRA gluons are vectorial (spin 1) bosons$%
^{(26)}$. These scalar bosons are:

\bigskip

{\normalsize 
\begin{eqnarray*}
p_{1}^{\alpha } &+&\overline{p_{1}^{\alpha }p_{3}^{\gamma }}p_{3}^{\gamma
}p_{2}^{\beta }{\rightarrow }p_{2}^{\beta } \\
p_{1}^{\alpha } &+&\overline{p_{1}^{\alpha }p_{4}^{\gamma }}p_{4}^{\gamma
}p_{2}^{\beta }{\rightarrow }p_{2}^{\beta }
\end{eqnarray*}
}

\pagebreak 
\begin{eqnarray*}
p_{2}^{\beta } &+&\overline{p_{2}^{\beta }p_{3}^{\gamma }}p_{3}^{\gamma
}p_{1}^{\alpha }{\rightarrow }p_{1}^{\alpha } \\
p_{2}^{\beta } &+&\overline{p_{2}^{\beta }p_{4}^{\gamma }}p_{4}^{\gamma
}p_{1}^{\alpha }{\rightarrow }p_{1}^{\alpha } \\
p_{2}^{\gamma } &+&\overline{p_{2}^{\gamma }p_{1}^{\beta }}p_{1}^{\beta
}p_{3}^{\alpha }{\rightarrow }p_{3}^{\alpha } \\
p_{2}^{\gamma } &+&\overline{p_{2}^{\gamma }p_{4}^{\beta }}p_{4}^{\beta
}p_{3}^{\alpha }{\rightarrow }p_{3}^{\alpha } \\
p_{3}^{\alpha } &+&\overline{p_{3}^{\alpha }p_{1}^{\beta }}p_{1}^{\beta
}p_{2}^{\gamma }{\rightarrow }p_{2}^{\gamma } \\
p_{3}^{\alpha } &+&\overline{p_{3}^{\alpha }p_{4}^{\beta }}p_{4}^{\beta
}p_{2}^{\gamma }{\rightarrow }p_{2}^{\gamma }.
\end{eqnarray*}

\bigskip

\noindent As we saw in the previous section these bosons are responsible for
the production of quark masses.

\noindent

In the interactions in (iii) the last 6 have to be multiplied by two since
they go both ways. We count a total of 82 interactions. Identifying the $q%
\bar{q}$'s we have: \newline
\noindent a) Colored $q\bar{q}$'s in the outer shell: \newline
\noindent $d_{r}\overline{c_{g}}$, $s_{r}\overline{t_{g}}$, $d_{r}\overline{%
c_{b}}$, $s_{r}\overline{t_{b}}$, $u_{b}\overline{u_{g}}$, $d_{b}\overline{%
d_{g}}$, $s_{b}\overline{s_{g}}$, and their antiparticles. In this shell
there are no colorless mesons. \newline
\noindent b) Colored $q\bar{q}$'s in the inner shell: \newline
\noindent $d_{b}\overline{c_{g}}$, $s_{b}\overline{t_{g}}$, $d_{b}\overline{%
u_{r}}$, $b_{b}\overline{t_{r}}$, $c_{g}\overline{u_{r}}$, $b_{g}\overline{%
s_{r}}$, and their antiparticles. Also, there are no

\noindent colorless mesons in this shell.\newline
\noindent c) Colored $q\bar{q}$'s between the two shells: \newline
\noindent $d_{r}\overline{c_{g}}$, $s_{r}\overline{t_{g}}$, $c_{b}\overline{%
u_{r}}$, $b_{b}\overline{s_{r}}$, $d_{b}\overline{c_{g}}$, $s_{b}\overline{%
t_{g}}$, $u_{r}\overline{u_{b}}$, $c_{r}\overline{c_{b}}$, $t_{r}\overline{%
t_{b}}$, $u_{b}\overline{u_{g}}$, $d_{b}\overline{d_{g}}$, $s_{b}\overline{%
s_{g}}$, $u_{r}\overline{c_{g}}$, $s_{r}\overline{b_{g}}$, and their
antiparticles. As we will see gluons are colored mesons. \newline
\noindent d) Colorless $q\bar{q}$'s: \newline
\noindent 2 $d\bar{u}$, 2 $u\bar{d}$ 2 $b\bar{t}$, 2 $t\bar{b}$, 2 $d\bar{c}$%
, 2 $c\bar{d}$, 2 $s\bar{t}$, 2 $t\bar{s}$, 4 $u\bar{u}$, 4 $d\bar{d}$, and
4 $s\bar{s}$. This is a very important result because it shows that some
mesons are exchanged between quarks in the proton. We will see shortly which
mesons are exchanged. Counting the $q$'s and the $\bar{q}$'s we obtain: 16 $%
\bar{u}$, 16 $u$, 12 $\bar{d}$, 12 $d$, 12 $\bar{s}$, 12 $s$, 8 $\bar{c}$, 8 
$c$, 4 $\bar{b}$, 4 $b$, 8 $\bar{t}$, and 8 $t$.

If we rotate the supercolors of the inner shell counterclockwise we obtain
similar colored mesons between the two shells and the following colorless
mesons: 8 $u\bar{u}$, 4 $c\bar{c}$, 4 $t\bar{t}$, 4 $d\bar{d}$, 4 $s\bar{s}$%
, 2 $c\bar{u}$, 2 $u\bar{c}$, 2 $b\bar{s}$, and 2 $s\bar{b}$. Depending on
the spins the combinations of these pairs together with those just seen
above can generate the mesons: $\pi ^{+}$, $\pi ^{-}$, $\pi ^{0}$, $\phi
(1020)$, $\rho (770)$, $\omega (782)$, $\eta $, $D^{+}$, $D^{-}$, $D^{0}$, $%
\bar{D}^{0}$, $D^{\ast }(2007)^{0}$, $D^{\ast }(2010)^{\pm }$, $\eta
_{c}(1S) $, $J/\Psi (1S)$, $b\bar{t}$, $t\bar{b}$, $s\bar{t}$, $t\bar{s}$, $%
B_{s}^{0}$, $\overline{B_{s}}^{0}$, and $t\bar{t}$. The exchange of pions
inside the proton dictates the proton size of about 1F. As we will see pions
are also exchanged inside the neutron. It is important to notice that there
are no $K^{\pm }$, and no $K^{\ast }(892)$. Let us take a closer look at the
interactions in the inner shell in the clockwise direction beginning with $%
p_{1}$. They are:

{\normalsize 
\begin{eqnarray*}
p_{1}^{\gamma } &+&\overline{p_{1}^{\gamma }p_{2}^{\beta }}p_{2}^{\beta
}p_{3}^{\alpha }{\rightarrow }p_{3}^{\alpha } \\
p_{1}^{\gamma } &+&\overline{p_{1}^{\gamma }p_{4}^{\beta }}p_{4}^{\beta
}p_{3}^{\alpha }{\rightarrow }p_{3}^{\alpha } \\
p_{3}^{\alpha } &+&\overline{p_{3}^{\alpha }p_{1}^{\gamma }}p_{1}^{\gamma
}p_{2}^{\beta }{\rightarrow }p_{2}^{\beta } \\
p_{3}^{\alpha } &+&\overline{p_{3}^{\alpha }p_{4}^{\gamma }}p_{4}^{\gamma
}p_{2}^{\beta }{\rightarrow }p_{2}^{\beta } \\
p_{2}^{\beta } &+&\overline{p_{2}^{\beta }p_{3}^{\alpha }}p_{3}^{\alpha
}p_{1}^{\gamma }{\rightarrow }p_{1}^{\gamma } \\
p_{2}^{\beta } &+&\overline{p_{2}^{\beta }p_{4}^{\alpha }}p_{4}^{\alpha
}p_{1}^{\gamma }{\rightarrow }p_{1}^{\gamma }.
\end{eqnarray*}
}

\bigskip

\noindent Taking a look at the $p\bar{p}$'s pairs, $\overline{p_{1}^{\gamma
}p_{2}^{\beta }}p_{2}^{\beta }p_{3}^{\alpha }$, $\overline{p_{1}^{\gamma
}p_{4}^{\beta }}p_{4}^{\beta }p_{3}^{\alpha }$, $\overline{p_{3}^{\alpha
}p_{1}^{\gamma }}p_{1}^{\gamma }p_{2}^{\beta }$, $\overline{p_{3}^{\alpha
}p_{4}^{\gamma }}p_{4}^{\gamma }p_{2}^{\beta }$, $\overline{p_{2}^{\beta
}p_{3}^{\alpha }}p_{3}^{\alpha }p_{1}^{\gamma }$, and $\overline{%
p_{2}^{\beta }p_{4}^{\alpha }}p_{4}^{\alpha }p_{1}^{\gamma }$ we notice that
the numbers of $p_{i}^{j}$'s and $\bar{p_{i}^{j}}$'s are the same. They are
supergluons created by vacuum and arranged in a certain way to make the
interactions possible. As we will see later on this inner shell is quite
stable.

\subsubsection{\noindent {\protect\large The Neutron Sea Content}}

Doing the same for the neutron (Fig 2.4) we obtain the same number of
transitions. After averaging over the two configurations we obtain similar
colored $q\bar{q}$'s and the following colorless $q\bar{q}$'s: \noindent 8 $u%
\bar{u}$, 12 $d\bar{d}$, 8 $s\bar{s}$, 2 $c\bar{u}$, 2 $b\bar{s}$, 2 $u\bar{d%
}$, 2 $d\bar{c}$, 2 $t\bar{s}$, 2 $t\bar{b}$, 4 $b\bar{b}$, 4 $c\bar{c}$,
and their antiparticles. Their combinations generate the mesons: $\pi ^{+}$, 
$\pi ^{-}$, $\pi ^{0}$, $\phi (1020)$, $\rho (770)$, $\omega (782) $, $\eta $%
, $D^{+}$, $D^{-}$, $D^{0}$, $\bar{D}^{0}$, $D^{\ast }(2007)^{0}$, $D^{\ast
}(2010)^{\pm }$, $\eta _{c}(1S)$, $J/\Psi (1S)$, $b\bar{t}$, $t\bar{b}$, $%
B_{s}^{0}$, $\overline{B_{s}}^{0}$, $\Upsilon (1S)$, $t\bar{b}$, and $b\bar{t%
}$. Again, the exchange of pions determines neutron's size of about 1F.
Counting the $q$'s and the $\bar{q}$'s we have: 12 $\bar{u} $, 12 $u$, 16 $%
\bar{d}$, 16 $d$, 12 $\bar{s}$, 12 $s$, 8 $\bar{c}$, 8 $c$, 8 $\bar{b}$, 8 $%
b $, 4 $\bar{t}$, and 4 $t$.

\subsubsection{\noindent {\protect\large The Contribution of Both Seas to
the Structure Function }$F_{2}(x)$}

According to the parton picture we have

{\normalsize \textbf{
\begin{eqnarray}
\frac{1}{x}F_{2}^{ep}(x) &=&\left( \frac{2}{3}\right) ^{2}\left[ u^{p}(x)+%
\bar{u}^{p}(x)\right] +\left( \frac{2}{3}\right) ^{2}\left[ c^{p}(x)+\bar{c}%
^{p}(x)\right] +  \notag \\
&&\left( \frac{2}{3}\right) ^{2}\left[ t^{p}(x)+\bar{t}^{p}(x)\right]
+\left( \frac{1}{3}\right) ^{2}\left[ d^{p}(x)+\bar{d}^{p}(x)\right] + 
\notag \\
&&\left( \frac{1}{3}\right) ^{2}\left[ s^{p}(x)+\bar{s}^{p}(x)\right]
+\left( \frac{1}{3}\right) ^{2}\left[ b^{p}(x)+\bar{b}^{p}(x)\right]
\end{eqnarray}
}}

\noindent for the proton, and

{\normalsize \textbf{
\begin{eqnarray}
\frac{1}{x}F_{2}^{en}(x) &=&\left( \frac{2}{3}\right) ^{2}\left[ u^{n}(x)+%
\bar{u}^{n}(x)\right] +\left( \frac{2}{3}\right) ^{2}\left[ c^{n}(x)+\bar{c}%
^{n}(x)\right] +  \notag \\
&&\left( \frac{2}{3}\right) ^{2}\left[ t^{n}(x)+\bar{t}^{n}(x)\right]
+\left( \frac{1}{3}\right) ^{2}\left[ d^{n}(x)+\bar{d}^{n}(x)\right] + 
\notag \\
&&\left( \frac{1}{3}\right) ^{2}\left[ s^{n}(x)+\bar{s}^{n}(x)\right]
+\left( \frac{1}{3}\right) ^{2}\left[ b^{n}(x)+\bar{b}^{n}(x)\right]
\end{eqnarray}
}}

\noindent for the neutron. In these equations $q(x)$ represents the
probability distribution of quark $q$. In general it has been considered in
the literature that $u^{p}(x)=d^{n}(x)$, $d^{p}(x)=u^{n}(x)$, $%
s^{p}(x)=s^{n}(x)$, $c^{p}(x)=c^{n}(x)$, $b^{p}(x)=b^{n}(x)$, and $%
t^{p}(x)=t^{n}(x)$, but as we saw above these equalities are not true.
According to what was calculated above, in both nucleon seas (the colorless
seas) $u=\bar{u}$, $d=\bar{d}$, $s=\bar{s}$, $c=\bar{c}$, $b=\bar{b}$, and $%
t=\bar{t}$. Substituing these results and the numbers obtained above we
arrive at

{\normalsize \textbf{
\begin{eqnarray}
\frac{1}{x}F_{2}^{ep}(x) &=&\left( \frac{2}{3}\right) ^{2}\left[ 2\times
16+2\times 8+2\times 4\right] +  \notag \\
&&\left( \frac{1}{3}\right) ^{2}\left[ 2\times 12+2\times 12+2\times 4\right]
=31.111
\end{eqnarray}
}}

\noindent for the proton, and

{\normalsize \textbf{
\begin{eqnarray}
\frac{1}{x}F_{2}^{en}(x) &=&\left( \frac{2}{3}\right) ^{2}\left[ 2\times
12+2\times 8+2\times 4\right] +  \notag \\
&&\left( \frac{1}{3}\right) ^{2}\left[ 2\times 16+2\times 12+2\times 8\right]
=29.33333
\end{eqnarray}
}}

\noindent for the neutron. These numbers represent the contributions of each
sea to the two structure functions. We obtain then the very important result

\begin{equation}
\frac{F_{2}^{en}}{F_{2}^{ep}}=\frac{29.333}{31.11111}=0.94286.
\end{equation}
\noindent The saturation, therefore, due to the sea of $q\bar{q}$'s never
reaches one. This agrees completely with SLAC result shown in Figs. 2.6 for
low $x$. For high $x$ the sea does not contribute and we expect to have this
ratio equal to 0.25 which is a well known result.

\vspace*{0.3in}\pagebreak

\noindent {\Large References}

\bigskip

\noindent 1. M. E. de Souza, in \textit{Proceedings of the XII Brazilian
National Meeting of the Physics of Particles and\ Fields}, Caxambu, Minas
Gerais, Brazil, September 18-22, 1991.\newline
\noindent 2. M.E. de Souza, IX Meeting of Physicists of the North and
Northeast, Macei\'{o}, Alagoas, Brazil, November 07 and 08, 1991. \newline
\noindent 3. M.E. de Souza, 13th Interantional Conference on General
Relativity and Gravitation, Huerta Grande, Cordoba, Argentina, June 28-July
4, 1992. \newline
\noindent 4. M.E. de Souza, in \textit{Proceedings of the XIII Brazilian
National Meeting of the Physics of Particles and Fields}, Caxambu, Minas
Gerais, Brazil, September 16-20, 1992.\newline
\noindent 5. M.E. de Souza, X Meeting of Physicists of the North/Northeast,
Recife, Pernambuco, Brazil, December 2-4, 1992. \newline
\noindent 6. M.E. de Souza, in \textit{Proceedings of the XIV Brazilian
National Meeting of the Physics of Particles and\ Fields}, Caxambu, Minas
Gerais, Brazil, September 29-October 3, 1993. \newline
\noindent 7. M.E. de Souza, XI Meeting of Physicists of the North/Northeast,
Jo\~{a}o Pessoa, Para\'{i}ba, Brazil, November 17-19, 1993. \newline
\noindent 8. M.E. de Souza, in \textit{The Six Fundamental Forces of Nature}%
, Universidade Federal de Sergipe, S\~{a}o Crist\'{o}v\~{a}o, Sergipe,
Brazil, February 1994. \newline
\noindent 9. M.E. de Souza, International Symposium Physics Doesn't Stop:
Recent Developments in Phenomenology, University of Wisconsin,
Madison(Wisconsin), USA, April 11-13, 1994. \newline
\noindent 10. M.E. de Souza, in \textit{Proceedings of the XV Brazilian
National Meeting of the Physics of Particles and\ Fields}, Angra dos Reis,
Rio de Janeiro, Brazil, October 4-8, 1994. \newline
\noindent 11. M.E. de Souza, XVI Brazilian National Meeting of the Physics
of Particles and Fields, Caxambu, MinasGerais, Brazil, October 24-28, 1995. 
\newline
\noindent 12. M.E. de Souza, XVII Brazilian National Meeting of the Physics
of \ Particles and Fields, Serra Negra, Minas Gerais, Brazil, October 24-28,
1996. \newline
\noindent 13. E.J. Eichten, K.D. Lane, and M.E. Peskin, Phys. Rev. Lett. 50,
811 (1983). \newline
\noindent 14. K. Hagiwara, S. Komamiya, and D. Zeppenfeld, Z. Phys. C29, 115
(1985). \newline
\noindent 15. N. Cabibbo, L. Maiani, and Y. Srivastava, Phys. Lett. 139, 459
(1984). \newline
\noindent 16. H. Fritzsch, in \textit{Proceedings of the twenty-second
Course of the International School of Subnuclear\ Physics}, 1984, ed. by A.
Zichichi (Plenum Press, New York, 1988). \newline
\noindent 17. G. 'tHooft, in \textit{Recent Developments in Gauge Theories},
eds. G. 'tHooft et al., Plenum Press, New York, 1980. \newline
\noindent 18. Y.N.Srivastava and A.Widom, hep-ph/0009032. \newline
\noindent 19. E.E. Chambers and R. Hofstadter, Phys. Rev. 103, 1454 (1956). 
\newline
\noindent 20. C. Kittel, W. D. Knight and M. A. Ruderman, in Mechanics,
Berkeley Physics Course, Vol. 1, pg. 451, McGraw-Hill Book Company, New
York(1965). \newline
\noindent 21. Particle Data Group, Review of Particle Properties, Phys. Rev.
D, 54, Part II, No. 1 (1966). \newline
\noindent 22. B. Povh and J. H\"{u}fner, Phys. Lett. B, 245, 653 (1990). 
\newline
\noindent 23. B.Povh, hep-ph/9908233. \newline
\noindent 24. D. terHaar, in \textit{Problems in Quantum Mechanics}, Pion
Limited, London, 1975. \newline
\noindent 25. C. Seife, Science, Vol. 292, No. 5518, p. 823 (2001). \newline
\noindent 26. D. H. Perkins, in \textit{Introduction to High Energy Physics}%
, Addison-Wesley, Menlo Park, CA, USA, 1987, p. 301.

\newpage

\bigskip \rule[0.5in]{0in}{0.17in}

\section{\noindent {\protect\LARGE The Strong, Superstrong and Higgs\newline
Bosons, Gluons and Interactions \newline
Between Primons}}

{\LARGE \bigskip \rule[1.5in]{0in}{0.17in}}

\subsection{\protect\Large The Higgs Bosons}

As we saw in the previous chapter the scalar colored bosons that are
exchanged in the $u(p_{1}p_{2})$ quark are $\eta _{ev}^{-}=\overline{%
p_{1}p_{2}}p_{2}p_{3}+\overline{p_{1}p_{4}}p_{4}p_{2}$ and $\eta _{ev}^{+}=%
\overline{p_{2}p_{3}}p_{3}p_{1}+\overline{p_{2}p_{4}}p_{4}p_{1}$. In the
same way in the $d(p_{2}p_{3})$ quark the boson $\eta _{ev}^{o}=\overline{%
p_{2}p_{1}}p_{1}p_{3}+\overline{p_{2}p_{4}p}_{4}p_{3}=\overline{p_{3}p_{1}}%
p_{1}p_{2}+\overline{p_{3}p_{4}p}_{4}p_{2}$ is exchanged. Doing the same for
the other quarks we obtain Table 3.1 below. As we saw before we can identify 
$\eta _{ev}^{+}$, $\eta _{ev}^{-}$, and $\eta _{ev}^{0}$ as the Higgs
bosons. That is, the Higgs bosons are scalar colored bosons and are
combinations of \ $p\overline{p}$ pairs.

Although $\eta _{ev}^{+}$ has three expressions in terms of $p\overline{p}$
pairs its mathematical expression has to be unique and is not expressed yet
in this and should be found.

When writing the Klein-Gordon equation 
\begin{equation}
\left[ \partial ^{\mu }\partial _{\mu }+\left( \frac{mc}{\hbar }\right) ^{2}%
\right] \phi =0
\end{equation}
for the uncharged Higgs field we should be careful because, actually, $m$ is
not an intrinsic(or bare) mass but an interaction mass because the Higss are
collections of $\overline{p_{j}p_{k}}p_{k}p_{i}$ and are colored mesons$.$
Of course, we should go on and find out what $m$ really is.

\vspace*{0.5in}

\begin{center}
\begin{tabular}{||c|c||}
\hline\hline
&  \\ 
Higgs Bosons & Expression in Terms of $p\overline{p}$ pairs \\ 
&  \\ \hline\hline
&  \\ 
$\eta _{ev}^{+}$ & 
\begin{tabular}{|c|}
\hline
$\overline{p_{2}p_{3}}p_{3}p_{1}+\overline{p_{2}p_{4}}p_{4}p_{1}$ \\ \hline
$\overline{p_{3}p_{2}}p_{2}p_{1}+\overline{p_{3}p_{4}}p_{4}p_{1}$ \\ \hline
$\overline{p_{4}p_{3}}p_{3}p_{1}+\overline{p_{4}p_{2}}p_{2}p_{1}$ \\ \hline
\end{tabular}
\\ 
&  \\ \hline
&  \\ 
$\eta _{ev}^{0}$ & 
\begin{tabular}{|c|}
\hline
$\overline{p_{2}p_{1}}p_{1}p_{3}+\overline{p_{2}p_{4}}p_{4}p_{3}$ \\ \hline
$\overline{p_{3}p_{1}}p_{1}p_{2}+\overline{p_{3}p_{4}}p_{4}p_{2}$ \\ \hline
$\overline{p_{2}p_{1}}p_{1}p_{4}+\overline{p_{2}p_{1}}p_{1}p_{4}$ \\ \hline
$\overline{p_{4}p_{1}}p_{1}p_{2}+\overline{p_{4}p_{1}}p_{1}p_{2}$ \\ \hline
$\overline{p_{3}p_{1}}p_{1}p_{4}+\overline{p_{3}p_{2}}p_{2}p_{4}$ \\ \hline
$\overline{p_{4}p_{1}}p_{1}p_{3}+\overline{p_{4}p_{2}}p_{2}p_{3}$ \\ \hline
\end{tabular}
\\ 
&  \\ \hline
&  \\ 
$\eta _{ev}^{-}$ & 
\begin{tabular}{|c|}
\hline
$\overline{p_{1}p_{3}}p_{3}p_{2}+\overline{p_{1}p_{4}}p_{4}p_{2}$ \\ 
$\overline{p_{1}p_{2}}p_{2}p_{3}+\overline{p_{1}p_{4}}p_{4}p_{3}$ \\ 
$\overline{p_{1}p_{3}}p_{3}p_{4}+\overline{p_{1}p_{2}}p_{2}p_{4}$ \\ \hline
\end{tabular}
\\ 
&  \\ \hline\hline
\end{tabular}
\end{center}

\vskip.2in

\begin{center}
\parbox{4in}
{Table 3.1. Expressions of the Higgs bosons in terms of primons.}
\end{center}

\bigskip

{\large Therefore, primons are very important fermions. Not only they form
quarks, they also form supergluons, gluons and Higgs bosons. And we are
seeing that the realm of mass is very strange for\emph{\ mass comes from
within.} }

\bigskip

\subsection{\protect\Large The Bosons of the Strong and Superstrong
Interactions}

{\normalsize \textbf{\ }}Doing the same for the other pairs of primons as we
did in the case of nucleons we find (for $i\neq j$) the other colorless $q%
\bar{q}$'s:

\noindent 
\begin{eqnarray*}
p_{1}^{i} &+&\overline{p_{1}^{i}p_{3}^{j}}p_{3}^{j}p_{4}^{i}{\rightarrow }%
p_{4}^{i} \\
p_{1}^{i} &+&\overline{p_{1}^{i}p_{2}^{j}}p_{2}^{j}p_{4}^{i}{\rightarrow }%
p_{4}^{i} \\
p_{2}^{i} &+&\overline{p_{2}^{i}p_{3}^{j}}p_{3}^{j}p_{4}^{i}{\rightarrow }%
p_{4}^{i}
\end{eqnarray*}

\pagebreak 
\begin{eqnarray*}
p_{2}^{i} &+&\overline{p_{2}^{i}p_{1}^{j}}p_{1}^{j}p_{4}^{i}{\rightarrow }%
p_{4}^{i} \\
p_{3}^{i} &+&\overline{p_{3}^{i}p_{2}^{j}}p_{2}^{j}p_{4}^{i}{\rightarrow }%
p_{4}^{i} \\
p_{3}^{i} &+&\overline{p_{3}^{i}p_{1}^{j}}p_{1}^{j}p_{4}^{i}{\rightarrow }%
p_{4}^{i} \\
p_{4}^{i} &+&\overline{p_{4}^{i}p_{1}^{j}}p_{1}^{j}p_{4}^{i}{\rightarrow }%
p_{4}^{i} \\
p_{4}^{i} &+&\overline{p_{4}^{i}p_{2}^{j}}p_{2}^{j}p_{4}^{i}{\rightarrow }%
p_{4}^{i} \\
p_{4}^{i} &+&\overline{p_{4}^{i}p_{3}^{j}}p_{3}^{j}p_{4}^{i}{\rightarrow }%
p_{4}^{i}
\end{eqnarray*}

\noindent and also in the opposite way. That is, we find the $q\bar{q}$'s: $b%
\bar{c}$, $c\bar{b}$, $s\bar{u}$, $u\bar{s}$, $b\bar{d}$, $d\bar{b}$, $t\bar{%
u}$, $u\bar{t}$, $s\bar{d}$, $d\bar{s}$, $t\bar{c}$, $c\bar{t}$, $t\bar{t}$, 
$s\bar{s}$, and $b\bar{b}$. Taking into account spin these pairs generate
the mesons: \noindent $K^{\pm }$, $K^{0}$, $\bar{K}^{0}$, $K^{\ast }(892)$, $%
\phi (1020)$, $b\bar{c}$, $c\bar{b}$, $B^{0}$, $\bar{B}^{0}$, $t\bar{u}$, $u%
\bar{t}$, $t\bar{c}$, $c\bar{t}$, $t\bar{t}$, and $b\bar{b}$.

Besides these we still have the mesons $p_{1}p_{2}\overline{p_{3}p_{4}}$, $%
p_{1}p_{3}\overline{p_{2}p_{4}}$, $p_{1}p_{4}\overline{p_{2}p_{3}}$, $%
p_{2}p_{3}\overline{p_{1}p_{2}}$, $p_{2}p_{4}\overline{p_{1}p_{2}}$, $%
p_{3}p_{4}\overline{p_{1}p_{2}}$ and their antiparticles. The first one, for
example, appears in the interaction $p_{3}p_{4}+\overline{p_{3}p_{4}}%
p_{1}p_{2}{\rightarrow }p_{1}p_{2}.$ The first two are the mesons $%
D_{s}^{\pm }$, $B^{\pm }$. These and the other ones calculated above are
quite familiar to us. We just have to classify them in terms of the two
interactions. In the interaction between two nucleons the dominant term is
due to the exchange of pions (we will discuss this subject later on), and as
we know pions are pseudoscalar mesons. Thus, we may assume that scalar
mesons mediate the strong interaction. On the other hand we may consider
that the vectorial mesons mediate the superstrong interaction. That is, the
superstrong interaction is caused by a vectorial field. We can, then,
separate the bosons according to these two interactions. We find, of course,
that spin plays a major role in the two interactions. According to the type
of interaction we have:\newline
\noindent a) Colorless Bosons of the strong interaction: $\ \pi ^{+}$, $\pi
^{-}$, $\pi ^{0}$, $\eta $, $K^{+}$, $K^{-}$, $K^{0}$, $\bar{K}^{0}$, $D^{+}$%
, $D^{-}$, $D^{0}$, $\bar{D}^{0}$, $D_{s}^{+}$, $D_{s}^{-}$, $B^{+}$, $B^{-}$%
, $B^{0}$, $\bar{B}^{0}$, $B_{s}^{0}$, $\overline{B_{s}}^{0}$, $\eta
_{c}(1S) $, etc. \newline
\noindent b) Colorless Bosons of the superstrong interaction: \noindent $%
\rho (770)$, $\omega (782)$, $\phi (1020)$, $K^{\ast }(892)$, $D^{\ast
}(2007)^{0}$, $D^{\ast }(2010)^{\pm }$, $J/\psi (1S)$, $\psi (2S)$, $\psi
(3700)$, $\psi (4040)$, $\Upsilon (1S)$, $\Upsilon (2S)$, $\Upsilon (3S)$, $%
\Upsilon (4S)$, etc.

Gluons are mediators of one of the properties of quarks which is color. As
we will see they are bosons of an effective field. Supergluons (mediators of
the supercolor field) are more fundamental.

Taking a look at the five fundamental interactions(weak, electromagnetic,
strong, superstrong and gravitational) we notice that only the gravitational
and electromagnetic interactions have just one boson, each, because they are
massless bosons. We also observe that only the weak interaction has three
vector bosons, which is a small and limited number of bosons. The strong and
superstrong interactions have many bosons and some of them have masses of
the same order. As was discussed in the first chapter there should still
exist another interaction called superweak interaction which was proposed by
de Souza in 1991$^{1,2}$. Actually it is a modified version of the fifth
force, proposed in 1987 by Fischbach$^{3}$. Since its range is supposed to
be infinite its boson should be massless. Therefore, the overall picture for
the fundamental interactions is the following: nature has six fundamental
interactions, three with infinite ranges and three with very small ranges,
that is, there are three interactions with massive bosons and three
interactions with massless bosons. The weak force is different from the
other two small range forces because it has only three very heavy bosons. As
we will see its bosons are the heaviest of all. Arranging the fundamental
forces in a table we obtain Table 3.2 below.

\bigskip

\ \ \ {\normalsize 
\begin{tabular}{||l|l|l||}
\hline
&  &  \\ 
Interaction & Bosons & Nature of the \\ 
&  & Field \\ 
&  &  \\ \hline\hline
Superstrong & supergluons, &  \\ \cline{2-3}
& $\rho (770)$, $\omega (782)$, $\phi (1020)$, $K^{\ast }(892)$ &  \\ 
\cline{2-3}
& $D^{\ast }(2007)^{0}$, $D^{\ast }(2010)^{\pm }$, $J/\psi (1S)$, &  \\ 
\cline{2-3}
& $\psi (2S)$, $\psi (3700)$, $\psi (4040)$, $\Upsilon (1S)$, &  \\ 
\cline{2-3}
& $\Upsilon (2S)$, $\Upsilon (3S)$, $\Upsilon (4S)$, .... & Vector \\ \hline
Strong & $\pi ^{\pm }$, $\pi ^{0}$, $\eta $, $K^{\pm }$, $K^{0}$, $\bar{K}%
^{0}$, $D^{\pm }$, $D^{0}$, &  \\ \cline{2-3}
& $\bar{D}^{0}$, $D_{s}^{+}$, $D_{s}^{-}$, $B^{+}$, $B^{-}$, $B^{0}$, &  \\ 
\cline{2-3}
& $\bar{B}^{0}$, $B_{s}^{0}$, $\overline{B_{s}}^{0}$, $\eta _{c}(1S)$, ....
& Pseudoscalar \\ \hline
Electromagnetic & $\gamma $ & Vector \\ \hline
Weak & $Z^{0}$, $W^{\pm }$ & Vector \\ \hline
Gravitational & $g?$ & Tensor? \\ \hline
Superweak & $\mathcal{N}$ & Scalar? \\ \hline\hline
\end{tabular}
}

\vskip.2in

\parbox{4.5in}
{Table 3.2. The Fundamental Forces of Nature and their Bosons. Observe 
that three interactions have massless bosons and the other three have massive
bosons.}

\bigskip

\subsection{\protect\bigskip {\protect\Large The Nature of Gluons}}

Let us analyze what the interaction $u_{b}+\bar{b}g\rightarrow u_{g}\;$
means in terms of primons. We will be able, then, to find out the nature of
gluons and to show that they are colored mesons and are massless because are
colored. Having in mind that supercolors can be interchanged the above
interaction is actually any of the four interactions:

\bigskip

\begin{eqnarray*}
a)\;p_{1}^{\alpha }p_{2}^{\beta }+\overline{bg}{} &\rightarrow
&{}p_{1}^{\alpha }p_{2}^{\gamma } \\
b)\;p_{1}^{\alpha }p_{2}^{\beta }+\overline{bg}{} &\rightarrow
&{}p_{1}^{\gamma }p_{2}^{\alpha } \\
c)\;p_{1}^{\beta }p_{2}^{\alpha }+\overline{bg}{} &\rightarrow
&{}p_{1}^{\alpha }p_{2}^{\gamma } \\
d)\;p_{1}^{\beta }p_{2}^{\alpha }+\overline{bg}{} &\rightarrow
&{}p_{1}^{\gamma }p_{2}^{\alpha }.
\end{eqnarray*}

\pagebreak

\noindent Let us consider the one-gluon interactions that are involved. The
first interaction is the result of any of the interactions 
\begin{eqnarray*}
p_{1}^{\alpha } &+&\overline{p_{1}^{\alpha }p_{2}^{\beta }}p_{2}^{\beta
}p_{1}^{\alpha }{\rightarrow }p_{1}^{\alpha } \\
p_{1}^{\alpha } &+&\overline{p_{1}^{\alpha }p_{2}^{\gamma }}p_{2}^{\gamma
}p_{1}^{\alpha }{\rightarrow }p_{1}^{\alpha } \\
p_{1}^{\alpha } &+&\overline{p_{1}^{\alpha }p_{3}^{\beta }}p_{3}^{\beta
}p_{1}^{\alpha }{\rightarrow }p_{1}^{\alpha } \\
p_{1}^{\alpha } &+&\overline{p_{1}^{\alpha }p_{3}^{\gamma }}p_{3}^{\gamma
}p_{1}^{\alpha }{\rightarrow }p_{1}^{\alpha } \\
p_{1}^{\alpha } &+&\overline{p_{1}^{\alpha }p_{4}^{\beta }}p_{4}^{\beta
}p_{1}^{\alpha }{\rightarrow }p_{1}^{\alpha } \\
p_{1}^{\alpha } &+&\overline{p_{1}^{\alpha }p_{4}^{\gamma }}p_{4}^{\gamma
}p_{1}^{\alpha }{\rightarrow }p_{1}^{\alpha }
\end{eqnarray*}
\noindent together with any of the following interactions 
\begin{eqnarray*}
p_{2}^{\beta } &+&\overline{p_{2}^{\beta }p_{1}^{\alpha }}p_{1}^{\alpha
}p_{2}^{\gamma }{\rightarrow }p_{2}^{\gamma } \\
p_{2}^{\beta } &+&\overline{p_{2}^{\beta }p_{3}^{\alpha }}p_{3}^{\alpha
}p_{2}^{\gamma }{\rightarrow }p_{2}^{\gamma } \\
p_{2}^{\beta } &+&\overline{p_{2}^{\beta }p_{4}^{\alpha }}p_{4}^{\alpha
}p_{2}^{\gamma }{\rightarrow }p_{2}^{\gamma }.
\end{eqnarray*}
\noindent The colored mesons with one-gluon exchange are in the last three
interactions. They are $u^{g}\overline{u^{b}}$, $d^{g}\overline{d^{b}}$, and 
$s^{g}\overline{s^{b}}$. Similarly, the one-gluon interactions involved in
(b), $p_{1}^{\alpha }p_{2}^{\beta }+\overline{bg}\rightarrow p_{1}^{\gamma
}p_{2}^{\alpha }\;$ are $p_{1}^{\alpha }\rightarrow p_{2}^{\alpha
},\;\;p_{2}^{\beta }\rightarrow p_{1}^{\gamma }\;\;$ which involve the
exchange of the colorless mesons $d\bar{c}\;$ and $s\bar{t}\;$, and the
colored mesons $c^{g}\overline{d^{b}}\;$ and $t^{g}\overline{s^{b}}$.

\noindent We also find that interaction (c) above involves the exchange of
the colored mesons $d^{g}\overline{c^{b}}\;$ and $s^{g}\overline{t^{b}}$.
Finally, in interaction (d) above there is the exchange of the colored
mesons $u^{g}\overline{u^{b}}$, $c^{g}\overline{c^{b}}$, and $t^{g}\overline{%
t^{b}}$. Summarizing we have that the gluon $\bar{b}g\;$ that acts between
two $u\;$ quarks is the overall effect of the action of the colored mesons: $%
u^{g}\overline{u^{b}}$, $d^{g}\overline{d^{b}}$, $s^{g}\overline{s^{b}}$, $%
c^{g}\overline{d^{b}}$, $t^{g}\overline{s^{b}}$, $d^{g}\overline{c^{b}}$, $%
s^{g}\overline{t^{b}}$, $u^{g}\overline{u^{b}}$, $c^{g}\overline{c^{b}}$,
and $t^{g}\overline{t^{b}}$. We see that it is a set of nine colored mesons.

Let us now see which mesons we have in the interaction $d_{b}+\bar{b}%
g\rightarrow d_{g}$. Doing in the same way as we did above we find that they
are the colored mesons: $c^{g}\overline{c^{b}}$, $d^{g}\overline{d^{b}}$, $%
b^{g}\overline{b^{b}}$, $c^{g}\overline{u^{b}}$, $u^{g}\overline{c^{b}}$, $%
b^{g}\overline{s^{b}}$, $s^{g}\overline{b^{b}}$, $u^{g}\overline{u^{b}}$, $%
s^{g}\overline{s^{b}}$. It also has nine members. We clearly see that this
set is different from that one just above and since QCD tells us that they
are equivalent, i.e., they are the same gluon, then each colored meson has
to be massless and equivalent to any other one. And they have to be massless
because they are colored mesons. In the calculation of the amplitude $M\;$
of the interaction $q_{b}+\bar{b}g\rightarrow q_{g}$, the effective coupling
constant $g_{s}=\sqrt{4{\pi }\alpha _{s}}\;$ is equal to nine times the
coupling constant of each colored mesons, $G_{c}$. That is, $g_{s}=9G_{c}$.
As we see, together with each one-gluon exchange there is also the exchange
of colorless mesons(scalar or vectorial mesons). We just stop at this point
because the calculations for all gluons are enormous, of course. This means
that gluons (that is, QCD) simplifies matters a lot. That is actually the
reason why QCD is so great and successful. We conclude then that the color
field is the field of a collective effective interaction such as other
collective interactions like magnons or phonons.

\subsection{\protect\Large The Interaction Matrix Between Primons of
Different Quarks with Different Supercolors}

As we saw in the sections above we need four primons and three supercolors
to generate quarks in the three \ \ colors. This means that in terms of
flavors primons can be represented by the Dirac spinors 
\begin{equation*}
\Psi _{1}=\left( 
\begin{array}{c}
1 \\ 
0 \\ 
0 \\ 
0
\end{array}
\right) ,\;\;\Psi _{2}=\left( 
\begin{array}{c}
0 \\ 
1 \\ 
0 \\ 
0
\end{array}
\right) ,\;\;\Psi _{3}=\left( 
\begin{array}{c}
0 \\ 
0 \\ 
1 \\ 
0
\end{array}
\right) ,\;\;\Psi _{4}=\left( 
\begin{array}{c}
0 \\ 
0 \\ 
0 \\ 
1
\end{array}
\right) .
\end{equation*}
\newline
As to supercolors we may represent them by the three-element columns 
\begin{equation*}
sc_{\alpha }=\left( 
\begin{array}{c}
1 \\ 
0 \\ 
0
\end{array}
\right) ,\;\;sc_{\beta }=\left( 
\begin{array}{c}
0 \\ 
1 \\ 
0
\end{array}
\right) ,\;\;sc_{\gamma }=\left( 
\begin{array}{c}
0 \\ 
0 \\ 
1
\end{array}
\right) .
\end{equation*}

\noindent The supercolor generators are the three-dimensional generators of
SU(2) 
\begin{equation*}
{\Theta }^{1}=\left( 
\begin{array}{ccc}
0 & 0 & 0 \\ 
0 & 0 & -i \\ 
0 & i & 0
\end{array}
\right) ,\;\;{\Theta }^{2}=\left( 
\begin{array}{ccc}
0 & 0 & i \\ 
0 & 0 & 0 \\ 
-i & 0 & 0
\end{array}
\right) ,\;\;{\Theta }^{3}=\left( 
\begin{array}{ccc}
0 & -i & 0 \\ 
i & 0 & 0 \\ 
0 & 0 & 0
\end{array}
\right) ,
\end{equation*}
which are three of the eight generators of $SU(3)$, and obey the relations 
\begin{equation}
\lbrack \Theta ^{j},\Theta ^{k}]=i\varepsilon ^{jkl}\Theta ^{l}.
\end{equation}

\noindent Let us call them supergluons. Such as gluons supergluons are
vectorial ($S=1$) and also massless. According to the ideas above developed
the combinations of equal supercolors do not produce a color. That is
exactly what we have: 
\begin{equation*}
{sc_{\alpha }^{\dagger }\Theta }{^{j}}sc_{\alpha }={sc_{\beta }^{\dagger
}\Theta ^{j}}sc_{\beta }={sc_{\gamma }^{\dagger }\Theta ^{j}}sc_{\gamma }=0,
\end{equation*}
where $j=1,2,3$. With different supercolors we have the sums: \newline
\begin{equation*}
\sum_{j=1}^{3}{sc_{\alpha }^{\dagger }\Theta }^{j}sc_{\beta
}=-i,\;\;\sum_{j=1}^{3}{sc_{\alpha }^{\dagger }\Theta }^{j}sc_{\gamma
}=i,\;\;\sum_{j=1}^{3}{sc_{\beta }^{\dagger }\Theta }^{j}sc_{\gamma }=-i.
\end{equation*}
\newline
\noindent Therefore, the substructure of $SU(3)$(color) is $SU(2)$%
(supercolor).

Between two primons of different quarks there is the exchange of colored
mesons which are part of the gluonic coupling between them. For example, in
the interaction between the two primons $p_{1}^{\alpha }$, and $p_{2}^{\beta
}\;$ we should have diagram 1 (listed as Fig 3.1). \noindent The mesons $%
d_{r}\overline{c_{g}}\;$ and $s_{r}\overline{t_{g}}\;$ are exchanged in this
case. The gluonic coupling $g_{pp}$, between two primons is, then, just 2/9
of the gluonic coupling between two quarks. Hence, the interaction matrix
for the above primons (with the above supercolors) is 
\begin{eqnarray}
M &=&i\sum_{jk}\left[ \overline{\Psi _{1}(3)}sc_{3}^{\dagger }\right] \left[
-i\frac{g_{pp}}{2}{\Theta }^{j}\gamma ^{\mu }\right] \left[ {\Psi _{1}(1)}%
sc_{1}\right] \left[ \frac{-ig_{\mu \nu }\delta ^{jk}}{q^{2}}\right] \left[ 
\overline{\Psi _{2}(4)}sc_{4}^{\dagger }\right]  \notag \\
&&\times \left[ -i\frac{g_{pp}}{2}{\Theta }^{k}\gamma ^{\nu }\right] {\times 
}\left[ {\Psi _{2}(2)}sc_{2}\right]
\end{eqnarray}
\noindent and hence we obtain

\begin{equation}
M=\sum_{j}\frac{-g_{pp}^{2}}{4}\frac{1}{q^{2}}\left[ \overline{\Psi _{1}(3)}%
\gamma ^{\mu }\Psi _{1}(1)\right] \left[ \overline{\Psi _{2}(4)}\gamma _{\mu
}\Psi _{2}(2)\right] \left( sc_{3}^{\dagger }{\Theta }^{j}sc_{1}\right)
\left( sc_{4}^{\dagger }{\Theta }^{j}sc_{2}\right) .
\end{equation}

\noindent Therefore, the potential between the two primons is a
Coulombian-like potential

\bigskip

\begin{equation}
V_{pp}=-F\frac{g_{pp}{\hbar }c}{4\pi }\frac{1}{r}
\end{equation}
\noindent where $F\;\;$ is a supercolor factor given by 
\begin{equation}
F=\frac{1}{4}\sum_{j}\left( sc_{3}^{+}{\Theta }^{j}sc_{1}\right) \left(
sc_{4}^{+}{\Theta }^{j}sc_{2}\right)
\end{equation}
\noindent in which $sc\;$ stands for the supercolor wavefunction. Let us
calculate the supercolor factor $F\;$ in the interaction between the two
primons above. In this case $F\;$ is given by 
\begin{equation}
F=\frac{1}{4}\frac{1}{\sqrt{2}}\frac{1}{\sqrt{2}}\left[ 2\sum_{j}\left( {%
\Theta }_{11}^{j}{\Theta }_{22}^{j}+{\Theta }_{12}^{j}{\Theta }%
_{21}^{j}\right) \right] =\frac{1}{4}(1)=\frac{1}{4}.
\end{equation}
\noindent where ${\Theta }_{jk}^{i}\;$ are the matrix elements of the
generators. Since $F>0$, there is a net attraction between the two primons
due to the supercolor field. This means that the inner shell of the nucleons
is very tightly bound because it has three different supercolors. That is
the reason why it has a mean radius of only $\approx 0.15$ fm.

We should still consider the repulsion and attraction due to the exchange of
colorless mesons between primons with equal supercolors.

\vspace*{0.5in}

\subsection{\protect\Large The Lagrangian of Quantum Superchromodynamics }

What we developed above in section 3.4, following the footsteps of QCD, is a
quantum superchromodynamics(QSCD) which deals with the interactions between
primons with different supercolors. According to what was established until
now primons should be very light fermions with a mass of about 1eV. Then, we
can propose that the free Lagrangian for primons is 
\begin{equation}
L=i{\hbar }c\overline{\Psi }\gamma ^{\mu }\partial _{\mu }\Psi -mc^{2}%
\overline{\Psi }\Psi
\end{equation}
\noindent in which $\Psi \;$ is the column 
\begin{equation}
\Psi =\left( 
\begin{array}{c}
\Psi _{\alpha } \\ 
\Psi _{\beta } \\ 
\Psi {\gamma }
\end{array}
\right)
\end{equation}

\noindent and $\Psi _{i}$ is a four-component Dirac spinor. \noindent In the
same way as is done in QCD we can construct the gauge invariant (under
supercolor SU(2)) QSCD Lagrangian 
\begin{equation}
L=i{\hbar }c\overline{\Psi }\gamma ^{\mu }\partial _{\mu }\Psi -mc^{2}%
\overline{\Psi }\Psi -\frac{1}{16{\pi }}\Gamma ^{\mu {\nu }}\Gamma _{\mu {%
\nu }}-g_{sc}\overline{\Psi }\gamma ^{\mu }{\Theta }{\Psi }A_{\mu }
\end{equation}
\noindent in which $g_{sc}$ is the supercolor coupling constant, and $\Gamma
^{\mu \nu }\;$ are the supergluon fields. The above Lagrangian should hold
for each primon because there are four different mass terms. That is, there
are four different Lagrangians.

Since primons are almost massless the Lagrangian (for each primon) can be
written as 
\begin{equation}
L=i{\hbar }c\overline{\Psi }\gamma ^{\mu }\partial _{\mu }\Psi -\frac{1}{%
16\pi }\Sigma ^{\mu \nu }\Sigma _{\mu \nu }-g_{sc}\overline{\Psi }\gamma
^{\mu }{\Theta }{\Psi }A_{\mu }.
\end{equation}
\noindent The above Lagrangians given by Eqs. 47 and 48 are invariant under
local SU(2) gauge transformations and describe the interaction of each
primon (that is, each flavor) with the three massless vector fields
(supergluons). The Dirac fields make the three supercolor currents 
\begin{equation}
I^{\zeta }=cg_{sc}\bar{\Psi}\gamma ^{\zeta }{\Theta }\Psi
\end{equation}
\noindent which are the sources of the supercolor fields.

\vspace*{0.3in}

\subsection{\protect\Large Primons and Weak Interactions}

As we saw $p_{1}\;$ and $p_{2}\;$ are stable, and $p_{3}\;$ and $p_{4}\;$
are unstable. Since $d\;$ decays into $u\;$ according to $d\rightarrow ue^{-}%
\bar{\nu _{e}}$, $p_{3}\;$ has to decay as $p_{3}\rightarrow p_{1}e^{-}\bar{%
\nu _{e}}$. Other weak decays of quarks cannot be explained in such a simple
way in terms of primons. It is a very hard task to build a whole new theory
of weak interactions of quarks taking into account the existence of primons.
It is a hard task but there are some clues to follow. As we saw above
primons have to exist: \textbf{there are many compelling evidences.}

\bigskip

\noindent {\Large References}

\bigskip

\noindent 1. M.E. de Souza, in \textit{Proceedings of the XII Brazilian
National Meeting of the Physics of Particles and\ Fields}, \ Caxambu, Minas
Gerais, Brazil, September 18-22, 1991. \newline
\noindent 2. M.E. de Souza, in \textit{The Six Fundamental Forces of Nature,}
p. 6, Universidade Federal de Sergipe, 1994. \newline
\noindent 3. E. Fischbach, in \textit{Proceedings of the NATO Advanced Study
Institute on Gravitational Measurements, \ Fundamental Metrology and
Constants}, 1987, ed. by V. de Sabbata and V. N. Melnikov (D. Reidel \
Publishing Company, Dordrecht, Holland, 1988).

\newpage

\bigskip \rule[0.5in]{0in}{0.17in}

\section{\noindent {\protect\LARGE Some Topics of QCD}}

{\LARGE \bigskip \rule[1.5in]{0in}{0.17in}}

\subsection{\protect\Large The Potential of a Quark Pair and the Usual QCD
Potential}

As two quarks($Q\bar{Q}$) are brought to a very close distance(below 0.5 fm,
presumably) from each other they should experience the strong force and also
the superstrong force. Since $Q\bar{Q}\;$ form bound states there should
exist a net molecular potential well between them. At large distances it
should be dominated by the strong force (Yukawa) potential 
\begin{equation}
V_{Q\bar{Q}}(r)=-\frac{(g_{s}^{Q})^{2}e^{-\mu _{s}{r}}}{r}.
\end{equation}
\noindent On the other hand QCD shows that there is an effective Coulombian
potential produced by the color field given by 
\begin{equation}
V_{QCD}(r)=-\frac{4}{3}\frac{\alpha _{s}}{r}+{\beta }r.
\end{equation}
\noindent

We do not know the value of $g_{s}^{Q}$, but we may assume that $%
(g_{s}^{Q})^{2}\;$ is of the order of $\alpha _{s}$. Then, it is easy to see
that for $\mu r\ll 1\;$ the two potentials may have the same order of
magnitude. When $r\;$ increases $V_{Q\bar{Q}}\;$ will be above the first
term of $V_{QCD}$, which decreases slowly to zero. The term $\beta r\;$
raises the potential and makes it get closer to $V_{Q\bar{Q}}$, as is shown
in Fig. 4.1.

It has been said in most textbooks on elementary particles that the data of
the experiments UA1 (Arnison et al.)$^{(1)}$ and UA2 (Bagnaia et al.)$^{(2)}$
at the CERN $p\bar{(p)}$ collider provide the best direct evidence that the
QCD potential at small $r\;$is proportional to $1/r$. But, the data show
much more than this simplistic conclusion. The data is shown in Fig. 4.2.
Parametrizing the data in the form $(sin\frac{\theta }{2})^{-n}\;$one
obtains $n=4.16\;$for the slope up to $\sin ^{4}\theta /2\approx 0.1$.
Notice that the center of the first point at the top is off the straight
line somewhat. This deviation may indicate that the differential cross
section tends to saturate as we go to small angles. A better fitting to the
data may be provided by a differential cross section of the form 
\begin{equation}
\frac{d\sigma }{d{\Omega }}{\propto }\frac{1}{(1+4(\frac{k}{\mu _{ss}})^{2}{%
sin}^{2}\frac{\theta }{2})^{2}}
\end{equation}
\noindent which means that the interacting potential for very short
distances is of the Yukawa type. Since $q^{2}=2000$GeV$^{2}$, $q\;$is about
45 GeV, and so, $k=1.56\times 10^{3}$fm$^{-1}$. For $sin^{4}\theta /2\approx
0.01$, $\theta \approx 37^{o}$. This is not a small angle, and if the
saturation is already beginning for such angles, then 
\begin{equation}
4(\frac{k}{\mu _{ss}})^{2}{sin}^{2}{\theta /2}{\sim }1.
\end{equation}
\noindent This means that $\mu _{ss}\sim 10-10^{3}fm^{-1}$, and thus the
order of magnitude of the range of the superstrong interaction is of about $%
0.1-10^{-3}$fm. Therefore, its bosons have masses in the range (1 - 1000)
GeV. Several experiments have, indeed, shown that the strong force becomes
repulsive at distances smaller than about 0.45 fm. Of course, it is not the
strong force, it is the superstrong force.

The success of the usual QCD potential is due to the use of several
adjustable parameters in the models and due to the existence of the two
primon shells described in chapter 2. As we saw the inner shell is quite
close to the center(mean radius of only $r_{1}=0.15$fm) while the outer
shell has quite a large mean radius $r_{2}\;$of about 0.65fm, that is, $%
r_{2}\approx 4r_{1}$. Therefore, it is almost a central potential for the
primons of the outer shell. That is, we can say that there is an approximate
central potential due to the existence of a strong charge $g_{1}\;$at the
center and another strong charge in the outer layer. Let us now estimate the
value of \noindent the coupling constant $\alpha _{s}\approx g_{1}g_{2}\;$%
between the two shells. Each quark has a strong charge of about 1/3. Thus, a
primon has a strong charge of about 1/6. But each shell has three primons,
and therefore, each shell has a strong charge of about $3\times \frac{1}{6}=%
\frac{1}{2}$. Then, the product $g_{1}g_{2}$, that is, $\alpha _{s}$, is
about 0.25 which is the experimental value of $\alpha _{s}$at $Q=3$GeV. As
discussed above, at very high Q the effective coupling should diminish due
to the action of the superstrong interaction. The lowest value of $\alpha
_{s}$(around 0.1) at $Q=100$GeV does include the effect of the superstrong
force. Please, find below a very important discussion on $\alpha _{s}$. It
is worth saying that the potential 
\begin{equation}
V_{QCD}(r)=-\frac{4}{3}\frac{\alpha _{s}}{r}+{\beta }r.
\end{equation}
\noindent has a very serious inconsistency when applied to mesons: it allows
an infinite number of bound states and this is not realistic at all.

Therefore, the effective potential between quarks should be a molecular
potential (due to the exchange of scalar and vectorial mesons) which can be
approximately described by a Coulombian potential (due to the effective
color field). As will be shown below the molecular potential also exhibits
asymptotic freedom.

As was seen in chapter 3 gluons are sets of colored mesons, that is, gluons
are effective collective bosons. In other words, they produce an overall
action which is the result of the actions of the exchanges of all the
different colored mesons, and they are, then, bosons of an effective field
which is produced by the collective interactions of supercolors and colors.
As we saw in chapter 3 the more basic interactions are governed by SU(2).
When we consider the interactions between quarks the overall effective
interaction mediated by gluons is described by SU(3). This means that QCD is
a great theory exactly because it simplifies matters a lot since gluons are
collections of colored mesons.

\bigskip

\subsection{\protect\Large The Confining Term of the Usual QCD \newline
Potential}

\bigskip Quarks are confined to distances shorter than 1F in nucleons, for
example. Let us try to understand why this is so. According to what was
developed above pions and other mesons are exchanged between quarks inside
nucleons. The mesons with the longest ranges are pions. That is why nucleons
have sizes of approximately 1F, and that is why quarks are confined to 1F.

In the light of what was developed up to now we see that when we try to free
the quarks of a proton with another proton we force the distance between
their quarks to be very small, that is, we make the quarks of the two
protons to get very close and, consequently, heavier scalar and vector
mesons (colorless) are exchanged among their primons, that is, among their
quarks, and therefore, they get more and more bound.

For improving our understanding on this issue let us assume that the overall
interaction with one scalar meson and a more massive vector meson produces a
sort of molecular-like potential, which to a good approximation can be
described by (not very far from equilibrium) 
\begin{equation}
V(r)=V_{o}+\frac{1}{2}kr^{2}-\gamma r^{3}.
\end{equation}
\noindent Depending on the values of the constants the last two terms of the
right side may be very close to $-\frac{4}{3}\frac{\alpha }{r}+\beta r$
(take a look at Fig. 4.1).

\bigskip

\subsection{\noindent {\protect\Large Asymptotic Freedom}}

It has been proven in the literature that the coupling constant of QCD$%
^{\prime }$s effective color field shows asymptotic behavior. That is 
\begin{equation}
\alpha _{s}(q^{2})=\frac{\alpha _{s}(q_{0}^{2})}{1+\frac{1}{4\pi }\alpha
_{s}(q_{0}^{2})\beta _{0}\ln {\frac{q^{2}}{q_{0}^{2}}}}
\end{equation}
\noindent in which $\beta _{0}\;$is a constant given by the number of
flavors.

Then, the molecular potential has to exhibit a similar asymptotic behavior,
and it actually does as will be proven below. As we go to higher
energies(i.e., to smaller $r$) there is more and more the influence of the
superstrong force, which being repulsive, diminishes the strength of the
strong force. As has been shown there should exist a molecular effective
potential between two quarks whose mathematical expression may be very
complicated. The effective force may even become zero at the bottom of each
well. Just to show that the effective coupling constant diminishes with $r\;$%
let us approximate the effective potential by 
\begin{equation}
V_{eff}=-\beta _{s}\frac{e^{-\mu _{s}r}}{r}+\beta _{ss}\frac{e^{-\mu _{ss}r}%
}{r}
\end{equation}
\newline
\noindent where $\beta _{s}=(g_{s}^{Q})^{2}$($Q\;$for quark) , and $\beta
_{ss}=(g_{ss}^{Q})^{2}\;$are the strong and superstrong couplings,
respectively. But according to QCD the effective potential for small $r\;$%
should be given by 
\begin{equation}
V_{eff}=-\frac{\alpha _{s}}{r}.
\end{equation}
\noindent We expect that $\beta _{ss}\gg \beta _{s}$, and $\mu _{ss}\gg \mu
_{s}$. Just to have a practical example, let us make $\beta _{ss}=10\beta
_{s}\;$and $\mu _{ss}=10\mu _{s}$. This means a boson with a mass of about
1.4 GeV. As we will see shortly this is very reasonable. Making $\beta
_{s}=1 $(GeV)(fm), and $\mu _{s}\approx 0.71$fm$^{-1}$, we obtain that $%
\alpha _{s}\;$should be 
\begin{equation}
\alpha _{s}=e^{-0.71r}-10e^{-10r}.
\end{equation}

\noindent The values of $\alpha _{s}\;$for different values of $r\;$are
shown in Table 4.1. We will see later on that the above values for $\beta
_{ss}\;$and $\mu _{ss}\;$agree quite well with supernovae data (see more on
this in chapter 6). Since a baryon has 6 primons, the $\beta _{ss}\;$of each
primon, $\beta _{ss}^{p}$, is about $\frac{1}{6}$10(GeV)(fermi)$\approx 2$%
GeVfm. As we will see in chapter 10 there are about 20 repulsive terms
between two nucleons. Therefore the effective superstrong coupling between
two baryons is about 20x2GeVfm=40 (GeV)(fm).

In momentum space the above effective molecular potential (Eq. 57) is$^{3}$%
\begin{equation}
V_{q}=-\beta _{s}\frac{1^{(1)}{\cdot }1^{(2)}}{q^{2}+\mu _{s}^{2}-i\eta }%
+\beta _{ss}\frac{\gamma ^{(1)}{\cdot }\gamma ^{(2)}}{q^{2}+\mu
_{ss}^{2}-i\eta }
\end{equation}
\noindent where $q_{\lambda }=(\vec{q},iq_{0})\;$is the four-momentum
transfer and $\gamma _{\lambda }\;$are gamma matrices, and $\mu _{ss}\;$and $%
\mu _{s}\;$are the inverse Compton wavelengths of the superstrong and
strong(pions) bosons, respectively. In the nonrelativistic limit we can make
the approximations $\gamma ^{(1)}\cdot \gamma ^{(2)}\rightarrow 1^{(1)}\cdot
1^{(2)}\;$and $|q_{0}|\ll |\vec{q}|$. In this case, for high momentum
transfer, that is, $q\gg \mu _{ss},\mu _{s}$, if we expand in powers of $%
\frac{\mu _{s}^{2}}{q^{2}}\;$and $\frac{\mu _{ss}^{2}}{q^{2}}\;$up to second
order we obtain 
\begin{equation}
V_{q}=\frac{\beta _{ss}-\beta _{s}}{q^{2}}-\frac{\beta _{ss}\mu
_{ss}^{2}-\beta _{s}\mu _{s}^{2}}{q^{4}}.
\end{equation}

\vskip.2in

\begin{center}
{\normalsize 
\begin{tabular}{||l|c||}
\hline
&  \\ 
$r$(fm) & $\alpha _{s}$(GeV$\cdot $fm) \\ 
&  \\ \hline\hline
0.50 & 0.634 \\ \hline
0.40 & 0.570 \\ \hline
0.30 & 0.310 \\ \hline
0.29 & 0.264 \\ \hline
0.28 & 0.212 \\ \hline
0.27 & 0.154 \\ \hline
0.26 & 0.089 \\ \hline
0.25 & 0.017 \\ \hline
0.24 & - 0.064 \\ \hline
... & ... \\ \hline\hline
\end{tabular}
}
\end{center}

\vskip.3in

\begin{center}
\parbox{4.5in}
{Table 4.1. An example of how the effective coupling constant(which is the 
result of the strong and superstrong interactions) varies with $r$. In this
case the superstrong coupling constant is chosen to be 10 times the strong 
coupling constant and the ranges of the strong and supertrong
interactions are 1.4 fm and 0.1 fm, respectively.}
\end{center}

\bigskip

If the second term is very small the effective potential will be very close
to QCD massless vectorial field. In order to achieve this it is enough to
have $\beta _{ss}\mu _{ss}^{2}\approx \beta _{s}\mu _{s}^{2}$, that is, 
\begin{equation}
\left( \frac{\mu _{ss}}{\mu _{s}}\right) ^{2}=\frac{\beta _{s}}{\beta _{ss}}.
\end{equation}
\noindent Using the above figures we obtain $\frac{\mu _{ss}}{\mu _{s}}%
\approx \frac{1}{22}$, that is, we find that the boson of the new
interaction has a mass of about 3GeV. Therefore, the higher the energy is,
the better QCD gets because the molecular potential tends more and more to a
constant, which is its bottom.

Let us now see the asymptotic behavior of the molecular potential. The
expansion of $V(q)\;$for high momentum transfer up to order $q^{6}\;$is 
\begin{equation}
V_{q}=\frac{\beta _{ss}-\beta _{s}}{q^{2}}-\frac{\beta _{ss}\mu
_{ss}^{2}-\beta _{s}\mu _{s}^{2}}{q^{4}}+\frac{\beta _{ss}\mu
_{ss}^{4}-\beta _{s}\mu _{s}^{4}}{q^{6}}-....
\end{equation}
\noindent Making $V(q)=\frac{\alpha }{q^{2}}\;$we have 
\begin{equation}
\alpha (q^{2})=\beta _{ss}-\beta _{s}-\frac{\beta _{ss}\mu _{ss}^{2}-\beta
_{s}\mu _{s}^{2}}{q^{2}}+\frac{\beta _{ss}\mu _{ss}^{4}-\beta _{s}\mu
_{s}^{4}}{q^{4}}-....
\end{equation}
Making $\delta _{0}=\beta _{ss}-\beta _{s}$, $\delta _{2}=\beta _{ss}\mu
_{ss}^{2}-\beta _{s}\mu _{s}^{2}$, $\delta _{4}=\beta _{ss}\mu
_{ss}^{4}-\beta _{s}\mu _{s}^{4}$, ..., we get 
\begin{equation}
\alpha (q^{2})=\delta _{0}-\frac{\delta _{2}}{q^{2}}+\frac{\delta _{4}}{q^{4}%
}-....
\end{equation}
\noindent and 
\begin{equation}
\alpha (q_{0}^{2})=\delta _{0}-\frac{\delta _{2}}{q_{0}^{2}}+\frac{\delta
_{4}}{q_{0}^{4}}-....
\end{equation}
\noindent Making $Q=q-q_{0}$, and dividing $\alpha (q^{2})\;$by $\alpha
(q_{0}^{2})$, we have 
\begin{equation}
\frac{\alpha (q^{2})}{\alpha (q_{0}^{2})}=\frac{1-\frac{a}{q_{0}^{2}}\left(
1+\frac{Q}{q_{0}}\right) ^{-2}+\frac{b}{q_{0}^{4}}\left( 1+\frac{Q}{q_{0}}%
\right) ^{-4}-....}{1-\frac{a}{q_{0}^{2}}+\frac{b}{q_{0}^{4}}-....}
\end{equation}
\noindent in which $a=\frac{\delta _{2}}{\delta _{0}}\;$and $b=\frac{\delta
_{4}}{\delta _{0}}$. For small $Q\;$we can expand the above expression and
obtain 
\begin{equation}
\frac{\alpha (q^{2})}{\alpha (q_{0}^{2})}=1+\frac{2Q}{q_{0}}\frac{1}{\Delta }%
\left( \frac{a}{q_{0}^{2}}-\frac{2b}{q_{0}^{4}}+....\right) +....
\end{equation}
\noindent where $\Delta =1-\frac{a}{q_{0}^{2}}+\frac{b}{q_{0}^{4}}-....$For $%
Q\;$small we also have 
\begin{equation}
\ln {\frac{q^{2}}{q_{0}^{2}}}=2\ln {\frac{q}{q_{0}}}=2\left( \frac{Q}{q_{0}}-%
\frac{Q^{2}}{2q_{0}^{2}}+\frac{Q^{3}}{3q_{0}^{3}}-....\right) {\approx }2%
\frac{Q}{q_{0}}.
\end{equation}

\noindent Finally we obtain 
\begin{equation}
\frac{\alpha (q^{2})}{\alpha (q_{0}^{2})}{\approx }1+f(q_{0})\ln {\frac{q^{2}%
}{q_{0}^{2}}}
\end{equation}
\noindent or 
\begin{equation}
\alpha (q^{2}){\approx }\frac{\alpha (q_{0}^{2})}{1-f(q_{0})\ln {\frac{q^{2}%
}{q_{0}^{2}}}}
\end{equation}
\noindent with 
\begin{equation}
f(q_{0})=\frac{\frac{a}{q_{0}^{2}}-\frac{2b}{q_{0}^{4}}+....}{1-\frac{a}{%
q_{0}^{2}}+\frac{b}{q_{0}^{4}}-....}
\end{equation}
\noindent That is, the effective coupling constant of the molecular
potential also shows an asymptotic behavior. And we can, then, ask which
asymptotic behavior we are really measuring, the asymptotic behavior of the
color field or the asymptotic behavior of the effective potential? Or are we
seing the same thing? \newline

\bigskip

\noindent {\Large References}

\bigskip

\noindent 1. G. Arnison, et al., Phys. Lett. 136B, 294(1984). \newline
\noindent 2. P. Bagnaia, et al., Phys. Lett. 138B, 430(1984). \newline
\noindent 3. R.C.Tolman, Phys.Rev. 55, 364(1939).

\bigskip \pagebreak

\rule[0.5in]{0in}{0.17in}

\section{\protect\LARGE The Energies of Hadrons and \newline
the Electric Charge in Baryons}

{\LARGE \bigskip \rule[1.5in]{0in}{0.17in}}

\subsection{\protect\bigskip {\protect\Large The Energies of Baryons}}

We will see below another evidence for the existence of the superstrong
interaction, which is the calculation of the energies of baryons. As was
shown above quarks are subject to attractive and repulsive forces.
Therefore, we can propose that a sort of effective molecular potential well
acts between any two of them. The expansion of such a potential about its
minimum yields a harmonic oscillator potential. Thus, we may consider that
every pair of quarks oscillates about an equilibrium distance $r_{q}$. For
small departures from equilibrium the potential must be of the form 
\begin{equation}
V(r)=V_{o}+K(r-r_{q})^{2}
\end{equation}
\noindent where $K$ is a constant and $V_{o}$ is a negative constant
representing the depth of the potential well. By doing so we are able to
calculate the energies of baryons. The present treatment is very different
from other calculations of baryon levels found in the literature. In those
calculations ad hoc central harmonic potentials have been used.

As is well known there are several important works that deal with this
problem. One of the most important is the pioneering work of Gasiorowicz and
Rosner$^{1}$ which has calculation of baryon levels and magnetic moments of
baryons using approximate wavefuncions. Another important work is that of
Isgur and Karl$^{2}$ which strongly suggests that non-relativistic quantum
mechanics can be used in the calculation of baryon spectra. Other very
important attempts towards the understanding of baryon spectra are the works
of Capstick and Isgur$^{3}$, Bhaduri et al.$^{4}$, Murthy et al.$^{5}$, \
Murthy et al$^{6}$, and Stassat et al.$^{7}$. An important work attempting
to describe baryon spectra is the recent work of Hosaka, Toki and Takayama$%
^{8}$ published in \textbf{1998}. This last work arrives at an important
equation which had already been deduced by \textbf{De Souza a long time ago,
in 1992}$^{9}$\textbf{. Other works by De Souza published before 1998
include it}$^{10,11}$\textbf{.}

\paragraph{\protect\large 5.1.1 In Cartesian Coordinates}

In the initial calculation we use normal cartesian coordinates which, of
course, does not consider the angular momentum of the system, that is, it
does not take into account the symmetries of the system. But this section is
very important because it calculates the energy levels. In the next section
we will link each level to its angular momentum. Considering the work of
Isgur and Karl$^{2}$ as to the use of non-relativistic quantum mechanics and
using a deformed harmonic oscillator potential$^{12,4.5,6}$ we can write the
Hamiltonian in normal cartesian coordinates as 
\begin{equation}
\sum_{i=1}^{6}\frac{{\partial }^{2}\psi }{{\partial }{\xi }_{i}^{2}}+\frac{2%
}{{\hbar }^{2}}\left( E-\frac{1}{2}\sum_{i=1}^{6}{\omega }_{i}{{\xi _{i}}^{2}%
}\right) \psi =0
\end{equation}
\noindent where we have used the fact that the three quarks are always in a
plane. The above equation may be resolved into a sum of 6 equations 
\begin{equation}
\frac{{\partial }^{2}\psi }{{\partial }{\xi }_{i}^{2}}+\frac{2}{{\hbar }^{2}}%
\left( E_{i}-\frac{1}{2}\omega _{i}{\xi _{i}}^{2}\right) \psi =0,
\end{equation}
\noindent which is the equation of a single harmonic oscillator of potential
energy $\frac{1}{2}\omega _{i}\xi _{i}^{2}$ and unitary mass with $%
E=\sum_{i=1}^{6}E_{i}$.

The general solution is a superposition of 6 harmonic motions in the 6
normal coordinates. The eigenfunctions $\psi _{i}(\xi _{i})$ are the
ordinary harmonic oscillator eigenfuntions 
\begin{equation}
\psi _{i}(\xi _{i})=N_{v_{i}}e^{-(\alpha _{i}/2)\xi _{i}^{2}}H_{v_{i}}(\sqrt{%
\alpha _{i}}\xi _{i}),
\end{equation}
\noindent where $N_{v_{i}}$ is a normalization constant, $\alpha _{i}=\nu
_{i}/\hbar $ and $H_{v_{i}}(\sqrt{\alpha _{i}}\xi _{i})$ is a Hermite
polynomial of the $v_{i}$th degree. For large $\xi _{i}$ the eigenfunctions
are governed by the exponential functions which make the eigenfunctions go
to zero very fast.

The energy of each harmonic oscillator is 
\begin{equation}
E_{i}=h\nu _{i}(v_{i}+\frac{1}{2}),
\end{equation}
\noindent where $v_{i}=0,1,2,3,...$ and $\nu _{i}$ is the classical
oscillation frequency of the normal ``vibration'' $i$, and $v_{i}$ is the
``vibrational'' quantum number. The total energy of the system can assume
only the values 
\begin{equation}
E(v_{1},v_{2},v_{3},...v_{6})=h\nu _{1}(v_{1}+\frac{1}{2})+h\nu _{2}(v_{2}+%
\frac{1}{2})+...h\nu _{6}(v_{6}+\frac{1}{2}).
\end{equation}

As was said above the three quarks in a baryon must always be in a plane.
Therefore, each quark is composed of two oscillators and so we may rearrange
the energy expression as 
\begin{equation}
E(n,m,k)=h\nu _{1}(n+1)+h\nu _{2}(m+1)+h\nu _{3}(k+1),
\end{equation}
\noindent where $n=v_{1}+v_{2},m=v_{3}+v_{4},k=v_{5}+v_{6}$. Of course, $%
n,m,k$ can assume the values, 0,1,2,3,... We may find the constants $h\nu $
from the ground states of some baryons. They are the known quark masses
taken as $m_{u}=m_{d}=0.31$Gev, $m_{s}=0.5$Gev, $m_{c}=1.7$Gev,$m_{b}=5$Gev
and $m_{t}=174$GeV.

The states obtained with the above Hamiltonian are degenerate with respect
to isospin (if the quark masses are the same) so that our calculation does
not distinguish between nucleonic and $\Delta $ states, or between $\Sigma $
and $\Lambda $ states. In the tables below the experimental values of baryon
masses were taken from reference 13.

Let us start the calculation with the states ddu(neutron), uud(proton) and
ddd($\Delta ^{-}$), uuu ($\Delta ^{++}$) and their resonances. All the
energies below are given in Gev. Because $m_{u}=m_{d}$, we have that the
energies calculated by the formula 
\begin{equation}
E_{n,m,k}=0.31(n+m+k+3)
\end{equation}
\noindent correspond to many energy states. The calculated values are
displayed in Table 5.1. The last column on the right is a rough
classification which will be cleared up in the next section. One observes in
Table 5.1 that the particles that belong to it are $N$ and $\Delta $, which
are particles that decay via the strong interaction either into $N$ or into $%
\Delta $(besides the electromagnetic decay, sometimes). For example

\begin{itemize}
\item  $\Delta (1232)\rightarrow N\pi $;

\item  $N(1440)\rightarrow N\pi $, $N\pi \pi $, $\Delta \rho $, $N\rho $;

\item  $\Delta (1600)\rightarrow N\pi $, $N\pi \pi $, $\Delta \pi $, $N\rho $%
, $N(1440)\pi $;
\end{itemize}

\noindent Therefore, with the help of Table 8 we can easily understand the
above decays. As we will see when a resonance decays into a particle of
another table, then the decay is weak. For example, $\Delta
(1905)\rightarrow \Sigma K$; $N(1650)\rightarrow \Lambda K$.

The energies of the particles $\Lambda $ and $\Sigma $, which are composed
of $uus$ and $uds$ are given by 
\begin{equation}
E_{n,m,k}=0.31(n+m+2)+0.5(k+1).
\end{equation}
\noindent The results are displayed in Table 5.2. The agreement with the
experimental values is excellent. As to the decay modes one observes the
same as for $N$ and $\Delta $, that is, decays via the strong interaction go
as $\Sigma \rightarrow \Lambda $ and $\Lambda \rightarrow \Sigma $. By means
of the weak interaction the two particles decay into $N$ and $\Delta $.

\bigskip \pagebreak

\vspace*{0.5in}

\begin{center}
\begin{tabular}{||c||c|l|c|l|c||}
\hline
&  &  &  &  &  \\ 
$n,m,k$ & $E_{C}(Gev)$ & $E_{M}$(Gev) & Error(\%) & $L_{2I.2J}$ & Parity \\ 
&  &  &  &  &  \\ \hline\hline
0,0,0 & 0.93 & 0.938($N$) & 0.9 & $P_{11}$ & + \\ \hline
$n+m+k=1$ & 1.24 & 1.232($\Delta $) & 0.6 & $P_{33}$ & + \\ \hline
$n+m+k=2$ & 1.55 & 1.52($N$) & 1.9 & $D_{13}$ & - \\ 
$n+m+k=2$ & 1.55 & 1.535($N$) & 1.0 & $S_{11}$ & - \\ 
$n+m+k=2$ & 1.55 & 1.6($\Delta $) & 3.1 & $P_{33}$ & + \\ 
$n+m+k=2$ & 1.55 & 1.62($\Delta $) & 4.5 & $S_{31}$ & - \\ \hline
$n+m+k=3$ & 1.86 & 1.90($N$) & 2.2 & $P_{13}$ & + \\ 
$n+m+k=3$ & 1.86 & 1.90($\Delta $) & 2.2 & $S_{31}$ & - \\ 
$n+m+k=3$ & 1.86 & 1.905($\Delta $) & 2.4 & $F_{35}$ & + \\ 
$n+m+k=3$ & 1.86 & 1.91($\Delta $) & 2.7 & $P_{31}$ & + \\ 
$n+m+k=3$ & 1.86 & 1.92($\Delta $) & 3.2 & $P_{33}$ & + \\ \hline
$n+m+k=4$ & 2.17 & 2.08($N$) & 4.1 & $D_{13}$ & - \\ 
$n+m+k=4$ & 2.17 & 2.09($N$) & 3.7 & $S_{11}$ & - \\ 
$n+m+k=4$ & 2.17 & 2.10($N$) & 3.2 & $P_{11}$ & + \\ 
$n+m+k=4$ & 2.17 & 2.15($\Delta $) & 0.9 & $S_{31}$ & - \\ 
$n+m+k=4$ & 2.17 & 2.19($N$) & 0.9 & $G_{17}$ & - \\ 
$n+m+k=4$ & 2.17 & 2.20($N$) & 1.4 & $D_{15}$ & - \\ 
$n+m+k=4$ & 2.17 & 2.20($\Delta $) & 1.4 & $G_{37}$ & - \\ 
$n+m+k=4$ & 2.17 & 2.22($N$) & 2.3 & $H_{19}$ & + \\ 
$n+m+k=4$ & 2.17 & 2.225($N$) & 5.5 & $G_{19}$ & - \\ \hline
$n+m+k=5$ & 2.48 & 2.39($\Delta $) & 3.6 & $F_{37}$ & + \\ 
$n+m+k=5$ & 2.48 & 2.40($\Delta $) & 3.2 & $G_{39}$ & - \\ 
$n+m+k=5$ & 2.48 & 2.42($\Delta $) & 2.4 & $H_{3,11}$ & + \\ \hline
$n+m+k=6$ & 2.79 & 2.7($N$) & 3.2 & $K_{1,13}$ & + \\ 
$n+m+k=6$ & 2.79 & 2.75($\Delta $) & 1.4 & $I_{3,13}$ & - \\ \hline
$n+m+k=7$ & 3.10 & 3.100($N$) & 0 & $L_{1,15}$ & ? \\ \hline
$n+m+k=8$ & 3.21 & ? & ? & ? & ? \\ \hline
$n+m+k=9$ & 3.72 & ? & ? & ? & ? \\ \hline
$n+m+k=9$ & 4.03 & ? & ? & ? & ? \\ \hline
... & ... & ... & ... & ... &  \\ \hline\hline
\end{tabular}

\vskip .2in

\parbox{4.5in}
{Table 5.1. Baryon states $N$ and $\Delta$. The energies $E_{C}$ were 
calculated according to the formula $E_{n,m,k} = 0.31(n+m+k+3)$ in which
$n,m,k$ are integers. $E_{M}$ is the measured energy.  The error means the 
absolute value of $(E_{C} - E_{M})/E_{C}$. We are able, of  course, to 
predict the energies of many other resonances.}
\end{center}

\bigskip \pagebreak

For the $\Xi ^{o}$($uss$) and $\Xi ^{-}$($dss$) baryons the energies are
expressed by 
\begin{equation}
E_{n,m,k}=0.31(n+1)+0.5(m+k+2).
\end{equation}
\noindent See Table 5.3 to check the agreement with the experimental data.
In this case the last column is almost empty due to a lack of experimental
data.

\bigskip

\begin{center}
\begin{tabular}{||c||c|l|c|c|c||}
\hline\hline
&  &  &  &  &  \\ 
State($n,m,k$) & $E_{C}(Gev)$ & $E_{M}$(Gev) & Error(\%) & $L_{2I,2J}$ & 
Parity \\ 
&  &  &  &  &  \\ \hline\hline
0,0,0 & 1.12 & 1.116($\Lambda $) & 0.4 & $P_{01}$ & + \\ 
0,0,0 & 1.12 & 1.193($\Sigma $) & 6.5 & $P_{11}$ & + \\ \hline
$n+m=1$, k=0 & 1.43 & 1.385($\Sigma $) & 3.2 & $P_{13}$ & + \\ 
$n+m=1$, k=0 & 1.43 & 1.405($\Lambda $) & 1.7 & $S_{01}$ & - \\ 
$n+m=1$, k=0 & 1.43 & 1.48($\Sigma $) & 3.5 & ? & ? \\ \hline
0,0,1 & 1.62 & 1.52($\Lambda $) & 6.2 & $D_{03}$ & - \\ 
0,0,1 & 1.62 & 1.56($\Sigma $) & 3.7 & ? & ? \\ 
0,0,1 & 1.62 & 1.58($\Sigma $) & 2.5 & $D_{13}$ & - \\ 
0,0,1 & 1.62 & 1.60($\Lambda $) & 1.2 & $P_{01}$ & + \\ 
0,0,1 & 1.62 & 1.62($\Sigma $) & 0 & $S_{11}$ & - \\ 
0,0,1 & 1.62 & 1.66($\Sigma $) & 2.5 & $P_{11}$ & + \\ 
0,0,1 & 1.62 & 1.67($\Sigma $) & 3.1 & $D_{13}$ & - \\ 
0,0,1 & 1.62 & 1.67($\Lambda $) & 3.1 & $S_{01}$ & - \\ \hline
$n+m=2$, k=0 & 1.74 & 1.69($\Lambda $) & 2.9 & $D_{03}$ & - \\ 
$n+m=2$, k=0 & 1.74 & 1.69($\Sigma $) & 2.9 & ? & ? \\ 
$n+m=2$, k=0 & 1.74 & 1.75($\Sigma $) & 0.6 & $S_{11}$ & - \\ 
$n+m=2$, k=0 & 1.74 & 1.77($\Sigma $) & 1.7 & $P_{11}$ & + \\ 
$n+m=2$, k=0 & 1.74 & 1.775($\Sigma $) & 2.0 & $D_{15}$ & - \\ 
$n+m=2$, k=0 & 1.74 & 1.80($\Lambda $) & 3.4 & $S_{01}$ & - \\ 
$n+m=2$, k=0 & 1.74 & 1.81($\Lambda $) & 4.0 & $P_{01}$ & + \\ 
$n+m=2$, k=0 & 1.74 & 1.82($\Lambda $) & 4.6 & $F_{05}$ & + \\ 
$n+m=2$, k=0 & 1.74 & 1.83($\Lambda $) & 5.2 & $D_{05}$ & - \\ \hline\hline
\end{tabular}
\end{center}

\centerline{Continues on next page}

\pagebreak

\vspace*{0.5in}

\begin{center}
\begin{tabular}{||c||c|l|c|c|c||}
\hline\hline
&  &  &  &  &  \\ 
State($n,m,k$) & $E_{C}$(Gev) & $E_{M}$(Gev) & Error(\%) & $L_{2I,2J}$ & 
Parity \\ 
&  &  &  &  &  \\ \hline\hline
$n+m=1$, k=1 & 1.93 & 1.84($\Sigma $) & 4.7 & $P_{13}$ & + \\ 
$n+m=1$, k=1 & 1.93 & 1.88($\Sigma $) & 2.6 & $P_{11}$ & + \\ 
$n+m=1$, k=1 & 1.93 & 1.89($\Lambda $) & 2.1 & $P_{03}$ & + \\ 
$n+m=1$, k=1 & 1.93 & 1.915($\Sigma $) & 0.8 & $F_{15}$ & + \\ 
$n+m=1$, k=1 & 1.93 & 1.94($\Sigma $) & 0.5 & $D_{13}$ & - \\ \hline
$n+m=3$, k=0 & 2.05 & 2.00($\Lambda $) & 2.5 & ? &  \\ 
$n+m=3$, k=0 & 2.05 & 2.00($\Sigma $) & 2.4 & $S_{11}$ & - \\ 
$n+m=3$, k=0 & 2.05 & 2.02($\Lambda $) & 1.5 & $F_{07}$ & + \\ 
$n+m=3$, k=0 & 2.05 & 2.03($\Sigma $) & 1.0 & $F_{17}$ & + \\ 
$n+m=3$, k=0 & 2.05 & 2.07($\Sigma $) & 1.0 & $F_{15}$ & + \\ 
$n+m=3$, k=0 & 2.05 & 2.08($\Sigma $) & 1.5 & $P_{13}$ & + \\ \hline
0,0,2 & 2.12 & 2.10($\Sigma $) & 0.9 & $G_{17}$ & - \\ 
0,0,2 & 2.12 & 2.10($\Lambda $) & 0.9 & $G_{07}$ & - \\ 
0,0,2 & 2.12 & 2.11($\Lambda $) & 0.5 & $F_{05}$ & + \\ \hline
$n+m=2$, k=1 & 2.24 & 2.25($\Sigma $) & 0.5 & ? & ? \\ \hline
$n+m=4$, k=0 & 2.36 & 2.325($\Lambda $) & 1.5 & $D_{03}$ & - \\ 
$n+m=4$, k=0 & 2.36 & 2.35($\Lambda $) & 0.4 & $H_{09}$ & + \\ \hline
$n+m=1$, k=2 & 2.43 & 2.455 & 2.5 & ? &  \\ \hline
$n+m=3$, k=1 & 2.55 & 2.585($\Lambda $) & 1.4 & ? & ? \\ \hline
0,0,3 & 2.62 & 2.62($\Sigma $) & 0 & ? & ? \\ \hline
$n+m=5$, k=0 & 2.67 & to be found & ? & ? &  \\ \hline
$n+m=2$, k=2 & 2.74 & to be found & ? & ? &  \\ \hline
$n+m=4$, k=1 & 2.86 & to be found & ? & ? &  \\ \hline
$n+m=1$, k=3 & 2.93 & to be found & ? & ? &  \\ \hline
$n+m=6$, k=0 & 2.98 & 3.00($\Sigma $) & 0.7 & ? & ? \\ \hline
$n+m=3$, k=2 & 3.05 & to be found & ? & ? &  \\ \hline
$n=m=0$, k=4 & 3.12 & to be found & ? & ? &  \\ \hline
$n+m=5$, k=1 & 3.17 & 3.17($\Sigma $) & 0 & ? & ? \\ \hline
$n+m=2$, k=3 & 3.24 & to be found & ? & ? &  \\ \hline
& ... & ... & ... & ... & ... \\ \hline\hline
\end{tabular}
\end{center}

\vskip.2in

\begin{center}
\parbox{4.5in}
{Table 5.2. Baryon states $\Sigma$ and $\Lambda$. The energies $E_{C}$ were
calculated according to the formula $E_{n,m,k}= 0.31(n+m+2) + 0.5(k+1)$.
$E_{M}$ is the measured energy. The error  means the absolute value of 
$(E_{C} - E_{M})/E_{C}$.}
\end{center}

\pagebreak

\vspace*{0.5in}

\begin{center}
\begin{tabular}{||c||l|l|l|c|c||}
\hline
&  &  &  &  &  \\ 
State($n,m,k$) & $E_{C}$(Gev) & $E_{M}$(Gev) & Error(\%) & $L_{2I,2J}$ & 
Parity \\ 
&  &  &  &  &  \\ \hline\hline
0,0,0 & 1.31 & 1.315 & 0.5 & $P_{11}$ & + \\ \hline
1,0,0 & 1.62 & 1.53 & 5.6 & $P_{13}$ & + \\ 
1,0,0 & 1.62 & 1.62 & 0 & ? & ? \\ 
1,0,0 & 1.62 & 1.69 & 4.3 & ? & ? \\ \hline
n=0, $m+k=1$ & 1.81 & 1.82 & 0.6 & $D_{13}$ & - \\ \hline
2,0,0 & 1.93 & 1.95 & 1.0 & ? & ? \\ \hline
n=1, $m+k=1$ & 2.12 & 2.03 & 4.2 & ? & ? \\ 
n=1, $m+k=1$ & 2.12 & 2.12 & 0 & ? & ? \\ \hline
n=3, $m=k=0$ & 2.24 & 2.25 & 0.5 & ? & ? \\ \hline
n=0, $m+k=2$ & 2.31 & 2.37 & 2.6 & ? & ? \\ \hline
n=2, $m+k=1$ & 2.43 & to be found & ? & ? & ? \\ \hline
n=4, $m=k=0$ & 2.55 & 2.5 & 2.0 & ? & ? \\ \hline
n=1, $m+k=2$ & 2.62 & to be found & ? & ? & ? \\ 
& ... & ... & ... & ... & ... \\ \hline\hline
\end{tabular}
\end{center}

\vskip.2in

\begin{center}
\parbox{4.5in}
{Table 5.3. Baryon states $\Xi$. The energies $E_{C}$ were
calculated according to the formula $E_{n,m,k}= 0.31(n+1) + 0.5(m+k+2)$.
$E_{M}$ is the measured energy. The error means the absolute 
value of $(E_{C} - E_{M})/E_{C}$. The state $\Xi(1530)P_{13}$
appears to be the lowest state of the composite $\Xi\biguplus\pi$. Its
decay is in fact $\Xi\pi$.}
\end{center}

\vskip .3in

In the same way the energies of $\Omega $($sss$) are obtained by 
\begin{equation}
E_{n,m,k}=0.5(n+m+k+3).
\end{equation}
\noindent The energies are displayed in Table 5.4. The discrepancies are
higher, of the order of 10\% and decreases as the energy increases. This is
a tendency which is also observed for the other particles. This may mean
that, at the bottom, the potential is less flat than the potential of a
harmonic oscillator. The decays occur as with the $\Xi $, that is, one sees
weak, electromagnetic and strong decays into other particles such as $\Xi $
and $\Lambda $.

\pagebreak

\vspace*{0.5in}

\begin{center}
\begin{tabular}{||c||l|c|l||}
\hline
&  &  &  \\ 
State($n,m,k$) & $E_{C}$(Gev) & $E_{M}$(Gev) & Error(\%) \\ 
&  &  &  \\ \hline\hline
0,0,0 & 1.5 & 1.672 & 11.7 \\ \hline
$n+m+k=1$ & 2.0 & 2.25 & 12.5 \\ \hline
$n+m+k=2$ & 2.5 & 2.47 & 1.2 \\ \hline
$n+m+k=3$ & 3.0 & to be found & ? \\ \hline
... & ... & ... & ... \\ \hline\hline
\end{tabular}
\end{center}

\vskip.2in

\begin{center}
\parbox{4.5in}
{Table 5.4. Baryon states $\Omega$. The energies $E_{C}$ were
calculated according to the formula $E_{n,m,k}= 0.5(n+m+k+3)$, and
$E_{M}$ is the measured energy.}
\end{center}

\vskip .3in

The energies of the charmed baryons($C=+1$) $\Lambda _{c}^{+}$, $\Sigma
_{c}^{++}$, $\Sigma _{c}^{+}$ and $\Sigma _{c}^{0}$ are given by 
\begin{equation}
E_{n,m,k}=0.31(n+m+2)+1.7(k+1).
\end{equation}
\noindent The levels are shown in Table 5.5.

For the charmed baryons($C=+1$) $\Xi _{c}^{+}$ and $\Xi _{c}^{0}$ we have 
\begin{equation}
E_{n,m,k}=0.31(n+1)+0.5(m+1)+1.7(k+1).
\end{equation}
\noindent The results are displayed in Table 5.6.

As for the $\Omega _{c}^{0}$, its energies are 
\begin{equation}
E_{n,m,k}=0.5(n+m+2)+1.7(k+1).
\end{equation}
\noindent Table 5.7 shows the results of the energy levels.

\pagebreak

\vspace*{0.5in}

\begin{center}
\begin{tabular}{||c||l|c|l||}
\hline
&  &  &  \\ 
State($n,m,k$) & $E_{C}$(Gev) & $E_{M}$(Gev) & Error(\%) \\ 
&  &  &  \\ \hline\hline
0,0,0 & 2.32 & 2.285($\Lambda _{c}$) & 1.5 \\ \hline
$n+m=1$, k=0 & 2.63 & 2.594($\Lambda _{c}$) & 0.1 \\ 
$n+m=1$, k=0 & 2.63 & 2.627($\Lambda _{c}$) & 0.01 \\ \hline
$n+m=2$, k=0 & 2.94 & to be found & ? \\ \hline
... & ... & ... & ... \\ \hline\hline
\end{tabular}
\end{center}

\vskip.2in

\begin{center}
\parbox{4.5in}
{Table 5.5. Baryon states $\Lambda_{c}$ and $\Sigma_{c}$. The 
energies $E_{C}$ were calculated according to the formula 
$E_{n,m,k}= 0.31(n+m+2) + 1.7(k+1)$. The state with energy 2.63 MeV 
had already been predicted
in another version of this work. The experimental levels 2.594 MeV
and 2.627 MeV have confirmed the theoretical values. It appears that
the level $\Sigma_{c}(2.455)$ is a composition of the level 
$(0,0,0)$(that is the 2.285 $\Lambda_{c}$) with a pion as is also 
inferred from its decay.}
\end{center}

\vskip.3in

\begin{center}
\begin{tabular}{||c||l|c|l||}
\hline
&  &  &  \\ 
State($n,m,k$) & $E_{C}$(Gev) & $E_{M}$(Gev) & Error(\%) \\ 
&  &  &  \\ \hline\hline
0,0,0 & 2.51 & 2.47($\Xi _{c}^{+}$ & 1.6 \\ \cline{2-3}
& 2.51 & 2.47($\Xi _{c}^{0}$ & 1.6 \\ \hline
1,0,0 & 2.82 & to be found & ? \\ \hline
0,1,0 & 3.01 & to be found & ? \\ \hline
... & ... & ... & ... \\ \hline\hline
\end{tabular}
\end{center}

\vskip.2in

\begin{center}
\parbox{4.5in}
{Table 5.6. Baryon states $\Xi_{c}$. The energies $E_{C}$ were 
calculated according to the formula $E_{n,m,k}= 0.31(n+1) + 0.5(m+1) 
+ 1.7(k+1)$. $E_{M}$ is the measured energy. The recently found level 
$\Xi_{c}(2645)$ is probably a composition of the regular level 
$\Xi_{c}^{+}$ with a pion as its decay confirms.}
\end{center}

\pagebreak

\vspace*{0.5in}

\begin{center}
\begin{tabular}{||c||l|c|l||}
\hline
&  &  &  \\ 
State($n,m,k$) & $E_{C}$(Gev) & $E_{M}$(Gev) & Error(\%) \\ 
&  &  &  \\ \hline\hline
0,0,0 & 2.7 & 2.704($\Omega _{c}^{0}$ & 0 \\ \hline
$n+m=1$, k=0 & 3.2 & to be found & ? \\ \hline
$n+m=2$, k=0 & 3.7 & to be found & ? \\ \hline
... & ... & ... & ... \\ \hline\hline
\end{tabular}
\end{center}

\vskip.2in

\begin{center}
\parbox{4.5in}
{Table 5.7. Baryon states $\Omega_{c}$. The energies $E_{C}$ were 
calculated according to the formula $E_{n,m,k}= 0.5(n+m+2) + 1.7(k+1)$.
The energy of the level 
$(0,0,0)$ above shown had been predicted in other versions of this work.}
\end{center}

\vskip .3in

We may predict the energies of many other baryons given by the formulas:

\begin{itemize}
\item  ucc and dcc, $E_{n,m,k}=0.31(n+1)+1.7(m+k+2)$;

\item  scc, $E_{n,m,k}=0.5(n+1)+1.7(m+k+2)$;

\item  ccc, $E_{n,m,k}=1.7(n+m+k+3)$;

\item  ccb, $E_{n,m,k}=1.7(n+m+2)+5(k+1)$;

\item  cbb, $E_{n,m,k}=1.7(n+1)+5(m+k+2)$;

\item  ubb and dbb, $E_{n,m,k}=0.31(n+1)+5(m+k+2)$ ;

\item  uub, udb and ddb, $E_{n,m,k}=0.31(n+m+2)+5(k+1)$;

\item  bbb, $E_{n,m,k}=5(n+m+k+3)$;

\item  usb and dsb, $E_{n,m,k}=0.31(n+1)+0.5(m+1)+5(k+1)$;

\item  sbb, $E_{n,m,k}=0.5(n+1)+5(m+k+2)$;

\item  scb, $E_{n,m,k}=0.5(n+1)+1.7(m+1)+5(k+1)$;

\item  ucb, $E_{n,m,k}=0.31(n+1)+1.7(m+1)+5(k+1)$;

\item  ttt, $E_{n,m,k}=(174\pm 17)(n+m+k+3)$;

\item  and all combinations of t with u, d, c, s and b.
\end{itemize}

The first state(0,0,0) of $udb$ which has an energy equal to 5.641GeV has
been recently found. The above formula for this state yields the energy
5.62GeV. The error is just 0.3\%.

\pagebreak

\paragraph{\protect\large 5.1.2 In Polar Coordinates}

In order to address the angular momentum we have to use spherical or polar
coordinates. Since the three quarks of a baryon are always in a plane we can
use polar coordinates. We choose the Z axis perpendicular to this plane. Now
the eigenfunctions are angular momentum eigenfunctions (of the orbital
angular momentum). Thus, we have three oscillators in a plane. Considering
that they are independent the radial Schr\"{o}dinger equation for the
stationary states of each oscillator is given by$^{14}$%
\begin{equation}
\left[ -\frac{\hbar ^{2}}{2\mu }\left( \frac{{\partial }^{2}}{{\partial \rho 
}^{2}}+\frac{1}{\rho }\frac{\partial }{\partial \rho }-\frac{m_{z}}{\rho ^{2}%
}\right) +\frac{1}{2}\mu \omega ^{2}\rho {^{2}}\right] R_{Em}(\rho
)=ER_{Em}(\rho )
\end{equation}
where $m_{z}$ is the quantum number associated to $L_{z}$. Therefore, what
we have is the following: three independent oscillators with orbital angular
momenta $\mathbf{L}_{\mathbf{1}},\ \mathbf{L}_{\mathbf{2}}$ and $\mathbf{L}_{%
\mathbf{3}}$ which have the Z components$\ L_{z1},L_{z2}$ and $L_{z3}$ in
the plane containing the quarks. Of course, the system has a total orbital
angular momentum $\mathbf{L=}$ $\mathbf{L}_{\mathbf{1}}+\ \mathbf{L}_{%
\mathbf{2}}+\mathbf{L}_{\mathbf{3}}$ and there is a quantum number $l_{i}$
associated to each $\mathbf{L}_{\mathbf{i}}$. The eigenvalues of the energy
are given by$^{14}$%
\begin{equation}
E=(2r_{1}+|m_{1}|+1)h\nu _{1}+(2r_{2}+|m_{2}|+1)h\nu
_{2}+(2r_{3}+|m_{3}|+1)h\nu _{3}
\end{equation}
in which $r_{1},r_{2},r_{3}=0,1,2,3,...$ and $|m_{i}|=0,1,2,3.....,l_{i}.$
Comparing the above equation with the equation 
\begin{equation*}
E(n,m,k)=h\nu _{1}(n+1)+h\nu _{2}(m+1)+h\nu _{3}(k+1),
\end{equation*}
we see that $n=2r_{1}+|m_{1}|$, $m=2r_{2}+|m_{2}|$, $k=2r_{3}+|m_{3}|$.

Let us recall that if we have three angular momenta $L_{1},L_{2}$ and $L_{3}$
described by the quantum numbers $l_{1},l_{2},l_{3}$ the total orbital
angular momentum $L$ will be described by the quantum number $l$ given by 
\begin{equation}
l_{1}+l_{2}+l_{3}\geq l\geq ||l_{1}-l_{2}|-l_{3}|
\end{equation}
where $l_{1}\geq |m_{1}|,$ $l_{2}\geq |m_{2}|,$ $l_{3}\geq |m_{3}|.$

Taking into account spin we form the total angular momentum given by $%
\mathbf{J=L+S}$ and the quantum numbers of $\mathbf{J}$ are $j=l\pm s$ where 
$s$ is the spin quantum number. As we will see we will be able to describe
all baryon levels.

\vspace*{0.3in}

\noindent {\large {a) Baryons N and $\Delta $}}

Let us begin the calculation with the particles $N$ and $\Delta $. We will
classify the levels by energy according to Table 5.1. The first state of $N$
is the state $(n=0,m=0,k=0)$ with energy 0.93 GeV. Therefore in this case $%
l_{1}=l_{2}=l_{3}=0$ and then $l=0$. Hence this is the positive parity state 
$P_{11}$ and we have

\smallskip

\begin{center}
\begin{tabular}{||c|c|c|c||}
\hline\hline
&  &  &  \\ 
$l$ & $N$ & $\Delta $ & Parity \\ 
&  &  &  \\ \hline
$0$ & $0.938P_{11}$ & $?$ & + \\ \hline\hline
\end{tabular}

\smallskip
\end{center}

The second energy level (1.24\nolinebreak\ GeV) which is the first state of $%
\Delta $ has $n+m+k=1.$ This means that $%
2r_{1}+|m_{1}|+2r_{2}+|m_{2}|+2r_{3}+|m_{3}|=1$. Thus, $%
|m_{1}|+|m_{2}|+|m_{3}|=1$ and $l_{1}+l_{2}+l_{3}\geq 1$, and we can choose
the sets $%
|m_{1}|=1,|m_{2}|=|m_{3}|=0;|m_{1}|=|m_{3}|=0,|m_{2}|=1;|m_{1}|=1,|m_{2}|=|m_{3}|=0 
$, and $l_{1}=2,l_{2}=l_{3}=0$, or $l_{2}=2,l_{1}=l_{3}=0$, or still $%
l_{3}=2,l_{1}=l_{2}=0$ which produce $l=2$ and thus the level

\begin{center}
\begin{tabular}{||c|c|c|c||}
\hline\hline
&  &  &  \\ 
$l$ & $N$ & $\Delta $ & Parity \\ 
&  &  &  \\ \hline
$2$ & ? & 1.232$P_{33}$ & + \\ \hline\hline
\end{tabular}

\smallskip
\end{center}

In the third energy level (1.55 Gev) $%
n+m+k=2=2r_{1}+|m_{1}|+2r_{2}+|m_{2}|+2r_{3}+|m_{3}|.$ This means that $%
|m_{1}|+|m_{2}|+|m_{3}|=2,0$ and we have the sets of possible values of $%
l_{1},l_{2},l_{3}$

\smallskip

\begin{center}
\begin{tabular}{||c||c|c|c|c|c|c|c||}
\hline\hline
$l_{1},l_{2},l_{3}$ & 2,0,0 & 0,2,0 & 0,0,2 & 1,1,0 & 1,0,1 & 0,1,1 & 0,0,0
\\ \hline
$l$ & 2 & 2 & 2 & 0,1,2 & 0,1,2 & 0,1,2 & 0 \\ \hline\hline
\end{tabular}

\smallskip
\end{center}

\noindent in which the second column presents the values of $l$ that satisfy
the condition $l_{1}+l_{2}+l_{3}\geq 2,0.$ There are thus the following
states

\begin{center}
\begin{tabular}{||c|c|c|c||}
\hline\hline
&  &  &  \\ 
$l$ & $N$ & $\Delta $ & Parity \\ 
&  &  &  \\ \hline
$0$ & $1.44P_{11}$, $1.71P_{11}$ & $?$ & + \\ \hline
$1$ & $
\begin{tabular}{c}
$1.535S_{11}$, $1.65S_{11}$ \\ 
$1.52D_{13}$, $1.70D_{13},1.675D_{15}$%
\end{tabular}
$ & $
\begin{tabular}{c}
$1.62S_{31}$ \\ 
$1.70D_{33}$%
\end{tabular}
$ & - \\ \hline
$2$ & $1.68F_{15},1.72P_{13}$ & $1.6P_{33}$ & + \\ \hline\hline
\end{tabular}

\smallskip
\end{center}

\noindent because we can have $%
j=1/2=0+1/2=1-1/2;j=3/2=1+1/2=2-1/2;j=5/2=1+3/2=2+1/2.$

The fourth energy level (1.86 Gev) has $%
n+m+k=3=2r_{1}+|m_{1}|+2r_{2}+|m_{2}|+2r_{3}+|m_{3}|$ which makes $%
|m_{1}|+|m_{2}|+|m_{3}|=3,1$ and $l_{1}+l_{2}+l_{3}\geq 3,1$. We have
therefore the possibilities

\smallskip

\begin{center}
\begin{tabular}{||c|c||}
\hline\hline
$l_{1},l_{2},l_{3}$ & $l$ \\ \hline\hline
3,0,0 & 3 \\ \hline
0,3,0 & 3 \\ \hline
0,0,3 & 3 \\ \hline
2,1,0 & 3,2,1 \\ \hline\hline
\end{tabular}
\begin{tabular}{||c|c||}
\hline\hline
$l_{1},l_{2},l_{3}$ & $l$ \\ \hline\hline
2,0,1 & 3,2,1 \\ \hline
1,0,2 & 3,2,1 \\ \hline
1,2,0 & 3,2,1 \\ \hline
0,1,2 & 3,2,1 \\ \hline\hline
\end{tabular}
\begin{tabular}{||c|c||}
\hline\hline
$l_{1},l_{2},l_{3}$ & $l$ \\ \hline\hline
0,2,1 & 3,2,1 \\ \hline
1,0,0 & 1 \\ \hline
0,1,0 & 1 \\ \hline
0,0,1 & 1 \\ \hline\hline
\end{tabular}
\end{center}

\noindent and the states

\begin{center}
\begin{tabular}{||c|c|c|c||}
\hline\hline
&  &  &  \\ 
$l$ & $N$ & $\Delta $ & Parity \\ 
&  &  &  \\ \hline
$1$ & $2.08D_{13}$ & $
\begin{tabular}{c}
$1.90S_{31}$ \\ 
$1.94D_{33}$%
\end{tabular}
$ & - \\ \hline
$2$ & $1.90P_{13}$,$2.00F_{15}$, $1.99F_{17}$ & $1.91P_{31}$, $1.92P_{33}$, $%
1.905F_{35}$, & + \\ 
& $2.00F_{35}$,$1.95F_{37}$ & ? &  \\ \hline
$3$ & ? & $1.93D_{35}$ & - \\ \hline\hline
\end{tabular}

\smallskip
\end{center}

In the fifth energy level (2.17 Gev) $n+m+k=4=$ $%
2r_{1}+|m_{1}|+2r_{2}+|m_{2}|+2r_{3}+|m_{3}|$ which yields $%
|m_{1}|+|m_{2}|+|m_{3}|=4,2,0$ and $l_{1}+l_{2}+l_{3}\geq 4,2,0$. We can
then have $l_{1}=l_{2}=l_{3}=0$ $(l=0)$ and also

\smallskip

\begin{center}
\begin{tabular}{||c|c||}
\hline\hline
$l_{1},l_{2},l_{3}$ & $l$ \\ \hline\hline
4,0,0 & 4 \\ \hline
0,4,0 & 4 \\ \hline
0,0,4 & 4 \\ \hline
3,1,0 & 4,3,2 \\ \hline
3,0,1 & 4,3,2 \\ \hline
1,3,0 & 4,3,2 \\ \hline
1,0,3 & 4,3,2 \\ \hline
0,3,1 & 4,3,2 \\ \hline
0,1,3 & 4,3,2 \\ \hline\hline
\end{tabular}
\begin{tabular}{||c|c||}
\hline\hline
$l_{1},l_{2},l_{3}$ & $l$ \\ \hline\hline
2,2,0 & 4,3,2,1,0 \\ \hline
2,0,2 & 4,3,2,1,0 \\ \hline
0,2,2 & 4,3,2,1,0 \\ \hline
2,0,0 & 2 \\ \hline
0,2,0 & 2 \\ \hline
0,0,2 & 2 \\ \hline
1,1,0 & 2,1,0 \\ \hline
1,0,1 & 2,1,0 \\ \hline
0,1,1 & 2,1,0 \\ \hline\hline
\end{tabular}
\end{center}

\noindent and hence the states

\begin{center}
\bigskip \rule[0.5in]{0in}{0.17in}

\begin{tabular}{||c|c|c|c||}
\hline\hline
&  &  &  \\ 
$l$ & $N$ & $\Delta $ & Parity \\ 
&  &  &  \\ \hline
$0$ & $2.10P_{11}$ & ? & + \\ \hline
$1$ & $2.08S_{11}$, $2.20D_{15}$ & $2.15S_{31}$, $2.35D_{35}$ & - \\ \hline
$2$ & ? & $2.39F_{37}?$ & + \\ \hline
$3$ & $2.19G_{17}$, $2.25G_{19}$ & $2.20G_{37}$ & - \\ \hline
$4$ & $2.22H_{19}$ & $2.3H_{39}$ & + \\ \hline\hline
\end{tabular}

\medskip
\end{center}

\smallskip

In the sixth energy level (2.48 Gev) $%
n+m+k=5=2r_{1}+|m_{1}|+2r_{2}+|m_{2}|+2r_{3}+|m_{3}|$ which produces $%
|m_{1}|+|m_{2}|+|m_{3}|=5,3,1$ and $l_{1}+l_{2}+l_{3}\geq 5,3,1$. We have
then the possibilities

\smallskip

\begin{center}
\begin{tabular}{||c|c||}
\hline\hline
$l_{1},l_{2},l_{3}$ & $l$ \\ \hline\hline
5,0,0 & 5 \\ \hline
0,5,0 & 5 \\ \hline
0,0,5 & 5 \\ \hline
4,1,0 & 5,4,3 \\ \hline
4,0,1 & 5,4,3 \\ \hline\hline
\end{tabular}
\begin{tabular}{||c|c||}
\hline\hline
$l_{1},l_{2},l_{3}$ & $l$ \\ \hline\hline
1,4,0 & 5,4,3 \\ \hline
1,0,4 & 5,4,3 \\ \hline
0,4,1 & 5,4,3 \\ \hline
0,1,4 & 5,4,3 \\ \hline\hline
\end{tabular}
\begin{tabular}{||c|c||}
\hline\hline
$l_{1},l_{2},l_{3}$ & $l$ \\ \hline\hline
3,0,0 & 3 \\ \hline
0,3,0 & 3 \\ \hline
0,0,3 & 3 \\ \hline
3,1,0 & 4,3,2 \\ \hline
3,0,1 & 4,3,2 \\ \hline\hline
\end{tabular}
\begin{tabular}{||c|c||}
\hline\hline
$l_{1},l_{2},l_{3}$ & $l$ \\ \hline\hline
1,3,0 & 4,3,2 \\ \hline
1,0,3 & 4,3,2 \\ \hline
0,3,1 & 4,3,2 \\ \hline
0,1,3 & 4,3,2 \\ \hline\hline
\end{tabular}

\bigskip
\end{center}

\noindent Thus we identify the states

\begin{center}
\begin{tabular}{||c|c|c|c||}
\hline\hline
&  &  &  \\ 
$l$ & $N$ & $\Delta $ & Parity \\ 
&  &  &  \\ \hline
$2$ & ? & $2.39F_{37}$ & + \\ \hline
$3$ & ? & $2.40G_{39}$ & - \\ \hline
$4$ & ? & $2.42H_{3,11}$ & + \\ \hline
$5$ & $2.60I_{1,11}$ & ? & - \\ \hline\hline
\end{tabular}

\smallskip
\end{center}

The seventh energy state (2.79 Gev) has $n+m+k=6$ $%
=2r_{1}+|m_{1}|+2r_{2}+|m_{2}|+2r_{3}+|m_{3}|$ which produces $%
|m_{1}|+|m_{2}|+|m_{3}|=6,4,2,0$ and $l_{1}+l_{2}+l_{3}\geq 6,4,2,0$. We
have then the possibilities\ below

\smallskip

\begin{center}
\begin{tabular}{||c||c|c|c|c|c|c|c||}
\hline\hline
$l_{1},l_{2},l_{3}$ & 6,0,0 & 0,6,0 & 0,0,6 & 5,1,0 & 5,0,1 & 1,5,0 & 1,0,5
\\ \hline
$l$ & 6 & 6 & 6 & 6,5,4 & 6,5,4 & 6,5,4 & 6,5,4 \\ \hline\hline
\end{tabular}
\end{center}

\smallskip

\noindent and the states

\smallskip

\begin{center}
\bigskip \rule[0.5in]{0in}{0.17in}

\begin{tabular}{||c|c|c|c||}
\hline\hline
&  &  &  \\ 
$l$ & $N$ & $\Delta $ & Parity \\ 
&  &  &  \\ \hline
$4$ & ? & ? & + \\ \hline
$5$ & ? & $2.75I_{3,13}$ & - \\ \hline
$6$ & $2.7K_{1,13}$ & $2.95K_{3,15}$ & + \\ \hline\hline
\end{tabular}
\end{center}

\bigskip

\noindent {\large {b) Baryons $\Sigma $ and $\Lambda $}}

Now let us do the calculation for $\Sigma $ and $\Lambda $. According to
Table 5.2 the first energy state (1.12 Gev) is $(n=0,m=0,k=0)$ and hence we
can have $l_{1}=0,l_{2}=0,l_{3}=0$ which yields $l=0$ and the states

\smallskip

\begin{center}
\begin{tabular}{||c|c|c|c||}
\hline\hline
&  &  &  \\ 
$l$ & $\Sigma $ & $\Lambda $ & Parity \\ 
&  &  &  \\ \hline
$0$ & $1.193P_{11}$ & $1.116P_{01}$ & + \\ \hline\hline
\end{tabular}

\smallskip
\end{center}

In the second energy level (1.43 Gev) $n+m=1,k=0$ which makes $%
2r_{1}+|m_{1}|+2r_{2}+|m_{2}|=1$ and $2r_{3}+|m_{3}|=0$. This actually makes 
$|m_{1}|+|m_{2}|=1$ and $|m_{3}|=0.$ That is, we have the condition $%
l_{1}+l_{2}\geq 1,$ $l_{3}\geq 0$ which allows us to choose the possibilities

\begin{center}
\begin{tabular}{||c||c|c|c||}
\hline\hline
$l_{1},l_{2},l_{3}$ & 1,1,0 & 1,0,1 & 0,1,1 \\ \hline
$l$ & 2,1,0 & 2,1,0 & 2,1,0 \\ \hline\hline
\end{tabular}
\end{center}

\noindent that produce the states

\begin{center}
\begin{tabular}{||c|c|c|c||}
\hline\hline
&  &  &  \\ 
$l$ & $\Sigma $ & $\Lambda $ & Parity \\ 
&  &  &  \\ \hline
$0$ & $1.385P_{13}$ & ? & + \\ \hline
$1$ & ? & $1.405S_{01}$ & - \\ \hline
$2$ & ? & ? & + \\ \hline\hline
\end{tabular}

\smallskip
\end{center}

\noindent and the state $1.48\Sigma $ is either $S_{13},S_{11}(l=1)$ or $%
F_{15}(l=2).$

In the third energy level (1.62 Gev) $n=m=0,k=1$ and we have $%
|m_{1}|=0,|m_{2}|=0$ and $|m_{3}|=1.$ That is, we have the condition $%
l_{1}\geq 0,l_{2}\geq 0,$ $l_{3}\geq 1$ which allows us to choose $%
l_{1}=l_{2}=0,l_{3}=1;l_{1}=l_{3}=1,l_{3}=0$; $l_{1}=0,l_{2}=l_{3}=1$, and
the states

\smallskip

\begin{center}
\begin{tabular}{||c|c|c|c||}
\hline\hline
&  &  &  \\ 
$l$ & $\Sigma $ & $\Lambda $ & Parity \\ 
&  &  &  \\ \hline
$0$ & $1.66P_{11}$ & $1.60P_{01}$ & + \\ \hline
$1$ & $
\begin{tabular}{c}
$1.62S_{11}$ \\ 
$1.58D_{13}$%
\end{tabular}
$ & 
\begin{tabular}{c}
$1.67S_{01}$ \\ 
$1.52D_{03}$%
\end{tabular}
& - \\ \hline
$2$ & ? & ? & + \\ \hline\hline
\end{tabular}
\end{center}

\smallskip

\noindent and then the state $1.56\Sigma $ is probably $F_{15}($ $l=2)$.

The fourth energy level (1.74 Gev) has $n+m=2=2r_{1}+|m_{1}|+2r_{2}+|m_{2}|$
and $k=2r_{3}+|m_{3}|=0$, and thus we obtain $|m_{1}|+|m_{2}|=2,0$ and $%
|m_{3}|=0.$ Hence we have the condition $l_{1}+l_{2}\geq 2,0$ and $l_{3}\geq
0$. We can then choose $%
l_{1}=2,l_{2}=l_{3}=0;l_{1}=l_{3}=0,l_{2}=2;l_{1}=l_{2}=1,l_{3}=0$ and thus
we can identify the states

\smallskip

\begin{center}
\smallskip 
\begin{tabular}{||c|c|c|c||}
\hline\hline
&  &  &  \\ 
$l$ & $\Sigma $ & $\Lambda $ & Parity \\ 
&  &  &  \\ \hline
$0$ & $1.77P_{11}$ & $1.81P_{01}$ & + \\ \hline
$1$ & $
\begin{tabular}{c}
$1.75S_{11}$ \\ 
$1.67D_{13},1.775D_{15}$%
\end{tabular}
$ & $
\begin{tabular}{c}
$1.80S_{01}$ \\ 
$1.69D_{03}$%
\end{tabular}
$ & - \\ \hline
$2$ & ? & $1.82F_{05}$ & + \\ \hline\hline
\end{tabular}

\smallskip
\end{center}

\noindent and then the level $1.69\Sigma $ is probably $F_{15}(l=2)$.

In the fifth energy level (1.93 Gev) $n+m=1=2r_{1}+|m_{1}|+2r_{2}+|m_{2}|$
and $k=1=2r_{3}+|m_{3}|$, and thus we obtain $|m_{1}|+|m_{2}|=1$ and $%
|m_{3}|=1.$ Hence we have the condition $l_{1}+l_{2}\geq 1$ and $l_{3}\geq 1$%
. We can then have the sets $l_{1}=1,l_{2}=0,l_{3}=1;$ $l_{1}=0,$ $%
l_{2}=1,l_{3}=1.$ Both yield $l=2,1,0$ and we can identify the states

\bigskip

\begin{center}
\begin{tabular}{||c|c|c|c||}
\hline\hline
&  &  &  \\ 
$l$ & $\Sigma $ & $\Lambda $ & Parity \\ 
&  &  &  \\ \hline
$0$ & $1.84P_{11},1.84P_{13}$ & $1.89P_{03}$ & + \\ \hline
$1$ & $1.94D_{13}$ & $1.83D_{05}$ & - \\ \hline
$2$ & $1.915F_{15}$ & $?$ & + \\ \hline\hline
\end{tabular}

\smallskip
\end{center}

The sixth energy level (2.05 GeV) has $n+m=3=2r_{1}+|m_{1}|+2r_{2}+|m_{2}|$
and $k=0=2r_{3}+|m_{3}|$, and thus we obtain $|m_{1}|+|m_{2}|=3,1$ and $%
|m_{3}|=0.$ Hence we have the condition $l_{1}+l_{2}\geq 3,1$ and $l_{3}\geq
0$. We can then have the sets $l_{1}=2,l_{2}=1,l_{3}=0;$ $l_{1}=1,$ $%
l_{2}=2,l_{3}=0$ which make $l=3,2,1,$ for $l_{1}+l_{2}\geq 3$ and the sets $%
l_{1}=1,l_{2}=1,l_{3}=0;$ $l_{1}=1,$ $l_{2}=1,l_{3}=0$ which make $l=2,1,0,$
for $l_{1}+l_{2}\geq 1$. We can identify the states

\smallskip

\begin{center}
\begin{tabular}{||c|c|c|c||}
\hline\hline
&  &  &  \\ 
$l$ & $\Sigma $ & $\Lambda $ & Parity \\ 
&  &  &  \\ \hline
$0$ & $2.08P_{13}$ & ? & + \\ \hline
$1$ & $2.00S_{11}$ & ? & - \\ \hline
$2$ & $2.07F_{15},2.03F_{17}$ & $2.02F_{07}$ & + \\ \hline
$3$ & $?$ & ? & - \\ \hline\hline
\end{tabular}

\smallskip
\end{center}

In the seventh energy level (2.12 GeV) $%
n=0=2r_{1}+|m_{1}|,m=0=2r_{2}+|m_{2}|,k=2=2r_{3}+|m_{3}|$ and thus $%
|m_{1}|=0,|m_{2}|=0,$ $|m_{3}|=2,0.$ Hence we have the condition $l_{1}\geq
0,l_{2}\geq 0$ and $l_{3}\geq 2,0$. We can then choose the sets $%
l_{1}=0,l_{2}=0,l_{3}=2;$ $l_{1}=0,$ $l_{2}=0,l_{3}=3$ which make $l=3,2$,
and the states

\smallskip

\begin{center}
\begin{tabular}{||c|c|c|c||}
\hline\hline
&  &  &  \\ 
$l$ & $\Sigma $ & $\Lambda $ & Parity \\ 
&  &  &  \\ \hline
$l=2$ & $?$ & $2.11F_{05}$ & + \\ \hline
$l=3$ & $2.10G_{17}$ & $2.10G_{07}$ & - \\ \hline\hline
\end{tabular}

\smallskip
\end{center}

Unfortunately, the angular momenta of the other energy levels have not been
found but they can surely be explained according to what was developed above.

\bigskip

\noindent {\large {c) Baryons $\Xi $}}

For these baryons only some angular momenta are known. The first energy
level (1.31 GeV) has $n=0,m=0.l=0$ which make $l_{1}=l_{2}=l_{3}=0$ and $l=0$
and is thus a $P$ state. Therefore we obtain

\smallskip

\begin{center}
\begin{tabular}{||c|c|c||}
\hline\hline
&  &  \\ 
$l$ & $\Xi $ & Parity \\ 
&  &  \\ \hline
$0$ & $1.318P_{11}$ & + \\ \hline\hline
\end{tabular}

\smallskip
\end{center}

In the second energy level (1.62 GeV) $%
n=1=2r_{1}+|m_{1}|,m=0=2r_{2}+|m_{2}|,k=0=2r_{3}+|m_{3}|$ and thus $%
|m_{1}|=1,|m_{2}|=0,$ $|m_{3}|=0.$ Hence we have the condition $l_{1}\geq
1,l_{2}\geq 0$ and $l_{3}\geq 0$. We can then have the sets $%
l_{1}=1,l_{2}=0,l_{3}=0;$ $l_{1}=1,$ $l_{2}=1,l_{3}=0$ which make $l=2,1,0$,
and the states

\smallskip

\begin{center}
\begin{tabular}{||c|c|c||}
\hline\hline
&  &  \\ 
$l$ & $\Xi $ & Parity \\ 
&  &  \\ \hline
$0$ & $1.53P_{13}$ & + \\ \hline
$1$ & ? & - \\ \hline
$2$ & $?$ & + \\ \hline\hline
\end{tabular}

\smallskip
\end{center}

\noindent and thus the two levels $1.62$ and $1.69$ are probably either $S$, 
$D$ or $F$ states.

The third energy level (1.81GeV) has $%
n=0=2r_{1}+|m_{1}|,m+k=1=2r_{2}+|m_{2}|+2r_{3}+|m_{3}|$ and thus $%
|m_{1}|=0,|m_{2}|+$ $|m_{3}|=1.$ Hence we have the condition $l_{1}\geq
0,l_{2}+$ $l_{3}\geq 1$. We can then have the sets $l_{1}=0,l_{2}=1,l_{3}=0;$
$l_{1}=0,$ $l_{2}=0,l_{3}=1$ which make $l=1$, and the state

\smallskip

\begin{center}
\begin{tabular}{||c|c|c||}
\hline\hline
&  &  \\ 
$l$ & $\Xi $ & Parity \\ 
&  &  \\ \hline
$l=1$ & $1.82D_{13}$ & - \\ \hline\hline
\end{tabular}

\smallskip
\end{center}

In the fourth energy level (1.93GeV) $%
n=2=2r_{1}+|m_{1}|,m=0=2r_{2}+|m_{2}|,k=0=2r_{3}+|m_{3}|$ and thus $%
|m_{1}|=2,0,|m_{2}|=0,$ $|m_{3}|=0.$ Hence we have the condition $l_{1}\geq
0,2,l_{2}\geq 0$ and $l_{3}\geq 0$. We can then choose the set $%
l_{1}=2,l_{2}=0,l_{3}=0$ which produces $l=2$, and the state 1.93GeV is
probably an $F$ state.

\paragraph{\noindent {\protect\large 5.1.3 Relation between energy and
angular momentum}}

From Eqs. 88 and 89 we have 
\begin{equation*}
\begin{array}{c}
E=(2r_{1}+|m_{1}|+1)h\nu _{1}+(2r_{2}+|m_{2}|+1)h\nu
_{2}+(2r_{3}+|m_{3}|+1)h\nu \\ 
l_{1}+l_{2}+l_{3}\geq l\geq ||l_{1}-l_{2}|-l_{3}|\text{ with }l_{1}\geq
|m_{1}|,l_{2}\geq |m_{2}|,l_{3}\geq |m_{3}|.
\end{array}
\end{equation*}

\noindent in which $l_{1}$, $l_{2}$ and $l_{3}$ are the quantum numbers of
the angular momenta $\overrightarrow{L_{1}}$, $\overrightarrow{L_{2}}$, and $%
\overrightarrow{L_{3}}$, and $m_{1}$,$m_{2}$, $m_{3}$ are the quantum
numbers of their projections on the Z axis, respectively. Therefore, we
clearly see that levels with large energies have large angular momenta as is
quite evident from the experimental data.

\bigskip

\paragraph{\protect\large 5.1.4 The sizes of baryons}

The solution in Cartesian coordinates is also useful for calculating in a
quite simple manner the average size of a baryon. As is known the average
potential energy of each oscillator is half of the total energy, that is,

\begin{equation}
<\frac{1}{2}k{{\xi _{i}}^{2}>=}\frac{h\nu _{i}}{2}(v_{i}+\frac{1}{2})
\end{equation}

\noindent but since there are two directions for each quark in the plane
there actually are two oscillators per quark and thus we have the potential
energy $E_{q}$ associated to each quark 
\begin{equation}
E_{q}=h\nu _{q}(n_{i}+1)
\end{equation}

\noindent where $n_{i}=0,1,2,3,...$ and $h\nu _{q}$ is the constituent quark
mass constant. Thus taking into account Eq. 06 and the above fact on the
relation between the total energy and the potential energy for an oscillator
it can be written that

\begin{equation}
\begin{array}{c}
E(n,m,k)=h\nu _{1}(n+1)+h\nu _{2}(m+1)+h\nu _{3}(k+1)= \\ 
=2\times \left( <\frac{1}{2}k_{1}\eta {{_{1}}^{2}>+}<\frac{1}{2}k_{2}\eta {{%
_{2}}^{2}>+}<\frac{1}{2}k_{3}\eta {{_{3}}^{2}>}\right) = \\ 
=\ <k_{1}\eta {{_{1}}^{2}>+}<k_{2}\eta {{_{2}}^{2}>+}<k_{3}\eta {{_{3}}^{2}>}
\end{array}
\end{equation}

\noindent where $\eta {{_{i}}^{2}=}$ ${{\xi _{ij}}^{2}+{\xi _{ik}}^{2}}$ in
which $j$ and $k$ are the two orthogonal directions of the two oscillators.
One can then make the association 
\begin{equation}
h\nu _{i}(n+1)=\,<k_{1}\eta {{_{1}}^{2}>}
\end{equation}

\noindent and hence the average radius $\mathcal{R\ }$\ of a baryon can be
given by

\begin{equation}
\begin{array}{c}
\mathcal{R}(n,m,k)=\left( \sqrt{<\eta {{_{1}}^{2}>}<\eta {{_{2}}^{2}>}<\eta {%
{_{3}}^{2}>}}\right) ^{1/3}= \\ 
=\left( \sqrt{\frac{h\nu _{1}(n+1)}{k_{1}}\frac{h\nu _{2}(m+1)}{k_{2}}\frac{%
h\nu _{3}(k+1)}{k_{3}}}\right) ^{1/3}.
\end{array}
\end{equation}

It is quite obvious that the application of the above formula should be
first done to the proton. In the fundamental level $n=m=k=0$ and $h\nu
_{1}=h\nu _{2}=h\nu _{3}=$ $0.31$GeV, and making the reasonable supposition
that $k_{1}=k_{2}=k_{3}=k$, thus 
\begin{equation}
\mathcal{R}_{0}=\mathcal{R}(0,0,0)=\sqrt{\frac{h\nu _{1}}{k}}.
\end{equation}

\noindent If one uses for the average size of a proton$^{15}$ the figure of $%
\sqrt{0.72}$fm = 0.85fm one has $k\approx 0.5$GeV/fm$^{2}$ which is a very
reasonable figure because if it is multiplied by the characteristic distance
of 1fm (of course!) the constant $k^{\prime }\approx 0.5$GeV/fm is obtained
which is quite close to the value of the constant $K$ used in the QCD
motivated potential$^{16,17}$ 
\begin{equation}
V_{QCD}=-CF\frac{\alpha _{s}}{r}+Kr
\end{equation}
which is assumed to be of the order of 1GeV/fm.

From Table 5.1 one has that for $n=m=k=2$ the energy of a proton is about
2.80GeV which gives an average radius of about 1.39fm and hence one sees
that the size of a baryon does not change much with the its energy.
Therefore it can be said that the smallest radius of a proton is about 0.8fm
and its largest radius is or the order of 1.4fm.

For the ground states of $\ \Sigma ^{-}$ and $\Xi ^{-}$ reference 15 gives,
respectively, the radii $\sqrt{0.54}$fm = 0.73fm and $\sqrt{0.43}$fm =
0.66fm. In terms of quarks $\Sigma ^{-}$ is $dus$ and therefore one should
have $k_{2}=k_{3}$ and $k_{1}\approx k\approx 0.5$GeV/fm$^{2}$ and 
\begin{equation}
\mathcal{R}(n,m,k)=\left( \sqrt{\frac{h\nu _{1}(n+1)\times h\nu
_{2}(m+1)\times h\nu _{3}(k+1)}{k_{1}(k_{3})^{2}}}\right) ^{1/3}.
\end{equation}
Using the above value it is obtained that 
\begin{equation*}
0.73=\left( \sqrt{\frac{0.31\times 0.31\times 0.5}{0.5\times (k_{3})^{2}}}%
\right) ^{1/3}
\end{equation*}

\noindent which yields $k_{3}=0.80$GeV/fm$^{2}$. From Table 5.2 it is seen
that for $n+m=5$, $k=1$, the energy is 3.17GeV which is the highest energy
level up to now. If one takes, for example, $n=2$, $m=3$ one has an average
radius of about 1.24fm. Also it is found that the average radii of \ $\Sigma
^{-}$\ are much smaller than those of the proton for levels with the same
quantum numbers $n$, $m$, $k$.

Now one can turn to $\Xi ^{-}$ which in terms of quarks is $dss$. Then it is
expected to have the same $k_{3}\approx 0.80$GeV/fm$^{2}$ (two of them)
above and a new $k$, which can be called $k_{ss}$. Using the ground state
radius of \ $\Xi ^{-}$ (0.66fm) one obtains $k_{ss}\approx 1.47$GeV/fm$^{2}$%
. For the excited states the average radius (in fm) is thus 
\begin{equation}
\mathcal{R}(n,m,k)=\left( \sqrt{\frac{0.31(n+1)\times 0.5(m+1)\times 0.5(k+1)%
}{1.47(0.8)^{2}}}\right) ^{1/3}
\end{equation}
which for the highest known excited state 2.55GeV ($n=4$, $m=k=0$) gives $%
\mathcal{R}(4,0,0)\approx 1.48$fm. Using the value $k_{ss}\approx 1.47$GeV/fm%
$^{2}$ the radius of the ground state of $\Omega $ is estimated to be about
0.58fm.

Putting together the above values the very important table below is obtained
for the constant $k$ (which is a sort of constant of confinement) in terms
of the pairs of interacting quarks.

\vskip .2in

\begin{center}
{\normalsize 
\parbox{2.3in}
{Table 8. The harmonic oscillator constant $k$ for some pairs of 
interacting quarks in terms of their reduced mass} }
\end{center}

\vskip .1in

\begin{center}
{\normalsize 
\begin{tabular}{c|ccc}
\hline\hline
&  &  &  \\ 
& u-u & u-s & s-s \\ 
&  &  &  \\ \hline\hline
&  &  &  \\ 
$k$(\text{GeV/fm}$^{2}$\text{)} & 0.5 & 0.8 & 1.47 \\ 
&  &  &  \\ 
$\mu$ \text{(in GeV/c}$^{2}$\text{)} & 0.15 & 0.188 & 0.25 \\ \hline
\end{tabular}
}
\end{center}

\vskip .2in

\noindent The table shows that $k$ increases with the reduced mass of the
pair of interacting quarks. When the data are fitted to a polynomial up to
second order in the reduced mass of the pair of interacting quarks the
following polynomial is obtained 
\begin{equation}
k(\mu )=0.1188-1.7561\mu +28.6508\mu ^{2}.
\end{equation}
It is interesting that the coeficient of the last term is quite large and
thus the first derivative increases very rapidly with $\mu $. As more
massive quarks are considered the degree of the polynomial may increase but
just to have a lower bound one can calculate the value of $k$ for the
interaction between two top quarks. The above formula gives $k=7416$GeV/fm$%
^{2}$. If the above data are fitted to a polynomial with a higher degree,
for example, $k(\mu )=A+B\mu ^{2}+C\mu ^{3}$, the following values are
obtained: $A=-0.0268$, $B=23.3794$, and $C=0.2278$. Since the value of $C$
is small and $B$ is of the same order of 28.6508, the first polynomial (Eq.
99) is a good approximation. If it is used for obtaining the $k$ between
quarks $u$ and $c$ one has $k(uc)\approx 1.53$GeV/fm$^{2}$, $k(sc)\approx
3.7 $GeV/fm$^{2}$, $k(cc)\approx 19$GeV/fm$^{2}$. And then one has that the
radii of the ground states of the charmed baryons $\Lambda _{c}^{+}$, $%
\Sigma _{c}^{++}$, $\Sigma _{c}^{+}$ and $\Sigma _{c}^{0}$ are about 
\begin{equation*}
R_{c}\approx \left( \sqrt{\frac{0.31\times 0.31\times 1.7}{0.5(1.53)^{2}}}%
\right) ^{1/3}\approx 0.7\text{fm}
\end{equation*}

\noindent which is not so small due to the influence of the interaction
between the two $u$ quarks. As to $\Omega _{c}$ its ground state has a radius

\begin{equation*}
R_{ssc}\approx \left( \sqrt{\frac{0.5\times 0.5\times 1.7}{1.53(3.7)^{2}}}%
\right) ^{1/3}\approx 0.5\text{fm}
\end{equation*}

\noindent and the ground state of the $ccc$ baryon has the quite small
radius of just

\begin{equation*}
R_{ccc}\approx \sqrt{\frac{1.7}{19}}\approx 0.3\text{fm.}
\end{equation*}

\noindent Since the value of $k(cc)\approx 19$GeV/fm$^{2}$ was obtained by
means of an extrapolation the above figure of $R_{ccc}$ should be taken as a
crude approximation.

In the case of the ${ttt}$ baryon an even cruder number is gotten for its
radius because its value for $k$ is expected to be larger than the above
figure of $7416$GeV/fm$^{2},$\ but it is instructive anyway to calculate its
order of magnitude which in this case produces an upper bound for its
radius. Therefore one can say that the radius of the ground state of the ${%
ttt}$ system is 
\begin{equation*}
R_{ttt}<\sqrt{\frac{174}{7416}}=0.15\text{fm.}
\end{equation*}

\noindent which is a very important number just because the top quark is the
most massive quark.

Since in this work the motion of the plane where quarks are sitting was not
taken into account conclusions can not be drawn on the shape of baryons
using the above figures.

\paragraph{\protect\large 5.1.5 Spin-Orbit Interaction}

We clearly notice that the splitting of some levels are caused by the
spin-orbit interaction. For example, consider the states $1.90S_{31}$ and $%
1.91P_{31}$ of $\Delta $ which differ by the values of $l=1$ and $l=0$,
respectively. Since we are assuming a harmonic potential and as the
spin-orbit term is proportional to $\frac{1}{r}\frac{dV}{dr}$ we can write

\begin{equation}
\Delta E_{SL}\approx C_{1}\overrightarrow{S}.\overrightarrow{L}=C\left[
j(j+1)-l(l+1)-s(s+1)\right]
\end{equation}
for $N$ and $\Delta $ baryons which have quarks with equal masses. Using for
the above case $j=1/2,s=1/2$ we find $C\approx 5$MeV which shows that the
influence of the spin-orbit interaction is small. Considering the levels $%
1.91P_{31}$ and $1.92P_{33}$ we find $C\approx 3.3$MeV which is of the same
order of the above $C$. The same holds in the case of the other baryons: for
example, consider the states $1.75S_{11}$ and $1.77P_{11}$ of $\Sigma $ or
the states $1.80S_{01}$ and $1.81P_{01}$ of $\Lambda .$ We see that there is
a small energy difference between these states.

\bigskip Concluding this section we can infer that the outstanding agreement
with the experimental data implies that quarks do not move at relativistic
speeds inside baryons. As has been argued by Lichtenberg$^{15}$ and others
it is hard to see how $SU(6)$ is a good approximate dynamical symmetry of
baryons if quarks move at relativistic velocities inside baryons.

We clearly see that the masses of baryons are expressed quite well by the
simple model above described. It lends support to the general framework of
having quarks as the basic building particles of baryons. Therefore, it
agrees well with QCD.

It is not an easy task to include the anharmonicity of the potential but it
is important to draw some conclusions on it because it is linked directly to
the number of bound states. When we add a negative anharmonic term to a
harmonic oscillator we obtain the energies$^{16}$ 
\begin{equation}
E_{n}=C+h{\nu }\left( n+\frac{1}{2}\right) -A\left( n+\frac{1}{2}\right) ^{2}
\end{equation}
\noindent in which the third term takes into account the anharmonicity of
the potential. This term can not be larger than the second term. Thus one
should always have 
\begin{equation}
n<\frac{1}{2}\left( \frac{h\nu }{A}-1\right) .
\end{equation}
\noindent This $n$ is the number of bound states. As we can see from the
tables the anharmonicity is below 5\% in general. Considering just one
harmonic oscillator,the maximum value of $n$, that is, the number of bound
states is about $\frac{1}{2}(20-1)\approx 10$. Of course this is just a
rough number because there are more oscillators to take into account. But,
anyway, the important conclusion is that a small anharmonicity allows many
bound states. The lack of a significant anharmonicity implies that the
potential is very symmetric, and this is expected since we are dealing with
interactions between two quarks. As we will see shortly in the case of $q%
\bar{q}$ the potential is not as symmetric.

\bigskip

\subsection{\protect\Large Generalization of the Gell-Mann-Okubo Mass Formula%
}

\noindent

The Gell-Mann-Okubo mass formula 
\begin{equation}
M=M_{0}+M_{1}Y+M_{2}\left( I(I+1)-\frac{Y^{2}}{4}\right)
\end{equation}
\noindent where $M_{0}$, $M_{1}$ and $M_{2}$ are suitable constants, $I$ is
the isospin, and $Y$ is the hypercharge, has been widely used as a relation
among the masses of baryon states belonging either to an octet or to a
decuplet. This is a phenomenological formula ``with no clear physical
reasons for the assumptions on which it is based''$^{20}$. As we will show
shortly the reason behind the above mass formula is the general formula for
the mass of a baryon 
\begin{equation}
E_{n,m,k}=\hbar {\nu _{1}}(n+1)+\hbar {\nu _{2}}(m+1)+\hbar {\nu _{3}}(k+1).
\end{equation}
\noindent For the decuplet of $SU_{3}$(u,d,s) Eq. (103) becomes 
\begin{equation}
M=M_{0}+M_{1}Y
\end{equation}
\noindent where $Y$ is the hypercharge. The relation among the masses of
baryons of the $SU_{3}$(u,d,s) decuplet is given by 
\begin{equation}
M_{\Sigma }-M_{\Delta }=M_{\Xi }-M_{\Sigma }=M_{\Omega ^{-}}-M_{\Xi }.
\end{equation}
\noindent According to Eq. (104) the equality of the first two terms of Eq.
(106) is given by 
\begin{eqnarray}
0.31(n+1) &+&0.5(m+k+2)+0.31(n+m+k+3)=  \notag \\
&&2\left( 0.31(n+m+2)+0.5(k+1)\right)
\end{eqnarray}
\noindent which is satisfied for any $n$, and $m=k$. Actually, instead of $%
\Delta $ we may have either $\Delta $ or $N$. For example(see Tables 5.1,
5.2 and 5.3),

\bigskip

\begin{itemize}
\item  $n=0$, $k=m=0$, $1.12 - 0.93 = 1.31 - 1.12 = 0.19$;

\item  $n=0$, $k=m=1$, $1.93 - 1.55 = 2.31 - 1.93 = 0.38$;

\item  $n=1$, $k=m=0$, $1.43 - 1.24 = 1.62 - 1.43 = 0.19$;

\item  $n=1$, $k=m=1$, $2.24 - 1.86 = 2.62 - 2.24 = 0.38$;

\item  $n=2$, $k=m=0$, $1.74 - 1.55 = 1.93 - 1.74 = 0.19$;

\item  $n=3$, $k=m=0$, $2.05 - 1.86 = 2.24 - 2.05 = 0.19$;

\item  ...
\end{itemize}

\noindent The equality of the first term with the third term of Eq. (106)
yields 
\begin{eqnarray}
0.31(n+m+2) &+&0.5(k+1)-0.31(n+m+k+3)=  \notag \\
0.5(n+m+k+3) &-&0.31(n+1)-0.5(m+k+2)
\end{eqnarray}
\noindent which is satisfied for any $n,m,k$. Again, instead of $\Delta $ we
may have $N$. For example(observe Tables 5.1, 5.2, 5.3 and 5.4),

\begin{itemize}
\item  $n=m=k=0$, $1.12 - 0.93 = 1.5 - 1.31 = 0.19$;

\item  $n=0$, $m+k=1$, $1.43 - 1.24 = 2.0 - 1.81 = 0.19$;

\item  $n=k=0$, $m=2$, $1.74 - 1.55 = 2.5 - 2.31 = 0.19$;

\item  ...
\end{itemize}

\noindent Finally, equating the second and third terms of Eq. (106) one
obtains 
\begin{eqnarray}
0.5(n+m+k+3) &+&0.31(n+m+2)+0.5(k+1)=  \notag \\
&&2\left( 0.31(n+1)+0.5(m+k+2)\right)
\end{eqnarray}
\noindent which is satisfied if $n=m$ for any value of $k$. As examples one
finds(see Tables 5.2, 5.3 and 5.4)

\begin{itemize}
\item  $n=m=k=0$, $1.31 - 1.12 = 1.5 - 1.31 = 0.19$;

\item  $n=m=0$, $k=1$, $1.81 - 1.62 = 2 - 1.81 = 0.19$;

\item  $n=m=1$, $k=0$, $2.12 - 1.74 = 2.5 - 2.12 = 0.38$;

\item  ...
\end{itemize}

\noindent For an octet of $SU_{3}$(u,d,s) one obtains 
\begin{equation}
3M_{\Lambda }+M_{\Sigma }=2M_{N}-M_{\Xi }
\end{equation}
\noindent which in terms of Eq. (104) becomes 
\begin{eqnarray}
2\left( 0.31(n+m+2)\right. &{+}&\left. 0.5(k+1)\right) =  \notag \\
0.31(n+m+k+3) &{-}&0.31(n+1)-0.5(m+k+2).
\end{eqnarray}
\noindent This equation is satisfied if $k=m$ for any $n$. For example, one
has(see Tables 5.1, 5.2 and 5.3)

\begin{itemize}
\item  $n=m=k=0$, $2{\times}1.12 = 0.93 + 1.31$;

\item  $n=1$, $m=k=0$, $2{\times}1.43 = 1.24 + 1.62$;

\item  $n=2$, $m=k=0$, $2{\times}1.74 = 1.55 + 1.93$;

\item  $n=3$, $m=k=0$, $2{\times}2.05 = 1.86 + 2.24$;

\item  $n=0$, $m=k=1$, $2{\times}1.93 = 1.55 + 2.31$;

\item  $n=k=m=1$, $2{\times}2.24 = 1.86 + 2.62$;

\item  ... .
\end{itemize}

Let us now try to relate the constants $M_{0}$ and $M_{1}$ to the quark
masses. Let us consider, for example, the decuplet of $SU_{3}$(u,d,s). In
terms of the hypercharge the masses of the particles are described by 
\begin{equation}
M_{\Omega ^{-}}=M_{0}-2M_{1};
\end{equation}
\begin{equation}
M_{\Xi }=M_{0}-M_{1};
\end{equation}
\begin{equation}
M_{\Sigma }=M_{0};
\end{equation}
\begin{equation}
M_{\Delta }=M_{0}+M_{1}.
\end{equation}
\noindent As we calculated above from the masses of $\Xi $, $\Sigma $ and $%
\Delta $ one finds that $m=k$(any $n$) and from the masses of $\Omega ^{-}$, 
$\Xi $ and $\Sigma $ one has $n=m$(any $k$). Therefore, in terms of $Y$ the
masses of $\Xi $, $\Sigma $ and $\Delta $ are given by 
\begin{equation}
M_{n,m}(Y)=0.31(n+m+2)+0.5(m+1)-0.19(m+1)Y
\end{equation}

\noindent and the mass of $\Omega ^{-}$ is the above formula with $n=m$,
that is, 
\begin{equation}
M_{\Omega ^{-}}(Y)=(1.12-0.19Y)(n+1).
\end{equation}

From the $SU(4)$ multiplets of baryons made of $u$, $d$, $s$, and $c$
quarks, and considering Eq. (104) one obtains, for example, 
\begin{equation}
M_{\Omega _{ccc}}-M_{\Xi _{cc}}=M_{\Xi _{cc}}-M_{\Sigma _{c}}=M_{\Sigma
_{c}}-M_{\Delta };
\end{equation}
\noindent 
\begin{equation}
M_{\Omega _{ccc}}-M_{\Omega _{cc}}=M_{\Omega _{cc}}-M_{\Omega
_{c}}=M_{\Omega _{c}}-M_{\Omega }
\end{equation}
\noindent and 
\begin{equation}
2M_{\Xi _{cc}}=M_{\Omega _{ccc}}+M_{\Sigma _{c}}
\end{equation}
\noindent or more generally, one obtains 
\begin{equation}
M_{q_{1}q_{1}q_{1}}-M_{q_{2}q_{1}q_{1}}=M_{q_{2}q_{1}q_{1}}-M_{q_{2}q_{2}q_{1}}=M_{q_{2}q_{2}q_{1}}-M_{q_{2}q_{2}q_{2}}
\end{equation}
\noindent and 
\begin{equation}
2M_{q_{1}q_{1}q_{2}}=M_{q_{1}q_{1}q_{1}}+M_{q_{1}q_{2}q_{2}}
\end{equation}
\noindent in which we can consider $SU(6)$, that is, $q_{i}$ may be $u$, $d$%
, $c$, $s$ $b$, and $t$. In the case of considering $u$ and $d$, we may have
the combinations $ud$, $uu$, and $dd$ for $q_{i}q_{i}$. We also may have 
\begin{equation}
M_{q_{1}q_{2}q_{3}}-M_{q_{4}q_{2}q_{3}}=M_{q_{1}q_{i}q_{j}}-M_{q_{4}q_{i}q_{j}}.
\end{equation}

We conclude this section saying that the Gell-Mann-Okubo mass formula is a
natural consequence of the pairwise interacting harmonic potential among
quarks.

We find in the literature several relations among the masses of baryons.
They are, actually, just special cases of the above formulas.

\vspace*{0.5in}

\subsection{{\protect\Large The Bound States of }$Q\overline{Q}$ 
{\protect\Large Mesons\newline
}}

\bigskip

According to QCD a meson is a colorless state which transforms under $SU_{3}$
as 
\begin{equation*}
q^{in}q_{jn}=\bar{q}_{in}q_{jn}.
\end{equation*}
\noindent Following the theory presented above it is reasonable to admit
that there is also a sort of molecular potential in the interaction between
a quark and an antiquark. Let us begin with the heavy mesons $c\bar{c}$ and $%
b\bar{b}$.

It is important to say that there are many important papers in the
literature on the meson spectrum. One of the pioneers is the work of Eichten
et al.$^{21}$. Other important works are those of Godfrey and Isgur$^{22}$,
Gupta et al.$^{23,24,25}$, and Itoh et al.$^{26,27}.$

The $q\bar{q}$ potential is not known and some \textit{ad hoc} potentials
have been used, especially in the description of the energy levels of
quarkonia. The most successful of all is a Coulomb-like potential with a
confining term, the Cornel potential$^{20,28}$ 
\begin{equation}
V(r)=C-\frac{K}{r}+\frac{r}{a^{2}}.
\end{equation}
\noindent The second term is completely justified in terms of the vectorial
color field of QCD and, hence, we can make an analogy with the
electromagnetic field. The confining term is not very consistent because it
produces an enormous attractive quark-interquark force independent of $r$,
but as we know the strong force decreases rapidly to zero when $r$ is larger
than a few fermi. Moreover it allows an infinite number of bound states
which is not true. Actually the number of bound states is quite small.

We will see shortly that a much more realistic potential is a molecular-type
potential like the Morse potential which is widely used in the description
of the vibrational and rotational spectra of diatomic molecules. Actually,
the Cornel potential is successful because it is close to Kratzer molecular
potential$^{29}$ 
\begin{equation}
V(r)=-2D\left( \frac{a}{r}-\frac{1}{2}\frac{a^{2}}{r^{2}}\right) .
\end{equation}
\noindent which at the bottom is less roundish than Morse potential.

Taking into account a centrifugal term 
\begin{equation}
V_{c}(r)=\frac{{\hbar }^{2}{\vec{L}}^{2}}{2Mr^{2}},
\end{equation}

\noindent where $M$ is the reduced mass of the system, and using
non-relativistic quantum mechanics one obtains that either Morse or Kratzer
potential produces the energy levels given by$^{30}$ 
\begin{eqnarray}
E_{nl} &=&C+h{\nu }\left( n+\frac{1}{2}\right) -A\left( n+\frac{1}{2}\right)
^{2}+Rl(l+1)  \notag \\
&&-VR\left( n+\frac{1}{2}\right) \left( l+\frac{1}{2}\right) ^{2}+....
\end{eqnarray}
\noindent where C, $h\nu $, A, R and VR are constants, and $n$ and $l$ are
the integers 0,1,2,3,4,5,... The first term is a constant related to the
depth of the potential, the second describes harmonic vibrations, the third
term takes into account the anharmonicity of the potential, the fourth term
describes rotations with constant moment of inertia, and the fifth term
shows the coupling between vibrations and rotations. In this work the last
term will be disregarded. The masses of all particles have been taken from
reference 13.

\bigskip

\subsubsection{\protect\bigskip {\protect\large Heavy Mesons}}

\paragraph{{\protect\large {a) $c\bar{c}$ Bound States}}}

Let us consider the $c\bar{c}$ system first. The first two levels, $\eta
_{c}(1S)(2979.8MeV)$ and $J/\Psi (1S)(3096.9MeV)$ are a hyperfine doublet,
that is, the result of a splitting caused by the spin-spin interaction. The
splitting can be calculated by 
\begin{equation}
{\Delta }E_{\vec{s_{1}}{\cdot }\vec{s_{2}}}=A\frac{\vec{s_{1}}{\cdot }\vec{%
s_{2}}}{{m_{1}}m_{2}}
\end{equation}
\noindent in which 
\begin{equation*}
\vec{s_{1}}{\cdot }\vec{s_{2}}=\left\{ 
\begin{array}{ll}
\frac{1}{4}{\hbar }^{2} & \mbox{for $S=1$} \\ 
-\frac{3}{4}{\hbar }^{2} & \mbox{for $S=0.$}
\end{array}
\right.
\end{equation*}
\noindent The difference between the two levels $S_{0}$ and $S_{1}$ is
117.1MeV. Removing the splitting one has the degenerate level $c\bar{c}_{0}$
with an energy of $3096.9MeV-\frac{1}{4}117.1MeV=2979.8MeV+\frac{3}{4}%
117.1MeV=3067.6MeV$. The levels \newline
$\Psi (2S)(3686.0MeV)$ and $\Psi (3769.9MeV)$ are vector mesons. Taking the
spin-spin energy off one obtains the energies $3686.0MeV-\frac{1}{4}%
117.1MeV=3656.7MeV$, and $3769.9MeV-\frac{1}{4}117.1MeV=3740.6MeV$. Applying
our model to these three levels one has

\begin{equation}
C_{c}+h\nu _{c}\left( 0+\frac{1}{2}\right) -A_{c}\left( 0+\frac{1}{2}\right)
^{2}=3067.6
\end{equation}
\begin{equation}
C_{c}+h\nu _{c}\left( 1+\frac{1}{2}\right) -A_{c}\left( 1+\frac{1}{2}\right)
^{2}=3656.7
\end{equation}
\begin{equation}
C_{c}+h\nu _{c}\left( 2+\frac{1}{2}\right) -A_{c}\left( 2+\frac{1}{2}\right)
^{2}=3740.6.
\end{equation}
\noindent From these equations one obtains $C_{c}=2583.6$MeV, $A_{c}=252.6$%
MeV, and $h\nu _{c}=1094.0$MeV. The constant $C_{c}$ is positive just
because we are actually considering the highest bound state with an energy
of 3740.6MeV, instead of zero because it is close to the charmonium
threshold. If we redefine the levels taking this into account the constant $%
C_{c}$ becomes about - 1186.3MeV.

The third term in Eq. (127) can not be larger than the second term. Thus one
should always have 
\begin{equation}
n<\frac{1}{2}\left( \frac{h\nu _{c}}{A_{c}}-1\right) .
\end{equation}
\noindent This $n$ is the number of bound states. In this case one obtains 
\begin{equation}
n<\frac{1}{2}\left( \frac{1094}{252.6}-1\right) =2.165
\end{equation}
\noindent which agrees quite well with reality since there are only three $%
S_{1}$ bound states($n=0,1,2$).

Let us now take into consideration the centrifugal term in Eq. (127 ), $%
Rl(l+1)$, and also the spin-orbit interaction $\vec{L}{\cdot }\vec{S}$.
Since 
\begin{equation}
\vec{L}{\cdot }\vec{S}=\frac{1}{2}\left\{ J(J+1)-L(L+1)-S(S+1)\right\} ,
\end{equation}
\noindent one obtains a total contribution in energy of 
\begin{equation}
B_{c}\left\{ J(J+1)-S(S+1)\right\} +(R_{c}-B_{c})l(l+1),
\end{equation}

\noindent which is responsible for the splitting of the (1P) level into the
three levels $\Xi _{co}(1P)$(3415.1 MeV), $\Xi _{c1}(1P)$(3510.53 MeV), and $%
\Xi _{c2}(1P)$(3556.17 MeV). One gets 
\begin{equation*}
B_{c}(0-1{\times }2)+(R_{c}-B_{c})1{\times }2=3415.1-29.3-3067.6=318.2
\end{equation*}
\begin{equation*}
B_{c}(1{\times }2-1{\times }2)+(R_{c}-B_{c})1{\times }%
2=3510.5-29.3-3067.6=413.6
\end{equation*}
\begin{equation*}
B_{c}(2{\times }3-1{\times }2)+(R_{c}-B_{c})1{\times }%
2=3556.2-29.3-3067.6=459.3.
\end{equation*}
\noindent From the first two equations one has $B_{c}=47.7$MeV and $%
R_{c}=254.5$MeV. Using these values in the third level one obtains(in MeV) $%
47.7(2{\times }3-1{\times }2)+206.8{\times }1{\times }2=604.4$, instead of
459.3MeV, and one arrives at an energy of $3067.6+604.4+29.3=3701.3$MeV for
the level $\Xi _{c2}(1P)$. It is just over 3\% off. The levels above the
threshold can not be explained by our simple model.

\bigskip

\paragraph{{\protect\large {b) $b\overline{b}$ Bound States}}}

Now let us consider the $b\overline{b}$ \ \ \ \ system. In this case there
is not the scalar level $\eta _{b}$(which corresponds to $\eta _{c}$ in
charmonium), but the corresponding splitting between $S_{0}$ and $S_{1}$
would have an energy of about 80.0MeV. The three levels $\Upsilon (1S)$, $%
\Upsilon (2S)$, and $\Upsilon (3S)$ correspond to $n=0,1,2$ in our potential
model. They are all vector states. Taking the spin-spin interaction energy
off, and applying Eq. (127) above one has 
\begin{equation}
C_{b}+h\nu _{b}\left( 0+\frac{1}{2}\right) -A_{b}\left( 0+\frac{1}{2}\right)
^{2}=9460.4-\frac{1}{4}80.0=9440.4
\end{equation}
\begin{equation}
C_{b}+h\nu _{b}\left( 1+\frac{1}{2}\right) -A_{b}\left( 1+\frac{1}{2}\right)
^{2}=10023.3-\frac{1}{4}80.0=10003.3
\end{equation}
\begin{equation}
C_{b}+h\nu _{b}\left( 2+\frac{1}{2}\right) -A_{b}\left( 2+\frac{1}{2}\right)
^{2}=10355.3-\frac{1}{4}80.0=10335.3,
\end{equation}
\noindent from which one obtains $C_{b}=9072.4$MeV, $h\nu _{b}=793.8$MeV,
and $A_{b}=115.5$MeV. Calculating the energy of the level with n=3 one gets
10435.8MeV. Adding the spin-spin interaction energy of about 20MeV one has
10455.8MeV which is quite close to the actual value of 10580.0MeV(error
below 1\%). It is important to notice that this level is actually slightly
above the threshold. As done above the number of bound states is about

\begin{equation*}
\frac{1}{2}\left( \frac{793.8}{115.5}-1\right) =3.39.
\end{equation*}
\noindent Again this agrees with reality because there are only four $S_{1}$
bound states($n=0,1,2,3$).

As in charmonium one can make the constant $C_{b}$ negative subtracting the
threshold energy from 9072.4MeV. One gets about $C_{b}= - 488$MeV.

Let us now consider the centrifugal term in Eq. (127), $Rl(l+1)$, and also
the spin-orbit interaction $\vec{L}{\cdot }\vec{S}$. As 
\begin{equation*}
\vec{L}{\cdot }\vec{S}=\frac{1}{2}\left\{ J(J+1)-L(L+1)-S(S+1)\right\} ,
\end{equation*}
\noindent one obtains a total contribution in energy of 
\begin{equation}
B_{b}\left\{ J(J+1)-S(S+1)\right\} +(R_{b}-B_{b})l(l+1),
\end{equation}
\noindent which is responsible for the splitting of the $\Upsilon (1S)$ into
the three levels $\Xi _{bo}(1P)$(9859.8 MeV), $\Xi _{b1}(1P)$ (9891.9 MeV),
and $\Xi _{b2}(1P)$(9913.2 MeV). One obtains 
\begin{equation*}
B_{b}(0-1{\times }2)+(R_{b}-B_{b})1{\times }2=9859.8-20.0-9440.4=399.4
\end{equation*}
\begin{equation*}
B_{b}(1{\times }2-1{\times }2)+(R_{b}-B_{b})1{\times }%
2=9891.9-20.0-9440.4=431.5
\end{equation*}
\begin{equation*}
B_{b}(2{\times }3-1{\times }2)+(R_{b}-B_{b})1{\times }%
2=9913.2-20.0-9440.4=452.8.
\end{equation*}
\noindent From the first two equations one obtains $B_{b}=16.1$MeV and $%
R_{b}=231.8$MeV. Using these values in the third level one gets(in MeV) $%
16.1(2{\times }3-1{\times }2)+215.8{\times }1{\times }2=496.0$, instead of
452.8MeV, and an energy of $9440.4+496.0+20.0=9956.4$MeV for the level $\Xi
_{b2}(1P)$. It is just 0.2\% off.

Doing the same for the splitting of the level $\Upsilon(2S)$ into the three
levels $\Xi_{bo}(2P)$(10232.1 MeV), $\Xi_{b1}(2P)$(10255.2 MeV), and $%
\Xi_{b2}(2P)$(10268.5 MeV), one gets 
\begin{eqnarray*}
B^{\prime}_{b}(0 - 1{\times}2) + (R^{\prime}_{b} - B^{\prime}_{b})1{\times}2
&=& 10232.1 - 20.0 - 10003.3 = 208.8
\end{eqnarray*}
\begin{eqnarray*}
B^{\prime}_{b}(1{\times}2 - 1{\times}2) + (R^{\prime}_{b} - B^{\prime}_{b})1{%
\times}2 &=& 10255.2 - 20.0 - 10003.3 = 231.9
\end{eqnarray*}
\begin{eqnarray*}
B^{\prime}_{b}(2{\times}3 - 1{\times}2) + (R^{\prime}_{b} - B^{\prime}_{b})1{%
\times}2 &=& 10268.5 - 20.0 - 10003.3 = 245.2.
\end{eqnarray*}

\noindent From the first two equations one gets $B^{\prime}_{b}=11.6$MeV and 
$R^{\prime}_{b}=127.6$MeV. Using these values in the third level one has(in
MeV) $11.6(2{\times}3 - 1{\times}2) + 127.6{\times}1{\times}2 = 301.6$,
instead of 245.2MeV, and an energy of $10003.3 + 301.6 + 20.0=10324.9$MeV
for the level $\Xi_{b2}(2P)$. It is just 0.2\% off.

Concluding one sees that this simple model gives very satisfactory results
and describes well the $c\bar{c}$ and $\Upsilon $ mesons. This model can not
be applied to the other heavy mesons due to the lack of experimental data,
but we can guess that their bound states are probably the levels of a
molecular potential. \newline

\subsubsection{\protect\bigskip {\protect\large Light Mesons}}

\paragraph{{\protect\large {a) Kaons}}}

The first two levels, $K(494)$ and $K^{\ast }(892)$ are a hyperfine doublet,
caused by the spin-spin interaction. Removing the spin-spin energy we get
the $K_{0}$ level with energy equal to $892MeV-\frac{1}{4}398MeV=494MeV+%
\frac{3}{4}398MeV=792.5MeV$.

Doing the same with the levels $K(1460)$ and $K^{*}(1680)$ we obtain a
degenerate level $K_{1}$ with energy equal to $1680MeV - \frac{1}{4}220MeV =
1460MeV + \frac{3}{4}220MeV = 1625MeV$.

Up to now no vectorial kaon with $J=1$(that is, a $1^{--}$) has been found
with energy close to 1800MeV but, as we see the splitting diminishes as the
energy increases, and, thus the error will be small if we consider the
energy of $K_{2}$ to be about 1830MeV.

Applying again the formula 
\begin{equation*}
E_{nl}=C+h{\nu }\left( n+\frac{1}{2}\right) -A\left( n+\frac{1}{2}\right)
^{2}+....
\end{equation*}
\noindent to the three levels $K_{0}$, $K_{1}$ and $K_{2}$, we have 
\begin{equation}
C_{K}+h\nu _{K}\left( 0+\frac{1}{2}\right) -A_{K}\left( 0+\frac{1}{2}\right)
^{2}=792.5
\end{equation}
\begin{equation}
C_{K}+h\nu _{K}\left( 1+\frac{1}{2}\right) -A_{K}\left( 1+\frac{1}{2}\right)
^{2}=1625
\end{equation}
\begin{equation}
C_{K}+h\nu _{K}\left( 2+\frac{1}{2}\right) -A_{K}\left( 2+\frac{1}{2}\right)
^{2}=1830.
\end{equation}

\noindent The solution yields $C_{K}=141$MeV, $A_{K}=313.8$MeV, and $%
h\nu_{K}=1460$MeV. The number of bound states is about 
\begin{eqnarray*}
n &{\approx}& \frac{1}{2}\left(\frac{1460}{313.8} - 1\right) = 1.8
\end{eqnarray*}
\noindent which agrees quite well with reality since there are only three $%
S_0$ bound states($n=0,1,2$).

It is quite hard to relate the $P$, $D$, and $F$ states to these levels. For
instance, we expected $^{3}D_{3}$ to have an energy larger than $^{3}D_{2}$,
but, actually, it is the opposite way, the former has an energy of about
1816MeV(the $K_{2}(1820)$ meson) while the latter has an energy of about
1770MeV(the $K_{3}^{\ast }(1780)$ state). Maybe in this case there is an
inverted spin-orbit interaction as in the case of nuclear levels.

\paragraph{{\protect\large {b) $\protect\eta $ and $\protect\omega $ Bound
States}}}

In the case of light mesons our calculation using non-relativistic quantum
mechanics is just a crude approximation because the velocities of the light
quarks $u$ and $d$ are very high for energies above 100MeV. Since the $q\bar{%
q}$ system rotates about a center of mass, most of the velocity should be a
transverse velocity, but the relative velocity along the line of the
particles should also be high.

The mesons that we will consider are the states $\eta(548)$, $\omega(782)$, $%
\eta(1295)$, $\omega(1420)$, and $\eta(1760)$.

The first two levels, $\eta(548)$ and $\omega(782)$ are a hyperfine doublet,
caused by the spin-spin interaction. Removing the spin-spin energy we get a
degenerate level with energy equal to $782MeV - \frac{1}{4}234MeV = 548MeV + 
\frac{3}{4}234MeV = 723.5MeV$. Let us call as $\eta\omega_{0}$.

Doing the same with the levels $\eta(1295)$ and $\omega(1420)$ we obtain the
degenerate level $\eta\omega_{1}$ with energy equal to $1420MeV - \frac{1}{4}%
125MeV = 1295MeV + \frac{3}{4}125MeV = 1388.7MeV$.

Up to now no vectorial $\omega$ with $J=1$(that is, a $1^{--}$) has been
found with energy close to 1800MeV but, as we see the splitting diminishes
as the energy increases, and, thus the error will be small if we consider
the energy of $\eta\omega_{2}$ to be about 1760MeV.

Applying the formula 
\begin{eqnarray*}
E_{nl} &=& C_{\eta\omega} + h{\nu}\left(n + \frac{1}{2}\right) - A\left(n + 
\frac{1}{2}\right)^{2} + ....
\end{eqnarray*}
\noindent to the three levels $\eta\omega_{0}$, $\eta\omega_{1}$ and $%
\eta\omega_{2}$, we have

\begin{equation}
C_{\eta \omega }+h\nu \left( 0+\frac{1}{2}\right) -A\left( 0+\frac{1}{2}%
\right) ^{2}=723.5
\end{equation}
\begin{equation}
C_{\eta \omega }+h\nu \left( 1+\frac{1}{2}\right) -A\left( 1+\frac{1}{2}%
\right) ^{2}=1388.7
\end{equation}
\begin{equation}
C_{\eta \omega }+h\nu \left( 2+\frac{1}{2}\right) -A\left( 2+\frac{1}{2}%
\right) ^{2}=1760.
\end{equation}
\noindent The solution produces $C_{\eta \omega }=289.7$MeV, $A=147$MeV, and 
$h\nu =959.1$MeV. The number of bound states $n$ is 
\begin{equation*}
n<\frac{1}{2}\left( \frac{959.1}{147}-1\right) =2.8
\end{equation*}
\noindent which agrees quite well with reality since there are only three $%
S_{0}$ bound states($n=0,1,2$).

Again it is quite hard to relate the $P$, $D$, and $F$ states to these
levels. For instance, we expected $^{3}P_{2}$ to have an energy larger than $%
^{3}P_{1}$, but, actually, it is the opposite way, the former has an energy
of about 1270MeV(the $f_{2}(1270)$ meson) while the latter has an energy of
about 1285MeV(the $f_{1}(1285)$ state). Maybe in this case there is again an
inverted spin-orbit interaction as in the case of nuclear levels.

\paragraph{\protect\large c) $\protect\pi $ and $\protect\rho $ Bound States}

As we said above this is a rough calculation, but it indicates quite well
the number of bound states.

First we will consider the states $\pi (140)$, $\rho (770)$, $\pi (1300)$, $%
\rho (1450)$, and $\pi (1770)$. The first two levels, $\pi (140)$ and $\rho
(770)$ are a hyperfine doublet, caused by the spin-spin interaction. The
removal of the spin-spin energy produces the degenerate level with energy
equal to $770MeV-\frac{1}{4}630MeV=140MeV+\frac{3}{4}630MeV=613MeV$. Let us
call as $\pi \rho _{0}$.

Doing the same with the levels $\pi(1300)$ and $\rho(1450)$ we get the
degenerate level $\pi\rho_{1}$ with energy equal to $1450MeV - \frac{1}{4}%
150MeV = 1300MeV + \frac{3}{4}150MeV = 1407MeV$.

No vectorial meson $\rho$ with $J=1$(that is, a $1^{--}$) has been found
with energy close to 1800MeV but, since the splitting diminishes as the
energy increases, the error will be small if we consider the energy of $%
\pi\rho_{2}$ to be about 1770MeV.

Applying the formula 
\begin{eqnarray*}
E_{nl} &=& C_{\pi\rho} + h{\nu}\left(n + \frac{1}{2}\right) - A\left(n + 
\frac{1}{2}\right)^{2} + ...
\end{eqnarray*}

\noindent to the three levels $\pi \rho _{0}$, $\pi \rho _{1}$ and $\pi \rho
_{2}$, we obtain 
\begin{equation}
C_{\pi \rho }+h\nu \left( 0+\frac{1}{2}\right) -A\left( 0+\frac{1}{2}\right)
^{2}=613
\end{equation}
\begin{equation}
C_{\pi \rho }+h\nu \left( 1+\frac{1}{2}\right) -A\left( 1+\frac{1}{2}\right)
^{2}=1407
\end{equation}
\begin{equation}
C_{\pi \rho }+h\nu \left( 2+\frac{1}{2}\right) -A\left( 2+\frac{1}{2}\right)
^{2}=1770.
\end{equation}
\noindent The solution gives $C_{\pi \rho }=54.3$MeV, $A=215$MeV, and $h\nu
=1225$MeV. The number of bound states $n$ is 
\begin{equation*}
n<\frac{1}{2}\left( \frac{1225}{215}-1\right) =2.3
\end{equation*}
\noindent which agrees quite well with reality since there are only three $%
S_{0}$ bound states($n=0,1,2$).

It is quite hard to relate the $P$, $D$, and $F$ states to these levels. Let
us try to describe only the states $a_{1}(1260)$ and $a_{2}(1320)$. In order
to do it we will take into consideration the centrifugal term $Rl(l+1)$, and
also the spin-orbit interaction $\vec{L}{\cdot }\vec{S}$. Since 
\begin{equation*}
\vec{L}{\cdot }\vec{S}=\frac{1}{2}\left\{ J(J+1)-L(L+1)-S(S+1)\right\} ,
\end{equation*}
\noindent we obtain a total contribution in energy of 
\begin{equation}
B\left\{ J(J+1)-S(S+1)\right\} +(R-B)l(l+1),
\end{equation}
\noindent which is responsible for the splitting of the (1P) levels into the
three levels $b_{1}(1235)$, $a_{1}(1260)$, and $a_{2}(1320)$. For the last
two we have 
\begin{equation*}
B(1{\times }2-1{\times }2)+(R-B)1{\times }2=1260-770=490,
\end{equation*}
\begin{equation*}
B(2{\times }3-1{\times }2)+(R-B)1{\times }2=1320-770=550.
\end{equation*}
\noindent Solving we obtain $B=15$MeV and $R=260$MeV. This shows that the
spin-orbit contribution is small while the rotational contribution is quite
significative. The centrifugal energy is 
\begin{equation}
V_{c}(r)=\frac{{\hbar }^{2}{\vec{L}}^{2}}{2Mr^{2}}.
\end{equation}

\noindent As we saw above it is about 260MeV, and $l(l+1)=2$. Using for $M$
a reduced mass of about 330/2=165MeV, for quarks $u$ and $d$ we obtain 
\begin{equation}
\left( \sqrt{<\frac{1}{r^{2}}>}\right) ^{-1}=\frac{\hbar }{\sqrt{MR}}=0.98F,
\end{equation}
\noindent which is quite consistent.

It is quite interesting that the potentials of all mesons are not symmetric
due to the anharmonic term which should be of the type $-{\beta }x^{3}$(with 
$\beta >0$). Such a term makes the potential to be above the harmonic
potential for $x<0$, and below that of a harmonic potential for $x>0$. This
asymmetric part of the potential conveys the idea of a very strong repulsion
for very close distances. How can we explain it with pointlike quarks?
Clearly we can not. This is another evidence for quark composition and the
superstrong interaction. As we saw in the case of baryons the potential is
much less asymmetric than in the case of mesons. This difference comes from
the fact that in the case of mesons the potential is between a quark and an
antiquark. This reveals some features of the superstrong interaction.

\subsubsection{\protect\bigskip {\protect\Large The Sizes of Mesons}}

The centrifugal energy is 
\begin{equation*}
V_{c}(r)=\frac{{\hbar }^{2}{\vec{L}}^{2}}{2Mr^{2}}=\frac{{\hbar }%
^{2}c^{2}e^{4}{\vec{L}}^{2}}{2e^{4}Mc^{2}r^{2}}=\frac{e^{4}{\vec{L}}^{2}}{2{%
\alpha }^{2}Mc^{2}r^{2}}
\end{equation*}
where $M$ is the reduced mass. Using the calculated values of \noindent $%
<V_{c}(r)>=R$ of section 5.3 we obtain the table below for $\left( \sqrt{<%
\frac{1}{r^{2}}>}\right) ^{-1}$ with $l=1$:

\begin{center}
\begin{equation*}
\begin{tabular}{||c|c|c|c||}
\hline\hline
&  &  &  \\ 
Meson & $R$(MeV) & $M$(MeV/c$^{2}$) & $\left( \sqrt{<\frac{1}{r^{2}}>}%
\right) ^{-1}$(F) \\ 
&  &  &  \\ \hline
$\pi \rho $ & $260$ & $150$ & $1.00$ \\ \hline
$c\bar{c}$ & $254.5$ & $850$ & $0.43$ \\ \hline
$b\overline{b}$ & $231.8$ & $2500$ & $0.26$ \\ \hline\hline
\end{tabular}
\end{equation*}
\end{center}

\noindent These values are quite consistent. For $l=0$ we expect much
smaller values. Recall that $r$ is the distance between $q$ and $\overline{q%
\text{.}}$

Taking a closer look at the above table we notice that $R$ does not change
much and that it decreases slowly as the reduced mass $M$ increases.
Therefore, making an extrapolation we can say that it is about $250$ MeV for
light mesons and mesons with intermediate masses, and for $t\overline{t}$ it
should be of the order of $200$ MeV. Defining as a measure of the radius of
the meson, $rad$, \ the expression 
\begin{equation*}
rad=\left( \sqrt{<\frac{1}{r^{2}}>}\right) ^{-1}
\end{equation*}

\noindent we can make an estimation of the radii of the other mesons for $%
l=1 $. Putting together the calculated values and the above values we have
the table shown below:

\bigskip

\begin{center}
\begin{tabular}{||c|c|c|c|}
\hline\hline
Mesons & $R$ (Mev) & $Mc^{2}$(Mev) & radius(F) \\ \hline\hline
$\pi \rho $ & 260 & 150 & 1.00 \\ \hline
$\eta \omega $ & $\sim 250$ & 150 & 1.00 \\ \hline
$K^{+},K^{-},K^{0},\overline{K^{0}}$ & $\sim 250$ & $\frac{0.3\times 0.5}{%
0.3+0.8}=0.188$ & 0.9 \\ \hline
$\phi $ & $\sim 250$ & 0.250 & 0.8 \\ \hline
$D^{+},D^{-},D^{0},\overline{D^{0}}$ & $\sim 250$ & $\frac{1.7\times 0.3}{%
1.7+0.3}=0.255$ & 0.8 \\ \hline
$B^{+},B^{-},B^{0},\overline{B^{0}}$ & $\sim 250$ & $\frac{5\times 0.3}{5+0.3%
}=0.283$ & 0.75 \\ \hline
$D_{s}^{+},D_{s}^{-}$ & $\sim 250$ & $\frac{1.7\times 0.5}{1.7+0.5}=0.386$ & 
0.6 \\ \hline
$c\bar{c}$ & 254.5 & 850 & 0.43 \\ \hline
$b\overline{b}$ & 231.8 & 2500 & 0.26 \\ \hline
$t\overline{t}$ & $\sim 200$ & 87000 & 0.05 \\ \hline\hline
\end{tabular}
\end{center}

\bigskip

\noindent The numbers of the above table are quite important. There are some
definite features. We see that light mesons and intermediate mesons have
about the same size, ranging from 1F to 0.5F. It shows that the $b\overline{b%
}$ system is quite small and that $t\overline{t}$ is extremely small. Its
radius agrees quite well with the approximate radius of the $ttt$ baryon
which is smaller than 0.15F. We now see that its radius should be of the
order of 0.05F, that is, three times smaller than previously calculated. If
we now go back to section 5.1.4 and use the radius of about 0.05F for $ttt$
we obtain 
\begin{equation*}
k\approx \frac{174}{0.05^{2}}=69600GeV/F^{2}
\end{equation*}

\noindent which multiplied by its approximate radius yields a QCD-like $K$
of about 3480 GeV/F. As we see the figure of 0.05F is close to the radius of
the electron which is known to be smaller than 0.01F. The crucial difference
is that the electron is very light while the $ttt$ is extremely heavy. And
since the top quark is the $p_{1}p_{4}$ system primons are, for sure,
point-like particles. {\large Therefore, we can make the strong statement
that more massive matter comes from less massive matter, that is, mass comes
from within. This means that there should exist primordial, noncreated
particles(with mass). They should be primons and and leptons. We will see
more on this in Chapter 7.}

\bigskip

\noindent {\Large References}

\bigskip

\noindent 1. S. Gasiorowicz and J.L.Rosner, Am. J. Phys., \textbf{49},
954(1981).

\noindent 2. N. Isgur and G. Karl, Phys. Rev. \textbf{D18}, 4187 (1978).

\noindent 3. S. Capstick and N. Isgur, \ Phys. Rev. \textbf{D34}, 2809
(1986).

\noindent 4. R.K. Bhaduri, B.K. Jennings and J.C. Waddington, Phys. Rev. 
\textbf{D29}, 2051 (1984).

\noindent 5. M.V.N. Murthy, M. Dey, J. Dey and R.K. Bhaduri, Phys. Rev. 
\textbf{D30}, 152 (1984).

\noindent 6. M.V.N. Murthy, M. Brack, R.K. Bhaduri and B.K. Jennings, Z.
Phys. \textbf{C29}, 385 (1985).

\noindent 7. P. Stassat, Fl. Stancu and J.-M. Richard, nucl-th/9905015.

\noindent 8. A. Hosaka, H. Toki, and M. Tokayama, Mod. Phys. Lett \textbf{A13%
}, 1699 (1998).

\noindent 9. M.E. de Souza, Proceedings of the XIV Brazilian National
Meeting of the Physics of Particles and Fields, Caxambu, Minas Gerais,
Brazil, September 29-October 3, p. 331(1993).

\noindent 10. M.E. de Souza, in The Six Fundamental Forces of Nature,
Universidade Federal de Sergipe, S\~{a}o Crist\'{o}v\~{a}o, Sergipe, Brazil,
February 1994. \newline
\noindent 11. M.E. de Souza, in Some Important Consequences of the Existence
of the Superstrong Interaction, Universidade Federal de Sergipe, S\~{a}o
Crist\'{o}v\~{a}o, Sergipe, Brazil, October, 1997.

\noindent 12. H. Toki, J. Dey and M. Dey, Phys. Lett, \textbf{B133}, 20
(1983).

\noindent \noindent 13. Particle Data Group, Phys. Rev. \textbf{D54 (Part I)}%
, (1996).

\noindent 14. R. Shankar, in Principles of Quantum Mechanics, 2nd ed.,
Plenum Press, New York, 1994, pp 316-317.

\noindent 15. B. Povh, hep-ph/9908233

\noindent 16. Fayyazuddin and Riazuddin, in A modern Introduction to
Particle Physics, World Scientific, Singapore, 1992, pp 249-256.

\noindent 17. D. H. Perkins, in Introduction to High Energy Physics, 3rd
ed., Addison-Wesley Publishing Company Inc., Menlo Park, 1987, pp. 177-178.\ 

\noindent 18. D.B. Lichtenberg, in Unitary Symmetries and Elementary
Particles, Academic Press, New York, USA(1970).

\noindent 19. S. Fl\"{u}gge, in Practical Quantum Mechanics, Vol. I,
Springer-Verlag, New York, USA, 1974, p. 186.

\noindent 20. D.H. Perkins, in Introduction to High Energy Physics,
Addison-Wesley, Menlo Park, CA, USA, 1987, p. 152.

\noindent 21. E. Eichten et al., Phys. Rev. Lett, \textbf{34}, 369(1975).

\noindent 22. S. Godfrey and N. Isgur, Phys. Rev. \textbf{D32}, 189(1985).

\noindent 23. S.N. Gupta, S.F. Radford, W.W. Repko, Phys. Rev. \textbf{D34},
201(1986).

\noindent 24. S.N. Gupta, J.M. Johnson, Phys. Rev. \textbf{D51}, 168(1995).

\noindent 25. S.N. Gupta, J.M. Johnson, Phys. Rev. \textbf{D53}, 312(1995).

\noindent 26. C. Itoh, T. Minamikawa, K. Miura, and T. Watanabe, Nuovo
Cimento \textbf{105A}, 1539(1992).

\noindent 27. C. Itoh, T. Minamikawa, K. Miura, and T. Watanabe, Nuovo
Cimento \textbf{109A}, 569(1996).

\noindent 28. Fayyazuddin and Riazuddin, in A Modern Introduction to
Particle Physics, World Scientific, Singapore(1992), p. 256.

\noindent 29. S. Fl\"{u}gge, in Practical Quantum Mechanics, Vol. I,
Springer-Verlag, New York, USA, 1974, p. 178.

\noindent 30\noindent . Ibidem, p. 180.

\noindent \pagebreak

\bigskip \rule[0.5in]{0in}{0.17in}

\section{\noindent {\protect\LARGE The Superstrong Force and \newline
the Universe}}

{\LARGE \bigskip \rule[1.5in]{0in}{0.17in}}

\subsection{\protect\Large The Supernovae Evidence for the Superstrong
Interaction}

Type II supernovae release the enormous energy$^{1}\;$ of about $10^{51}\;$
ergs, which corresponds to initial shock velocities of $5{\times }10^{7}$%
m/s. Several different models have attempted without success$^{1}\;$ to
explain how the gravitational energy released during collapse could reach
the outer layers of the star. Moreover it is not clear at all why the core
explodes in the first place. No known theoretical model to date has been
able to make the core collapse and ``to produce both a gas remnant and a
neutron star''$^{1}$. We may shed some light onto this issue by simply
proposing that the explosion is caused by the superstrong interaction. The
mechanism of the explosion may be as follows: Due to gravitational collapse
all nucleons(and electrons) of the star get more and more squeezed up to the
point that the repulsion caused by the superstrong interaction among the
nucleons begins to play an important role because of the very small range of
this new interaction. With further collapse a point is reached when the
repulsion overcomes the gravitational attraction and a rapid expansion takes
place in the core of the star while the outer layers are still collapsing.
We will have, then, the shock of the outer envelope of the core with the
inner envelope of the outer layer. If, at the moment of the shock, the
velocities of the hydrogen atoms of the outer envelope of the core are much
larger than those of the inner envelope of the outer layer, then there will
not be any neutron star, and the gas will just expand and forms the gas
remnant. If it happens the other way around or if the velocities of the two
envelopes are similar then there will be the formation of a neutron star
because in this case the core bounces back. It is worth noting that the core
may be formed only if all nucleons of its envelope collide at the same
time(or almost) with the nucleons of the other envelope. This is only
possible if the nucleons of the core form a gigantic spherical lattice. When
the collision happens its effects in the inner layers of nucleons in the
core are damped.

The energy of $10^{51}\;$ ergs corresponds to an energy of about $0.6$MeV
per baryon. A nucleon has a radius of about 1fm and the superstrong
interaction will be dominant only if the nucleons are very close to each
other. Taking the distance from their centers we may consider that when they
are very squeezed they are separated by about 2fm(from center to center).
This is consistent with nuclear physics data. Therefore, we can write 
\begin{equation}
40(GeV)\frac{e^{-2\mu _{ss}}}{2}=0.6(MeV)
\end{equation}
\noindent where and $\mu _{ss}\;$ is given in $fm^{-1}$. The factor of 40
was taken from section 4.3. We obtain $\mu _{ss}\approx {5}$ $fm^{-1}$,
which means that the mediator of the new interaction has a mass of about 0.7
GeV. If nucleons get even more squeezed, separated by just
1fm(center-to-center) we obtain $\mu _{ss}{\approx }11$ fm$^{-1}\;$ and a
mediator with a mass around 1.5GeV. This is quite in line with what we
calculated in chapter 3. Mediators with masses in this range abound: $\rho
(770)$, $\omega (782)$, $\phi (1020)$, $D^{\ast }(2007)^{0}$,$D^{\ast
}(2010)^{\pm }$, $D_{1}(2420)^{0}$, etc. It is important to say that the
search for quark composition has been aimed at too high energies, in the TeV
region. But, as we see, the superstrong bosons are not as massive.

These considerations are also in line with the repulsion which is one of the
features attributed to the strong force at very small distances. Walecka$%
^{2}\;$ has developed a theory of highly condensed matter in the domain of
the strong force assuming that the strong repulsion is due to the exchange
of $\omega $. He constructed a relativistic Lagrangian that allows nucleons
to interact attractively by means of scalar pion exchange and repulsively by
means of the more massive vector meson $\omega $. At very high densities he
finds that the vector meson field dominates and one recovers Zel$^{\prime }$%
dovich result 
\begin{equation}
\displaystyle P\rightarrow {\rho }c^{2};v_{s}{\rightarrow }c.
\end{equation}
\noindent where $v_{s}\;$ is the thermodynamic speed of sound in the medium, 
$P\;$ is the pressure, and $\rho \;$ is the density. In his article he
defines the two dimensionless coupling constants 
\begin{equation*}
\displaystyle{c_{s}}^{2}=\frac{{g_{s}}^{2}}{{\hbar }c^{3}}\frac{M^{2}}{\mu
^{2}},\;\;{c_{v}}^{2}=\frac{{g_{v}}^{2}}{{\hbar }c}\frac{M^{2}}{m^{2}},
\end{equation*}
\noindent in which $\frac{{g_{s}}^{2}}{{\hbar }c^{3}}\;$ and $\frac{{g_{v}}%
^{2}}{{\hbar }c}\;$ are, respectively, the coupling constants of the
strong(pionic) and vectorial fields, and $M$, $\mu \;$ and $m\;$ are the
inverse Compton wavelengths 
\begin{equation*}
M{\equiv }\frac{m_{b}c}{{\hbar }},\;\;\mu {\equiv }\frac{m_{s}c}{{\hbar }}%
,\;\;m{\equiv }\frac{m_{v}c}{{\hbar }}
\end{equation*}
\noindent where $m_{b}\;$ is the proton mass. Using data of nuclear matter
he obtains ${c_{s}}^{2}=266.9\;$ and ${c_{v}}^{2}=195.7$. Considering that
the vector field is actually caused by the superstrong interaction we
observe that the value of $c_{v}\;$ is consistent with the figures that we
obtained for $\mu _{ss}\;$ and $\beta _{ss}^{2}$. Therefore, Walecka$%
^{\prime }$s theory is essentially right.

Claims of the experimental discovery of a new interaction have been made by
Baurov and Kopajev$^{3}$(and references therein). According to them the new
interaction is manifested by the magnetic activity of solar flares on the
surface of the Sun. As they say ``The new interaction must be very strong in
that case because the vector potential $\vec{A}\;$ of the tubes is of the
order of ${\sim }10^{11}$Gs.cm ...'' It may actually be the same superstrong
interaction we have discussed above. Although it has been manifested on the
surface of the Sun, its origin may be traced to its center where the density
is of the order of the nuclear density. Moreover we may explain why it
happens in flares in the following way: Due to gravitational contraction the
density may increase momentarily to such a point that the superstrong
interaction becomes important. This is especially true right at the center
of the Sun. Then, the very squeezed baryons(nucleons) are expelled to the
outer layers of the Sun due to their mutual repulsion, and, in such a
process, we expect that the magnetic activity will increase and, thus, the
solar flares are generated.

\subsection{\protect\bigskip {\protect\Large The Formation and Evolution of
Galaxies}}

\subsubsection{\protect\bigskip {\protect\large The Formation of Galaxies
and Primordial Stars}}

The superstrong force explains the formation of galaxies in a quite easy
way. As M\'{a}rquez et al.$^{4}\;$ and Yahil$^{5}\;$ have shown, high
redshift galaxies are very small, having diameters smaller than 1kpc. This
is so because they are born as quasars which have sizes much smaller than
1kpc. In the beginning of the Universe, because of the repulsion caused by
the superstrong interaction, the nucleons attained high velocities, of the
order of supernovae$^{\prime }$, in the range of $10^{4}$km/s, but due to
the action of the strong force this velocity diminished, and probably went
down to $10^{3}$km/s. Gravity diminished further this velocity to the range $%
10^{2}$-$10^{3}$km/s, which is the range of the peculiar velocities of
galaxies. Assuming $v{\approx }10^{3}$km/s, and using the virial theorem, we
obtain that a newborn quasar had a radius 
\begin{equation}
R{\approx }\frac{GM}{2v^{2}}
\end{equation}
\noindent which is about 500 light years, where $M$ is the typical mass of a
galaxy($10^{11}$ suns). This is quite consistent with the data of
M\'{a}rquez et al.$^{4}\;$ and Yahil$^{5}$. In the data of M\'{a}rquez et
al. we see that some quasars have companion galaxies just some kpc away from
them. For example, there is a galaxy only 28kpc away from the quasar 3C 215
and it is surrounded by 14 galaxies within ${\pm }30$''. If we assume that
they are going away from each other with velocities around $10^{3}$km/s,
going backwards to the time when they were formed(touching each other), they
had sizes of approximately 100pc, which is very consistent with the above
calculation. This means that quasars(galaxies) were formed when the
temperature was about $10^{9}$K, just after the formation of the first light
nuclei. The Universe was very young, less than a second old. Considering the
typical number of baryons of a galaxy in a sphere with a radius of about 500
light years we obtain that the average distance among nucleons was only $%
10^{-4}$m and corresponds to an average density of $\rho {\approx }10^{-16}$%
kg/m$^{3}$. For $v{\approx }10^{3}$km/s we find that the Universe was about $%
10^{-11}$s old. The size of an atom is about $1\AA $, so that, atoms were
formed $10^{-17}$s after the Big Bang. The Big Bang was then a sort of
supernova explosion. If we form an enormous squeezed nucleus with all
baryons of the Universe we obtain a radius of about 50000km, which is
approximately equal to Jupiter radius. It was not a black hole because it
expanded due to the action of the superstrong interaction.

Let us now use Jeans criterion. The mass contained in a sphere of radius $%
\lambda _{J}\;$ is the Jeans mass 
\begin{equation}
M_{J}=\frac{4}{3}{\pi }{\lambda _{J}}^{3}\rho
\end{equation}
\noindent where $\rho $ is the density and $\lambda _{J}\;$ is given by 
\begin{equation}
\lambda _{J}=\sqrt{\frac{\pi }{G\bar{\rho}}}c_{s}
\end{equation}
\noindent where $c_{s}\;$ is the sound velocity. Using $c_{s}=10^{3}$km/s
and the above figure for $\rho \;$ we obtain $\lambda _{J}{\approx }10^{19}$%
m and $M_{J}=10^{41}$kg which is quite consistent with the virial theorem
calculation. This $M_{J}\;$ is the mass of a typical galaxy like the Milky
Way, and $10^{19}$m is about 300pc, which is about the size of a newborn
quasar, as was shown above.

Very close to the beginning of the Universe, when the density was just above
nuclear density ($10^{18}$kg/m$^{3}$) and the nucleons were still with
supernova velocities of about 50000km/s we obtain $\lambda _{J}{\approx }%
10^{4}$m and $M_{J}{\approx }10^{30}$kg. This $M_{J}\;$ is the typical mass
of a star like our Sun. This calculation means that quasars and stars were
formed almost at the same time. These were the primordial stars. This is in
line with the arguments and data of Shaver et al.$^{6}\;$and Pettini et al$%
^{7}$. Shaver et al. show in their work that the formation rate of stars and
the space density of quasars peak at the same redshift($z{\approx }2.5$) and
have the same redshift dependence. This fact links the formation of quasars
to the formation of primordial stars and rules out the existence of black
holes. It is the superstrong interaction that avoids the formation of these
hypothetical objects. Therefore, we should expect to have old stars and also
neutron stars in the bulges of galaxies. Of course this is a quite known
fact. For instance, very close to the center of the Milky Way there are very
old stars.

This picture formation of stars and quasars allows the possibility of having
neutron quasars which we may call \textit{quasons}. They could provide the
energy for the very energetic cosmic rays and would provide quite a lot of
dark mass. \newline

\subsubsection{\protect\bigskip {\protect\large The Evolution of Galaxies }}

The remarkable work of M\'{a}rquez et al.$^{4}\;$ has shown that very high
redshift elliptical galaxies harbor quasars. They have also shown that such
galaxies are very small(diameters smaller than 1kpc) and all of them are
ellipticals. All the studied objects(about 15 quasars) have extended
structures of ionized gas around them(this fact had already been presented
by other researchers). They have found other galaxies in the fields of the
studied objects only a few kpc away from them. Some of the quasars present
asymmetric radio sources with collimated one-sided jets of extended ionized
gas. This means that galaxies are born as quasars which become galaxies by
means of the shedding of matter(ionized gas) from their cores as a result of
the strong repulsion among their baryons caused by the superstrong
interaction. The authors have found that the quasar 3C 281 is a double radio
source. They also confirm the results of Miley and Hartsuijker$^{8}\;$ that
found that the quasar 3C 206 is also a classical double radio source. At
such high redshifts it is very unlikely that this double source was caused
by merging. It probably was caused by the breaking of the core of these
quasars into two cores, separated by a very small distance. This breaking
was caused by repulsion due to the superstrong interaction. The same kind of
phenomenon has been observed in galaxies. For example, our well behaved
normal galaxy M31 has two nuclei separated by just 5 light years$^{9}$. Very
recent data$^{10}\;$ of NGC 6240, which is considered a typical protogalaxy,
show that ``approximately 70\% of the total radio power at 20cm originates
from the nuclear region ($\leq $1.5kpc), of which half is emitted by two
unresolved (R$\leq $30pc) cores and half by a diffuse component. Nearly all
of the other 30\% of the total radio power comes from an arm-like region
extending westward from the nuclear region''. A very important property of
many quasars is their brightness which can vary from night to night. This
flickering may have its origin in the outward motion of large quantities of
matter from their cores. This brightness variability is also present in
Seyfert galaxies which are powerful sources of infrared radiation. Many of
them are also strong radio emitters. For example, over a period of a few
months, the nucleus of the Seyfert galaxy M77(or NGC1068) switches on and
off a power output equivalent to the total luminosity of our galaxy$^{11}$.
It is also worth noting that the nuclei of Seyfert galaxies are very bright
and have a general starlike appearance. Researchers have found that some
Seyfert galaxies exhibit explosive phenomena$^{11}$. For example, M77 and
NGC4151 expel huge amounts of gas from their nuclei. The spectra of both
galaxies show strong emission lines, just as quasars$^{\prime }$. Shaver et
al.$^{12}\;$ have found that there is a redshift cutoff in the number of
quasars around $z=2$, there is no quasar for $5<z<7$, and almost no quasar
for $z<0.5$. This clearly shows that quasars evolve into galaxies.

We can show a long list of similar phenomena that evidences the superstrong
interaction. Let us mention just some of them. NGC 2992 presents a jet-like
structure and a circum-nuclear ring$^{13}$. Falcke and Biermann$^{14}\;$
report that there is a large scale emission-like jet going outward from the
core of NGC 4258 with a mass of about 4${\times }10^{35}$kg and with a
kinetic power of approximately $10^{42}$ergs/s and expansion velocity of
2000km/s. This is of the same order of supernovae velocities. It is well
known that BL Lacertae objects are powerful sources of radio waves and
infrared radiation. They share with quasars the fact of exhibiting a
starlike appearance and of showing short-term brightness fluctuations. As
some quasars do, they also have a nebulosity around the bright nucleus.
Researchers$^{15}\;$ have managed to obtain the spectrum of their
nebulosity. \textit{The spectrum of the nebulosity is strikingly similar to
the spectrum of an elliptical galaxy}(M32$^{\prime }$s spectrum, in this
case). In terms of the evolution above described they are simply an
evolutionary stage of a quasar towards becoming a normal galaxy.

Radio galaxies share with BL Lacertae objects many of the properties of
quasars. As Heckman et al.$^{15}\;$ have shown, in the middle and far
infrared(MFIR) quasars are more powerful sources of MFIR radiation than
radio galaxies. Also, there have been investigations showing that the
emission from the narrow-line region(NLR) in radio-loud quasars is stronger
than in radio galaxies of the same radio power$^{16,17,18}$. Goodrich and
Cohen$^{19}\;$ have studied the polarization in the broad-line radio galaxy
3C 109. After the intervening dust is taken into account the absolute
V-magnitude of this galaxy becomes $-26.6\;$ or brighter, which puts it in
the quasar luminosity range. The investigators suggest that ``many radio
galaxies may be quasars with their jets pointed away from our direct line of
sight''. It has also been established that radio galaxies are found at
intermediate or high redshifts and that they are clearly related to galactic
evolution because as the redshift increases cluster galaxies become bluer on
average, and contain more young stars in their nuclei. This is also valid
for radio galaxies: the higher the redshift, the higher their activity. All
these data show that a radio galaxy is just an evolutionary stage of a
galaxy towards becoming a normal galaxy, i.e., it is just a stage of the
slow transformation by means of an overall expansion of a quasar into a
normal galaxy.

In the light of the above considerations the nuclei of old spirals must
exhibit a moderate activity. This is actually the case. The activity must be
inversely proportional to the galaxy$^{\prime }$s age, i.e., it must be a
function of luminosity. The bluer they are, the more active their nuclei
must be. As discussed above there must also exist a relation between this
activity and the size of the nucleus(as compared to the disk) in spiral
galaxies. Our galaxy has a mild activity at its center. Most of the activity
is concentrated in a region called Sagittarius A, which includes the
galactic center. It emits synchroton and infrared radiations. Despite its
large energy output Sagittarius A is quite small, being only about 40 light
years in diameter. Besides Sagittarius A our galaxy exposes other evidences
showing that in the past it was a much more compact object: 1) Close to the
center, \textit{on opposite sides of it}, there are two enormous expanding
arms of hydrogen going away from the center at speeds of 53km/s and 153km/s;
b) Even closer to the center there is the ring called Sagittarius B2 which
is expanding at a speed of 110km/s$^{(11,20)}$. It is worth noting that the
speeds are low(as compared to the velocities of relativistic electrons from
possible black holes). This phenomenon is not restricted to our galaxy.
Recent high-resolution molecular-line observations of external galaxies have
revealed that galactic nuclei are often associated with similar expanding
rings$^{21}$.

A new born quasar, as discussed above, must have most of its mass as
hydrogen, the rest being the primordial helium. But close to its center
there should also exist heavy elements. Therefore, it is mainly constituted
of protons. We expect that different parts of it will be subjected to the
superstrong force, especially close to its center where the gravitational
field is small. The expansion of the quasar has to be, thus, from within,
that is, from its center to its outer layers. The repulsion makes the quasar
increase in size and go through the intermediate stages which may include
radio galaxies and BL Lacertae objects. Far from the center big clumps of
hydrogen and helium gasses form stars. Considering what is exposed above we
may propose the following evolutionary scenarios described below.

\paragraph{\noindent {\protect\normalsize Elliptical galaxy \newline
}}

A quasar may become an elliptical galaxy by expanding slowly as a whole.
Because of rotation we may have several types of ellipticals according to
their oblateness. As is well known ellipticals do not exhibit much
rotation(as compared to spiral galaxies). This is explained as follows: As a
quasar expands its angular velocity decreases because of angular momentum
conservation. For example, the angular velocity of an EO must be given
by(disregarding mass loss) 
\begin{equation}
{\omega }_{EO}={\omega }_{Q}\left( {\frac{R_{Q}}{R_{EO}}}\right) ^{2}
\end{equation}
\noindent where $\omega _{Q}\;$ is the angular velocity of the quasar which
gave origin to the galaxy; $R_{EO}\;$ and $R_{Q}\;$ are the radii of the
elliptical galaxy and the quasar, respectively. Because $R_{EO}{\gg }R_{Q}$, 
${\omega }_{EO}{\ll }{\omega }_{Q}$. This is consistent with the slow
rotation of ellipticals. There is also the following consistency to be
considered. Most galaxies in the Universe are ellipticals(about 60\%) and as
was shown above this means that most quasars expand slowly. Therefore, most
quasars must not show rapid variability and must also be radio quite. This
is exactly what has been reported$^{22}$. Another evidence to be taken into
account is the reported nebulosity around some quasars. Boroson and Person$%
^{23}$ have studied this nebulosity spectroscopically. The emission lines
they found are of the same type as the emissions from a plasma.

\paragraph{\noindent {\protect\normalsize Spiral Galaxy \newline
}}

There are two possibilities in this case: normal spiral and barred spiral.
This happens when, at some point in its expansion towards becoming a galaxy,
a quasar expands rapidly by pouring matter outwards from its center, mainly
in opposite sides accross a diameter. This pouring will give origin to two
jets which will wind up around the central bulge because of rotation and
will create the two spiral arms. A possible mechanism is the following: Due
to rotation we expect to have some bulging in the spherical shape, and
because of angular momentum conservation the outpouring of matter may only
happen in a plane perpendicular to the axis of rotation. Because of rotation
the core of the quasar becomes also an ellipsoid. This core(which has a
higher concentration of baryons) may be broken into two parts, going to a
state of lower potential energy(of the superstrong interaction). These two
parts repel each other and form two centers(lobes) in the equator of the
quasar(or young galaxy). The quasar 3C 281, for example, is a double radio
source and has extended ionized gas around it$^{4}$. These two lobes are
also seen in many radio galaxies. As a consequence of the outpouring of
matter from each center there must exist all kinds of radiations covering
the whole electromagnetic spectrum, especially in the form of synchroton
radiation caused by collisions among atoms. Because of these collisions we
expect to have electrons stripped from hydrogen and helium atoms. These
electrons create the observed synchroton radiation which is associated with
jets in very active galaxies. These collisions provide also the enormous
energy output observed in quasars.

Besides the slow outpouring of matter from their centers, their bulges
should also expand as a whole by means of large amounts of matter which give
origin to the globular clusters. This expansion probably happens in the
beginning of the formation of the spiral. That is why the globular clusters
are so old. As the galaxy ages the activity at the galactic center
diminishes due to decrease of mass in the nucleus.

Barred spirals are galaxies that expel matter more vigorously. That is
exactly why their arms are not tightly wound. As the galaxy ages the arms
will curl up more and more and the bar will disappear because of the
ejection of matter outwards. It is worth noting that the more
spirals(including barred ones) are wound up the smaller are their nuclei
and, conversely, the larger are their bulges, the less they are wound up.
This happens because of the shedding of matter outwards from their nuclei
throughout the galaxy$^{\prime}$s life due to the action of the superstrong
force. The bar can be explained in terms of a more vigorous shedding of
matter outwards as compared to the shedding that takes place in normal
spirals. Therefore, as a spiral evolves its nucleus diminishes and the two
arms become more and more tightly wound up. In summary, the evolution of
galaxies probably follows one of the eight branches: \newline
i) Quasar(without jets) $\rightarrow\;$ BL Lacertae or radio galaxy $%
\rightarrow \linebreak \rightarrow \left\{ 
\begin{array}{l}
\mbox{Seyfert Galaxies} \\ 
\mbox{Elliptical Galaxies$\rightarrow$Spiral Galaxies}
\end{array}
\right. $ \newline
ii) Quasar(without jets) $\rightarrow\;$ BL Lacertae or radio galaxy $%
\rightarrow \left\{ 
\begin{array}{l}
\mbox{Seyfert Galaxies} \\ 
\mbox{Elliptical Galaxies}
\end{array}
\right. $ \newline
iii) Quasar(without jets) ${\rightarrow}\; \left\{ 
\begin{array}{l}
\mbox{Seyfert Galaxies} \\ 
\mbox{Elliptical Galaxies$\rightarrow$Spiral Galaxies}
\end{array}
\right. $ \newline
iv) Quasar(without jets) ${\rightarrow}\; \left\{ 
\begin{array}{l}
\mbox{Seyfert Galaxies} \\ 
\mbox{Elliptical Galaxies}
\end{array}
\right. $ \newline
v) Quasar(without jets) ${\rightarrow}\;$ Elliptical Galaxies \newline
vi) Quasar(with jets) $\rightarrow\;$ radio galaxy $\rightarrow \left\{ 
\begin{array}{l}
\mbox{Seyfert Galaxies} \\ 
\mbox{Spiral Galaxies}
\end{array}
\right. $ \newline
vii) Quasar(with jets) $\rightarrow\; \left\{ 
\begin{array}{l}
\mbox{Seyfert Galaxies} \\ 
\mbox{Spiral Galaxies}
\end{array}
\right. $ \newline
viii) Quasar(with jets) $\rightarrow\;$ Spiral Galaxies. \vskip .2in

Let us, now, make a general analysis including all kinds of galaxies.
Considering the evolution above proposed we do not expect to have very small
spiral galaxies because spirals must come from strong expulsion of matter
from quasars nuclei, and this must happen only when the number of baryons is
sufficiently large. This is the case, indeed, because dwarf galaxies are
either irregular or elliptical galaxies. Ellipticals have masses in the
range between $10^{5}\;$ and $10^{13}\;$ solar masses while spirals$%
^{\prime} $ masses are comprised between $10^{9}\;$ and $10^{11}\;$ solar
masses. Also, we expect that spirals have faster rotations than ellipticals
and, indeed, they do. This is just because the nuclei of spirals are smaller
than the nuclei of ellipticals(for the same mass, of course). Therefore,
spirals should have faster rotational velocities. It is also expected that,
since spirals shed gas to their disks throughout their lifetimes, their
disks must have young stars. This is an established fact. Our galaxy$%
^{\prime}$s disk, for example, has very hot, young(O-,B-, and A-type) stars,
type-I Cepheids, supergiants, open clusters, and interstellar gas and dust.
Each of these types represent young stars or the material from which they
are formed. Conversely, the globular clusters and the nucleus contain older
stars, such as RR Lyrae, type-II Cepheids, and long-period variables. This,
of course, is a general characteristic of all spirals. For example, Young O-
and B-type stars are the stars which outline the beautiful spiral arms of
the Whirlpool galaxy. Because of the lack of gas(i.e., because of the lack
of a disk) ellipticals also have primarily very old stars.

A very important support to the above evolution scheme is provided by the
number-luminosity relation $N(>l)$. When expanded in terms of the apparent
luminosity, $l$, the first term(Euclidean term) is given by$^{24}$ 
\begin{equation}
N(>l)=\frac{4{\pi }n(t_{o})}{3}{\left( \frac{L}{4{\pi }l}\right) }^{1.5}
\end{equation}
\noindent where $n(t_{o})\;$ is the present density of galaxies and $L$ is
the absolute magnitude. The correction term is always negative, so that the
number of faint objects($l\;$ small) should always be less than the number
that the $l^{-1.5}\;$ predicts. This conclusion is strongly contradicted by
observations on radio sources: many surveys of radio sources agree that
there are more faint sources than the $l^{-1.5}\;$ law predicts. The fitting
of the experimental data provides a law of the form$^{24}$ 
\begin{equation}
N(>l){\approx }\frac{constan}{l^{1.8}}.
\end{equation}
\noindent Since the formula breaks down for small $l$(i.e., faint distant
sources), we must conclude that in the past radio sources were brighter
and/or more numerous than they are today. This, of course, lends support to
the above evolutionary scheme.

It is worth mentioning that there is a very important drawback against the
traditional view of explaining the formation of arms in spirals by the
bulging effect of rotation. If this were the case we would find a higher
proportion of pulsars off the galactic equator of our galaxy. But the real
distribution reveals that these sources are mostly concentrated in the
galactic equator. The traditional view does not explain either why all
spirals have large amounts of gas in their disks. Besides, within the
traditional framework quasars are just exotic objects. Evolution is clearly
out of question without a repulsive short range force.

\subsection{\protect\Large The Formation of Structure: Bubbles, Sheets and
Clusters}

Only clumps of matter exceeding the Jeans mass stabilize and virialize.
Therefore, clumps of different sizes going apart from each other populated
the young Universe. These clumps were quasars, young galaxies and groups of
them with different numbers. Let us now turn to the present (local)
Universe. The sizes of the biggest clumps can be calculated using Jean$%
^{\prime }$s criterion. With an overall average density of about $10^{-29}$%
g/cm$^{3}\;$and a sound velocity around $10^{3}$km/s we obtain from Eq. 156 $%
\lambda _{J}{\approx }10^{24}m{\approx }67$Mpc which is of the order of
magnitude of the largest voids. Solving for the ratio $v=d/t=67Mpc/10^{17}s%
\; $ we find $v=2{\times }10^{7}$m/s, which is quite reasonable with the
above arguments and figures. As the density diminishes larger and larger
structures are formed. When the density was higher the structures were
smaller. For simplicity let us consider that the progenitor of a void was a
spherical volume with a radius $r$. Since the volume of the void increases
its density decreases and galaxies from different voids become clumped. This
process generates underdense and overdense regions and the overall effect is
the formation of clusters and superclusters. Of course, galaxies in clusters
are accelerated due to their mutual attraction.

\subsection{\protect\Large No Need of \ (for) Dark Matter}

It has generally been argued that most of the matter of the Universe, from
80\% to 90\%, is dark matter which has quite unusual properties. For
example, although it has to be very heavy it does not scatter radiation
which is very contradictory. It is shown below that for closing the Universe
there is no need of (for) dark matter.

Let us consider that the wall of a certain bubble of radius $R$ \ has a
thickness $\Delta t$. Assuming that $R>>\Delta t$ the total mass of the wall
is 
\begin{equation}
M\approx \rho 4\pi R^{2}\Delta t
\end{equation}
where $\rho $ is the density of matter in the wall and is at least $3\times
10^{-28}$kg/m$^{3(25)}$. Therefore, the potential energy of a galaxy in the
wall is 
\begin{equation}
E_{p}\approx -\frac{Gm\rho 4\pi R^{2}\Delta t}{R}.
\end{equation}

\noindent Hence the total energy of the galaxy is 
\begin{equation}
E\approx \frac{1}{2}mv^{2}-\frac{Gm\rho 4\pi R^{2}\Delta t}{R}=\frac{1}{2}%
mv^{2}-Gm\rho 4\pi R\Delta t
\end{equation}
where we have used Newtonian mechanics since space is flat on large scales.
In order to have a closed Universe the maximum value $V$ of $v$ is then 
\begin{equation}
V\approx \sqrt{\rho 8\pi GR_{M}\Delta t}
\end{equation}
in which $R_{M}$ is the maximum radius that the bubble can attain. If we
take $R_{M}\approx 40$Mpc/h$^{(26)}$, $\Delta t\approx 5/h$Mpc$^{(27)}$ and
for $\rho $ the matter density from galaxies, we obtain $V\approx (350-600)$%
km/s which is a quite reasonable figure and is of the order of the peculiar
velocities of galaxies. If the Universe had nine times more mass $V$ would
be in the range ($1050-1800)km/s$ which would be quite off the mark. {\large %
Hence, the mass that we have measured is approximately all the mass that
actually exists and it does close the Universe. }It may be slightly larger
just because of the mass of brown stars. The above formula also shows that
the velocities of galaxies in the wall increases with $\sqrt{R}$ which is in
good agreement with observations.

\subsection{\protect\Large The Expansion of the Universe and the Background
Radiation}

It is quite remarkable the similarity between a supernova explosion and the
Big Bang. In supernova debris we find sheets and filaments of gas, and
underdense and overdense regions. We find the same in the large scale
structure of the Universe: sheets, filaments and voids. There are more
similarities. In supernova debris we find shells of gas expanding at speeds
in the range $10^{3}-10^{4}$km/s. There are also shells in the Universe. As
di Nella and Paturel$^{28}\;$ show ``The distribution of galaxies up to a
distance of 200 Mpc (650 million light-years) is flat and shows a structure
like a shell roughly centered on the Local Supercluster (Virgo Cluster).
This result clearly confirms the existence of the hypergalactic large scale
structure noted in 1988. This is presently the largest structure ever
seen.'' This is so because both explosions, either in supernovae or in the
Universe, are caused by the same force: THE SUPERSTRONG FORCE. It is worth
mentioning that the above picture of the universal expansion maintains
nucleosynthesis untouched. We only have to reinterpret the cosmic background
radiation(CBR).

The recent data of Mather et al.$^{29}\;$ on the CBR indicate a temperature $%
T_{0}=2.7$K. As we konw the frequency at the peak of the spectrum, $\nu
_{max}$, is related to $T\;$ by $\nu _{max}/T=5.88{\times }10^{10}$Hzdeg$%
^{-1}$. On the other hand, during collapse, the temperature and the density
of a collapsing mass(supernova) obeys the equation$^{30}$ 
\begin{equation}
T=T_{c,i}\left( \frac{\rho }{\rho _{c,i}}\right) ^{\frac{1}{3}}
\end{equation}
\noindent where $T_{c,i}=8.0{\times }10^{9}$K, and $\rho _{c,i}=3.7{\times }%
10^{9}$gcm$^{-3}$, are the temperature and the density at the onset of
collapse. Using the above equation for a density slightly higher than
nuclear density, around $10^{15}$gcm$^{-3}$, we obtain $T=5.2{\times }%
10^{11} $K, and $\nu _{max}=3{\times }10^{22}$Hertz, which corresponds to an
energy of 124 MeV. This is quite close to the mass of pions. For example,
the annihilation of $\pi ^{+}\;$ with $\pi ^{-}\;$ produce photons with
energies of about 140 MeV. Therefore, the primordial photons that produced
the CMB may have been created by pion annihilation.

\bigskip

\subsection{\protect\Large The Planetary Evidence for the Superstrong
Interaction}

As McCaughrean and Mac Low$^{31}\;$ say ``Mass outflow is known to be a
common and perhaps inevitable part of star formation''. Edwards et al.$%
^{32}\;$ also states that observations of young low-mass stars at optical,
near-infrared, and milimeter wavelengths often reveal highly collimated
bipolar jets and molecular outflows. And jets carry large amounts of energy
and momentum from the central regions of young stellar objects$^{33}$(YSOs).
Moreover, between 25\% to 75\% of YSOs in the Orion nebula appear to have
disks$^{34}$.

It is very important to point out that no theory of planet formation is able
to offer a reasonable explanation for the origin of the large amount of iron
that is found in the cores of all planets, and the heavy elements(such as
uranium) found on Earth and for sure in other planets. Of course, the jets
and outflows mentioned above contain the planetary iron. How was it formed
and expelled? We may explain it as follows: Because of inhomogeneities, when
the solar nebular collapsed some parts of it got so squeezed that all heavy
elements were formed and expelled due to the action of the superstrong
interaction. That is, a small part of the sun suffered a supernova-like
explosion. We easily observe that it was just a small ejection since the
mass of all planets is only about one thousandth of the mass of the Sun. We
expect that more massive stars eject more mass from their centers.\newline

\subsection{\protect\Large The Rotation of Spiral Galaxies}

The rotational curve of spiral galaxies is one of the biggest puzzles of
nature. It is possible to give a reasonable explanation for this puzzle in
terms of the action of the superstrong force. In the process we will also
explain the formation of the spiral structure of the arms.

First, let us consider the central nucleus(or bulge). The whole bulge
expands slowly throughout the lifetime of a galaxy. For simplification let
us consider a uniform density for the bulge. Because mass varies as $r^{3}\;$
and the gravitational force varies as $r^{-2}\;$ we expect the tangential
velocity to be proportional to $r$.

Now, let us consider the tangential velocities of stars in the disk. As was
shown above the disk was formed by the shedding of matter from the center of
the galaxy where a denser core existed. The mass is expelled with speeds in
the range $10^{2}-10^{3}$km/s. Let us consider that the bulge has a radius $%
R_{B}\;$ and also that, because of the action of the superstrong force, a
certain mass of gas $m\;$ is expelled from the center(Fig. 6.1). Because of
its radial velocity, the mass $m$ will continue to distance itself from the
bulge, but its tangential velocity is kept fixed because of the action of
repulsion and because of the transfer of angular momentum from the bulge to
the mass. This may be shown in the following way: As the mass goes away from
the center it increases its angular momentum. At a distance $r$ the angular
momentum is given by 
\begin{equation}
J=mrv_{t}
\end{equation}
\noindent where $v_{t}\;$ is the tangential velocity. Because $J$(of the
mass $m$) increases with time(and with $r$) we have 
\begin{equation}
\frac{dv_{t}}{v_{t}}>-\frac{dr}{r}.
\end{equation}
\noindent Integrating, we obtain 
\begin{equation}
ln{\frac{v_{t}}{v_{to}}}>ln{\frac{r_{o}}{r}}
\end{equation}
\noindent where $r_{o}\;$ is the position of the mass at a time $t_{o}\;$
and $r\;$ is its position at a later time. Both positions are measured from
the center. Because the logarithm is an increasing function of the argument,
we must have 
\begin{equation}
\frac{v_{t}}{v_{to}}>\frac{r_{o}}{r}.
\end{equation}
\noindent We clearly see that $v_{t}=v_{t_{o}}\;$ is a solution of the above
inequality because $r\;$ is always larger than $r_{o}$. Thus the mass $m$
gains angular momentum. Because of conservation of angular momentum the
galactic nucleus must decrease its angular momentum by the same amount. A
recent study shows that the arms of spirals ``transport angular momentum
radially within galactic disks''$^{36}$. If we consider that the angular
velocity of the nucleus does not diminish(which is more plausible than
otherwise), then its mass must diminish, i.e., the nucleus needs to shed
more matter outwards. Since $v_{t}$ remains the same the angular velocity
must decrease as the mass goes away from the center. This generates the
differential rotation observed in all spiral galaxies. The formation of the
spiral structure is, therefore, directly connected with the evolution of the
galaxy.

We can easily show that the beautiful spiral arms are described by a
logarithmic spiral(in an inertial frame). The angle $\theta\;$ measures the
angular position of $m\;$ with respect to the center of the bulge and $%
\phi\; $ measures the angle in the bulge at position $R\;$ where the mass
left it. The angular velocity of the bulge is $\Omega$. Let us consider that
the tangential velocity of the mass $m\;$ is a constant. Therefore, we
obtain 
\begin{eqnarray}
r{\omega} &=& r\frac{d{\theta}}{dt} = R\frac{d{\phi}}{dt} = R{\Omega} =
v_{t} = constant
\end{eqnarray}
\noindent where $R\;$ is the radius of the galactic bulge and $v_{t}\;$ is
the tangential velocity of the mass $m$. We have that 
\begin{eqnarray}
d{\theta} &=& {\omega}dt = \frac{R{\Omega}}{r}dt = \frac{R{\Omega}}{rv_{r}}dr
\end{eqnarray}
\noindent where we have used the fact that $v_{r} = \frac{dr}{dt}$.
Considering that $v_{r}\;$ varies slowly with $r$(or $t$) we may integrate $d%
{\theta}\;$ and obtain 
\begin{eqnarray}
r &{\approx}& Re^{\frac{v_{r}}{v_{t}}\theta}.
\end{eqnarray}
\noindent \textit{This is the equation of the logarithmic spiral}. We
imediately obtain that 
\begin{eqnarray}
\omega &{\approx}& {\Omega}e^{-\frac{v_r}{v_t}\theta}.
\end{eqnarray}
\noindent We may also calculate $\phi$. It is given by 
\begin{eqnarray}
\phi &{\approx}& \kappa\left(e^{\frac{\theta}{k}} - 1\right)
\end{eqnarray}
\noindent where $\kappa\;$ is given by $v_{t}/v_{r}$.

The ratio $\kappa=v_{t}/v_{r}\;$ distinguishes between the two types of
spiral galaxies. If $\kappa{\ll}1$, then $\omega\;$ diminishes rapidly with $%
\theta$. This corresponds to spirals with bars. Conversely, if $\kappa{\gg}1$%
, then $\omega\;$ diminishes slowly and only reaches a very low value for
large $\theta$. This is consistent with the data on spiral galaxies. The
middle ground $\kappa{\approx}1\;$ corresponds to intermediate cases. A
typical spiral without bars should have $\kappa{\gg}1$.

Let us now consider the problem from the point of view of a frame fixed in
the galactic bulge and rotating with it(Fig. 6.2). We may define an angle $%
\psi \;$ related to $\theta \;$ and $\phi \;$ by $\psi =\phi -\theta $.
Therefore, $\psi \;$ is given by 
\begin{equation*}
\psi =\kappa \left( e^{\frac{\theta }{\kappa }}-1\right) -\theta .
\end{equation*}
\noindent For small $\theta \;$ one has $\psi {\approx }{\theta }^{2}/{%
2\kappa }\;$ and $r{\approx }Re^{\sqrt{\frac{2}{\kappa }\psi }}$ and for
large $\theta \;$ we have $\psi {\approx }{\kappa }e^{\frac{\theta }{\kappa }%
}\;$ and $r{\approx }R\frac{\psi }{\kappa }$. Therefore, in this rotating
frame the mass $m$ also describes a spiral curve as it moves away from the
center.

Let us now estimate the order of magnitude of the radial velocity of gas(and
stars) in the galactic disk. The radius of our galaxy is 50000 light years
and the age of our galaxy is of the order of magnitude of the age of the
Universe, $10^{17}$s. The gas which is at the edge of the disk must have
moved from the center with a mean velocity of about 5km/s. Since the gas was
expelled with much higher velocities the mean velocity of stars far from the
center are very small because the gravitational attraction slowed down the
mass to very small velocities. The above figure is just a rough estimation.
It is very important to obtain the mean radial velocities of stars in the
spiral arms of the Milky Way to compare with it.

\subsection{\protect\Large Black Holes Do Not Exist}

There have been several reports in the literature attributing to black holes
the infall of matter towards the centers of galaxies. Actually, as we will
see below the infall of matter towards the centers of galaxies is an
evidence of the existence of the superstrong interaction. And moreover, the
infall shows that the superstrong force has a very small range.

A possible explanation on the subject is the following: due to the action of
the superstrong force(repulsive) matter is expelled from the centers of
galaxies, but it may fall back if it does not acquire enough speed to escape
from the gravitational field of the center of the galaxy. And the fallback
mimics the existence of a black hole.

The black hole is avoided because of the existence of the superstrong
interaction. We can argue that such interaction has to exist to avoid such a
thing as a black hole which violates many physical laws simply because it
represents a cutoff in space time. Actually it is a cutoff from reality. Let
us take a simple example. Let us suppose that a certain number of hydrogen
atoms, for instance, fall into a black hole and disappears inside of it.
What is done with conservation of energy and momentum? And what is then done
with conservation of baryon number and lepton number? A more crucial
question is the following: do particles exist at all inside black holes?
From these questions and from several other ones we suspect that these
exotic and \textit{mathematical} objects do not exist. Nature has not
reserved a role for them.

It is very important to have in mind that mathematical solutions have a
broader scope than physical solutions simply because Nature is restrictive.
That is, reality is restrictive. A mathematical theory can be wonderful from
the intelectual point of view but it may be unphysical. \textit{We can not
force Nature to obey our theories and hypotheses.} We have to accept Nature
as it is.

Let us make some calculations having in mind what has been argued and
calculated in the previous sections and chapters. According to general
relativity an object with a surface potential $V_{S}\gtrsim (GM)/(Rc^{2})$
is a black hole. Let us consider, for example, an object with a mass of
about a solar mass confined to a volume with a very small radius of just
1km. It has a density of about $\rho \approx 10^{17}g/cm^{3}$. This object
has thus $10^{57}$ nucleons separated from each other by just 0.2 F. The
superstrong force halts the collapse by means of the exchange of vector
mesons. According to \ section 5.3 $b\overline{b}$ mesons are the main
mediators in this case for their Compton wavelengths are about 0.126 F and
their size is about 0.2F. We can also estimate the order of magnitude of the
superstrong coupling $(g_{ss}^{2})_{b\overline{b}}$ for $b\overline{b}$
mesons. In order to halt the collapse the overall repulsion due to the
superstrong interaction should be approximately equal to the gravitational
energy, that is, 
\begin{equation}
G\frac{M^{2}}{R}\approx n^{2}(g_{ss}^{2})_{b\overline{b}}\frac{4\pi }{\mu _{b%
\overline{b}}^{2}}\mathcal{V}.
\end{equation}
The formula on the right is calculated in reference 37. As $\mathcal{V}%
=(4\pi /3)R^{3}$ and $n=N/\mathcal{V}$, we obtain 
\begin{equation}
(g_{ss}^{2})_{b\overline{b}}\approx \frac{GM^{2}\mu _{b\overline{b}}^{2}}{%
4\pi RVn^{2}}=\frac{Gm_{p}^{2}\mu _{b\overline{b}}^{2}R^{2}}{3}
\end{equation}
where $m_{p}$ is the proton mass. The above formula gives 
\begin{equation}
(g_{ss}^{2})_{b\overline{b}}\approx 6.2\times 10^{-21}`Jm=3.9\times
10^{4}GeV\ F.
\end{equation}
In the case of $t\overline{t}$ mesons (size approximately equal to 0.007F)
we obtain 
\begin{equation}
(g_{ss}^{2})_{t\overline{t}}\approx 10^{7}GeV\ F.
\end{equation}
The above figures are completely attainable.

A serious and definite blow on the black hole idea was the recent finding of
an initial mass for the Universe$^{38}$ because at some point this mass was
concentrated in a very small volume and, according to general relativity, it
was a black hole. But the measurements are showing that it was not a black
hole, and was rather just a soup of protons, neutrons and electrons.

\bigskip

\noindent {\Large References}

\bigskip

\noindent 1. S.L. Shapiro and S.A. Teukolsky, in Black Holes, White Dwarfs
and Neutron Stars, John Wiley \& Sons, New York(1983), p. 513.

\noindent 2. J.D. Walecka, Annals of Phys. \textbf{83}, 491 (1974).

\noindent 3. Yu. A. Baurov and A.V. Kopajev, hep-ph/9701369.

\noindent \noindent 4. I. M\'{a}rquez, F. Durret, and P. Petitjean,
astro-ph/9810012.

\noindent 5. A. Yahil, astro-ph/9803052.

\noindent 6. P.A. Shaver, L.M. Hook, C.A. Jackson, J.V. Wall, and K.I.
Kellermann, astro-ph9801211.

\noindent 7. M. Pettini, C.C. Steidel, M. Dickinson, M. Kellogg, M.
Giavalisco, and K.L. Adelberger, in \textit{The Ultraviolet Universe at Low
and High Redshift: Probing the Progress of Galaxy Evolution} (Eds. W.H.
Waller et al.) AIP,1997.

\noindent \noindent 8. G. K. Miley and A.P. Hartsuijker, \textit{A \& AS} 
\textbf{291}, 29 (1978).

\noindent 9. T.S. Slatler, I.R. King, P. Krane, and R.I. Jedrzejwski,
astro-ph/9810264.

\noindent 10. E.J. M. Colbert, A. S. Wilson, and J. Bland-Hawthorn, \textit{%
The Radio Emission from the Ultra-Luminous Far-Infrared Galaxy NGC 6240},
preprint network astro-ph; astro-ph/9405046, May 1994.

\noindent 11. W. J. Kaufmann,III, in \textit{Galaxies and Quasars}%
(W.H.Freeman and Company, San Francisco, 1979).

\noindent 12. P.A. Shaver, L.M. Hook, C.A. Jackson, J.V. Wall, and K.I.
Kellermann, astro-ph9801211.

\noindent 13. S.C. Chapman, G.A.H. Walker and S.L. Morris, astro-ph/9810250.

\noindent 14. H. Falcke and P.L. Biermann, astro-ph/9810226.

\noindent 15. T. M. Heckman, K. C. Chambers and M. Postman, \textit{Ap.J.}, 
\textbf{391}, 39(1992).

\noindent 16. S. Baum and T. M. Heckman, \textit{Astrophys. J} \textbf{336},
702(1989).

\noindent 17. N. Jackson and I. Browne, \textit{Nature}, \textbf{343},
43(1990).

\noindent 18. A. Lawrence, \textit{Mon. Not. R. Astr. Soc.}, 1992, in press.

\noindent 19. R. W. Goodrich and M. H. Cohen, \textit{Astrophys. J.} \textit{%
391}, 623(1992).

\noindent 20. Y. Sofue, \textit{Astro. Lett. Comm.} \textbf{28}, 1(1990).

\noindent 21. N. Nakai, M. Hayashi, T. Handa, Y. Sofue, T. Hasegawa and M.
Sasaki, \textit{Pub. Astr. Soc. Japan} \textbf{39}, 685(1987).

\noindent 22. K. L. Visnovsky, C. D. Impey, C. B. Foltz, P. C. Hewett, R. J.
Weymann and S. L. Morris, \textit{Astrophys. J.} \textbf{391}, 560(1992).

\noindent 23. T. A. Boroson and S. E. Persson, \textit{Astrophys. J.} 
\textbf{293}, 120(1985).

\noindent 24. M. V. Berry, in \textit{Principles of Cosmology and
Gravitation, }Adam Hilger, Bristol (1991), p.116.

\noindent 25. Ibidem, p. 4.

\noindent 26. H. El-Ad, T. Piran, and L.N. da Costa, Mon. Not. R. Astron.
Soc. \textbf{287}, 790(1997).

\noindent 27. M. Roos, in Introduction to Cosmology, John Wiley \& Sons,
Chichester(1994), p. 5.

\noindent 28. H. di Nella et G. Paturel, \textit{C.R.Acad.Sci. Paris}, 
\textbf{t.319}, S\'{e}rie II, p. 57-62, 1994

\noindent 29. J.C.\ Mather et al., Astrophys. J.(Letters) \textbf{354},
L37(1990).

\noindent 30. S.L. Shapiro and S.A. Teukolsky, in Black Holes, White Dwarfs
and Neutron Stars, John Wiley \& Sons, New York(1983), p. 536.

\noindent \noindent 31. M.J.McCaughrean and M.-M. Mac Low, astro-ph/9611058.

\noindent 32. S. Edwards, T.P. Ray, and R. Mundt, in \textit{Protostars and
Planets III}, eds. E.H. Levy and J.I.Lunine (Tucson: University of Arizona
Press), p. 567,1993.

\noindent 33. G. Mellema and A. Frank, astro-ph/9710255.

\noindent 34. C.F. Prosser, J.R. Stauffer, L. Hartmann, D.R. Soderblom, B.F.
Jones, M.W. Werner, and M.J. McCaughrean, \textit{Ap.J.} \textbf{421},
517(1994).

\noindent 35. S.L. Shapiro and S.A. Teukolsky, in Black Holes, White Dwarfs
and Neutron Stars, John Wiley \& Sons, New York(1983), pages 209, 210.

\noindent 36. O. Y. Gnedin, J. Goodman, and Z. Frei, Princeton University
Observatory preprint \textit{Measuring Spiral Arm Torques: Results for M100}%
, Astro-ph/9501112.

\noindent 37. S.L. Shapiro and S.A. Teukolsky, in Black Holes, White Dwarfs
and Neutron Stars, John Wiley \& Sons, New York(1983), pp. 209,210.

\noindent 38. C. Seife, Science, Vol. 292, No. 5518, p. 823 (2001).

\pagebreak

\section{\protect\bigskip {\protect\LARGE Associated Fermions and the Hidden
Realm of Gravity}}

{\LARGE \bigskip \rule[1.5in]{0in}{0.17in}}

\subsection{\protect\Large Associated Fermions and the Dual Role of Neutrinos%
}

Let us take a closer look at the table below which was considered in chapter
I. At the center we find the structured states which are the different
media. Let us begin discussing what happens in ordinary matter, for example,
in a solid. In a solid the transport properties are mainly related to the
motion of electrons(fermions) which are the carriers of the electric charge.
Besides the carrier of the charge there is the carrier of the
electromagnetic interaction which is the photon (boson). In the nucleus
there is a similar pattern: the carriers of the strong charge(isospin) are
the nucleons (fermions) and the carriers of the strong force are bosons
(scalar mesons). In the quark we have that primons are the carriers of the
superstrong charge and the carriers of the superstrong force are vector
mesons(bosons). What about the galactic medium? Before answering this
question let us recall that neutrinos are copiously produced in galaxies and
fill in the Universe. And they are also produced in the weak decay of a
neutron into a proton. Hence, we can say that an electron neutrino carries a
superweak charge $\mathcal{G}_{\mathcal{N}N}\mathcal{\ }$which is
transferred to it from the neutron in the decay. Therefore electron
neutrinos should be the charge carriers of the superweak charge and a new
boson called \textit{numeron (}derived from the latin \textit{numerus }%
(number))\textit{\ , }$\mathcal{N}$, is the carrier of the superweak
interaction. It is chosen with this name because it is related to the number
of protons with respect to the number of neutrons. Hence we have found out
why electron neutrinos fill in the universe.

We may call these fermions of \textit{associated fermions} because they are
associated to the fundamental forces. And there is always a quantized
current linked to each associated fermion. \emph{The four associated
fermions {\large primon, electron, nucleon and neutrino,} are the only
fermions that are stable.}

\pagebreak

\vspace*{0.5in}

\begin{center}
\begin{tabular}{ccccccc}
\hline\hline
&  &  &  &  &  &  \\ 
& ? &  & quark &  & nucleon &  \\ 
&  &  &  &  &  &  \\ 
& nucleon &  & nucleus &  & atom &  \\ 
&  &  &  &  &  &  \\ 
& atom &  & gas &  & galaxy &  \\ 
&  &  & liquid &  &  &  \\ 
&  &  & solid &  &  &  \\ 
&  &  &  &  &  &  \\ 
& galaxy &  & galactic medium &  & ? &  \\ 
&  &  &  &  &  &  \\ \hline\hline
&  &  &  &  &  & 
\end{tabular}
\end{center}

\vskip .2in

\begin{center}
\parbox{4in}
{Table 1.1. The table is arranged in such a way to show the links between the 
structured states and the units of creation. The interrogation marks above 
imply the existence of prequarks and of the Universe itself as units of
creation}
\end{center}

\bigskip

\bigskip

The generalized fermionic current is therefore 
\begin{equation}
j_{V}^{\mu }=q\overline{\psi }\gamma ^{\mu }\psi
\end{equation}
for the two vectorial fields (superstrong and electromagnetic) where $q$ is
the generalized charge (electromagnetic or superstrong), and $\psi $ is the
fermion Dirac spinor for electrons or primons.

For the pionic coupling between nucleons we have

\begin{equation}
j_{S,\pi N}=g_{\pi N}\overline{\psi }\gamma ^{5}(\mathbf{\tau .\Phi })\psi
\end{equation}
in which $g_{\pi N}$ is the strong charge, $\mathbf{\tau }$ are the isospin
Pauli matrices, $\psi $ is the nucleonic isospinor 
\begin{equation}
\psi =\left( 
\begin{array}{c}
\varphi _{p} \\ 
\varphi _{n}
\end{array}
\right) .
\end{equation}

\pagebreak

\vspace*{0.5in}

\noindent and $\mathbf{\Phi }$ is the isovector 
\begin{equation}
\mathbf{\Phi }=\left( 
\begin{array}{c}
\phi _{\pi ^{+}} \\ 
\phi _{\pi ^{0}} \\ 
\phi _{\pi ^{-}}
\end{array}
\right)
\end{equation}
where $\phi $ are pseudoscalar functions. We will deal with the superweak
field in the next chapter. For now we can write just a general and
unpretentious expression for the superweak current such as 
\begin{equation}
j_{S,\mathcal{N}N}=\mathcal{G}_{\mathcal{N}N}\overline{\psi }\gamma ^{5}%
\mathbf{\Phi }_{\mathcal{N}}\psi
\end{equation}
where $\mathcal{G}_{\mathcal{N}N}$ is the superweak coupling constant, $\psi 
$ is the same as above, a nucleonic isospinor 
\begin{equation}
\psi =\left( 
\begin{array}{c}
\varphi _{p} \\ 
\varphi _{n}
\end{array}
\right) .
\end{equation}
and $\mathbf{\Phi }_{\mathcal{N}}$ is a scalar function. We see that 
\begin{equation}
j_{S,\mathcal{N}N}=\mathcal{G}_{\mathcal{N}N}\phi _{p}\mathbf{\Phi }_{%
\mathcal{N}}\phi _{n}+\mathcal{G}_{\mathcal{N}N}\phi _{n}\mathbf{\Phi }_{%
\mathcal{N}}\phi _{p}
\end{equation}
since the superweak neutrino current happens only when a proton interacts
with a neutron or vice versa.\ 

As we saw in chapters 2 and 3 primons also carry scalar currents. Therefore,
there is then a strong scalar current of primons similar to that given by
Eq.(182).

The existence of the fermionic currents above mentioned means that the
corresponding charges are quantized, that is, there is always a charge unit.
For example, for the electric charge the electron's is the minimum one.

We can then construct the table on the next page for the associated fermions
and the fundamental forces. In it we notice that the weak force is different
from the four forces above mentioned because it has neutrino weak currents
and electronic weak currents as well. \ Therefore neutrinos have a dual role
since they are one of the charge carriers of the weak force and also the
carrier of the superweak charge. The gravitational force appears to belong
to a very special class as will be shown below. The strong force is also
special in the sense that it has an associated fermion (the primon) in the
confined world and another associated fermion\ (the nucleon) for the
ordinary world. Maybe the Higgs bosons should be classified as a strong
boson.

The table points towards an important trend: primons, nucleons and electron
have masses, and thus, the neutrino should have a mass too.

\pagebreak

\vspace*{0.5in}

\begin{center}
\begin{tabular}{||c|c|c||}
\hline\hline
&  &  \\ 
Fundamental Force & Interaction Carriers & Charge Carriers \\ 
& (Bosons) & (Stable Fermions) \\ 
&  &  \\ 
&  &  \\ \hline\hline
&  &  \\ 
Superstrong & 
\begin{tabular}{l}
supergluons, $\rho (770)$, $\omega (782)$, \\ 
$\phi (1020)$, $K^{\ast }(892)$, $D^{\ast }(2007)^{0}$, \\ 
$D^{\ast }(2010)^{\pm }$, $J/\psi (1S)$, $\psi (2S)$, \\ 
$\psi (3700)$, $\psi (4040)$, $\Upsilon (1S)$, \\ 
$\Upsilon (2S)$, $\Upsilon (3S)$, $\Upsilon (4S)$, ....
\end{tabular}
& primons \\ 
&  &  \\ \hline
&  &  \\ 
Strong & 
\begin{tabular}{l}
$\pi ^{\pm }$, $\pi ^{0}$, $\eta $, $K^{\pm }$, $K^{0}$, $\bar{K}^{0}$, $%
D^{\pm }$, \\ 
$D^{0}$, $\bar{D}^{0}$, $D_{s}^{+}$, $D_{s}^{-}$, $B^{+}$, $B^{-}$, \\ 
$B^{0}$, $\bar{B}^{0}$, $B_{s}^{0}$, $\overline{B_{s}}^{0}$, $\eta _{c}(1S)$
....
\end{tabular}
& 
\begin{tabular}{c}
primons and \\ 
nucleons
\end{tabular}
\\ 
&  &  \\ 
&  &  \\ \hline
&  &  \\ 
Electromagnetic & $\gamma $ & electron \\ 
&  &  \\ \hline
&  &  \\ 
Superweak & $\mathcal{N}$ & neutrino($\nu _{e}$) \\ 
&  &  \\ \hline
&  &  \\ 
Weak & $Z^{0}$, $W^{\pm }$ & neutrino and \\ 
&  & electron \\ 
&  &  \\ \hline
&  &  \\ 
Gravity & graviton? & ? \\ 
&  &  \\ \hline\hline
\end{tabular}
\linebreak
\end{center}

\vskip .2in

\begin{center}
\parbox{4.5in}
{Table 7.1. Table of the Fundamental Forces, Interaction Bosons, and 
Associated Fermions. We notice that the weak interaction is special and 
that the gravitational interaction is more than special and very 
strange if its associated fermion does not exist.}
\end{center}

\pagebreak

\vspace*{0.5in}

\subsection{\protect\Large The Hidden Realm of Gravity }

Following the reasoning above developed we may ask if gravity has a charge
carrier. If it has one it is not known yet. But we suspect that such fermion
does not exist (except inside hadrons) because massless particles are
attracted gravitationally by a massive particle, as general relativity shows
it, and as the experiments have revealed. According to general relativity
the 4-momentum of a freely moving photon is written as$^{1}$%
\begin{equation}
\nabla _{p}p=0
\end{equation}
where the four-momentum of the photon is $p=\frac{d}{d\lambda }$ and $%
\lambda $ is an affine parameter. This geodesic equation can be written as 
\begin{equation}
\frac{dp^{\alpha }}{d\lambda ^{\ast }}+\Gamma _{\beta \gamma }^{\alpha
}p^{\beta }p^{\gamma }=0
\end{equation}
from which we calculate that a photon(\emph{zero mass}) suffers a deflection
given by the angle 
\begin{equation}
\Delta \phi =4M/b=1^{\prime \prime }.75(R_{\odot }/b)
\end{equation}
in which $M$ is the sun's mass, $R_{\odot \text{ }}$is the sun's radius and $%
b$ is the impact parameter. This means that even particles with
gravitational \ \ ``charge '' equal to zero suffer the influence of gravity.
On the other hand in Newtonian gravity we have that the gravitational
potential energy between two massive bodies is 
\begin{equation}
E_{p}=-G\frac{m_{1}m_{2}}{r}
\end{equation}
which is of Yukawa type. According to this equation the two masses are the
two gravitational charges. But since we do not have a fermionic mass carrier
we cannot write each mass as a multiple of the fermion mass. Therefore, mass
cannot be quantized. Without the fermionic mass carrier we cannot have mass
currents. How can gravity be quantized without quantizing mass and without
fermionic currents? \ 

Let us see what we arrive at if we admit the existence of such a mass
carrier. Let us call it \emph{masson}. Following section 7.1 we may suppose
it is a 1/2 spin fermion. Since it is a fermion it has to satisfy Dirac
equation which written in covariant form is (a free fermion) 
\begin{equation}
(i\hbar \gamma ^{\mu }\partial _{\mu }-mc)\psi =0.
\end{equation}
But it should also satisfy a four-vector mass current 
\begin{equation}
j^{\mu }=cm\overline{\psi }\gamma ^{\mu }\psi
\end{equation}

\pagebreak

\vspace*{0.5in}

\noindent because the mass is also the charge. Multiplying Dirac equation
above from the left by $\overline{\psi }\gamma ^{\mu }$ we have 
\begin{equation}
i\hbar \overline{\psi }\gamma ^{\mu }\gamma ^{\mu }\partial _{\mu }\psi
=j^{\mu }=cm\overline{\psi }\gamma ^{\mu }\psi .
\end{equation}
which shows that the masson mass comes from the variation of $\psi $.

The masson has to be extremely light and we expect it to interact with
vacuum. Let us consider that it suffers the action of an effective vacuum
potential $\Phi _{V}$ which is capable of creating mass, that is, 
\begin{equation}
\Phi _{V}\psi =nmc\psi
\end{equation}
where $n$ is an integer because $m$ is supposedly the quantum of mass. Thus
we can write 
\begin{equation}
i\hbar \gamma ^{\mu }\partial _{\mu }\psi -mc\psi +\Phi _{V}\psi =i\hbar
\gamma ^{\mu }\partial _{\mu }\psi -mc\psi +nmc\psi =0.
\end{equation}
In the case of minimum creation $n=1$, that is, we obtain 
\begin{equation*}
i\hbar \gamma ^{\mu }\partial _{\mu }\psi =0
\end{equation*}
and 
\begin{equation}
j^{\mu }=0
\end{equation}
which is quite contradictory because it means that the creation of mass does
not generate any mass current. {\large We see then that mass creation from
pure vacuum does not make any sense.}

From Dirac equation we have $i\hbar \gamma ^{\nu }\partial _{\nu }\psi
=cm\psi $ and 
\begin{equation*}
i\hbar \overline{\psi }\gamma ^{\mu }\gamma ^{\nu }\partial _{\nu }\psi =cm%
\overline{\psi }\gamma ^{\mu }\psi =j^{\mu }
\end{equation*}
As $\gamma _{\mu }\gamma ^{\mu }=4$, we can write 
\begin{equation}
i\hbar \overline{\psi }\gamma ^{\mu }\gamma ^{\nu }\partial _{\nu }(\gamma
_{\mu }\gamma ^{\mu }\psi )=i\hbar \overline{\psi }\gamma ^{\mu }\gamma
^{\nu }\gamma _{\mu }\partial _{\nu }(\gamma ^{\mu }\psi )=4mc\overline{\psi 
}\gamma ^{\mu }\psi .
\end{equation}
Since $\gamma ^{\mu }\psi $ is also a solution of Dirac equation we obtain 
\begin{equation}
i\hbar \overline{\psi }\gamma ^{\nu }\partial _{\nu }(\gamma ^{\mu }\psi )=mc%
\overline{\psi }\gamma ^{\mu }\psi
\end{equation}
and 
\begin{equation}
i\hbar \overline{\psi }\gamma ^{\nu }\gamma ^{\mu }\gamma _{\mu }\partial
_{\nu }(\gamma ^{\mu }\psi )=4mc\overline{\psi }\gamma ^{\mu }\psi
\end{equation}
Since $\gamma ^{\nu }\gamma ^{\mu }+\gamma ^{\mu }\gamma ^{\nu }=2g^{\nu
}{}^{\mu }$, summing up equations (195) and (197) we obtain

\begin{equation}
i\hbar \overline{\psi }g^{\nu }{}^{\mu }\gamma _{\mu }\partial _{\nu }\gamma
^{\mu }\psi =4mc\overline{\psi }\gamma ^{\mu }\psi
\end{equation}

\pagebreak

\vspace*{0.5in}

\noindent where $g^{\nu }{}^{\mu }$ is the metric 
\begin{equation}
g^{\nu \mu }{}=\left( 
\begin{array}{cccc}
1 & 0 & 0 & 0 \\ 
0 & -1 & 0 & 0 \\ 
0 & 0 & -1 & 0 \\ 
0 & 0 & 0 & -1
\end{array}
\right) .
\end{equation}

\noindent Therefore, we obtain the fermionic mass operator (of the \emph{%
masson}) 
\begin{equation}
m=\frac{i}{4c}\hbar g^{\nu }{}^{\mu }\gamma _{\mu }\partial _{\nu }
\end{equation}
and the mass current 
\begin{equation}
j^{\mu }=\frac{i\hbar }{4}\overline{\psi }g^{\nu }{}^{\mu }\gamma _{\mu
}\partial _{\nu }\gamma ^{\mu }\psi =\frac{i\hbar }{4}\overline{\psi }g^{\mu
\nu }{}\partial _{\nu }\gamma _{\mu }\gamma ^{\mu }\psi =i\hbar \overline{%
\psi }g^{\mu \nu }{}\partial _{\nu }\psi .
\end{equation}
These two equations clearly show that the \emph{masson }mass depends on the
metric. In curved space-time we can always choose a small region where
space-time is approximately flat. Hence, we can extend the meaning of $%
g^{\nu }{}^{\mu }$ to include curved space-time. Doing this we notice that
since the \emph{masson } mass depends on the metric it can not be unique,
that is, it has different values in different curved space-times. Since flat
space time is a local approximation of curved space-time its mass has only a
local meaning. Therefore, we stumbled into another obstacle in quantizing
gravity. We can do this formally. Let us take an orthogonal metric, that is
a metric in which $g^{\nu }{}^{\mu }=0$, for $\nu \neq \mu $. We have then
the metric 
\begin{equation}
g^{\nu \mu }{}=\left( 
\begin{array}{cccc}
g_{00} & 0 & 0 & 0 \\ 
0 & g_{11} & 0 & 0 \\ 
0 & 0 & g_{22} & 0 \\ 
0 & 0 & 0 & g_{33}
\end{array}
\right) .
\end{equation}
If we are in a very small region of curved space-time (without large
curvature) we can say that $g_{00}\approx 1+f_{00}$, $g_{11}\approx
-1+f_{11} $, $g_{22}\approx -1+f_{22}$ and $g_{33}\approx -1+f_{33}$, and we
have for small $f_{jj}$ 
\begin{equation}
\delta m=\frac{i}{4c}\hbar \Delta ^{\nu }{}^{\mu }\gamma _{\mu }\partial
_{\nu }
\end{equation}
with 
\begin{equation*}
\Delta ^{\nu }{}^{\mu }=\left( 
\begin{array}{cccc}
f_{00} & 0 & 0 & 0 \\ 
0 & f_{11} & 0 & 0 \\ 
0 & 0 & f_{22} & 0 \\ 
0 & 0 & 0 & f_{33}
\end{array}
\right)
\end{equation*}

\pagebreak

\vspace*{0.5in}

\noindent where $\delta m$ is non-Euclidean. This is the mass acquired by
the masson directly from curvature.

{\large We arrive at a quite strong and disturbing conclusion which is: if
the associated fermion of the gravitational field does not exist then
gravitons do not exist either in ordinary matter since bosons make the
interaction between fermions. And this leads us to say that the
gravitational field is not propagated at all in ordinary matter, that is, it
is always a static field.} {\large This is in line with the null results of
gravitational waves. }

The only solution to this puzzle is to assume that the primon is also the
fermion of the gravitational field but this would mean that gravitational
waves and gravitons only exist inside hadrons. There would then exist a
connection between the Higgs and the graviton. In this case confinement
would be related to the hidden realm of gravity. It is worth noticing that
there are models of hadrons that consider that quarks do not have definite
masses and that is why there is the concept of constituent mass. Also, let
us refresh our minds at this point and recall that we had to resort to a
special way (via Higgs bosons) for creating the masses of quarks and let us
also remember that leptons and primons appear to have inherent masses. The
special recourse shows us that mass is created from within, hidden by
confinement, that is, hidden from the ordinary world. This is in line,
actually, with having an initial mass for the Universe, which reveals that
the initial mass as well as the initial space were not created at all. And a
final reminder is more than appropriate: the confined world is also the
world of fractional charge which is in line with what we found before. As
was shown before the masses of quarks are connected to their charges and to
the Higgs bosons. And a last point is also important: the confined world
inside nucleons is also the world where spin has strange properties. We now
see that all this is interconnected and begins to be disentangled.

{\large Therefore, it looks like that gravitational waves (with quanta) in
ordinary matter are impossible and may only be possible inside hadrons (or
in neutron stars).}

Some critcs may say that Dirac equation cannot be applied at all in this
case. And I say that it has to be applied in the present Universe since it
has been proven to be .\textit{flat} and if it is flat it obeys Minkowski
space. Of course Dirac equation cannot be applied in the Planck scale, but
as we deduce from this work the Planck scale never happened.

{\large \bigskip }

\noindent {\Large References}

\bigskip

\noindent 1. C.W. Misner, K.S. Thorne and J.A.\ Wheeler, in Gravitation,
W.H. Freeman and Company, San Francisco, 1973, p. 446.

\pagebreak \vspace*{0.5in}

\section{\protect\bigskip {\protect\LARGE Properties of the Galactic
Structured State}}

{\LARGE \bigskip \rule[1.5in]{0in}{0.17in}}

\subsection{\protect\Large The Superweak Force and Its Connection to
Neutrinos}

From Noether's theorem$^{1}$ it follows that any fundamental force is linked
to a conservation law. That is why Fischbach proposed the fifth force based
on baryon number conservation. According to Fischbach the E\"{o}tv\"{o}s
experiment presented some discrepancies that could be eliminated by
supposing the existence of a composition dependent force.

The potential energy of such hypothetical force is usually represented by a
Yukawa potential which, when added to the standard Newtonian potential
energy, becomes$^{2}$ 
\begin{equation}
V(r)=-\frac{Gm_{1}m_{2}}{r}\left( 1+\alpha \exp {(-r/\lambda )}\right) ,
\end{equation}
\noindent where $\alpha $ is the new coupling in units of gravity and $%
\lambda $ is its range. The dependence on composition can be made explicit
by writing ${\alpha }=q_{i}q_{j}\zeta $ with 
\begin{equation}
\displaystyle q_{i}=cos{\theta }(N+Z)_{i}/\mu _{i}+sin{\theta }(N-Z)_{i}/\mu
_{i},
\end{equation}
\noindent where the new effective charge has been written as a linear
combination of the baryon number and the nuclear isospin per atomic mass
unit, and $\zeta $ is the coupling constant in terms of $G$.

Until now most experimental results have not confirmed the existence of this
force$^{3}$, although they do not rule it out because its coupling
constant(s) may be

\pagebreak

\vspace*{0.5in} \noindent smaller than previously thought. Adelberger et al.$%
^{3}$ have found the upper limit of $10^{-13}$ms${-2}$ in the acceleration
which means that the fifth force is at least $10^{10}$ smaller than the
gravitational force. Of course, this force may only exist if there is a
violation in the weak equivalence principle, which has been proven to hold$%
^{3}$ to a precision of one part in $10^{12}$. But it may be violated if the
precision is improved. They may, then, reveal the existence of the fifth
force.

The superweak force proposed in this work has the same character of that of
the fifth force, but has an infinite range. This means that the mass of the
mediating boson is zero. From the above expression for the fifth force
potential energy we may express the potential energy of the superweak force
in terms of the baryon numbers and isospins of two very large bodies i and j
as 
\begin{eqnarray}
V(r,N,Z) &=&\pm \left( A_{B}(N+Z)_{i}+A_{I}(N-Z)_{i}\right) {\times }  \notag
\\
&&\left( A_{B}(N+Z)_{j}+A_{I}(N-Z)_{j}\right) \mathcal{G}_{\mathcal{N}N}{^{2}%
}\frac{1}{r}
\end{eqnarray}
\noindent where $A_{B}$ and $A_{I}$ are constants and $\mathcal{G}_{\mathcal{%
N}N}$ is the coupling constant. The charge $Q=\mathcal{G}_{\mathcal{N}%
N}[A_{B}(N+Z)+A_{I}(N-Z)]$ poses a problem because mass is proportional to
its first term $\mathcal{G}_{\mathcal{N}N}A_{B}(N+Z)$. Let us take a look if
the second term is good enough for the intended purpose. \ First let us
absorb the constant $A_{I}$ into $\mathcal{G}_{\mathcal{N}N}$ and write $Q=%
\mathcal{G}_{\mathcal{N}N}(N-Z)=\mathcal{G}_{\mathcal{N}N}(2N-B)$. Take a
volume of space where radioactive decay happens in a certain body which has
initially $N_{1}$ neutrons, and $B$ neutrons and protons. At a later time
the body will have $N_{2}$ neutrons and still $B$ neutrons and protons.
Therefore, baryon number is conserved but isospin, of course, is not. The
difference $N_{1}-N_{2}=n_{e}=n_{\nu _{e}}$. Hence we can say that such
decay generates a neutrino current given by 
\begin{equation}
j_{SW,\mathcal{N}N}=\mathcal{G}_{\mathcal{N}N}\overline{\psi }\gamma ^{5}%
\mathbf{\Phi }_{\mathcal{N}}\psi
\end{equation}
as was seen in section 7.1. The smallest superweak charge is thus 
\begin{equation}
Q_{SW}=\mathcal{G}_{\mathcal{N}N}
\end{equation}
and means that besides carrying a lepton number the neutrino carries a
superweak charge. And since it carries a charge it probably has a mass. This
neutrino picture casts doubt on the existence of Majorana neutrinos.

The superweak potential energy between two bodies can then be described as 
\begin{equation}
V_{ij}(r_{ij},N_{i},Z_{j})=\pm (N-Z)_{i}{\times }(N-Z)_{j}\mathcal{G}_{%
\mathcal{N}N}{^{2}}\frac{1}{r_{ij}}.
\end{equation}
Let us make the $\pm 1$ equal to a constant $b$. Then 
\begin{equation}
V_{ij}(r_{ij},N_{i},Z_{j})=b(N-Z)_{i}{\times }(N-Z)_{j}\mathcal{G}_{\mathcal{%
N}N}{^{2}}\frac{1}{r_{ij}}.
\end{equation}

\pagebreak

\vspace*{0.5in}

\noindent The energy of the galaxies of the whole Universe is thus given by 
\begin{equation}
E=\underset{i=1}{\overset{N}{\sum }}\left( K_{i}\right) +\underset{i=1,i>j}{%
\overset{N}{\sum }}V_{ij}^{G}+\underset{i=1,i>j}{\overset{N}{\sum }}%
V_{ij}^{SW}
\end{equation}
where $V_{ij}^{G}$ and $V_{ij}^{SW}$ are the gravitational and superweak
potential energies of the pair $ij$. We can write 
\begin{eqnarray}
E &=&\underset{i=1}{\overset{N}{\sum }}\frac{1}{2}m_{i}v_{i}^{2}-\underset{%
i,j(i>j)}{\overset{N}{\sum }}m_{i}{\times m}_{j}\left( \sqrt{G}\right) ^{2}%
\frac{1}{r_{ij}}+  \notag \\
&&b\underset{i=1,i>j}{\overset{N}{\sum }}(2N-B)_{i}{\times }(2N-B)_{j}%
\mathcal{G}_{\mathcal{N}N}{^{2}}\frac{1}{r_{ij}} \\
&=&\underset{i=1}{\overset{N}{\sum }}\frac{1}{2}m_{i}v_{i}^{2}-\underset{%
i,j(i>j)}{\overset{N}{\sum }}m_{i}{\times m}_{j}\left( \sqrt{G}\right) ^{2}%
\frac{1}{r_{ij}}+  \notag \\
&&bB_{i}B_{j}\underset{i=1,i>j}{\overset{N}{\sum }}(2\eta _{i}-1){\times }%
(2\eta _{j}-1)\mathcal{G}_{\mathcal{N}N}{^{2}}\frac{1}{r_{ij}}
\end{eqnarray}
subject to the condition 
\begin{equation}
B=\underset{i=1}{\overset{N}{\sum }}B_{i}=\text{constant.}
\end{equation}

\bigskip

\bigskip It is important to notice that the number of neutrons and protons
vary from galaxy to galaxy for a given time(measured since the Big Bang).
This means that the superweak field among galaxies is not homogeneous. The
two potentials together produce an effective potential, $V(r)$, 
\begin{equation}
V(r)=-\underset{i=1,i>j}{\overset{N}{\sum }}m_{i}{\times m}_{j}\left( \sqrt{G%
}\right) ^{2}\frac{1}{r_{ij}}+b\underset{i=1,i>j}{\overset{N}{\sum }}%
B_{i}B_{j}(2\eta _{i}-1){\times }(2\eta _{j}-1)\mathcal{G}_{\mathcal{N}N}{%
^{2}}\frac{1}{r_{ij}}
\end{equation}
which is responsible for the spatial configuration of galaxies into bubbles
and sheets. Since $m=Bm_{p}$ we have 
\begin{equation}
V(r)=-\underset{i=1,i>j}{\overset{N}{\sum }}\left( \frac{B_{i}B_{j}}{r_{ij}}%
\right) \left[ \left( m_{p}\sqrt{G}\right) ^{2}-b(2\eta _{i}-1){\times }%
(2\eta _{j}-1)\mathcal{G}_{\mathcal{N}N}{^{2}}\right] .
\end{equation}

\pagebreak

\vspace*{0.5in} \noindent If we choose $b=-1$ we obtain 
\begin{equation}
V(r)=-\underset{i=1,i>j}{\overset{N}{\sum }}\left( \frac{B_{i}B_{j}}{r_{ij}}%
\right) \left[ \left( m_{p}\sqrt{G}\right) ^{2}+(2\eta _{i}-1){\times }%
(2\eta _{j}-1)\mathcal{G}_{\mathcal{N}N}{^{2}}\right]
\end{equation}
which accounts for the acceleration of galaxies simply because the product $%
(2\eta _{i}-1){\times }(2\eta _{j}-1)$ increases as the galaxy moves towards
the wall since the number of neutrons increases with time. The choice $b=1$
suggests that the superweak force should be, actually, an attractive force.
Hence its potential energy is 
\begin{equation}
V_{ij}(r_{ij},N_{i},Z_{j})=-(N-Z)_{i}{\times }(N-Z)_{j}\mathcal{G}_{\mathcal{%
N}N}{^{2}}\frac{1}{r_{ij}}.
\end{equation}
In the begining of the Universe $N_{i}\approx N_{j}\approx N_{0}$; $%
Z_{i}\approx Z_{j}\approx Z_{0}$; $r_{ij}\approx r_{0}$. We obtain then 
\begin{eqnarray}
V_{ij}^{0}(r_{0},N_{0},Z_{0}) &=&-(N-Z)_{0}{\times }(N-Z)_{0}\mathcal{G}_{%
\mathcal{N}N}{^{2}}\frac{1}{r_{0}}\varpropto  \notag \\
-g_{SF}^{2}\frac{\exp (-\mu _{SF}r_{0})}{r_{0}} &\sim &-g_{SF}^{2}\frac{1}{%
r_{0}},
\end{eqnarray}
for\ $\mu _{SF}r_{0}<<1,$that is, the superweak force is unified to the
strong force in the beginning of the Universe. This is a preliminary
unification and a more profound (in terms of field theory) is needed.

\subsection{\protect\Large The Galactic Medium is a Strange \ `Metal of
Clusters of Galaxies and Neutrinos'}

Since the `beginning$^{\prime }$ of the universe, after the formation of
primordial stars, the number of neutrons, $N,$ has increased at the expense
of the number of protons which has decreased. This increase in $N $ takes
place in the cores of stars by means of the fusion process that happens in
the whole Universe. As the Universe ages stars become white dwarfs, brown
dwarfs and neutron stars. During the aging process the core density of a
star increases and the high electron Fermi energy drives electron capture
onto nuclei and free protons. This last process, called neutronization,
happens via the weak interaction. The most significant neutronization
reactions$^{4}$ are:

\begin{itemize}
\item  {Electron capture by nuclei,}
\end{itemize}

\begin{eqnarray}
& & e^{-} + (Z,A)\;\;\overset{W}{\longrightarrow} \;\;\nu_{e} + (Z-1,A),
\end{eqnarray}

\pagebreak

\vspace*{0.5in}

\begin{itemize}
\item  {Electron capture by free protons,}
\end{itemize}

\begin{eqnarray}
& & e^{-} + p \;\;\overset{W}{\longrightarrow} \;\;\nu_{e} + n,
\end{eqnarray}
\vskip .1in \noindent where $W$ means that both reactions proceed via
charged currents of the electroweak interaction.

Of course, neutronization takes place in the stars of all galaxies, and
thus, the number of neutrons increases relative to the number of protons as
the universe ages. For example, a white dwarf in the slow cooling stage(for $%
T{\leq }10^{7}$K) reaches a steady proton to neutron density of about 1/8,
and takes about $10^{9}$ years to cool off completely, which is a time close
to the present age of the universe. By then, most stars have become white
dwarfs(or neutron stars). This steady increase is expected to be very slow.

Therefore we can say that galaxies release neutrinos and doing so increase
their superweak charge deficits. Making the analogy with the electric charge
we can say that galaxies are \emph{superweak ions}, that is, they have a
neutrino deficit. And neutrinos in the Universe are like electrons in a
metal, they are almost free and abundant. On the other hand observations
have shown that the medium formed by galaxies has quite a large degree of
order since voids have an average size of about 40$h{-1}$Mpc$^{5}$.
Therefore we can say that the Universe is a sort of metal of clusters of
galaxies in which neutrinos are the conductors.

\bigskip

\subsection{\protect\Large Properties of the Neutrino Gas of the Universe}

Neutronization is a very slow process. Less than 10\% of hydrogen has been
converted to helium. We also know that there is today only one atom of
helium for every 10 atoms of hydrogen. Since the total number of baryons of
the Universe is approximately $10^{69}$, there are, therefore, at least
about $10^{68}$ neutrinos wandering about in the Universe. As we will see
the neutrino density is much larger. We can say, then, that the Universe is
approximately an ideal neutrino gas. Taking into account the present radius
of the Universe ($\sim 3\times 10^{27}cm$) the lower bound for the neutrino
density is about $10^{-14}$ neutrinos/cm$^{3}$. The neutrino current (or
superweak current) is, then, \ 
\begin{equation}
j_{\nu }=n_{\nu }c\mathcal{G}_{\mathcal{N}N}\text{.}
\end{equation}

\pagebreak

\vspace*{0.5in}

\noindent Let us first consider free neutrinos. In this case their energies
are 
\begin{equation}
E(k)=c\hbar k=|\overrightarrow{p}|c=c\hbar \sqrt{%
k_{x}^{2}+k_{y}^{2}+k_{z}^{2}}\text{.}
\end{equation}
Using Born-von Karman boundary conditions the wave function solution for
stationary states is 
\begin{equation}
\Psi _{k}(\overrightarrow{r},t)=\frac{1}{V}\exp i(\overrightarrow{k}\cdot 
\overrightarrow{r}-\omega t)
\end{equation}
where $k_{x}=\frac{2\pi n_{x}}{L}$, $k_{y}=\frac{2\pi n_{y}}{L}$, $k_{z}=%
\frac{2\pi n_{z}}{L}$($n_{x},n_{y},n_{z}$ integers). Therefore, a region of
k-space of volume $\Omega $ contains $\frac{\Omega L^{3}}{8\pi ^{3}}$
allowed values of $\overrightarrow{k}$. Let us have in mind that in our case
we are dealing only with the neutrino $\nu _{e}$, that is, we are dealing
with only one helicity. The neutrino density of neutrinos with $%
\overrightarrow{k}$ within the Fermi sphere is therefore 
\begin{equation}
n_{F}^{\nu }=\frac{4\pi k_{F}^{3}}{3}\frac{1}{8\pi ^{3}}=\frac{k_{F}^{3}}{%
6\pi ^{2}}.
\end{equation}
Since the agglomerates are separated by about 50Mpc for the present Universe
we should have $\lambda _{F}\sim 2\pi /k_{F}\sim 50Mpc\sim 10^{24}m$ and
then $n_{F}^{\nu }\sim 10^{-78}$ neutrinos/cm$^{3}$. We also obtain $%
E_{F}\sim 10^{-49}J\approx 10^{-30}eV.$ Such a small energy only makes sense
for massless (or almost massless) particles. In the beginning of the
Universe $\lambda _{F}$ was much smaller and $E_{F}$ was then much larger.
This means that in the beginning more neutrinos were below the Fermi level.
We can thus say that $k_{i}($or $L$) depends on the red shift.

The energy levels of a neutrino confined in a box with sides equal to $L$
are given by 
\begin{equation}
E=c\hbar \sqrt{k_{x}^{2}+k_{y}^{2}+k_{z}^{2}}=\frac{2\pi c\hbar }{L}\sqrt{%
n_{x}^{2}+n_{y}^{2}+n_{z}^{2}}.
\end{equation}
The number of energy levels $N(E)$ per unit volume with energies between
zero and $E$ \ is 
\begin{equation}
N(E)=\frac{1}{8}\left( \frac{4}{3}\pi k^{3}\right) =\frac{4\pi ^{4}}{%
3h^{3}c^{3}}E^{3}.
\end{equation}
This expression has also been calculated by Kubo$^{6}$ in a different way.
Therefore, the neutrino density of levels at the Fermi energy is 
\begin{equation}
g(E_{F})=\frac{4\pi ^{4}}{h^{3}c^{3}}E_{F}^{2}.
\end{equation}
As $E_{F}$ is very small almost all neutrinos produced in stars attain
states above the Fermi energy. Therefore, most neutrinos in the Universe are
conductors of the superweak charge.

\pagebreak

\vspace*{0.5in}

\subsection{\protect\Large Neutrino Levels in a Weak Periodic Potential}

In the beginning of this section we will follow the footsteps of Ashcroft \&
Mermin$^{7}$ (but we also could follow chapter 3 of Ziman$^{8}$) on electron
levels in solids. Before considering the neutrinos that wander about in the
Universe let us consider the general problem of neutrinos subject to a
periodic potential 
\begin{equation}
U(\mathbf{r}+\mathbf{R})=U(\mathbf{r})
\end{equation}
where $\mathbf{R}$ is a Bravais lattice vector that we assume to exist. For
the Universe $\mathbf{R}$ depends on the redshift, that is, it depends on
the age of the Universe. In the young Universe, for example, $\mathbf{R}$
was small. In the present Universe (or local Universe) it is of the order of
the distance between agglomerates. The potential $U(\mathbf{r})$ is an
effective potential which should be the result of the gravitational and
superweak fields and its mathematical expression may be very complicated. We
can consider that the effective field is weak so that we can disregard pair
creation. We can also disregard the interaction of neutrinos with each
other, that is, we can use an independent neutrino approximation. Since we
are dealing with massless neutrinos they should satisfy Dirac equation 
\begin{equation}
\left( -ic\hbar \mathbf{\alpha }\cdot \mathbf{\nabla }+U(\mathbf{r})\right)
\Psi =i\hbar \frac{\partial \Psi }{\partial t}.
\end{equation}
For stationary solutions we have 
\begin{equation}
\Psi (\mathbf{r},t)=\psi (\mathbf{r})\exp (-i\frac{Et}{\hbar })
\end{equation}
and 
\begin{equation}
\left( -ic\hbar \mathbf{\alpha }\cdot \mathbf{\nabla }+U(\mathbf{r})\right)
\psi (\mathbf{r})=E\psi (\mathbf{r}).
\end{equation}

\bigskip We can also assume that Bloch's theorem is valid for neutrinos and
we can thus consider that $\Psi $ is a plane wave times a function of the
Bravais lattice, 
\begin{equation}
\Psi (\mathbf{r})=\exp (i\mathbf{k}\cdot \mathbf{r})u(\mathbf{r})
\end{equation}
where 
\begin{equation}
u(\mathbf{r}+\mathbf{R})=u(\mathbf{r})
\end{equation}
and thus 
\begin{equation}
\psi (\mathbf{r}+\mathbf{R})=\exp (i\mathbf{k}\cdot \mathbf{R})\psi (\mathbf{%
r}).
\end{equation}

\pagebreak

\vspace*{0.5in} \noindent Bloch's theorem is valid for neutrinos simply
because Floquet's theorem$^{9} $ can also be applied to Dirac's equation.

Now we expand $\psi (\mathbf{r})$ as 
\begin{equation}
\psi (\mathbf{r})=\underset{\mathbf{q}}{\sum }c_{\mathbf{q}}e^{i\mathbf{q.r}}
\end{equation}
and 
\begin{equation}
U(\mathbf{r})=\underset{\mathbf{K}}{\sum }U_{\mathbf{K}}e^{i\mathbf{K.r}}.
\end{equation}
The coefficients $U_{\mathbf{K}}$ are given by 
\begin{equation}
U_{\mathbf{K}}=\underset{cell}{\frac{1}{V}\int }d\mathbf{r}U(\mathbf{r})e^{-i%
\mathbf{K.r}}
\end{equation}
and satisfy the condition $U_{\mathbf{-K}}=U_{\mathbf{K}^{\mathbf{\ast }}}$.
By changing the reference level for the potential energy we can make 
\begin{equation}
U_{0}=\underset{cell}{\frac{1}{V}\int }d\mathbf{r}U(\mathbf{r})=0.
\end{equation}
Assuming symmetry inversion and making $U(\mathbf{r})=U(-\mathbf{r})$ we
have 
\begin{equation}
U_{\mathbf{-K}}=U_{\mathbf{K}}=U_{\mathbf{K}^{\ast }}
\end{equation}
Placing the corresponding expansions into Dirac equation we find 
\begin{equation}
\underset{\mathbf{q}}{\sum }e^{i\mathbf{q.r}}\left[ \left( c\hbar \mathbf{%
\alpha .q}-E\right) c_{\mathbf{q}}+\underset{\mathbf{K}^{\prime }}{\sum }U_{%
\mathbf{K}^{\prime }}c_{\mathbf{q-K}^{\prime }}\right] =0
\end{equation}
and since the plane waves form an orthogonal set we obtain 
\begin{equation}
\left( c\hbar \mathbf{\alpha .q}-E\right) c_{\mathbf{q}}+\underset{\mathbf{K}%
^{\prime }}{\sum }U_{\mathbf{K}^{\prime }}c_{\mathbf{q-K}^{\prime }}=0.
\end{equation}
Writing $\mathbf{q}$ in the form $\mathbf{q=k-K}$ so that $\mathbf{K}$ is a
reciprocal lattice vector chosen so that $\mathbf{k}$ lies in the first
brillouin zone we can write 
\begin{equation}
\left( c\hbar \mathbf{\alpha .(k-K)}-E\right) c_{\mathbf{k-K}}+\underset{%
\mathbf{K}^{\prime }}{\sum }U_{\mathbf{K}^{\prime }}c_{\mathbf{k-K-K}%
^{\prime }}=0
\end{equation}

\pagebreak

\vspace*{0.5in}

\noindent and making the change of variables $\mathbf{K}^{\prime
}\longrightarrow $ $\ \ \mathbf{K}^{\prime }-\mathbf{K}$ we finally obtain 
\begin{equation}
\left( c\hbar \mathbf{\alpha .(k-K)}-E\right) c_{\mathbf{k-K}}+\underset{%
\mathbf{K}^{\prime }}{\sum }U_{\mathbf{K}^{\prime }\mathbf{-K}}c_{\mathbf{k-K%
}^{\prime }}=0.
\end{equation}

The neutrino energy bands are revealed in a better way if we look for all
solutions to the Dirac equation that have the form $\Psi _{n\mathbf{k}}(%
\mathbf{r})=\exp (i\mathbf{k}\cdot \mathbf{r})u_{\mathbf{k}}(\mathbf{r})$.
Doing this we find 
\begin{eqnarray}
H_{\mathbf{k}}\exp (i\mathbf{k}\cdot \mathbf{r})u_{\mathbf{k}}(\mathbf{r})
&=&  \notag \\
\left( -ic\hbar \mathbf{\alpha }\cdot \mathbf{\nabla }+U(\mathbf{r})\right)
\exp (i\mathbf{k}\cdot \mathbf{r})u_{\mathbf{k}}(\mathbf{r}) &=& \\
\exp (i\mathbf{k}\cdot \mathbf{r})\left[ c\hbar \mathbf{\alpha }\cdot \left(
-i\mathbf{\nabla }+\mathbf{k}\right) +U(\mathbf{r})\right] u_{\mathbf{k}}(%
\mathbf{r}) &=&E_{\mathbf{k}}\exp (i\mathbf{k}\cdot \mathbf{r})u_{\mathbf{k}%
}(\mathbf{r})  \notag
\end{eqnarray}
and 
\begin{equation}
H_{\mathbf{k}}u_{\mathbf{k}}(\mathbf{r})=\left( -ic\hbar \mathbf{\alpha }%
\cdot \mathbf{\nabla }+U(\mathbf{r})\right) u_{\mathbf{k}}(\mathbf{r})=E_{%
\mathbf{k}}u_{\mathbf{k}}(\mathbf{r})
\end{equation}
which shows that we can find $E_{\mathbf{k}}$ solving this eigenvalue
problem. Of course, there is an infinite family of solutions with discretely
spaced eigenvalues, which are labeled by the band index \emph{n}, that is,
we have to write $E_{\mathbf{k}}=E_{n}(\mathbf{k}).$ Hence, for each \emph{n}
there is a set of levels specified by $E_{n}(\mathbf{k})$, called an energy
band.

From Davydov$^{10}$ we find that the mean velocity (effective velocity) of $%
\nu _{e}$ is 
\begin{equation}
\mathbf{v}_{m}=\text{ \ }\frac{c^{2}}{E}\int d\mathbf{r}\ \psi _{n\mathbf{k}%
}^{\ast }(\mathbf{r})\ \mathbf{p}\ \psi _{n\mathbf{k}}(\mathbf{r})=
\end{equation}
which may be smaller than $c$ and shows that the interaction of the electron
neutrino with the galactic lattice lowers its effective velocity, and thus,
makes the neutrino to acquire an \emph{effective mass}.

Neutrinos in the Universe are mainly produced in the cores of stars during
the process of fusion or during the creation of novae and supernovae. These
neutrinos have energies much larger than the Fermi energy, and therefore,
most neutrinos of the Universe are not in the \ ground state. This means
that as the Universe ages it becomes a better neutrino conductor. \ 

Let us follow the calculations of Ashcroft \& Mermin$^{11}$ from page 152 to
160, modified for the case of neutrinos. As is shown on p. 158, for $\mathbf{%
q}$ vectors on a Bragg plane 
\begin{equation}
E=E_{\mathbf{q}}^{0}\pm |U_{\mathbf{K}}|=c\hbar \mathbf{\alpha }\cdot 
\mathbf{q}\pm |U_{\mathbf{K}}|.
\end{equation}
Since the interacting lattice potential is small the second term above is
small. That is, the band gap is small. Since the $\mathbf{K}$ vectors depend
on the lattice,

\pagebreak

\vspace*{0.5in} \noindent and the lattice depends on the expansion of the
Universe, we can say that $\mathbf{K}$ depends on the redshift and as a
consequence $E$ depends on the redshift, $\mathsf{Z}$. As we go to higher
redshifts the $\mathbf{K}$ vectors increase, and thus the band gaps increase
too. Therefore, the band gap is not unique. For the present Universe (local
Universe), since $\mathbf{q}$ is much larger than $\mathbf{K}$ there is an
incredible number of repeated zones. Taking into account the redshift
dependence we should write 
\begin{equation}
E(\mathbf{q},\mathsf{Z})=c\hbar \mathbf{\alpha }\cdot \mathbf{q}\pm |U_{%
\mathbf{K}}(\mathsf{Z})|.
\end{equation}
In the beginning of the Universe the number of neutrinos was much smaller
than it is today and also the density was very high so that neutrinos were
not transparent. Therefore, in the beginning the Universe was a poor
neutrino conductor. As it aged it became a better conductor.

In the Universe the periodicity is not exact, that is, there is an \emph{%
average} distance between any two cluster which is about $50\ Mpc$. That is,
the lattice is not that of a crystalline solid. This means that we can only
define an average reciprocal lattice vector, $\mathbf{K}$, and then we may
not have forbidden bands but just bands having a smaller number of
neutrinos. But since most neutrinos in the Universe are very energetic their 
$\mathbf{k}$ vectors are much larger than the average reciprocal lattice
vector. That is, the Universe is a very good neutrino conductor. It is quite
important therefore to determine the crystallographic parameters of the
Universe, that is, the approximate direct lattice and the inverse lattice,
as well.

The superweak force between pairs of galaxies depend on the number of
neutrinos that are exchanged between them, that is, depend on their
superweak charges. As we see it is an unsurmountable problem. Please, see
calculation below for neutrino bands in one dimension.

\subsubsection{\protect\large Neutrino Energy Bands in One Dimension}

\bigskip Splitting up Dirac equation into two pairs we can write for the
neutrino 
\begin{equation}
\psi =\left( 
\begin{array}{c}
\psi _{a} \\ 
\psi _{b}
\end{array}
\right)
\end{equation}
where $\psi _{b}=\lambda \psi _{a}(\lambda =\pm 1)$. The wavefunction
satisfies the equation 
\begin{equation}
\left[ s_{3}\partial _{3}-\lambda i\frac{E-V(z)}{\hbar c}\right] \psi _{a}=0
\end{equation}
Using a wavefunction propagating in the z direction we have 
\begin{equation}
\psi _{a}(z)=Ce^{ikz}u_{k}(z)=\left( 
\begin{array}{c}
a \\ 
b
\end{array}
\right) e^{ikz}u_{k}(z)
\end{equation}

\pagebreak

\vspace*{0.5in}

\noindent and we obtain therefore 
\begin{equation}
\frac{\hbar }{2}\left( 
\begin{array}{cc}
1 & 0 \\ 
0 & -1
\end{array}
\right) \left( 
\begin{array}{c}
a \\ 
b
\end{array}
\right) \frac{du_{k}}{dz}+i\frac{\hbar }{2}ku_{k}\left( 
\begin{array}{cc}
1 & 0 \\ 
0 & -1
\end{array}
\right) \left( 
\begin{array}{c}
a \\ 
b
\end{array}
\right) =\lambda i\frac{E-V(z)}{\hbar c}\left( 
\begin{array}{c}
a \\ 
b
\end{array}
\right) u_{k},
\end{equation}
that is, the two equations 
\begin{eqnarray}
\frac{\hbar }{2}\frac{du_{k}}{dz}a+i\frac{\hbar }{2}ku_{k}a &=&\lambda i%
\frac{E-V(z)}{\hbar c}u_{k}a  \notag \\
-\frac{\hbar }{2}\frac{du_{k}}{dz}b-i\frac{\hbar }{2}ku_{k}b &=&\lambda i%
\frac{E-V(z)}{\hbar c}u_{k}b.
\end{eqnarray}
from which we get $ab=0$, that is, either $a=0$ or $b=0$. The pair of
equations above shows that the solution $a=1,$ $b=0$, $\lambda =\hbar /2$%
(neutrinos) is equivalent to the solution $b=1,a=0,\lambda =-\hbar /2$%
(antineutrinos), so that the two helicity states continue being possible and
with definite values. That is, we can write 
\begin{equation}
C=\left( 
\begin{array}{c}
1 \\ 
0
\end{array}
\right) \text{ for }\lambda =\hbar /2
\end{equation}
and 
\begin{equation}
C=\left( 
\begin{array}{c}
0 \\ 
1
\end{array}
\right) \text{ for }\lambda =-\hbar /2.
\end{equation}

Making $b=0$, that is, dealing with neutrinos, we are left with only the
first equation 
\begin{equation}
\frac{du_{k}}{dz}+iku_{k}=i\frac{E-V(z)}{\hbar c}u_{k}
\end{equation}
which can be easily solved (it is a particular case of Bernoulli's equation)
depending on the form of $V(z)$. For example, for a periodic potential of
the form $V(z)=V_{0}\cos Kz$, we have 
\begin{equation}
\frac{du_{k}}{dz}=i\left( \frac{E}{\hbar c}-k-\frac{V_{0}}{\hbar c}\cos
Kz\right) u_{k}
\end{equation}
whose solution is 
\begin{equation}
u_{k}(z)=u_{0}e^{i\left( \frac{E}{\hbar c}-k\right) z}e^{-i\frac{V_{0}}{%
\hbar c}\sin Kz}.
\end{equation}
Let us consider now that $K$ is a reciprocal lattice vector, that is, $%
Ka=2n\pi $, and let us find the condition that satisfies the relation 
\begin{equation}
u_{k}(z+a)=u_{k}(z).
\end{equation}

\pagebreak

\vspace*{0.5in}

\noindent It is satisified if 
\begin{equation}
\left( k-\frac{E}{\hbar c}\right) a=2l\pi ,l=\func{integer}
\end{equation}
that is, we find 
\begin{equation}
k(l)=\frac{E}{\hbar c}+2l\pi ,
\end{equation}
clearly showing the existence of neutrino energy bands. The wavefunction $%
\Psi (z,t)$ is 
\begin{equation*}
\Psi _{l}(z,t)=\left( 
\begin{array}{c}
1 \\ 
0 \\ 
1 \\ 
0
\end{array}
\right) u_{0}\ e^{i\left[ k(l)z-\frac{E}{\hbar }t\right] }\ e^{-i\frac{V_{0}%
}{\hbar c}\sin Kz}.
\end{equation*}

What we developed above can be applied to any (well-behaved potential)
because we can always express any function $V(z)$ in terms of a Fourier
series and, therefore, the general solution is always of the form 
\begin{equation}
u_{k}(z)=\left( 
\begin{array}{c}
1 \\ 
0 \\ 
1 \\ 
0
\end{array}
\right) u_{0}\ e^{i\left[ k(l)z-\frac{E}{\hbar }t\right] }\ \exp \left( -i%
\underset{j=1}{\overset{\infty }{\sum }}a_{j}\sin (K_{j}z+\phi )\right) .
\end{equation}

If neutrinos have a very small mass, then we can just use the well-known
theory of electron bands having in mind that there is always a factor of 2
which should be dropped because in the case of neutrinos we have either one
helicity state or the other one(neutrino or antineutrino) and in the case of
electrons the spin states do not mean particle and antiparticle.

\bigskip

\noindent {\Large References}

\bigskip

\noindent 1. E. Noether, in \textit{Nach. Ges. Wiss. G\"{o}ttingen,} 171,
1918.

\noindent 2. E. Fischbach, in \textit{Proceedings of the NATO Advanced Study
Institute on Gravitational Measurements, Fundamental Metrology and
Constants, 1987}, ed. by V. de Sabbata and V. N. Melnikov(D. Reidel
Publishing Company, Dordrecht, Holland, 1988).

\noindent 3. E. G. Adelberger, B. R. Heckel, C. W. Stubbs and W. F. Rogers, 
\textit{Annu. Rev. Nucl. Part. Sci.} \textbf{41}, 269(1991).

\noindent 4. S.L. Shapiro and S.A. Teukolsky, in Black Holes, White Dwarfs
and Neutron Stars, John Wiley \& Sons, New York(1983), p. 521.

\pagebreak

\vspace*{0.5in}

\noindent 5. H. El-Ad, T. Piran, and L.N. da Costa, \textit{Mon. Not. R.
Astron. Soc.} \textbf{287}, 790 (1997).

\noindent 6. R. Kubo, in Ststistical Physics, North-Holland Publishing
Company, Amsterdam(1965), p. 288.

\noindent 7. N.W. Ashcroft and N.D. Mermin, in Solid State Physics, Saunders
College, Philadelphia(1976), pp. 132-141.

\noindent 8. J.M. Ziman, in Principles of the Theory of Solids, Cambridge
University Press, London, 1972, pp. 77-112.

\noindent 9. S. Fl\"{u}gge, in Practical Quantum Mechanics, Vol. I,
Springer-Verlag, New York, USA, 1974, p. 62.

\noindent 10. A.S. Davydov, in Quantum Mechanics, 2nd ed., Pergamon Press,
Oxford(1965).

\noindent 11. N.W. Ashcroft and N.D. Mermin, in Solid State Physics,
Saunders College, Philadelphia(1976), pp. 152-160.

\pagebreak

\vspace*{0.5in}

\section{\protect\LARGE Another Solution to the Solar \newline
Neutrino Problem}

\bigskip \rule[1.5in]{0in}{0.17in}

\subsection{\protect\Large The Solar Neutrino Problem and Its Current
Solution}

Electron neutrinos are copiously produced in the interiors of stars. In the
Sun they are produced in the pp cycle which dominates the fusion process in
cool stars 
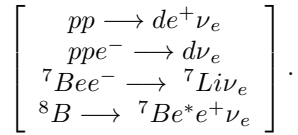
\begin{equation*}
\left[ 
\begin{array}{c}
pp\longrightarrow de^{+}\nu _{e} \\ 
ppe^{-}\longrightarrow d\nu _{e} \\ 
^{7}Bee^{-}\longrightarrow \text{ }^{7}Li\nu _{e} \\ 
^{8}B\longrightarrow \text{ }^{7}Be^{\ast }e^{+}\nu _{e}
\end{array}
\right] .
\end{equation*}
This process generates a certain neutrino flux \ $\Phi _{E}$ that would be
expected to be detected at the Earth. The data of super-Kamiokande$^{1,2}$
show that only about half of the predicted flux $\Phi _{E}$ is actually
detected. This constitutes the solar neutrino problem. There are three main
solutions to the problem without changing the standard solar model: a)
Vacuum Neutrino Oscillations; \ b) Resonant Matter Neutrino Oscillations;
and c) Neutrino with Magnetic Moment.

The first solution requires a quite large vacuum mixing angle which appears
to be unreal. The third solution is not very reasonable because according to
it the neutrino flux would depend on the solar activity related to sun
spots, but Kamiokande experiments have ruled out such dependence. Therefore,
only the second solution remains reasonable and that is why it is the most
acceptable solution. It considers that in their journey towards the Earth
electron neutrinos are transformed into $\nu _{\mu }$ and $\nu _{\tau }$. Of
course, it is only possible if $\nu _{e}$ has a mass

\pagebreak

\vspace*{0.5in}

\noindent diferent from zero. The main drawback is the following: \emph{the
other leptons are not transformed into each other, except in decaying
processes. \ }And an important \ question should be asked about this: why
not having the opposite, the transformation of $\nu _{\mu }$ and $\nu _{\tau
}$ into $\nu _{e}$? Another important drawback is that if neutrinos have
nonzero masses we expect that the masses of $\nu _{\mu }$ and $\nu _{\tau }$
are larger than that of $\nu _{e}.$ How can a less massive particle be
transformed into a more massive particle without the action of a third
particle? Let us recall that the only transformation that we know is that of 
$K^{0}$ and $\overline{K^{0}}$ into each other via an intermediate $\pi
^{+}\pi ^{-}$ pair. First of all it is not a transformation between
fermions, it is actually, a transformation between quarks due to a CP
transformation. In the case of neutrinos which pair of particles could be
the agent of the transformation since neutrinos are not composed of other
particles?

The recent data of SNO$^{3}$ has reduced the MSW oscillation solution to
only the Large Mixing Angle (LMA) solution according to the SNO
Collaboration itself, to Bahcall et al.$^{4}$ and other researchers. But
taking into account the data of SN1987A, Kachelriess et al.$^{5}$restricts
also the use of LMA-MSW saying that ``On the other hand the LMA-MSW solution
can easily survive as the best overall solution, although its size is
generally reduced when compared to fits to the solar data only ''.

\bigskip 

\subsection{\protect\Large Another Solution to the Solar Neutrino Problem}

As we saw in chapter 8 neutrinos are the charge carriers of the superweak
charge. Therefore, we can say that the solar model should be modified and
should take into account the superweak interaction. By means of it neutrinos
could interact with baryons and scatter inelastically off electrons and
muons before leaving the star. In this way the neutrino transparency in a
star would be reduced, that is, the neutrino flux would be reduced. At this
point it is quite hard to work out some numbers because we are in the dark
but we can say that the overall transparency should be substantially
reduced. Modifying what was developed by Shapiro and Teukolsky$^{6}$ we can
say that the effective mean free path, should be 
\begin{equation}
\lambda _{eff}=\left( \lambda _{e}\lambda _{n}\lambda _{N}\right) ^{1/3}
\end{equation}
where $\lambda _{e}$ is the neutrino mean free path due to the cross section
in $\nu _{e}-e^{-}$ scattering, $\lambda _{n}$ is the neutrino mean free
path due to $n-\nu _{e}$(caused by the weak interaction) scattering and $%
\lambda _{N}$ is the neutrino mean free path caused by the superweak current
due to the presence of nucleons. We expect that $\lambda _{N}$ is smaller
than $\lambda _{e}$ or $\lambda _{n}$ in order to diminish the neutrino
transparency. If we write 
\begin{equation}
\lambda _{N}=\frac{1}{\sigma _{N}\ n_{N}}
\end{equation}
where $\sigma _{N}$ is the cross section due to the superweak interaction
and $n_{N}$ is the nucleon density. Of course $\lambda _{N}$ should be a
function of $\frac{\rho _{nuc}}{\rho }$ and of $\frac{1}{E_{\nu }}$. That is,

\pagebreak

\vspace*{0.5in}

\noindent we expect to have a law of the form 
\begin{equation}
\lambda _{N}=C_{1}\left( \frac{\rho _{nuc}}{\rho }\right) ^{q}\left( \frac{%
C_{2}}{E_{\nu }}\right) ^{r},
\end{equation}
that is, a law similar in form to $\lambda _{e}$ and $\lambda _{n}.$ The
constants can be inferred from the experimental data and from solar models
because a closed theory on this new interaction will take a long time, of
course. The above equation means that a part of the neutrino cross section
measured on Earth has to be attributed to the superweak interaction. With
more neutrino data we should be able to measure it.

As is discussed above we should have $\lambda _{N}=\frac{1}{\sigma _{N}\
n_{N}}<(\lambda _{e}\lambda _{n})^{1/2}$. On the other hand for $\rho
\lesssim 2\rho _{nuc}$, $n_{N}\approx n_{p}=9.6\times 10^{35}\left( \frac{%
\rho _{nuc}}{\rho }\right) ^{2}cm^{-3}$, and hence we obtain$^{7}$ 
\begin{equation}
\frac{1}{\sigma _{N}}<1.92\times 10^{38}\left( \frac{\rho _{nuc}}{\rho }%
\right) ^{19/6}\left( \frac{10^{5}ev}{E_{\nu }}\right) m^{-2}.
\end{equation}
For B neutrinos $E=14MeV$ and $\rho $ in the core of the Sun is about $%
100g/cm^{3}$ and hence we have 
\begin{equation}
\sigma _{N}>1.5\times 10^{-74}cm^{2}
\end{equation}
whose lower bound is extremely small. In the calculation above we used $\rho
_{nuc}=2.8\times 10^{14}g/cm^{3}$.

We can have a better estimate making the following considerations. In order
to lower substantially the neutrino flux we should have $\lambda _{eff}\sim
R_{c}\approx 0.25R_{S}=1.7\times 10^{5}km$ ($R_{c}$ is the core radius of
the Sun) and thus we obtain $\lambda _{e}\lambda _{n}\lambda _{N}=\lambda
_{eff}^{3}\approx 4.9\times 10^{24}m^{3}.$ As $(\lambda _{e}\lambda
_{n})^{1/2}=2\times 10^{5}km\left( \frac{\rho _{nuc}}{\rho }\right) ^{7/6}$ $%
\left( \frac{10^{5}ev}{E_{\nu }}\right) ^{2.5}$ (ref. 6 above) we get 
\begin{equation}
\sigma _{N}=2\times 10^{-63}cm^{2}
\end{equation}
which is quite small. Just for comparison let us recall that weak processes
(inverse $\beta $-decay) have cross sections of the order of $10^{-43}cm^{2}$
(ref. 8)$.$

\bigskip

\noindent {\Large References}

\bigskip

\noindent 1) \ S. Fukuda et al., hep-ex/0103032.

\noindent 2) \ S. Fukuda et al., hep-ex/0103033.

\noindent 3) \ SNO Collaboration, nucl-ex/0204009; Phys. Rev. Lett. 89
(2002), 011302

\noindent 4) \ J.N. Bahcall, M. C. Gonzalez-Garcia, C. Pe\~{n}a-Garay,
CERN-TH/2002-094, IFIC-02-18.

\noindent 5) \ M. Kachelriess, A. Strumia, R. Tomas, J.W.F. Valle, Phys.Rev.
D65 (2002) 073016.

\noindent 6) \ S.L. Shapiro and S.A. Teukolsky, in Black Holes, White Dwarfs
and Neutron Stars, John Wiley \& Sons, New York(1983), pp. 326, 327.

\noindent 7) \ Idem, p 310.

\noindent 8) \ D.H.\ Perkins, Introduction to High Energy Physics, 3rd ed.,
Addison-Wesley, Menlo Park, California, 1987, p. 213.

\pagebreak

\vspace*{0.5in}

\section{\protect\bigskip {\protect\LARGE Some Topics in Nuclear Physics}}

\bigskip \rule[1.5in]{0in}{0.17in}

\subsection{\protect\Large The Nuclear Potential and the Stability of the
Deuteron, Triton and Alpha Particle}

The most accurate empirical nuclear potential to date is the Paris potential$%
^{1}$. It has two expressions: one for the antisymmetric states(with respect
to spin), allowed for two protons, two neutrons, as well as a proton and a
neutron, and one for the symmetric states(with respect to spin), accessible
only for the n-p system. In any case, when $S=0$, there is only a central
potential between any two nucleons($V_{C0}$). The Paris group has found that
the potential has four different terms and is described by $^{1,2}$ 
\begin{equation}
V(r)=V_{C1}(r)+V_{T}(r)\Omega _{T}+V_{S0}(r)\Omega _{S0}+V_{S02}(r)\Omega
_{S02}
\end{equation}
\noindent where 
\begin{equation}
\Omega _{T}=3\frac{(\overrightarrow{\sigma _{1}}.\vec{r})(\vec{\sigma _{2}}.%
\vec{r})}{r^{2}}-\vec{\sigma _{1}}.\vec{\sigma _{2}},
\end{equation}
\begin{equation}
\hbar \Omega _{S0}=(\vec{\sigma _{1}}+\vec{\sigma _{2}}).\vec{L},
\end{equation}
\begin{equation}
\hbar ^{2}\Omega _{S02}=(\vec{\sigma _{1}}.\vec{L})(\vec{\sigma _{2}}.\vec{L}%
)+(\vec{\sigma _{2}}.\vec{L})(\vec{\sigma _{1}}.\vec{L}).
\end{equation}
\noindent In these equations $\vec{L}\;$ is the total orbital angular
momentum of the nucleons, ${\hbar /2}\sigma \;$ is the spin operator of each
nucleon, the subscripts 1 and 2 in $\sigma \;$ refer to

\pagebreak

\vspace*{0.5in}

\noindent the two nucleons, and the subscript 1 in the first term refers to $%
S=1\;$ (it is $V_{C0}\;$ for S=0). The first three terms are responsible for
binding the deuteron. The term $V_{T}(r)\;$ is associated also with the
large electric quadrupole moment of the deuteron$^{2}$. We clearly see that
the spatial part of the wavefunction must be antisymmetric. Of course, the
spin wavefunctions $|S,S_{z}>$($|1,-1>$, $|1,0>$ and $|1,1>$) are symmetric
under particle exchange. The different terms of the potential are presented
in Fig. 10.1.

According to the ideas above discussed there is more repulsion for $S=0\;$
because in this case more vectorial mesons(colorless) are exchanged. Let us
verify this. In Fig. 10.2, which is for $S=0\;$ we should consider the
scalar mesons which are exchanged between the two primons of each pair below
(each primon belongs to a different nucleon): $\downarrow p_{3}^{\gamma
}p_{1}^{\gamma }\downarrow $, $\downarrow p_{1}^{\gamma }p_{1}^{\gamma
}\downarrow $, $\downarrow p_{2}^{\beta }p_{2}^{\beta }\downarrow $, $%
\uparrow p_{2}^{\beta }p_{2}^{\beta }\uparrow $, $\uparrow p_{3}^{\alpha
}p_{3}^{\alpha }\uparrow $, $\uparrow p_{3}^{\alpha }p_{2}^{\alpha }\uparrow 
$. The corresponding $q\bar{q}^{\prime }$s exchanged are: 01 $d\bar{u}$, 01 $%
u\bar{d}$, 01 $b\bar{t}$, 01 $t\bar{b}$, 02 $u\bar{u}$, 02 $c\bar{c}$, 02 $t%
\bar{t}$, 04 $u\bar{u}$, 04 $d\bar{d}$, 04 $s\bar{s}$, 02 $c\bar{c}$, 02 $d%
\bar{d}$, 02 $b\bar{b}$, 01 $u\bar{c}$, 01 $c\bar{u}$, 01 $s\bar{b}$, and 01 
$b\bar{s}$. This means that the following scalar mesons are exchanged: 01 $%
\pi ^{+}$, 01 $\pi ^{-}$, 02 $\pi ^{0}$, 04 $\eta $, 04 $\eta _{c}$, 01 $%
D^{0}$, 01 $\bar{D}^{0}$, 01 $B_{s}^{0}$, 01 $\bar{B}_{s}^{0}$, 01 $b\bar{t}$%
, 01 $t\bar{b}$, 02 $t\bar{t}$, and 02 $b\bar{b}$. Then, there are 22
attractive terms. This, of course, is for a certain configuration of the
supercolors of the two nucleons. In other configurations we can also have
scalar mesons exchanged between the two primons $\downarrow
p_{1}^{j}p_{2}^{j}\downarrow $. They are, respectively, the $q\bar{q}%
^{\prime }$s: $d\bar{c}$, $c\bar{d}$, $s\bar{t}$, and $t\bar{s}$. Then, for
other configurations there can also exist the exchange of the scalar mesons: 
$D^{+}$, $D^{-}$, $s\bar{t}$, and $t\bar{s}$.

Let us take care now of the vector mesons. They should be exchanged between
the primons of the pairs (for the supercolor configuration shown in the
figure): $\downarrow p_{3}^{\gamma}p_{1}^{\gamma}\uparrow $, $\downarrow
p_{1}^{\gamma}p_{1}^{\gamma}\uparrow $, $\downarrow
p_{2}^{\beta}p_{2}^{\beta}\uparrow $, $\uparrow
p_{2}^{\beta}p_{2}^{\beta}\downarrow $, $\downarrow
p_{2}^{\alpha}p_{2}^{\alpha}\uparrow $, $\downarrow
p_{2}^{\alpha}p_{3}^{\alpha}\uparrow $, The following $q\bar{q}^{\prime}$s
are exchanged: 01 $d\bar{u}$, 01 $u\bar{d}$, 01 $b\bar{t}$, 01 $t\bar{b}$,
02 $u\bar{u}$, 02 $c\bar{c}$, 02 $t\bar{t}$, 06 $u\bar{u}$, 06 $d\bar{d}$,
06 $s\bar{s}$, 02 $c\bar{c}$, 02 $d\bar{d}$, 02 $b\bar{b}$, 01 $u\bar{c}$,
01 $c\bar{u}$, 01 $s\bar{b}$, and 01 $b\bar{s}$. 
\noindent Therefore, there are the exchange of the vector mesons: 08 $\omega$
(or 08 $\rho\;$ or both of them), 06 $\phi$, 04 $J/{\Psi}$, 01 $%
D^{*}(2010)^{+}$, 01 $D^{*}(2010)^{-}$, and the vectorial mesons 01 $d\bar{u}
$, 01 $u\bar{d}$, 01 $b\bar{t}$, 01 $t\bar{b}$, 02 $t\bar{t}$, 02 $b\bar{b}$%
, 01 $s\bar{b}$, and 01 $b\bar{s}$. There are, then, 30 repulsive terms.

Taking into account the overall effect the repulsion probably overcomes the
attraction, and that is why there is no binding for $S=0$. Although it was
done for a certain supercolor configuration this is a general result. That
is, for other configurations the number of repulsive (vectorial) terms is
always much larger than the number of attractive (scalar) terms. The former
is about 1.5 the latter.

Let us now consider $S=1$. In this case let us take a look at Fig. 10.3. The
scalar mesons which are exchanged between the two primons (of different
nucleons) of each pair are: $\downarrow p_{2}^{\alpha }p_{2}^{\alpha
}\downarrow $, $\uparrow p_{2}^{\alpha }p_{2}^{\alpha }\uparrow $, $\uparrow
p_{2}^{\beta }p_{2}^{\beta }\uparrow $, $\uparrow p_{1}^{\beta }p_{1}^{\beta
}\uparrow $, $\uparrow p_{1}^{\beta }p_{2}^{\beta }\uparrow $, $\uparrow
p_{2}^{\beta }p_{1}^{\beta }\uparrow $. $\uparrow p_{3}^{\gamma
}p_{3}^{\gamma }\uparrow $, $\downarrow p_{3}^{\gamma }p_{1}^{\gamma
}\downarrow $. \noindent Thus, the corresponding $q\bar{q}^{\prime }$s are:
06 $u\bar{u}$, 06 $d\bar{d}$, 06 $s\bar{s}$, 02 $u\bar{u}$, 02 $c\bar{c}$,
02 $t\bar{t}$, 02 $c\bar{d}$, 02 $d\bar{c}$, 02 $t\bar{s}$, 02 $s\bar{t}$,
02 $c\bar{c}$, 02 $d\bar{d}$, 02 $b\bar{b}$, 01 $d\bar{u}$, 01 $u\bar{d}$,
01 $b\bar{t}$, and 01 $t\bar{b}$. There are, then, the following scalar
mesons: 06 $\eta $, 02 $\pi ^{0}$, 01 $\pi ^{+}$, 01 $\pi ^{-}$, 04 $\eta
_{c}$, 02 $D^{-}$, 02 $D^{+}$, 02 $t\bar{t}$, 02 $t\bar{s}$, 02 $s\bar{t}$,
02 $b\bar{b}$, 01 $b\bar{t}$, and 01 $t\bar{b}$. There are, then, 28
attractive terms. In other supercolor configurations we can also have scalar
mesons exchanged between the primons $\downarrow
p_{2}^{j}p_{3}^{j}\downarrow $. They are the $q\bar{q}$'s $u\bar{c}$, $c\bar{%
u}$, $s\bar{b}$, and $b\bar{s}$. Then, we can have the scalar mesons $D^{0}$%
, $\bar{D}^{0}$, $B_{s}^{0}$, and $\bar{B}_{s}^{0}$. Let us consider now the
vector mesons. They should be exchanged between the primons of the pairs: $%
\downarrow p_{2}^{\alpha }p_{2}^{\alpha }\uparrow $, $\uparrow p_{2}^{\alpha
}p_{2}^{\alpha }\downarrow $, $\downarrow p_{3}^{\gamma }p_{3}^{\gamma
}\uparrow $, $\uparrow p_{3}^{\gamma }p_{1}^{\gamma }\downarrow $. They are,
respectively: 04 $u\bar{u}$, 04 $d\bar{d}$, 04 $s\bar{s}$, 01 $d\bar{u}$, 01 
$u\bar{d}$, 01 $b\bar{t}$, 01 $t\bar{b}$, 02 $c\bar{c}$, 02 $d\bar{d}$, 02 $b%
\bar{b}$. There are, then, the following vector mesons: 04 $\omega $ (or 04 $%
\rho $ or both of them), 04 $\phi $, 02 $J/{\Psi }$, 01 $d\bar{u}$, 01 $u%
\bar{d}$, 01 $b\bar{t}$, 01 $t\bar{b}$, 02 $d\bar{d}$, 02 $b\bar{b}$. Hence,
there are 18 repulsive terms. For other supercolor configurations the number
of attractive (scalar) terms is either larger or equal to the number of
repulsive terms, and since the scalar mesons have shorter ranges than the
vectorial mesons (for the same $q\bar{q}$), there can exist an equilibrium
position, and then, the system with $S=1\;$ can be stable. In other
supercolor configurations we can have vectorial mesons exchanged between the
primon pairs $\downarrow p_{1}^{j}p_{1}^{j}\uparrow $, $\downarrow
p_{1}^{j}p_{2}^{j}\uparrow $, $\downarrow p_{2}^{j}p_{3}^{j}\uparrow $. They
are the $q\bar{q}$'s: $\;u\bar{u}$, $c\bar{c}$, $t\bar{t}$, $d\bar{c}$, $c%
\bar{d}$, $s\bar{t}$, $t\bar{s}$, $c\bar{u}$, $u\bar{c}$, $s\bar{b}$, and $b%
\bar{s}$. Then, we can also have the following vector mesons: $\;D^{\ast
}(2007)^{0}$, $\bar{D}^{\ast }(2007)^{0}$, $D^{\ast }(2010)^{+}$, and $%
D^{\ast }(2010)^{-}$.

This also shows us the way of analysing the stability of other compound
hadrons. It is clear then that the color field is not responsible for
binding the deuteron. We have then established the connection between quarks
and the exchange of scalar and vector mesons in the nucleons.

Therefore, in the deuteron (that is, for $S=1$), the most important scalar
term is the pionic term, and the most important vectorial term is that
involving $\omega $. This means that a quite good approximate nuclear
potential energy is 
\begin{equation}
V(r)=-g_{\pi }^{2}\frac{e^{\mu _{\pi }r}}{r}+g_{\omega }^{2}\frac{e^{\mu
_{\omega }r}}{r}
\end{equation}
\noindent which agrees with Walecka$^{\prime }$s theory of highly condensed
matter$^{3}$. It corresponds to the $V_{C1}(r)\;$\noindent term of the Paris
potential.

An improved potential energy with the first four scalar terms and the first
four vectorial terms is 
\begin{eqnarray}
V(r) &=& - g_{\pi}^{2}\frac{e^{\mu_{\pi}r}}{r} - g_{\eta}^{2}\frac{%
e^{\mu_{\eta}r}}{r} - g_{D}^{2}\frac{e^{\mu_{D}r}}{r} - g_{B}^{2}\frac{%
e^{\mu_{B}r}}{r} + g_{\omega}^{2}\frac{e^{\mu_{\omega}r}}{r} + g_{\rho}^{2}%
\frac{e^{\mu_{\rho}r}}{r}  \notag \\
& & + g_{\phi}^{2}\frac{e^{\mu_{\phi}r}}{r} + g_{D^*}^{2}\frac{e^{\mu_{D^*}r}%
}{r}.
\end{eqnarray}

When we consider the interaction between two protons or two neutrons we have
the same mesons. We can, then, propose a more complete Walecka$^{\prime }$s
Lagrangian (using Walecka$^{\prime }$s notation and constants) for the
nucleon-nucleon interaction: 
\begin{eqnarray}
L &=&-{\hbar }c\left[ \bar{\psi}\left( \gamma _{\lambda }\frac{\partial }{%
\partial x_{\lambda }}+M\right) \psi \right]  \notag \\
&&-\frac{c^{2}}{2}\left[ \left( \frac{\partial \phi ^{j}}{\partial
x_{\lambda }}\right) ^{2}+\left( \mu ^{j}\right) ^{2}\left( \phi ^{j}\right)
^{2}\right] -\frac{1}{4}F_{\lambda \nu }^{j}F_{\lambda \nu }^{j}  \notag \\
&&-\frac{(m^{j})^{2}}{2}V_{\lambda }^{j}V_{\lambda }^{j}+ig_{v}^{j}\bar{\psi}%
\gamma _{\lambda }{\psi }V_{\lambda }^{j}+g_{s}^{j}\bar{\psi}{\gamma ^{5}}{%
\psi }\phi ^{j},
\end{eqnarray}

\noindent where a summation on $j\;$ is implied. The nucleon field of mass $%
m_{N}\;$ is $\psi \;$, $\phi ^{j}\;$ is a neutral scalar meson field of mass 
$m_{s}^{j}$, $V_{\lambda }^{j}\;$ is a neutral vector meson field of mass $%
m_{v}^{j}$, and $F_{\lambda }^{j}\;$ is the field tensor associated to the
field $V_{\lambda }^{j}$. The quantities $M$, $\mu ^{j}$, and $m^{j}\;$ are
the inverse Compton wavelengths $M=\frac{m_{N}c}{\hbar }$, $\mu ^{j}=\frac{%
m_{s}^{j}c}{\hbar }$, and $m^{j}=\frac{m_{v}^{j}c}{\hbar }$. The index $j\;$
runs from 1 to 4 in order to take into account the scalar mesons $\pi $, $%
\eta $, $D$, and $B$, and the vector mesons $\omega $, $\rho $, $\phi $, and 
$D^{\ast }$. The masses of these mesons are within one order of magnitude.

On the other hand the $p_{3}\;$ of the neutron outer shell does not decay
because the deuteron is stable. Its stability happens due to the binding
with $p_{1}\;$(or $p_{2}\;$) by means of supergluons and by the exchange of
mesons depending on the supercolors. As we see in Fig. 10.4 the deuteron
exhibits a large quadrupole moment. There are two cores with equal positivie
charges, + 1/2 each. Because of the pair $p_{1}-p_{3}\;$ there is a net
positive charge of about +2/3 in the middle, between the two cores. And
there is a negative charge cloud of -1/3 on each side, around each core.
Such distribution produces two opposite electric dipole moments of about
(1/3)x0.6e(fm) and, therefore, a quadrupole moment of about 0.2e(fm)x1(fm)=2x%
$10^{-3}$e(barn), which is close to the experimental value of 2.82x$10^{-3}$%
e(barn)$^{(4)}$. \textit{Pointlike quarks moving randomly can not produce
such moments}. Without considering the above model, since the deuteron has
three d quarks we would expect it to be a very unstable system that would
decay very fast.

In the light of what was discussed above we can understand the large decay
constant of triton. We know that the spins of the two neutrons cancel each
other so that the spin of triton comes from the proton. The configuration of
primons(and quarks) of the system is described below in Fig. 10.5. There is
a net binding between $p_{1}\;$ and the two ${p_{3}}^{\prime }s$. Actually,
it must be an alternate binding between $p_{1}\;$ and each $p_{3}$. This
binding makes $p_{3}\;$ more stable so that instead of decaying in 920s it
decays in about $3.87{\times }10^{8}$s. The addition of another $p_{1}\;$
would make the system completely stable. Therefore, the alfa particle primon
configuration should be given by Fig. 10.6 and \textbf{\ it has a planar
configuration and not a piramidal one}. Due to the attraction of the four
inner shells the system is very tightly bound and, of course, very stable.
The eight ${p_{2}}^{\prime }$s of the outer layers will tend to stay away
from each other. We infer, thus, that the system has the following electric
charge distribution: the center(the region where the two ${p_{1}}^{\prime }$%
s and the two ${p_{3}}^{\prime }$s are) has a net charge of about $2{\times }%
(+5/6)-2{\times }(-1/6)=+4/3$; a middle region(corresponding to the position
of the four inner shells) with a charge of about $4{\times }(+1/2)=+2$; and
an outer region(corresponding to the positions of the eight ${p_{2}}^{\prime
}s$) with a charge of $8{\times }(-1/6)=-4/3$. The system, of course, as we
see, has no quadrupole moment. It is interesting to notice that an alpha
particle is not, therefore, a system of two deuterons. In this way we
explain that the saturation of the nuclear force is quite similar to the
saturation of chemical bonds. We can also understand the reason behind the
tensorial character of the nuclear force, which arises simply due to the
spatial arrangement of primons.

The spatial arrangements of primons(and quarks) in the triton and in the
alpha particle are quite in line with the work of Abbas$^{5}$. According to
his work the alpha particle has a hole at the center with a size of about 1
fm. The hole, of course, exists because of the repulsive and attractive
forces between the four closest primons (Fig. 10.6).

\bigskip

\subsection{{\protect\Large The Absence of Nuclides with A=5 and the
Instability of }$Be^{8}$ \newline
}

It is well known that there is no nuclide with A=5. It simply does not form,
even for a brief time. Why is it so? Taking a look at the primon
configuration of the alpha particle we can understand why. As we saw above
the binding happens in the middle among the four primons: the two pairs of $%
p_{1}-p_{3}$. Besides, it is a planar structure. Thus, there is no room for
another nucleon, that is, there is no bond left. We have a strong binding if
we put a neutron on one side and a proton on the other side (with opposite
spins) because in this case there will be another bond $p_{1}-p_{3}$. That
is why $Li^{6}\;$ is stable. It is also possible to put two protons(with
opposite spins) to form $He^{6}$. The binding between them occurs via the
bonding $p_{1}-p_{2}$. In this case the whole binding is not as strong as in 
$Li^{6}$, and $He^{6}\;$ is unstable and decays in about 0.81s.

We can also see that it is impossible to bind two alpha particles since
there is no bond left in any of them. Actually, the bonding could occur only
by means of the $p_{2}^{\prime}$s of the outer layers, but there is no
bonding between equal primons and, therefore, the binding does not take
place. We know that $Be^{8}\;$ is formed only for an extremely brief
time(about $10^{-23}$s) and breaks up into two alpha particles.

We should investigate how we can form all nuclides by arranging nucleons in
a way to form the bondings among primons, and try to relate the total spin
to each arrangement of primons. In the case of $Li^{6}\;$ and $He^{6}\;$ we
clearly see that $J=0$.

\bigskip

\noindent {\Large References}

\bigskip

\noindent 1. M. Lacombe, \textit{et al., Phys. Rev.} \textbf{C21}, 861
(1980).

\noindent 2. W. N. Cottingham and D. A. Greenwood, in \textit{An
Introduction to Nuclear Physics}, Cambridge University Press,
Cambridge(1992).

\noindent 3. J.D. Walecka, Annals of Phys. \textbf{83}, 491(1974).

\noindent 4. H. G. Kolsky, T.E. Phipps, Jr., N.F. Tamsey, and H.B. Silsbee,
Phys. Rev. \textbf{87}, 395(1952); recalculated by E.P. Auffray, Phys. Rev.
Letters \textbf{6}, 120(1961).

\noindent 5. A. Abbas, nucl-th/9911074.A. Abbas, nucl-th/9911074.

\bigskip

\pagebreak

\bigskip \rule[0.5in]{0in}{0.17in}

{\Huge Appendix}{\LARGE : Brief Vita}

\bigskip

BSc in Physics: Universidade Federal de Pernambuco,

\ \ \ \ \ \ \ \ \ \ \ \ \ \ \ \ \ \ \ \ \ \ \ Recife, Pernambuco, Brazil

MSc.(courses only) in Nuclear Engineering:

\ \ \ \ \ \ \ \ \ \ \ \ \ \ \ \ \ \ \ \ \ \ \ Universidade Federal de
Pernambuco,

\ \ \ \ \ \ \ \ \ \ \ \ \ \ \ \ \ \ \ \ \ \ \ \ Recife, Pernambuco, Brazil

MSc in Physics: Universidade Federal de Pernambuco,

\ \ \ \ \ \ \ \ \ \ \ \ \ \ \ \ \ \ \ \ \ \ \ \ \ Recife, Pernambuco, Brazil

PhD in Physics: University of Illinois at Chicago,

\ \ \ \ \ \ \ \ \ \ \ \ \ \ \ \ \ \ \ \ \ \ \ \ Chicago, Illinois, USA

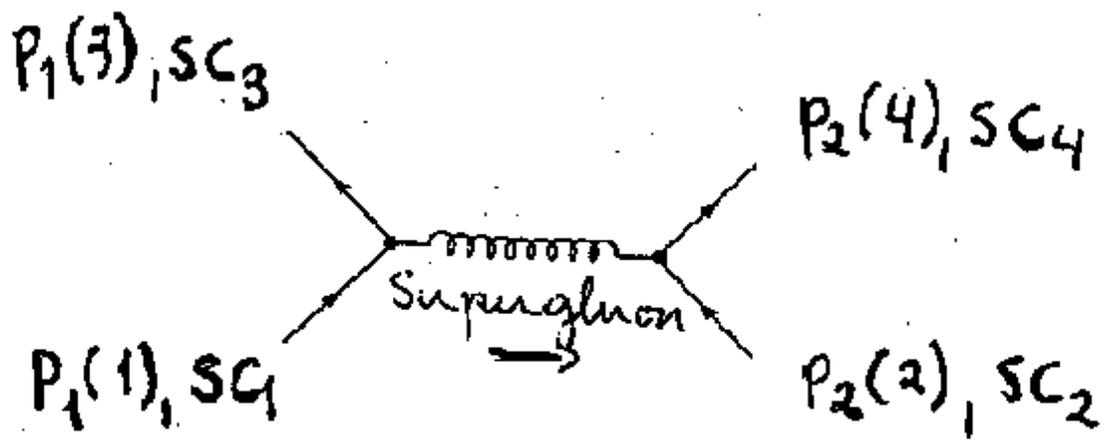
\end{document}